# Security, Fault Tolerance, and Communication Complexity in Distributed Systems

A thesis presented

by

## Donald Rozinak Beaver

to

The Division of Applied Sciences

in partial fulfillment of the requirements
for the degree of Doctor of Philosophy
in the subject of Computer Science
Harvard University
Cambridge, Massachusetts

May 1990

# Abstract

We present efficient and practical algorithms for a large, distributed system of processors to achieve reliable computations in a secure manner. Specifically, we address the problem of computing a general function of several private inputs distributed among the processors of a network, while ensuring the correctness of the results and the privacy of the inputs, despite accidental or malicious faults in the system.

Communication is often the most significant bottleneck in distributed computing. Our algorithms maintain a low cost in local processing time, are the first to achieve optimal levels of fault-tolerance, and most importantly, have low communication complexity. In contrast to the best known previous methods, which require large numbers of rounds even for fairly simple computations, we devise protocols that use small messages and a constant number of rounds *regardless* of the complexity of the function to be computed. Through direct algebraic approaches, we separate the *communication complexity of secure computing* from the *computational complexity* of the function to be computed.

We examine security under both the modern approach of computational complexity-based cryptography and the classical approach of unconditional, information-theoretic security. We develop a clear and concise set of definitions that support formal proofs of claims to security, addressing an important deficiency in the literature. Our protocols are provably secure.

In the realm of information-theoretic security, we characterize those functions which two parties can compute jointly with absolute privacy. We also characterize those functions which a weak processor can compute using the aid of powerful processors without having to reveal the instances of the problem it would like to solve. Our methods include a promising new technique called a *locally random reduction*, which has given rise not only to efficient solutions for many of the problems considered in this work but to several powerful new results in complexity theory.

# Contents

## Acknowledgments

I would very much like to acknowledge the guidance and advice of my advisor, Michael Rabin, and of my unofficial advisor, Joan Feigenbaum, whose help and efforts on my behalf have been instrumental. I would also like to thank Roger Brockett, Shafi Goldwasser, Mei Hsu, and Les Valiant. Several institutions besides Harvard have provided formal and informal support and a productive working environment, including Williams College, Cal Tech, the Hebrew University of Jerusalem, MIT, the Deutsches Forschungszentrum für Künstliche Intelligenz, and the New York Public Library. This work was supported in part by NSF grant CCR-870-4513.

The great many colleagues and graduate students who have made the past four years enjoyable, intellectually stimulating, or just bearable are too many to mention, but in particular I would like to thank Judit Bar-Ilan, Stuart Haber, Christos Kaklamanis, Philip Klein, Danny Krizanc, Lisa Neal, Rafail Ostrovsky, Nir Shavit, and Thanasis Tsantilas.

I especially want to thank Susan Greenbaum, Amanda Lathroum, Beralda Conceiçao de Lima, and Gerry Waggett for their unending support and inspiration.

And, of course, without two very important Professors, Don and Ollie Beaver, this effort would not have been possible. I dedicate this work to them, to all of my family, and especially to my brother, James.

## Coauthors

Some of the protocols appearing here represent joint work: Chapter 5 with Dr. Judit Bar-Ilan and Prof. Michael Rabin, Chapter 9 with Prof. Shafi Goldwasser, Chapter 10 with Prof. Silvio Micali and Mr. Phillip Rogaway, Chapters 11 and 12 with Dr. Joan Feigenbaum, and Chapters 13 and 14 with Dr. Joan Feigenbaum, Dr. Joe Kilian, and Mr. Phillip Rogaway.

# Chapter 1

# Introduction

It is true that you can fool all the people some of the time;
you can even fool some of the people all the time;
but you can't fool all of the people all the time.

— Abraham Lincoln

In an ideal world, a single, reliable, and trusted computer would take care of every computational need without delay. In practice, however, no single computer could be powerful enough, reliable enough, or even sufficiently accessible to satisfy such an ideal. For reasons of efficiency, reliability, security, and for the immense benefits of interaction, distributed computing is the only solution, and indeed it is rapidly being realized. The tremendous increase in dependence on large, interconnected computer systems makes a thorough analysis of their reliability of paramount importance. The issues of fault-tolerance and security are essential to capitalizing on the advantages of scale and interaction.

At the same time, the need for practical methods to ensure reliability and security can be satisfied only by efficient techniques. Unwieldy and inefficient methods are difficult to implement correctly and, even worse, are likely to be ignored altogether.

The protection of *communications* among various parties has a long and rich history, by no means restricted to the age of computers. Cryptographers have long concerned themselves with developing codes to encipher and decipher messages. Similarly, limiting and authorizing access to important resources (whether they are the headquarters of a military command or the files in a centralized computer) has also been the focus of centuries of analysis. Often, the methods developed for one problem are applicable to the other; for instance, the $\textsc{Unix}^{tm}$ operating system implements password schemes to authorize logins by using encryption functions originally designed to protect communications.

Security for *distributed computations,* however, is a more recent concern. The direct approach is to use classical methods for security that treat a collection of interacting computers as a group of individuals, protecting each computer individually and then protecting the communications between each pair of computers. Often, methods for security and reliability depend on the invulnerability of some central host which takes care of authorization, exclusive access to files, and so on.

Research into distributed system reliability (without regard to security) addresses the problems of failures in individual processors, and develops methods such as transaction processing and process migration to recover from individual failures. Together, the collection of processors can carry on computations despite local problems. Like the methods developed to protect communications, though, these methods still rely on one or a few central hosts, such as name-servers, for essential operations.

The very presence of a large number of somewhat independent processors affords a much greater degree of reliability and security than that provided by a single, central host. A large, distributed network of computers is in some ways very much like a community of people. Some elements are reliable, some are not; some are prone to accidental mishaps, others behave maliciously. On the whole, most are reliable most of the time, but there are few whom one would trust or depend upon completely. Yet with the proper "laws," individuals function together despite an imperfect world, and societies prosper despite individual flaws.

In this work, we take advantage of the synergy provided by the interaction among many processors in order to ensure that useful computations continue despite failures in sizable and *arbitrary* portions of a network. Interestingly and importantly, our methods apply to ensure both reliability and security. In retrospect this is not surprising: reliability and security are two sides of the same coin. A reliable system should be resilient against the worst-case failure, which is best regarded as a malicious attack. On the other hand, ensuring that an adversary cannot obtain information means that the nature of its attack is independent of that information. This has the advantage that, in an intuitive sense, the attack can be treated more like a random or accidental failure.

In addition to these intuitive connections between reliability and security, it turns out that the two have interesting and deeper connections when defined formally. Essentially, the two goals are unified by the single purpose of simulating a trusted and reliable central host, even though none is available and it would be imprudent even with assurances to depend on a single, supposedly reliable party.

A standard formal method to ensure that unspecified information is not leaked during an interaction is to demonstrate that the course of the interaction can be simulated, in a certain formal sense, based only on the information that is supposed to be leaked. This information is measured with respect to an ideal setting in which a trusted host is available; the trusted host leaks only specified facts, by definition.

We examine interaction in a deeper sense, investigating not simply the *information* flow from one processor to another but the *influence* that one processor may have over the computation. We develop broad definitions that compare the information and the influence that a participant or adversary has in one protocol to that which it would have in another protocol. The meaning of *correctness* in distributed computation is closely tied to the measure of *influence* over outputs, in the same way that *privacy* is tied to the measure of *information*. This observation forms the basis of a fundamentally new and unified approach to distributed security.

The focus of this dissertation is a collection of *efficient* methods whereby a collection of individuals can achieve reliable computations in a *provably* secure manner, despite the failure or corruption of some minority of the individuals.

Consider the following example. The board of trustees of a company would like to take a vote on a pressing issue, yet none has time to meet in a particular location. At the same time, because of the sensitivity of the issue, each trustee would prefer to keep his vote private. The everyday solution of writing a vote on a piece of paper and shuffling the sheets before counting the votes is physically impossible in this setting; the trustees can only communicate over telephone lines, perhaps using their workstations as well. An electronic analog of the paper-solution will not work: there may be no trusted party to count the votes; electronic messages may be easily traced or duplicated. We shall see how to take a secret ballot, and how to conduct more general distributed computations, in a simple, efficient, reliable, and secure manner.

The unusual and key aspect to our methods is that we do not assume the reliability of any particular processor or the availability of a central, trusted host. A cynical and paradoxical synopsis of our approach might be, "Trust everyone, but trust no-one." In less ambiguous terms, the community as a whole is reliable, but there is no particular individual whose reliability is guaranteed. Through "democratic" means, we ensure that the majority rules, and as long as the majority is reliable, the particular individuals who fail do not matter. In fact, reliable operations are ensured even without the need to identify faulty elements in advance, or to restart entire computations when failures occur (malicious or otherwise) and the offending components are cast out.

We treat a distributed computation in a general manner as follows. Say that each of $n$ processors in a network holds a private input value $x_i$. Together, they must compute some function $F(x_1, \ldots, x_n)$, without revealing anything about the inputs other than what could be deduced from learning $F(x_1, \ldots, x_n)$ alone. In the secret ballot example, a unanimous vote clearly reveals all the individual votes, but this is allowable since it is leaked by the *result*, not as a side effect of a protocol used to compute the result. In other words, the network must simulate a trusted external party who receives $x_1, \ldots, x_n$, computes $F(x_1, \ldots, x_n)$, and returns only the result.

Since some *arbitrary* collection of $t$ of the processors may be faulty, a centralized solution will not suffice. Our goal is to develop multiparty protocols to compute the result even though some of the processors or communication lines may be unreliable in a random or malicious manner.

Research into secure multiparty computations can be divided roughly along the same lines as the rest of cryptographic research: classical, information-theoretic security *vs.* the modern approach of complexity-based cryptography. The classical approaches to security examine secure communications under conditions where an eavesdropper has unlimited computational resources at his disposal. Few methods withstand such strong requirements for security. Fortunately, requiring perfect, information-theoretic security may be unnecessary in practical situations, if one is assured of some limitation on an eavesdropper's or adversary's resources. Diffie and Hellman pioneered the modern approach of computational complexity-based cryptography, in which the processors, reliable or otherwise, are assumed to have bounded amounts of time and memory to achieve their purposes, honest or otherwise [DH76].

Each approach has its own appeal and its own disadvantages. Information-theoretic cryptography guarantees protection of the information against any adversary (random or malicious in the worst possible way), but it may require unreasonable amounts of communication or computation. Complexity-based cryptography broadens the range of secure information processing since it permits only weaker adversaries, but most complexity-based results are based on unproven assumptions about the intractability of problems like factoring large numbers.

Based on unproven complexity-theoretic assumptions, inefficient methods for performing a variety of multiparty computations have been developed [Yao82a, Yao86, GMW86, GMW87, GHY87, Hab88, BG89]. Yao introduced the idea of private function computation by presenting a method for two cooperating individuals to compute a function $F(x, y)$ without having to reveal $x$ and $y$, given that the individuals are restricted to polynomial-time computations [Yao86]. Goldreich, Micali, and Wigderson showed that any function $F$ described by a Boolean circuit $C_F$ can be computed reliably and securely as long as a majority of the processors are reliable [GMW87]. Galil, Haber, and Yung improved the efficiency and implemented other desirable aspects [GHY87, Hab88]. Figure 1.1 describes some of the cryptographic results, and some of the "non-cryptographic" ones, about multiparty computations.

| Reference | Fault-Tol. | Crypto | Rounds | Message Size | Local Time |
|-----------|-----------|--------|--------|-------------|-----------|
| [Yao86] | $t = 1, n = 2$ | yes | const | poly | poly |
| [GMW87] | $t < n/2$ | yes | $Dn^2$ | poly | poly |
| [GHY87] | $t < n/2$ | yes | $D$ | poly | poly |
| [BGW88, CCD88] | $t < n/3$ | no | $D$ | poly | poly |
| [BB88] | $t < n/3$ | no | $D/\log n$ | poly | poly |
| [BB88] | $t < n/3$ | no | const | poss exp | poss exp |
| [RB89, Bea88] | $t < n/2$ | no | $D$ | poly | poly |
| [BFKR89] | $t < \log n$ | no | const | poly | poss exp |
| [BMR90] | $t < n/2$ | yes | const | poly | poly |

Figure 1.1: Various protocols for 2-party and $n$-party secure computations. "Crypto" indicates unproven complexity assumptions; $D$ is the depth of a circuit $C_F$ for the function $F$ to be computed; $n$ represents the number of players and the sizes of the inputs (for clarity of presentation); "poly" means polynomially-bounded; "poss exp" means possibly exponential (depending on the complexity of $F$). The cost measures include fault-tolerance, number of rounds of interaction, message size, and local computation time. Some of these results appear here.

The disadvantages to the early, cryptographic solutions are twofold. First, because these methods are based on evaluating a circuit $C_F$ representing the function $F$ gate by gate, the amount of interaction may become prohibitive. We shall say more about this issue momentarily.

The second disadvantage lies in the set of assumptions on which these solutions are founded. Many assume that factoring large numbers is difficult, or determining the discrete logarithm of a number modulo some prime is hard; others rely on the general assumption that one-way trapdoor functions exist, *i.e.* functions which are easy to compute but hard to invert (one-way), but which are easy to invert given additional, trapdoor information. Though these protocols are apparently secure given the current lack of efficient methods for factoring or computing discrete logarithms and the like, an advance in the techniques for solving these apparently intractable problems could destroy the foundations on which the protocols are based.

Ben-Or, Goldwasser, Wigderson, Chaum, Crépeau, and Damgaardhave recently introduced a new, "non-cryptographic" approach which circumvents the use of unproven assumptions, replacing them with certain reasonable assumptions about the network [BGW88, CCD88]. Namely, they assume that at most a third of the processors between the processors are faulty, and that secure pairwise communication lines are available. In other words, let $t$ be a bound on the number of faulty processors (for simplicity, regard a processor as faulty if its communication lines are faulty). As long as $3t < n$, any function $F(x_1, \ldots, x_n)$ described by a Boolean circuit $C_F$ can be computed reliably and securely in a complete network of $n$ processors with private communication lines. Because the protocols are based on circuit simulation using threshold schemes, the number of rounds of interaction and the sizes of the messages involved are related to the depth and size of $C_F$.

Since threshold schemes form the basis for many of these protocols, let us digress for a moment to describe them. Say that one processor holds a private item of information, $s$. It distributes this secret among the network, sending values (known as "pieces") to each processor $i$, so that provided that at most $t$ processors fail, the secret can be put together later. The secret, however, must remain a secret until then; no group of $t$ or fewer processors, even colluding and malicious ones, can glean any information about $s$ from their pieces. Shamir gave an elegant solution, known as secret sharing, based on properties of random polynomials [Sha79].

The recent methods for secret computation are based on combining secretly shared values to create new secretly shared values. Say $x$ and $y$ have been secretly shared among the system. Using a protocol for addition or multiplication, new secrets $u$ and $v$ can be constructed such that their hidden values are $u = x + y$ and $v = xy$. The new secrets have the same reliability and secrecy properties as secrets which have been distributed by some dealer, but there no longer need be a dealer who knows the hidden values of $u$ and $v$.

By evaluating the Boolean circuit $C_F$, a new secret $w$ whose hidden value is $F(x_1, \ldots, x_n)$ is constructed. The use of secret sharing is essential to hiding the intermediate values, which could reflect information about the inputs which should not be revealed. Most multiparty protocols, cryptographic or not, revolve around interactively simulating a circuit gate by gate, following a procedure introduced in [GMW87]. Thus, the communication complexity of most protocols is directly related to the com-putational (circuit) complexity of $F$. In concrete programming terms, the number of rounds of interaction is proportional to the time that a centrally-run program would take to compute $F$.

Even for useful functions having reasonable circuit depth, the number of rounds of interaction becomes prohibitive in a practical sense. Because communication is a bottleneck in distributed computations, protocols with high communication complexity are at a great disadvantage. The overhead of ensuring security by using these methods counteracts the advantages of computing distributively, even to the point of making simple computations too expensive to perform.

## 1.1 The Communication Complexity of Secure Computation

We examine a new measure of communication complexity: the *communication complexity of secure computation*, which measures the number of rounds of interaction and the number of bits sent during a protocol which computes a function *reliably* and *securely*. Developing secure and practical protocols with small communication complexity is the primary focus of this dissertation.

We may contrast the communication complexity of secure computation with the standard measure of communication complexity as follows. If we relax the demand for security, any function has a "small" communication complexity: each processor simply broadcasts its input $x_i$, and then individually computes $F(x_1, \ldots, x_n)$ from the messages it received. (Technically, we have ignored an important issue, namely how to achieve broadcast reliably using private communications, but this problem, known as Byzantine Agreement, has efficient solutions.) Thus, few rounds and small messages would suffice.

It is reasonable to expect that more and longer messages may be required to hide information otherwise leaked by a direct approach. This expectation holds for the particular protocols mentioned earlier, and even led some authors to conjecture that the circuit depth of a function is a lower bound on the number of rounds needed for secret computation.

We disprove that conjecture with our first result [BB88], which is the first to address the issue of communication complexity in secure computation: any circuit of depth $D$ can be evaluated in $O(D/\log n)$ rounds using small messages. For all functions in $NC^1$ (or more generally for any function admitting a circuit of polynomial size and logarithmic depth), this result gives protocols which operate in a small, constant number of rounds.

For example, consider a secret ballot over a legal measure which will pass only if there is a two-thirds majority. There is a simple and efficient protocol which operates in constant rounds regardless of the number of voters, to determine if the law is passed, without revealing the individual votes or even the exact tally. A variety of useful functions becomes efficiently computable using our techniques.

An even stronger and more surprising statement can be made if one is willing to permit fewer than $O(\log n)$ faults in the network, as opposed to the most general bound of $n/2$ or $n/3$ :

> *Any* function, *regardless* of its circuit depth or size, can be computed in a *constant* number of rounds of interaction and using small messages.

This is a very surprising result: the communication complexity of secure computation need not be related to the computational complexity of the function being computed. It lends great hope to the practical implementation of methods for distributed security, since communication is an expensive resource.

At the risk of basing protocols on unproven complexity-theoretic assumptions, we show also that the goal of achieving constant-round protocols to compute functions of arbitrary circuit depth can be achieved at the higher levels of fault-tolerance ($2t < n$).

The common idea underlying many of our protocols is the following: to compute a function in a distributed manner, convert it to several problems that can be computed locally and then combined to give the solution, despite errors in some of the local computations. The key, however, is to ensure that each locally computed problem is independent of the original to a certain degree, which hides information about the original problem and its answer and simultaneously provides a degree of resilience against local errors. The brunt of the computation is placed on local computations rather than communications.

These are important tools for practical methods for distributed security and reliability. With ample levels of fault-tolerance, and using small messages, the number of rounds of interaction is reduced greatly.

## 1.2 Fault Tolerance

The non-cryptographic protocols of [BGW88, CCD88] tolerate faults in a third of the network, namely $3t < n$. It is not difficult to show that a majority of faults is intolerable (*e.g.* the AND function cannot be computed by a half-faulty network [BGW88]). This left a gap between the achievable $3t < n$ bound and the $2t \leq n$ impossibility result.

We close the gap, showing that as long as the number of faults satisfies $2t < n$, any circuit can be evaluated securely and reliably. The crucial discovery that supports the tight bound on fault-tolerance relies on a simple problem which we call the *ABC Problem*: one processor must share three secrets, $a$, $b$, and $c$, and then prove that $ab = c$ without revealing any other information.

Our solution is simple and efficient, and supports secret computations based on arithmetic over exponentially large ($n$-bit) fields (such as the set of integers modulo $m$, where $m$ is an $n$-bit prime). A similar but less efficient result was obtained independently by Ben-Or [RB89]; in contrast, that solution uses bitwise operations which would require $O(n \log n)$ times as many rounds for the same arithmetic. Both methods rely on a recent technique by Rabin [RT88] for verifiable secret sharing without using cryptographic assumptions. (*Verifiable* secret sharing has the added property that all nonfaulty processors are assured that the secret is well-defined and reconstructable even when the dealer of the secret may be faulty.)

**Zero-Knowledge Proof Systems**   Our solution to the ABC problem also provides a fast and efficient way for one processor to prove that a general property holds on a string of bits, without having to reveal the proof.

Goldwasser, Micali, and Rackoff [GMR89] and Babai [BM88] pioneered the concept of zero-knowledge proof systems, in which one party, the prover, attempts to convince another party, the verifier, that a string $x$ is in a particular language $L$. The verifier learns whether $x \in L$, but as in the multiparty protocols discussed above, he learns nothing more than that result. Presumably the prover has far greater power to prove language membership, while the verifier is limited to polynomial time computations.

A similar but distinct kind of proof system involves committed strings, and is similar to the ABC problem described above. Say that the prover can place bits in envelopes and seal them, so that the verifier cannot read the bits but the prover cannot change them before opening the envelopes at some future time. In this case, the prover commits to a string of bits $x$, and then proves to the verifier that $x \in L$, without revealing the bits in $x$.

We give a formulation of both types of zero-knowledge proofs for a network of processors. Assuming that a majority of individuals are reliable, we show that a prover can prove membership in *any* language in IP (the class of languages with polynomial-time 2-player proof systems) without having to know which processors are reliable. Our result applies to both flavors of proof systems, proving statements about known strings and proving statements about hidden but committed strings.

Thus, if the prover, the verifier, and a third party are present, the prover is assured that his proof leaks no additional information, as long as *either* the verifier *or* the third party is reliable.

Multiparty zero-knowledge proof systems are distinct from the related idea of Goldwasser *et al* [BGKW88], who examine two-*prover* proof systems in which provers are kept physically separate. Here, we allow communication between all processors, where just one of them is a prover.

## 1.3   Locally Random Reductions

We develop a powerful new tool called a *locally random reduction* to achieve many of the results mentioned above, along with other results we shall describe in a moment. A *random reduction* from the problem of computing some function $f(x)$ to the problem of computing another function $g(y)$ is a pair of probabilistic, polynomial-time algorithms $P$ and $Q$. By computing $P(x)$, one obtains a list of values $(y_1, \ldots, y_m)$ at which to evaluate $g$. The results are interpolated to determine $f(x)$ by evaluating $Q(g(y_1), \ldots, g(y_m))$. The reduction is $(k, m)$-*locally random* if for any subset $S$ of the $y$'s such that $|S| \leq k$, the distribution on $\{y_i \mid i \in S\}$ induced by $P$ is the same, regardless of $x$.

This new tool has widespread applications. It forms the basis of a theory for program testing developed by Lipton [Lip89], who also observed that, using our general technique for LRR's, the PERMANENT function has a locally random reduction. Blum, Levin, and Luby [BLL89] used it to show that if the PERMANENT function is computable on average in polynomial time with a uniform distribution, then $\#P \subseteq \text{ZPP}$. Nisan [Nis89] showed that any language in the polynomial time hierarchy admits a two-prover interactive proof system; subsequently, Fortnow, Lund, Karloff, and Nisan [LFKN89] showed that $\text{PH} \subseteq \text{IP}$. Previously, it was not known whether even coSAT admitted interactive proofs. Finally, Shamir [Sha89] solved a fundamental open question, showing $\text{IP} = \text{PSPACE}$. The algebraic approach of locally random reductions has sparked quite a bit of interesting results in cryptography and complexity theory.

We apply locally random reductions to several problems in cryptography including the secure multiparty computations and zero-knowledge proof systems described above. As mentioned, the communication complexity of secure computation can be made independent of the computational complexity of the function. Locally random reductions serve this purpose by allowing us to move the computational effort needed to compute the function from the communication lines to the local processors, thus removing a serious burden of secure protocols.

Locally random reductions were inspired by the problem of *instance-hiding schemes* [AFK89, BF90]: a weak processor, $A$, wishes to utilize the computational resources of powerful oracles $B_1, \ldots, B_m$, in order to compute an intractable function $f(x)$. The weak processor does not, however, want to reveal anything about $x$. We show that, using a $(1, m)$-LRR (defined in Chapter 12), any function of $m$ bits has an instance-hiding scheme.

## 1.4 Extreme Fault-Tolerance

**Privacy Without Cryptographic Assumptions** We have mentioned that general computations are impossible when fault levels constitute a majority of the network. To understand better the nature of privacy in distributed computations, we examine protocols tolerating greater numbers but strictly weaker types of faults. In fact, let us assume that no processor fails at all, but that some number $t$ of them may pool their information to attempt to glean information that should remain private. What sorts of functions can be computed *privately* when $2t \geq n$, without making unproven assumptions?

We take a first step in this direction by completely characterizing the functions $f$ which can be computed privately by $n = 2$ parties, any $t = 1$ of which may "collude." That is, for general functions of two arguments, we give a straightforward method to determine if $F$ is privately computable. This result was obtained independently (with an incorrect proof) by Kushilevitz [Kus89]. For the case of $n \geq 2$, Chor and Kushilevitz [CK89] showed that a very limited subset of the class of Boolean valued functions are privately computable for $2t \geq n$. The generalization to general-valued functions of several arguments remains open.

**Privacy and Reliability With Cryptographic Assumptions**  Not only are most functions impossible to compute with perfect security when $2t \geq n$, but an issue of fairness arises. In general, if a majority of participants halt, the whole computation must come to a halt. A faulty majority can therefore halt just after learning the result of the computation, and just before the nonfaulty participants learn the result.

By restricting the processors and the adversarial processors to polynomial-time computations, we give new protocols which solve both of these problems, making the weakest possible unproven cryptographic assumption in this case, that a protocol for two-player Oblivious Transfer exists. A processor with unbounded resources could break encryption schemes; thus we do not violate the impossibility of perfect secrecy.

The important aspect of computations with a faulty majority is the issue of fairness. Yao [Yao86] and Galil, Haber, and Yung [GHY87] proposed a definition which did not allow the adversary to have access to the programs of the nonfaulty processors, and gave cryptographic solutions under these circumstances. We formulate a stronger definition, allowing the adversary access to the programs of the reliable processors, and which we believe best captures the notion of fairness in a quantitative and intuitive sense. Furthermore, we are able to achieve this stronger property of fairness.

In addition to developing extremal protocols based on cryptographic assumptions, we consider instead a network equipped with "noisy" channels, namely channels which allow messages to pass with a 50-50 chance. We give alternative protocols which utilize these channels instead of depending on unproven assumptions, showing that fairness, privacy, and correctness are achievable even when faults abound.

## 1.5   Definitions and Formal Proofs

In order to prove that protocols are correct and private, a formal model for distributed protocols and the adversaries that represent the faulty behavior of the participants is needed.

While the current literature contains many new techniques, it often omits any mention of an underlying formal model for security, often giving only intuitive arguments for the claimed results. While intuitive arguments are satisfactory to the cursory reader, and ideal for the fast dissemination of results, the lack of formal proofs undermines any assurance that the methods do provide security and fault-tolerance. Furthermore, with the variety of assumptions and network characteristics used by the various solutions, it is difficult to compare the advantages of one over another.

The formulation of a rigorous and coherent model is a primary contribution of this dissertation. The computational power of the participants and the adversaries, the available communication channels, the assumptions made to ensure secrecy, the nature of the adversaries, the definitions of correctness, privacy, and fairness, all require a formal and explicit treatment. We give a framework to define and prove various properties of the networks and the protocols, enabling one to compare what is achievable across the range of network characteristics and assumptions made. In keeping with the historical trend of cryptography, we provide a model which supports an analysis of perfect, information-theoretic security, yet is adaptable to the modern approach of computational complexity-based cryptography.

But our work goes far beyond the mere clarification of mechanical specifications of interactions — which, though necessary for formal proofs of resilience, are less than stimulating. We address the difficult and often subtle problem of giving formal definitions for properties such as correctness and privacy, issues that seem so intuitively easy that common sense often leads down the wrong path. The research literature is filled with *ad hoc* properties and with attempts to formalize each of them. Few definitions are satisfactory, and because of their *ad hoc* nature, few attempts to actually prove claims to security and reliability appear.

We give a surprisingly simple and coherent definition of *resilience* (see 3.4 in §3.4), a combination of security and reliability that unifies the various properties that one desires in a multiparty computation. The subtle and key aspect of our approach is to avoid analyzing each desired property of a protocol separately, instead giving a single definition from which all properties arise. In a nutshell: a real multiparty protocol is intended to achieve the results that an *ideal* protocol having a trusted host would achieve, and the information and influence of an adversary must be no greater than that in the ideal case. We define rigorously what it means to "achieve the same results," giving a general and powerful tool for measuring the accomplishments of one protocol against those of another.

## 1.6 Applications

The protocols we provide achieve secure and reliable distributed computations of arbitrary functions $F(x_1, \ldots, x_n)$. We demonstrate how this abstraction can fit in with practical goals.

We develop some efficient applications to practically-motivated problems. For example, we give a fast method for a processor to authenticate itself and to authorize its transactions. We provide a quick protocol to deliver mail anonymously, protecting the identity of the sender and even of the receiver. We describe a reliability mechanism which resists traffic analysis, by hiding not just the results of its computations but the nature of the computations themselves (*e.g.* whether it is idle or processing a request for a file, etc.).

| Faults | Rounds | Bits | Local Time | Function Class |
|---|---|---|---|---|
| $t < n/2$ | $\text{depth}(C_F)/\log n$ | poly | poly | P (or poly-size circuits) |
|  | constant | poly | poly | $NC^1$ (or log-depth circuits) |
|  |  | exp | $\text{size}(C_F)$ | any $C_F$ |
| $t = O(\log n)$ | constant | poly | exp | any $C_F$ |

Figure 1.2: Various tradeoffs in communication complexity, fault-tolerance, and function-complexity for the results developed here.

## 1.7  Summary

We develop a host of communication-efficient protocols to achieve optimally fault-tolerant and secure distributed computations. Figure 1.2 describes the various combinations and trade-offs in rounds, message sizes, and the computational complexity of the desired computations.

We give the first protocol to achieve a secure multiparty computation using a number of rounds less than the circuit depth of the function to be computed. In fact, we give the first protocols that use only a constant number of rounds.

Our protocols are proven secure and reliable using a concise and precise set of formal definitions. The formulation of these definitions is a major contribution of this dissertation and provides a clear and general basis for proving the resilience of other protocols in the literature.

We investigate fault tolerance when faults constitute a majority of the network. We characterize the class of functions which are privately computable against a passive adversary. We give a strong formal definition for fairness when the faults are Byzantine, and show how to achieve it using additional assumptions.

We develop a powerful new tool, the *locally random reduction*, which has broad applications. It reduces the communication complexity of many secure protocols drastically, regardless of the computational complexity of the computation the protocol attempts to achieve. It provides a solution to the problem of utilizing powerful public resources without compromising privacy. It has inspired applications to program testing and interactive proof systems and led to several fundamental new results.

# Chapter 2

# Preliminary Definitions

> The rest of the pages he picked up from the floor, bunched together, and threw down between his legs into the bowl. Then he pulled the chain. Down went the helping names, the influential numbers, the addresses that could mean so much, into the round, roaring sea and onto the rails. Already they were lost a mile behind, blowing over the track now, over the glimpses of hedges into the lightning-passing fields.
>
> Home and help were over. He had eight pounds ten and Lucille Harris' address. "Many people have begun worse," he said aloud.
>
> — Dylan Thomas, *Adventures in the Skin Trade*

We shall require a modest repertoire of definitions before the results of this dissertation can be stated and proved. Many of these definitions are standard and we sketch them here for completeness; others have not appeared or are unsatisfying for many reasons, and the precise and concise formalization of their intuitive meanings is a novel, necessary, and unifying contribution of this dissertation. In this chapter we present general definitions and describe the mechanical nature of a protocol – *i.e.* the nature of the participants and adversaries, the network, and the sequence of computations and communications involved. In Chapter 3 we examine the precise nature of security and reliability for multiparty protocols, a more subtle and significant issue.

Let us emphasize once again the fundamental abstraction of a multiparty computation:

> Each of $n$ parties holds a private input value $x_i$, and together the $n$ parties wish to compute a (finite and multi-valued "probabilistic") function $F(x_1, \ldots, x_n)$ without revealing anything but the result. That is, they wish to simulate the following situation: a trusted and reliable party receives each of their inputs privately, computes the function $F(x_1, \ldots, x_n)$, and returns the results to the players.

In order to appreciate and to compare the variety of protocols under the broad and varied set of assumptions and network models, a unifying model is needed. The important mechanical parameters of a model and solution can be listed in a few dimensions. The participants may be resource-unbounded automata, having a finite or even an infinite number of states (the "information-theoretic" model), or Turing machines with limited resources (the "complexity-based" model). The network may connect the participants completely, and it may also support a range of communication channels, broadcast and private channels being the most important but by no means exclusive. Unproven complexity theoretic assumptions or cryptographic assumptions may be made; often, for example, protocols are based on the unproven existence of one-way or trapdoor functions. Normally, one allows an adversary to observe and perhaps to change the communications to and from various participants. The adversary may be unbounded or bounded, static or dynamic in its choice of participants to corrupt, and passive or malicious in its actions.

## 2.1 Notation

We use a vector notation to represent a labelled set of items: $\vec{x} = \{(1, x_1), \ldots, (n, x_n)\}$; we shall normally omit the labels, writing $\vec{x} = \{x_1, \ldots, x_n\}$. A subscript denotes a subset of those values: $\vec{x}_T = \{x_i \mid i \in T\}$. Bars indicate complements: $\overline{T} = [n] - T$, where $[n] = \{1, \ldots, n\}$. The operation $a$ div $b$ is defined as $b\lfloor \frac{a}{b} \rfloor$.

We adopt the standard alphabet $\Sigma = \{0, 1\}$. We may extend it through a trivial encoding to include other characters, especially the delimiter, #, and a symbol $\Lambda$ indicating a null message. Let $\Sigma^{\leq c} = \bigcup_{i=0}^{c} \Sigma^c$. If $\sigma$ is a string then $\sigma[a..b]$ denotes the substring of $\sigma$ from the $a^{th}$ character to the $b^{th}$, inclusive. The symbol $\oplus$ denotes exclusive or of bits, or the bitwise exclusive-or when applied to a pair of strings. The exclusive-or of the bits $b_i$ as $i$ ranges over a specified set is denoted $\oplus_i b_i$. The symbol $\circ$, when used with strings, denotes concatenation. For notational convenience, we define $H_m = (\Sigma^\star)^m$, the set of $m$-tuples of strings, and we let $H = \bigcup_{i=1}^{\infty} H_i$. We assume a uniquely decodable encoding of $m$-tuples of strings into single strings.

We assume a natural encoding of sets of strings when written using $\Sigma$ : if $S = \{s_1, \ldots, s_n\}$ is a set of strings (where, without loss of generality, $s_1, \ldots, s_n$ are in lexicographic order), then $S$ is written as $s_1 \# s_2 \# \cdots \# s_n$. If $S'$ is a set of objects each of which has a natural encoding as a string, then the encoding of $S'$ is the same as the encoding of the set $S$ of encodings of each of its members. When we speak of a set of messages or a state being written on a tape, we implicitly assume a natural and uniquely decodable (*e.g.* prefix-free) way to tag various items such as messages (include the sender, receiver, communication channel, contents, and time of sending) or Turing machine states.

The notation "$i$ :" in a protocol description indicates a local computations and variable assignments of player $i$. The notation "$i \to j : m$" indicates that $i$ sends $m$ to $j$, and $i \to [n] : m$ means that $i$ broadcasts $m$. The notation $(1 \leq i \leq n)$ indicates that the succeeding text is performed in parallel for $i$ in the range from 1 to $n$.

Let $\mathtt{dist}(X)$ denote the set of all distributions on a set $X$. The probability of event $Y$ with respect to distribution $P$ is denoted $\Pr_P[Y]$. A *probabilistic function* $F$ is a function mapping some domain $X$ to a set of *distributions* on a set $Y$, that is, $F : X \to \mathtt{dist}(Y)$. The output of a probabilistic function is a distribution, but sometimes we may abuse notation and refer to a the output as a *sample* taken according to that distribution.

The difference of two distributions $P$ and $Q$ on a set $X$ is defined by $|P - Q| = \max_{W \subseteq X} |\mathrm{Pr}_P [W] - \mathrm{Pr}_Q [W]|$. We denote a sample $x$ taken according to $P$ by $x \leftarrow P$; a distribution may also be written as $P = \{x \leftarrow P\}$. Let $\{x \leftarrow P \mid Y(x)\}$ be a sample $x$ taken according to the distribution $P$ subject to the condition that $Y(x)$ holds. Let $f$ be a function from some number $j$ of arguments to a range $X$; we denote by

$$\{x_1 \leftarrow P_1; x_2 \leftarrow P_2(x_1); x_3 \leftarrow P_3(x_1, x_2); \dots; x_j \leftarrow P_j(x_1, \dots, x_{j-1}) : f(x_1, x_2, \dots, x_j)\}$$

the distribution on $X$ induced by running the $j$ experiments in order and applying $f$ to the results. Where clear from context, we use the same notation to describe a sample drawn from that distribution. The composition of two probabilistic functions $F : X \to \mathtt{dist}(Y)$ and $G : Y \to \mathtt{dist}(Z)$ is the probabilistic function $H : X \to \mathtt{dist}(Z)$ given by

$$H(x) = \{y \leftarrow F(x); z \leftarrow G(y) : z\}.$$

Some often used distributions are: the uniform distribution on a set $X$, denoted $\boldsymbol{uniform}(X)$; the uniform distribution on the set $\mathtt{Polyn}(t, s)$ of polynomials $f(u)$ of degree $t$ satisfying $f(0) = s$, denoted $\boldsymbol{UPoly}(t, s)$; and the set of $n$ values obtained by selecting such a polynomial at random and evaluating it at $n$ points, defined by

$$\mathtt{Pieces}(n, t, s) = \{\vec{y} \in E^n \mid (\exists f \in \mathtt{Polyn}(t, s))(\forall i) y_i = f(\alpha_i)\}$$
$$\boldsymbol{UPieces}(n, t, s) = \boldsymbol{uniform}(\mathtt{Pieces})$$

Here, $E$ is some finite field and the $\alpha_i$ values are implicitly fixed, normally either $\{1, \dots, n\}$ or $\{1, \omega, \dots, \omega^{n-1}\}$ where $\omega$ is a primitive $n^{th}$ root of unity. A coin biased to 0 by $b$ is the distribution $\mathtt{bias}(b) \in \mathtt{dist}(\{0, 1\})$ given by $\mathrm{Pr}[0] = \frac{1}{2} + b$.

An ensemble is a family $\mathcal{P} = \{P(z, k)\}$ of distributions on $\Sigma^{\leq p(|z|, k)}$, parametrized by $z \in \Sigma^\star$ and $k \in \mathbf{N}$, where $p(\cdot, \cdot)$ is polynomially bounded.

**Definition 2.1** *Ensembles $\mathcal{P}$ and $\mathcal{Q}$ are $\delta(k)$-**indistinguishable** if*

$$(\forall k)(\forall z) \qquad |\mathcal{P}(z, k) - \mathcal{Q}(z, k)| \leq \delta(k).$$

We write this as $\mathcal{P} \approx^{\delta(k)} \mathcal{Q}$. Under the standard $O$-notation, $f(k) = O(g(k))$ means $(\exists c, k_0)(k \geq k_0 \Rightarrow f(k) \leq c \cdot g(k))$. We call $k_0$ the convergence parameter.

**Definition 2.2** *Ensembles $\mathcal{P}$ and $\mathcal{Q}$ are $O(\delta(k))$-**indistinguishable** if*

$$(\exists \Delta : \mathbf{N} \to \mathbf{N}) \quad \mathcal{P} \approx^{\Delta(k)} \mathcal{Q} \qquad \text{and} \quad \Delta(k) = O(\delta(k)).$$

We use the following adjectives to delineate various abilities to discriminate between

ensembles:

| | | | | |
|---|---|---|---|---|
| **perfect** | (written $\mathcal{P} \approx \mathcal{Q}$) | if | | $\mathcal{P} \approx {}^0 \mathcal{Q}$. |
| **exponential** | (written $\mathcal{P} \approx^e \mathcal{Q}$) | if | $(\exists c > 1)$ | $\mathcal{P} \approx {}^{O(c^{-k})} \mathcal{Q}$. |
| **statistical** | (written $\mathcal{P} \approx^s \mathcal{Q}$) | if | $(\forall c)$ | $\mathcal{P} \approx {}^{O(k^{-c})} \mathcal{Q}$. |
| **computational** | (written $\mathcal{P} \overset{\sim}{\approx} \mathcal{Q}$) | if | | (see below) |

Computational distinguishability is based on the notion of a resource-bounded machine that tries to distinguish strings sampled from one distribution or the other. A *distinguisher* $A$ is a probabilistic polynomial size circuit which outputs either 0 or 1. Let $A_{\mathcal{P}(z,k)}$ denote the probability that $A$ outputs a 0 on an input selected according to $\mathcal{P}(z,k)$. Let PPC denote the class of probabilistic polynomial size circuit families.

**Definition 2.3** *Ensembles $\mathcal{P}$ and $\mathcal{Q}$ are $\delta(k)$-computationally indistinguishable if*

$$(\forall A \in \text{PPC})(\exists k_0)(\forall k \geq k_0)(\forall z) \quad |A_{\mathcal{P}(z,k)} - A_{\mathcal{P}(z,k)}| \leq \delta(k).$$

We say they are simply *computationally indistinguishable* if $\delta(k) = O(k^{-c})$ for some fixed $c$, and write $\mathcal{P} \overset{\sim}{\approx} \mathcal{Q}$.

We shall soon consider ensembles induced by running a protocol with particular parameters $n$ (number of players) and $m$ (size of inputs). Two *families* of ensembles $\{\mathcal{P}(i,j)\}$ and $\{\mathcal{Q}(i,j)\}$ are $\delta(k)$-indistinguishable if for all $i$ and $j$, $\mathcal{P}(i,j) \approx^{\delta(k)} \mathcal{Q}(i,j)$. We write $\{\mathcal{P}(i,j)\} \asymp^{\delta(k)} \{\mathcal{Q}(i,j)\}$.

An ensemble $\mathcal{P}$ is *polynomially generable* if there exists a polynomial size circuit family with random inputs that, on input $z$ and $k$, produces a distribution identical to $\mathcal{P}(z,k)$.

## 2.2   Two-Party Protocols

Two-party protocols are programs for a pair of machines that must jointly compute a result. In its most general form, the definition of a two-party protocol specifies only that the machines take turns computing and communicating, eventually placing an output on each of their output tapes. In fact, the two participants need not be machines but can be arbitrary, even non-recursive functions mapping inputs to a random variable describing the output. We shall postpone discussion in full generality, however, and for the moment consider a pair of machines.

Security for two-party protocols addresses what a given protocol accomplishes when one of the participants is replaces by a different machine, namely a *corrupted* machine. Protocols are designed to withstand or at least to detect such changes and to achieve the same results as when neither party is replaced.

The two common types of machines used to model interactive computations are Turing machines and circuits. Depending on the degree and the nature of the faults, one model has advantages over the other. Non-uniform circuits have greater computational power than Turing machines, so a protocol which is secure against faulty circuits is stronger than one secure against faulty Turing machines. On the other hand, Turing machines present a more natural model for real-world computations, in some senses, and thus the power of nonfaulty members of a system is modelled more accurately. The protocols presented in this work generally require the power of a polynomial time Turing machine at most, but they are stated with respect to polynomial size circuits or more powerful players, to provide maximal security. A slightly different way to state this is that any uncorrupted player is in general a Turing machine, whereas an adversary that corrupts it may replace it by a more powerful, non-uniform circuit.

### 2.2.1 Interactive Turing Machines

An *interactive Turing machine* [GMR89] is a probabilistic Turing machine having a read-only input tape $x$ (which may be public or private), a private work tape $W$, a private random tape $R$ with a unidirectional read-head, a read-only communication tape $C_{in}$, and a writable communication tape $C_{out}$. A random tape is a convenient intuitive tool to describe the generalized computation of a probabilistic Turing machine: in particular, a probabilistic Turing machine is a probabilistic function that maps input $x$ to the distribution induced by choosing uniformly at random a mapping from $\mathbf{N}$ to $\{0, 1\}$, regarding this as the random tape of a standard, deterministic, multitape Turing machine, and operating the machine in the usual way. of bits, regarding this as A machine makes a random decision, *i.e.* flips a coin, by reading its random tape.

The interactive machine has a special *inactive* state, which it enters after having performed a computation and writing on its communication tape. It is awoken from the inactive state when the other machine enters its inactive state. Local computation time is the maximal number of steps used between entering inactive states; total local computation is the total number of steps executed during the interaction.

An *interactive protocol* is a pair of interactive Turing machines [A,B] sharing the same public input tape and sharing communication tapes in the standard sense. They take turns being active. Normally, one or both of the machines uses polynomial local computation time, and often one of the machines is more powerful. For instance, in the zero-knowledge proof model, a machine which proves a complicated theorem may be permitted a polynomial space computation to assist it in convincing a weak , polynomial-time verifier of the truth of the theorem.

The *communication complexity* of a protocol is measured by two factors: first, the number of rounds, *i.e.* the number of times each Turing machine enters its inactive state, and second, the message complexity, *i.e.* the total number of bits sent during the protocol. Each measure is a function of the size $m$ of the inputs.

### 2.2.2 Non-Uniform Interaction

For many reasons, a non-uniform model is superior to the uniform Turing machine approach. For instance, the security of an encryption scheme is often stated with respect to circuits rather than Turing machines. A canonical example is given in [GM84]. Consider a public-key cryptosystem in which a user A publishes his encryption key $E_A$. In order for the system to be secure, there should be no way for an adversary to choose a polynomial-time Turing machine which inverts the encryption, even if the adversary can choose the machine after the encryption key is published. The solution to this problem as presented in [GM84] is to allow non-uniform adversaries, such as circuits, which can have such knowledge "wired in."

There are other, more important reasons to allow non-uniformity. In order to combine protocols in a modular way, a way to model information obtained in earlier interactions is needed [TW87, Ore87, GMR89]. Furthermore, because external information may be available, even a protocol composed of many subprotocols may not start with a *tabula rasa*.

Rather than using circuits as the model, though, it is more natural to employ "non-uniform" Turing machines; the extensions to the circuit model to allow interaction are somewhat inelegant. Since the power of polynomial size circuits is achieved by polynomial time Turing machines with polynomial size advice strings (the class P/poly), there is no loss of generality in defining security and correctness with respect to Turing machines that take advice.

A *non-uniform interactive Turing machine* is an interactive Turing machine having a private read-only auxiliary input tape $\sigma$. A two-player *non-uniform interactive protocol* is a pair of non-uniform interactive Turing machines. Normally, each machine is taken to be a polynomial-time machine, though for some purposes the machines may be allowed unbounded computation time and unbounded advice.

In general, we shall use either the non-uniform Turing machine model just described, or a more general model in which each player's local computation is described by some arbitrary, perhaps non-recursive function, The former corresponds to the modern trend of "complexity-based" cryptography, whereas the latter corresponds to the classical approach of "information-theoretic" security, where issues of computability and complexity are irrelevant.

### 2.2.3 Two-Party Interactive Proof Systems

Goldwasser, Micali, and Rackoff [GMR89] and Babai [BM88] pioneered similar definitions of what it means for one machine to convince another that a string is in a language, even when the doubting party does not have the resources to decide the answer directly. As defined in [GMR89], an *interactive proof system* [P,V] for a language $L$ is a two-party protocol for machines called the *prover* (P) and the *verifier* (V). The verifier must use polynomial time, while the prover has unbounded computation time and space. Both machines employ probabilistic computation. At the end of the protocol, the verifier outputs *accept* or *reject*. In order to be a proof system, the protocol must have the following properties:

- **Completeness:** For $x \in L$, if the protocol is run with correct P and V, then for every $c > 0$ and sufficiently large $n = |x|$, V accepts $x$ with probability at least $1 - n^{-c}$.

- **Soundness:** For $x \notin L$, if the protocol is run with any $P^\star$ and V, then for every $c > 0$ and sufficiently large $n = |x|$, V rejects $x$ with probability at least $1 - n^{-c}$.

This definition generalizes in a straightforward way to cover proofs that $f(x) = y$ for a function $f$ and values $x$ and $y$.

### 2.2.4   Zero-Knowledge

It is natural to examine properties of the protocol under situations in which one machine or the other is replaced by a different, "faulty" machine, which tries to obtain more information than it truly deserves. If the transcript of messages send during a protocol can be generated by a machine given only the output of the protocol, then intuitively that machine learns nothing apart from the output by participating in the protocol. In an ideal case, a trusted prover would state whether or not $x \in L$, and the verifier would accept the statement, learning only that fact regardless of whether it were itself corrupt. We wish to address the situation in which the prover might be corrupt, while ensuring that no information is leaked beyond that of the ideal case.

Let $L \subseteq \Sigma^\star$ be a language. An ensemble $\mathcal{P}(x, a)$ is *perfectly (exponentially, statistically, computationally) approximable on $L$* if, for $x \in L$, there is a probabilistic expected polynomial time Turing machine $M$ whose output $M(x, a)$ is perfectly (exponentially, statistically, computationally) indistinguishable from $\mathcal{P}(x, a)$. The machine $M$ is called a *simulator*.

Notice the "one-sidedness" of the previous definitions. If $x$ is not in the language, there is no restriction that the distribution be generable by some machine.

Let $V^\star$ be a non-uniform interactive Turing machine with input $x$ and auxiliary (advice) input $a$. A protocol [P,V] is *perfectly (statistically, exponentially, computationally) zero-knowledge on $L$ for $V^\star$* if the ensemble induced by the message history of the interaction between $P$ and $V^\star$ is perfectly (statistically, exponentially, computationally) approximable on the language

$$L^\star = \{(x, a) \mid x \in L\}.$$

The protocol [P,V] is zero-knowledge on $L$ if it is zero-knowledge on $L$ for any such $V^\star$.

A more powerful version of zero-knowledge considers a single, universal simulator that simulates the conversation between $P$ and any $V^\star$, using $V^\star$ as a "black box" whose contents cannot be examined or reset but whose input/output behavior is accessible. A good discussion of this and other flavors of zero-knowledge appears in [Ore87].

### 2.2.5   Zero-Knowledge Proof Systems

With the definitions given above, it is easy to describe a zero-knowledge proof system [GMR89]:

A protocol [P,V] is a perfect (statistical, computational) zero-knowledge proof system for a language $L$ if it is an interactive proof system for $L$ and it is perfectly (statistically, computationally) zero-knowledge on $L$.

## 2.3   Multiparty Protocols: The Participants

### 2.3.1   Player ex Machina

A *player* is an interactive automaton, namely an automaton with an input tape, an auxiliary input tape, an output tape, and a random tape. Let us first consider the familiar case of interactive Turing machines.

The definitions for the multiparty scenario are straightforward extensions of those given for two-party protocols. An interactive Turing machine with an input and an output communication tape is easily regarded as one having $n - 1$ input tapes and $n - 1$ output tapes; technically, the bits of virtual tape $j$ are encoded in positions $j + k(n - 1)$ on the single tape, where $k$ ranges over the integers and specifies the position on the virtual tape. As in the two-party case, each has a read-only input tape, a private auxiliary tape, a private work tape, and a private output tape.

A *multiparty Turing-machine protocol* is a set of $n$ interactive Turing machines $\{M_1, \ldots, M_n\}$ each having $n - 1$ pairs of communication tapes: the read-only communication tapes of machine $M_i$ are labelled $\{R_{ij}\}_{j=1,\ldots,n; j \neq i}$, and its exclusive-write tapes are labelled $\{W_{ij}\}_{j=1,\ldots,n; j \neq i}$.

A *round* in a synchronous protocol operating on a complete network consists of the following events. Each machine $M_i$ enters its active state, computes, and returns to its inactive state. Each tape $W_{ij}$ is then copied to $R_{ji}$. We shall soon describe these mechanics in more detail, including how adversaries affect the execution.

The computation time of machine $M_i$ is the sum of the steps it takes while it is in an active state. The message complexity of the protocol is the total length of all messages sent.

### 2.3.2   General Players

Before specifying the sequence of events and computations occurring during the execution of a protocol in more detail, let us first describe a player in the most general terms. We allow a broader model than Turing machines for a few reasons. Turing machines are restricted to computing recursive functions. If they compute a probabilistic function using a random tape, then the resulting distribution is a recursive function of the input and random tape. Non-uniform functions and general distributions would not be achievable. In fact, even simple distributions, such as generating random numbers modulo 3, are not computable in bounded time.

Even though the protocols we design will not require non-recursive functions or distributions, we would like them to be secure against the most powerful adversary. By using a broad computational model, we cover all the participants — players and adversaries — in the protocol, and do not need to pay attention to *how* a new set of messages is computed (*e.g.* whether by Turing machine or non-recursive function) unless we wish to do so under particular circumstances. This corresponds to the "information-theoretic" model, in which one is concerned with "perfect" secrecy; any distinguishable difference or dependence among distributions on messages sent from player to player is considered to give away information, regardless of the resources needed to detect it. The information-theoretic model does not restrict the power of the players and malicious adversaries to compute recursive or even polynomial-time functions on the messages they see; in fact, each player is defined more generally as a probabilistic function from input messages to output messages.[1]

Thus, for full generality, we use a very general format: we describe players as automata that may have a finite or infinite state set, and we set bounds on available resources only under specific circumstances.not set bounds on available resources.[2] Our informal use of the term "automaton" extends beyond the common usage describing a finite-state machine, since we allow for infinite state sets. A Turing machine with infinite work and random tapes is an automaton having an infinite general-state set, where each state describes the finite control and the contents of the tapes.

Each player has a state set $Q$, a standard input set $X \subseteq \Sigma^\star$, and an auxiliary input set $Z$. For Turing machines, each input in $X$ includes a string $x$ and a security parameter $k$. The initial state of a player is selected according to the input, auxiliary input, and security parameter $k$; that is, each player has a function $q^0 : X \times Z \times \mathbf{N} \to Q$.

As in the case of *non-uniform interactive protocols* (§2.2.2), the auxiliary inputs serve a few purposes. They introduce and encapsulate non-uniformity in a uniform model; that is, when giving the conceptually simple specification that each player is a Turing machine, they are a convenient means to introduce non-uniformity. They may also represent information held jointly by various players, such as shared private encryption keys. Their most common use, however, is not as part of the protocol *per se* but as a record of information obtained in other or previous computations and interactions. For example, auxiliary input $a_i$ may represent the history of conversations in which player $i$ participated during the previous protocol. As in the composition of zero-knowledge proof systems, examining resilience for arbitrary auxiliary inputs allows one to demonstrate that secure protocols remain secure when composed sequentially (*cf.* [Ore87, Hab88]).

---

[1] Even though a function from strings to strings can be specified in a discrete manner, a *distribution* on strings need not have some discrete specification. Often a probabilistic machine is regarded as having a discrete and uniformly distributed input (*e.g.* a random tape of bits). Using *functions* rather than discrete specifications of functions to describe the allowable transitions and distributions associated with an "automaton" is more general and more powerful.

[2] In the "cryptographic" model, for example, we may assume that players and adversaries are polynomial-time Turing machines. It boils down to the comment made earlier: one can either regard a corrupted participant as a player gone awry, in which case it is better to regard each player as powerful, or one can describe several different computational models, one for uncorrupted players, one for an adversary, one for corrupted players, and so on.

During each round of a protocol, each player performs a probabilistic computation and sends messages distributed according to the results. The sets of incoming and outgoing messages depend on the types of communication channels and the number of other players in the network. In particular, each player is described by a *probabilistic* transition function $\delta$ that produces a *distribution* on new states and new messages based on the current state and incoming messages:

$$\delta : Q \times H \to \mathtt{dist}(Q \times H).$$

In other words, given a state and the messages of the previous round, $\delta$ produces a distribution on new states and messages from which distribution the subsequent state and outgoing messages of the player are drawn.

Note that the transition function produces a *distribution* on outputs, in analogy to the computation of a probabilistic Turing machine, which need not compute a *function* of its inputs, but rather produces a distribution on outputs based on its input and induced by the uniformly random bits on its random tape. A Turing machine can also be regarded as producing a *sample* from that distribution (in particular, a Turing machine *is* a function of its input and random input). We consider a general player in a similar, dual fashion: as producing a distribution, or as producing a sample from that distribution. This will allow us to describe protocol executions in terms of the *distributions* on messages as well as in terms of *particular message histories* generated according to those distributions.

**Definition 2.4** *A* **player** *is a tuple* $(Q, X, Z, q^0, \delta)$ *whose components are as described above. A* **polynomial-time Turing machine player** *is a player whose transition function* $\delta$ *and initial-state function* $q^0$ *are computable in time polynomial in the size of its input, auxiliary input, incoming messages, and an additional parameter,* $k$. *A* **polynomial-size circuit player** *is a player whose transition and initial-state functions are computable by a circuit family of size polynomial in the above parameters.*

## 2.4 Protocols

### 2.4.1 Networks and Communication Channels

A *channel* is a probabilistic function $C : \Sigma^\star \to \mathbf{N} \times \mathbf{N} \times \Sigma^\star$ that takes a message $m$ as input and produces a distribution on sets of deliverable messages. A deliverable messages is a triple of the form $(s, r, m)$, where $1 \leq s, r \leq n$ and $m \in \Sigma^\star$. After all channel functions are applied and the resulting distributions are sampled, the collection of resulting triples forms the set $\mathcal{M}$ of messages to be delivered. The subset $\mathcal{M}_s$ defined as $\{(s, r, m) \mid (s, r, m) \in \mathcal{M}, 1 \leq r \leq m\}$ is placed on the input tape of player $r$. If player $s$ writes $(m, C_{sr})$ on its communication tape (meaning "send $m$ on channel $C_{sr}$.") then channel $C_{sr}$ is applied to $m$, producing $(s, r, m)$, which is then placed on the communication tape of player $r$. As mentioned in §2.1, the set of messages appear in lexicographic order on the tape.

For example, a private channel $C_{sr}$ from player $s$ to player $r$ is the function that on input $m$ produces $\{(s, r, m)\}$. A broadcast channel $C_s$ on input $m$ produces $\{(s, 1, m), (s, 2, m), \dots, (s, n, m)\}$. An oblivious transfer channel $C_{sr}^{\mathrm{OT}}$ gives weight $1/2$ to messages of the form $\{(s, r, m)\}$ and $1/2$ to messages of the form $\{(s, r, \Lambda)\}$. An anonymous channel is modelled by supplying the recipient simply with $m_j$.

### 2.4.2   Protocols

An $n$-player protocol is essentially a set of players, channels, and output functions:

$$<\Pi> = \{(P_1, \mathcal{C}_1, \mathcal{O}_1, Y_1), \dots, (P_n, \mathcal{C}_n, \mathcal{O}_n, Y_n)\},$$

where $\mathcal{O}_i : Q_i \to Y_i$ is an output function mapping the (final) state of player $i$ to an output value in $Y_i$. This can be taken to be the contents of a work tape or output tape at the end of the protocol. The set of outputs depends on the purpose of the protocol but we shall take it to be a set of $m$-bit strings. Player $P_i$ is described by $P_i = (Q_i, X_i, Z_i, q_i^0, \delta_i)$. Each $\mathcal{C}_i$ denotes a set $\{C^j\}_{j \in \mathbf{N}}$ of channels on which player $P_i$ sends its messages. We sometimes omit the square brackets and write $\Pi(n, m)$ instead; when other parties such as adversaries or trusted hosts are discussed, the square brackets are intended to delimit the list of participants.

A *network* is a set of channels. A *protocol* $\Pi(n, m)$ is implementable on a given network only if its channels are a subset of those in the network. A *protocol family* $\Pi = \{\Pi(n, m, k)\}_{n,m \in \mathbf{N}}$ specifies a protocol for each $n$ and $m$. For each number $n$ of players and $m$ of bits in the inputs, $\Pi(n, m)$ denotes a series of protocols $\Pi(n, m)$. Finally, $\Pi(n)$ denotes a sequence of sets of $n$ players, each with $m$ built in, but for purposes of asymptotic complexity analysis, each of these players can be regarded as Turing machines that read $n$ and $m$ from its input tape.

### 2.4.3   Protocol Compilers for Turing Machines

At times we may wish to regard the participants as Turing machines even though the function $F(x_1, \dots, x_n)$ itself may not be recursive.[3] One reason for regarding the players as circuits or non-uniform Turing machines with respect to their power to "break" the protocol is that the specification of a non-uniform function $F$ might itself lend power to an otherwise uniform or polynomial-time machine. We therefore turn our attention to protocol "compilers," which produce the specifications of the players for the various possible network sizes, input sizes, and security parameters.

---

[3]How can recursive machines compute a non-recursive function? As we shall shortly discuss, it is often convenient to consider a universal machine that receives a written specification of $F$ as an input. The specification, which may be a circuit or formula, need not be generable by a Turing machine.

A protocol compiler $\mathcal{C}$ is a circuit family $\{C_{n,m,k}\}$ where $n$ specifies network size, $m$ specifies the number of bits in the arguments, and $k$ is a security parameter for protocols which may ensure security and reliability with high probability, although not certainty. The compiler $\mathcal{C}$ produces a description of $n$ Turing machines, each of which has an input $x_i$ and $n$ communication tapes. The computational resources of the compiler (its size and depth) may or may not be of importance to the protocol designer, though usually they are related to the complexity of $F$. As with circuit families in general, the compiler itself may be *uniform* (generated by a Turing machine) or non-uniform, which would be necessary to specify protocols computing general, perhaps non-recursive functions $F$.

A *universal protocol* specifies a single Turing machine $M$ which takes a special input $(1^n, 1^m, 1^k, i, C_F)$, where $i$ indicates the identity of the machine in the network and $C_F$ is a circuit description of the function $F$ for the given size and number of inputs. In this case, the protocol for $n$ players consists of $n$ copies of $M$, each of which is supplied with the appropriate parameters and a unique identification. Our protocols are generally of this type.

We measure the *message complexity* `MesgSize`$(\Pi)$, (maximal number of bits in any execution), *round complexity* `Rounds`$(\Pi) = R$, *local computational complexity* `LocalCpx`$(\Pi)$, (resources required by a Turing machine to compute the player's transitions between rounds), *computational complexity of $F$* `FnCpx`$(\Pi)$, (Turing machine complexity of computing the outputs of all the players) as functions of $n$, $m$, and $k$, unless otherwise specified. Stating that a protocol uses polynomial size messages means, for example, that it communicates at most $p(n, m, k)$ bits in any execution, where $p(n, m, k) = O((nmk)^c)$ for some fixed $c$.

Usually, it is not difficult to design universal protocols in which the communication complexity is polynomial in $n, m, k$ and the description of $C_F$. The trick is to design a compiler whose protocols have communication complexity polynomial in $n$, $m$, and $k$ only, regardless of $F$ and its complexity.

### 2.4.4 Reliable Protocol Execution

Let us formally describe the sequence of steps that occur during a protocol when all the participants are reliable.

Since our purpose is to specify *precisely* the nature of a protocol and its execution rather than to become immersed in formalism, let us use some shorthand to relieve the burden of the notation. As mentioned, during each round of a protocol, a player performs a computation and then sends messages over its various channels. Let $\mathtt{mesg}^{out}(U, V, r)$ denote the set of messages sent from players in set $U$ to players in set $V$ during round $r$. Let $\mathtt{mesg}(U, V, r)$ denote the set of messages actually delivered from players in set $U$ to players in set $V$, that is, those messages drawn according to the various channel distributions acting on $\mathtt{mesg}^{out}(U, V, r)$. Denote the messages received by $V$ from $U$ and $U$ from $V$ by $\mathcal{M}(U, V, r) = \mathtt{mesg}(U, V, r) \cup \mathtt{mesg}(V, U, r)$. (As stated in §2.1, we assume a natural and uniquely decodable encoding of these sets of messages as strings.) We sometimes add a label to describe a substring of a message, so that $\mathtt{mesg}(i, j, r; x)$ indicates the message $i$ sends to $j$ about variable $x$.

Let $\mathtt{send}(\{m_1, \ldots, m_k\}, \{C_1, \ldots, C_k\})$ denote the set of messages generated by applying each message $m_i$ to channel $C_i$.

An execution of a synchronous $R(n, m, k)$-round protocol $\Pi(n, m, k)$ on inputs $\vec{x}$ and auxiliary inputs $\vec{a}$ is the sequence of states and messages obtained from the following experiment. Initialize all states according to $\vec{x}$ and $\vec{a}$. At each round, select the new states and outgoing messages using the transition functions of each player, apply the messages to the channels, and generate the messages actually sent. Finally, after round $R = R(n, m, k)$, allow one final computation. The output of player $i$ is a function of its final state; in the case of a Turing machine, the output is written on the output tape.

More formally, but still using a somewhat loose notation, consider the case of a network with private lines and broadcast channels. We denote the messages sent over private lines by $\mathtt{mesg}_{priv}(U, V, r)$ and those sent over broadcast lines by $\mathtt{mesg}_{broad}(U, [n], r)$. Let $\mathcal{C}_{priv}$ and $\mathcal{C}_{broad}$ denote the private and broadcast channels. An execution of the protocol is generated by the experiment shown in Figure 2.1.

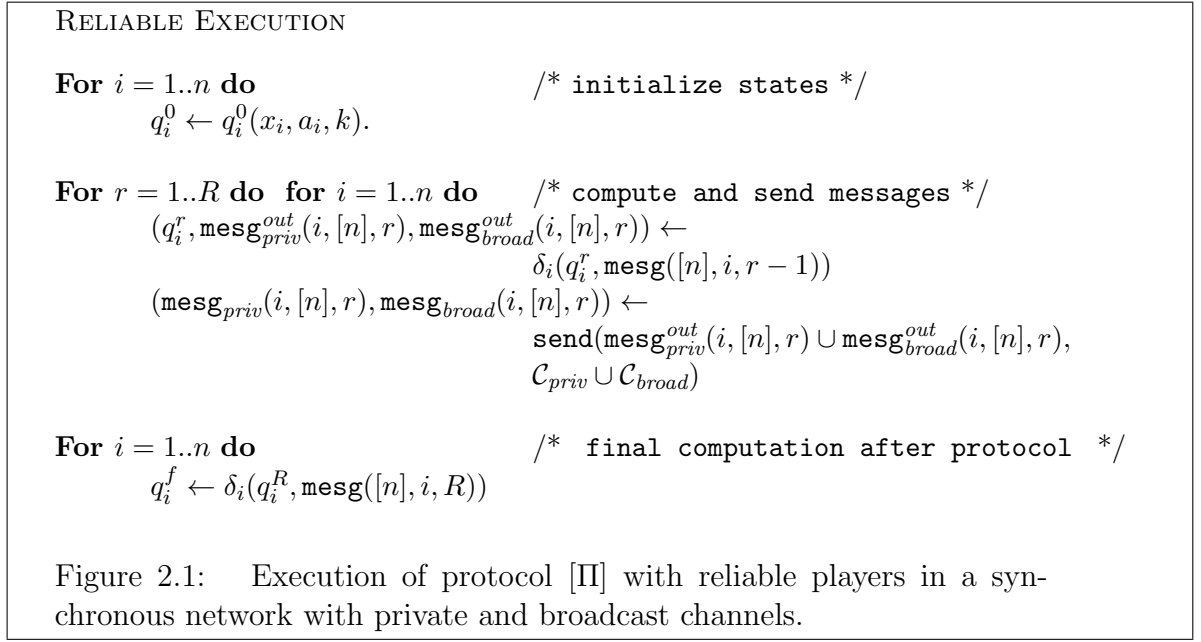

Figure 2.1: Execution of protocol $[\Pi]$ with reliable players in a synchronous network with private and broadcast channels.

## 2.4.5 Views

A global state vector is a sequence of $n$ states, one for each player. In general, we may consider state vectors for arbitrary subsets of the $n$ players. Let $\tilde{Q}$ be the set of all partial state vectors:

$$\tilde{Q} = \{\vec{q}_B \mid \vec{q} \in Q_1 \times \cdots \times Q_n, B \subseteq [n]\}.$$

Similarly, in order to consider sets of inputs and auxiliary inputs for arbitrary subsets of the players, we define

$$\tilde{X} = \{(\vec{x}_B, \vec{a}_B) \mid \vec{x} \in X_1 \times \cdots \times X_n, \vec{a} \in Z_1 \times \cdots \times Z_n, B \subseteq [n]\},$$

and we use this notation to describe the set $\tilde{Y}$ of outputs for arbitrary subsets of the players.

We call the state vector and messages seen by a subset $B$ of the players at a given round $r$ a *view*:

$$\texttt{view}_B^r = (\vec{q}_B^r, \mathcal{M}(B, [n], r)).$$

The set of all possible views is

$$\tilde{V} = \tilde{Q} \times H.$$

A *history* or *execution* of the protocol is thus a member of the set $\texttt{exec}$ of inputs, auxiliary inputs, initial states, sequences of views, and final states:

$$\texttt{exec} = \tilde{X} \times \tilde{Z} \times \tilde{Q} \times (\tilde{V})^R \times \tilde{Q}.$$

A protocol induces a distribution on executions according to the experiment described formally in §2.4.4. We may regard a protocol as a *probabilistic function* from inputs to executions; the function is the composition of several probabilistic functions as described.

### 2.4.6 Memoryless Protocols and Players

As described thus far, a protocol is *memoryless*; the current states and messages are all that matter to the execution of the protocol. The final states of the players do not necessarily reflect the history of the protocol.

In practical situations it is often impossible to assure that no record is kept of previous operations. Backups are kept; transactions must be rolled back to consistent states when failures occur; and a wide variety of reasons make it extremely risky to assume that a history of previous behavior is erased before an adversary gains control of an installation.

In order to assure maximal security, we consider a stronger model in which each player is replaced by a modified player that records its view of the history of the protocol in its current state. In other words, the state $q_i^r$ of player $i$ at round $r$ determines the sequence $(q_i^0, (q_i^1, \mathcal{M}(i, [n], 1)), (q_i^2, \mathcal{M}(i, [n], 2)), \ldots, (q_i^r, \mathcal{M}(i, [n], r)))$. The transition function $\delta_i$ is easily modified to accommodate this record, giving rise to a player whose behavior is identical but whose state records the sequence of states and messages seen by that player. In the case of Turing machines, this means that each Turing machine is equipped with an extra, history tape on which it records all its computations (communications, random bits, contents of work tapes, etc.).

We adopt this as the *standard* model, and assume that all players are transformed in this manner. In the standard model, the set of states of players in some coalition $T$ at round $r$ defines not only their current view, $\mathtt{view}_T^r$, but the history of the protocol up to that point: $\vec{x}_T, \vec{a}_T, \vec{q}_T^{\,0}, \mathtt{view}_T^1, \ldots, \mathtt{view}_T^r$. We distinguish the weaker model, in which players may forget previous state information, by specifically noting it as the *memoryless* model. The memoryless model has advantages in *proving* privacy against dynamic adversaries in cryptographic models [BHY89] because it permits easier simulations of the knowledge of corrupted players.

## 2.5 Adversaries and Fault Models

This dissertation examines the security and reliability of multiparty protocols under various fault models. The types of allowable faults are described below. The definitions for the multiparty case also apply to the two-party scenario as a special case.

Failures are modelled by an adversary which "chooses" to substitute new messages for the messages otherwise computed according to the protocol. This "choice" is simply an interpretation of the string output by the adversary, which is simply a player that computes an output string based on some transition function and its current state; the description of a protocol execution states how that output string is interpreted as a request to corrupt individual participants.

A *static* adversary must choose the subset it will corrupt before the protocol is executed. A *dynamic* adversary may choose at the end of a round the set of new processors it will corrupt for the next round. In either case, the adversary may be allowed to *rush* messages, that is, to wait for nonfaulty processors to send their messages, see the messages to which it has access, and then specify the messages of the corrupted processors for that round. Finally, a *strongly dynamic* adversary can rush all messages, examine them, and decide to corrupt machines of its choice before the messages are sent; the set of corrupted machines may encompass every player in the network over the course of the execution, though at any particular time the current coalition must be an allowable one. We shall consider dynamic but not strongly dynamic adversaries.

We measure security with respect to an *adversary class*, namely a set of adversaries. A typical example is the class of all polynomial-time Turing machines. A more powerful class allows unlimited computing time, though still requiring messages to be recursive functions of the inputs. Another powerful, non-uniform adversary class is the set of all polynomial-size circuit families (equivalently, polynomial-time Turing machines that take advice). The most powerful class, corresponding to "information-theoretic" security, allows the adversary to be a general player, namely to have an arbitrary transition function with an arbitrary set of states.

An important parameter of the adversary is the number $t$ of machines it is allowed to corrupt. The three ranges of primary importance are $t < n/3$, $n/3 \leq t < n/2$, and $n/2 \leq t \leq n - 1$. Different levels of fault-tolerance and security are achievable for each of these ranges.

In addition to specifying the manner in which the adversary can choose processors to corrupt, there are various sets of restrictions on the types of messages it is allowed to substitute. In order of increasing power, the types of faults are as follows.

- **Passive:** The adversary cannot substitute different messages for any of the original, uncorrupted machines. It does, however, have access to the tapes and states of the machines it "corrupts." This model is sometimes called the *gossip* model.

- **Fail-Stop:** The corrupted processors cannot write improper messages (*i.e.* different from the protocol specifications), but may halt at some round, sending only null ($\Lambda$) messages thereafter.

- **Omission:** The corrupted processors cannot write improper messages but may omit some of them periodically (replacing them by $\Lambda$). This model is useful in examining faulty communication lines, where occasional messages are lost.

- **Byzantine:** The adversary may compute any message of its choice (with restrictions, of course, if the adversary is computationally bounded), whether proper or improper according to the protocol, and it may omit messages as well.

A protocol attacked by an adversary is formally nothing more than a protocol with $n + 1$ players, each evaluating a probabilistic transition function and receiving and producing messages. When a specific adversary is concerned, the protocol is denoted $[A, \Pi]$; when an adversary may be chosen from some adversary class $\mathcal{A}$, the set of resulting protocols is denoted $<A, \Pi>$.

## 2.5.1 Passive Adversaries

A *fault class* $\mathcal{T}$ is a collection of allowable coalitions, namely a collection of subsets of $[n]$. We require that if $T \in \mathcal{T}$ and $U \subseteq T$, then $U \in \mathcal{T}$. The standard fault class is the *t-fault class*, $\mathcal{T} = \{T \subseteq [n] \mid |T| \leq t\}$, allowing any set $T$ of size $t$ or less to be corrupted.

Formally, a passive adversary is a tuple $(Q_A, Z_A, q_A^0, \delta_A, T)$, where $Q_A$ is a set of states, $q_A^0$ is a function mapping auxiliary inputs in $Z_A$ to initial states in $Q_A$, $\delta_A$ is a transition function to be described shortly, and $T$ is a function mapping $Q_A$ to $[n]$, describing the current coalition that the adversary chooses to corrupt. The passive adversary does not generate messages but it does choose a new state and new coalition based on its current state and the view of the coalition it currently corrupts:

$$\delta_A : Q_A \times \tilde{V} \to \mathtt{dist}(Q_A).$$

A passive adversary is *static* if $T$ is constant, and thus fixed before the protocol. A passive adversary is *dynamic* if $T$ varies with the state of the adversary. Note that if the number $n$ of players is fixed, then there is no essential difference between static and dynamic adversaries. There are a finite number, $\binom{n}{t}$ of fixed coalitions; the chance of selecting any one of them at random is a constant $c_n = 1/\binom{n}{t}$. Any dynamic adversary chooses some coalition $T$ with probability at least $c_n$. Thus there is some static adversary that starts with coalition $T$ and therefore has a constant fraction ($c_n$) of the probability of the adversary to corrupt the protocol successfully. We have not yet presented definitions of security, but it turns out that constant factors do not matter; essentially, there is a static adversary having the same power as the dynamic one. If $n$ grows, then the probability of a particular coalition $T$ vanishes, and there is a difference between static and dynamic.

A passive $\mathcal{T}$-adversary for a fault class $\mathcal{T}$ is a passive adversary for which

(1) for all $q_A \in Q_A$, $T(q_A) \in \mathcal{T}$, and

(2) it always chooses a new coalition that is a superset of the current one (*i.e.*, for all $q_A \in Q_A$, for all partial views $\mathtt{view}^r_{T(q_A)}$, for all states $q'_A$ having nonzero probability weight in $\delta_A(q_A, \mathtt{view}^r_{T(q_A)})$, $T(q_A) \subseteq T(q'_A)$).

Since we shall be concerned with fault classes containing all coalitions of size $t$ or less, we refer simply to a passive $t$-adversary.

Figure 2.2 shows how a $R$-round protocol is executed synchronously in the presence of a passive adversary. The reliable execution of a protocol defines a view seen by a subset of players; here, it is straightforward to include the state of the adversary and the messages it sees in the definitions of views and histories. For clarity, we abuse notation for the final computation, omitting mention of the last messages generated by the transition function $\delta$ because they are ignored.

### 2.5.2 Byzantine Adversaries

A Byzantine or malicious adversary $(Q_A, Z_A, q^0_A, \delta_A, T)$, has the added capability of overwriting the messages sent by players in the coalition it has corrupted. We thus extend its transition function to allow it to generate malicious messages:

$$\delta_A : Q_A \times \tilde{V} \to \mathtt{dist}(Q_A \times H).$$

As before, a Byzantine adversary is static if $T$ is a constant function and dynamic if $T$ depends on the state. A Byzantine $t$-adversary is allowed the fault class $\mathcal{T} = \{T \mid |T| \leq t\}$.

The most severe sort of Byzantine adversary can *rush* messages, meaning it receives the messages sent from reliable players before other messages are delivered, allowing it to choose a larger coalition to corrupt and to choose new messages depending on the good messages it has seen. Rather than consider an assortment of transition functions, one of which is used to generate new adversarial messages and others which determine new choices of coalitions after seeing rushed messages, we simply apply the same transition function $\delta_A$ after each set of rushed messages is received, allowing the adversary to change its state and select new players to corrupt. Formally, the messages generated by the transition function are ignored, until the

---

**PASSIVE EXECUTION**

**For** $i = 1..n$ **do**
$\qquad q_i \leftarrow q_i^0(x_i, a_i, k)$.
Set
$\qquad q_A \leftarrow q_A^0(a_A, n, m, k)$.

**For** $r = 1..R$ **do**
$\qquad$ **For** $i = 1..n$ **do** $\qquad$ /* each computes and sends messages */
$\qquad\qquad (q_i, \mathtt{mesg}^{out}(i, [n], r)) \leftarrow \delta_i(q_i, \mathtt{mesg}([n], i, r-1))$
$\qquad\qquad \mathtt{mesg}(i, [n], r) \leftarrow \mathtt{send}(\mathtt{mesg}^{out}(i, [n], r), \mathcal{C}_1 \cup \cdots \cup \mathcal{C}_n)$
$\qquad$ Set $\qquad\qquad$ /* adversary computes */
$\qquad\qquad q_A \leftarrow \delta_A(q_A, \mathtt{view}_{T(q_A)}^{r-1})$.

**For** $i = 1..n$ **do** $\qquad\qquad$ /* final computation */
$\qquad q_i^f \leftarrow \delta_i(q_i, \mathtt{mesg}([n], i, R))$
Set
$\qquad q_A^f \leftarrow \delta_A(q_A, \mathtt{view}_{T(q_A)}^R)$.

---

Figure 2.2: Execution of protocol $<A, \Pi>$ with a passive adversary $A$.

point at which the adversary stops enlarging the coalition it has chosen to corrupt. At that point, whatever messages the adversary generated are delivered to the appropriate recipients *via the channels from the corrupted players*, and the still-unsent messages between reliable players are also delivered.

The steps involved in an execution of the protocol with a rushing, Byzantine adversary on inputs $\vec{x}$, auxiliary inputs $\vec{a}$, and adversary auxiliary input $a_A$ are shown in Figure 2.3. For clarity of proofs, we sometimes regard the **Repeat** step (2.1.2) as a **For** loop that is repeated $t$ times; the adversary can increase the coalition no more than $t$ times, of course. The state of player $i$ or the adversary at round $r$ is referred to as $q_i^r$ or as $q_A^R$, respectively. If we wish to dissect the execution even further, examining the states within the **Repeat** loop, we refer to the $\rho^{th}$ repetition of the repeat loop using the notation $(r, \rho)$. Hence we may speak of $q_A^{(r, \rho)}$ or of $q_A^r = q_A^{(r,0)}$.

We shall have reason to speak not simply of the specific states and messages occurring during the protocol but of the distributions on those states. Let

$$\begin{aligned}
\mathtt{Vnames} \quad = \quad & \{x_1, \ldots, x_n, a_1, \ldots, a_n, a_A, q_1, \ldots, q_n, q_A, \\
& \mathtt{mesg}^{out}(1, 1), \ldots, \mathtt{mesg}^{out}(n, n), \mathtt{mesg}(1, 1), \ldots, \mathtt{mesg}(n, n), \\
& y_1, \ldots, y_n, y_A\}.
\end{aligned}$$

(Semantically, this is a set of "labels" for random variables.) The distribution on, say, the state $q_i^r$ at round $r$ will then be described by $\mathtt{V}(q_i, r)$. In general, the distribution on the state or message $v \in \mathtt{Vnames}$ at round $r$ is described by $\mathtt{V}(v, r)$. Each distribution $V(v, r)$ is a probabilistic function of the results of earlier samples as specified by Figure 2.3. For example, $V(q_i, r)$ is a probabilistic function (namely, the transition function $\delta_i$) of $V(q_i, r-1), V(\texttt{mesg}(1, i), r-1), V(\texttt{mesg}(2, i), r-1), \ldots, V(\texttt{mesg}(n, i), r-1)$. Or, for example, the variable $V(\texttt{mesg}(1, 1), r)$ is in fact a probabilistic function (namely, the channel function) $V(\texttt{mesg}^{out}(1, 1), r)$. The "results" or "outputs" of a distributed computation are described by $Y_1 = V(y_1, R), \ldots, Y_n = V(y_n, R), Y_A = V(y_A, R)$. Without loss of generality, the random variable $V(y_A, R)$ is "completely dependent" on the variables $V(q_A, r)$ and $V(\texttt{mesg}(i, j), r)$ for $i, j \in T_{V(q_A, R)}$ and $1 \leq r \leq R$, in the sense that the output of the adversary specifies every state and message it has seen.

Thus, the execution of a protocol is simply a sample taken from a set of inter-dependent random variables $V(v, r)$. The text of this chapter shows how to generate this joint distribution by sampling probabilistic functions in a specified order. Our proofs will fundamentally show that two probabilistic functions are the same, where each is defined by a different sequence of application of probabilistic functions.

## 2.6   Outputs and Output Distributions

At the end of a general protocol the adversary and each player writes an output string $Y_A = \mathcal{O}_A(q_A^f)$ or $Y_i = \mathcal{O}_i(q_i^f)$ (respectively) on an output tape. A corrupted player's tape contains $\Lambda$.

Let adversary $A$ have auxiliary input $a$. An execution of a protocol $\Pi(n, m)$ on inputs $\vec{x} = (x_1, \ldots, x_n)$, auxiliary inputs $\vec{a} = (a_1, \ldots, a_n)$, and security parameter $k$ induces a *distribution* on $(\Sigma^\star)^{2n+2}$, that is, on the outputs and views of $A$ and of the players. The distribution is denoted:

$$[A, \Pi]^{hist}(n, m)(\vec{x} \circ \vec{a} \circ a, k) = (Y_A, Y_1, Y_2, \ldots, Y_n, \texttt{view}_A, \texttt{view}_1, \texttt{view}_2, \ldots, \texttt{view}_n).$$

For fixed $n$ and $m$, parametrizing over $z = \vec{x} \circ \vec{a} \circ a$ and security parameter $k$ represents an ensemble, $[A, \Pi]^{hist}(n, m)$. A universal protocol that computes a particular function $F$ may include a circuit description $C_F$ in the $z$ parameter. For readability, we occasionally write $\Pi(\vec{x} \circ \vec{a} \circ a, k)$ when $n$ and $m$ are understood in the context. The family of ensembles when $n$ and $m$ vary is written $[A, \Pi]^{hist} = \{[A, \Pi]^{hist}(n, m)\}_{n,m \in \mathbf{N}}$, and is induced by the family of protocols $\Pi = \{\Pi(n, m)\}_{n,m \in \mathbf{N}}$.

When analyzed separately, the issues of privacy and correctness examine the views apart from the outputs. We denote the output (which, without loss of generality, includes the view) of the adversary generated in a protocol by:

$$[A, \Pi]^{Y_A}(\vec{x} \circ \vec{a} \circ a, k) = Y_A.$$

We use a similar notation for the outputs of the players:

$$[A, \Pi]^{\vec{Y}}(\vec{x} \circ \vec{a} \circ a, k) = (Y_1, \ldots, Y_n).$$

Privacy concerns $[A, \Pi]^{Y_A}$, whereas correctness concerns $[A, \Pi]^{\vec{Y}}$. The ensemble specifying the view of the adversary and the outputs of the players (not their views) is:

$$[A, \Pi](\vec{x} \circ \vec{a} \circ a, k) = (Y_A, Y_1, \ldots, Y_n).$$

We shall see it is advantageous to consider $[A, \Pi]$ as a whole, not to break it down into components describing privacy and correctness.

## 2.7   The Function to Compute

For full generality, we wish to consider families of probabilistic functions mapping strings to strings (equivalently, integers to integers, encoded in a natural fashion). Let $F = \{F^{n,m}\}_{n,m \in \mathbf{N}}$ be a family of probabilistic functions with $n$ inputs of length $m$ and $n$ outputs of length $m$ :

$$F^{n,m} : X_1 \times X_2 \times \cdots \times X_n \to Y_1 \times Y_2 \times \cdots \times Y_n$$

where each $X_i, Y_i \subseteq \{0, 1\}^m$.

Computing probabilistic functions allows us to accomplish a wide range of tasks. It is often the case that a given set of inputs is not mapped in a 1-1 manner to a set of outputs and thus is not described by a function *per se*. A coin toss, useful for tasks such as leader election and symmetry breaking, is a primary example.

Lest our analysis become too complicated, however, we consider distributions described by probabilistic circuits, namely distributions induced by setting certain bits of a Boolean circuit uniformly at random. This covers all probabilistic Turing machine computations and is a natural case to consider.

Let $\mathcal{F}^{det,poly}$ be the set of all deterministic function families $\hat{F} = \{\hat{F}^{n,m}\}_{n,m \in \mathbf{N}}$ satisfying

$$\hat{F}^{n,m} : \{0, 1\}^{nm+\rho(n,m)} \to \{0, 1\}^{nm}$$

where $\rho(n, m) = O((nm)^c)$ for some $c$. Each $F^{n,m}$ can be described as a circuit with $nm$ inputs and a polynomial number of supplementary inputs, and $nm$ outputs (or, if one prefers, as $nm$ such circuits each having a one-bit output).

Let $\mathcal{F}$ be the set of all probabilistic function families $F = \{F^{n,m}\}$ such that there exists an $\hat{F} \in \mathcal{F}^{det,poly}$ satisfying

$$F^{n,m}(\vec{x}) = \{\vec{r} \leftarrow \mathbf{uniform}(\{0, 1\}^{\rho(n,m)}) : \hat{F}^{n,m}(\vec{x}, \vec{r})\}$$

Denote the restriction of $F$ to its $i^{th}$ output bit by $F_i(x_1, \ldots, x_n)$.

A protocol for $F \in \mathcal{F}$ should compute a vector $(y_1, y_2, \ldots, y_n)$ selected according to the distribution $F(x_1, x_2, \ldots, x_n)$, and should provide player $i$ with the correct value $y_i$.

We make two simplifying observations based on the ability of each player to generate uniformly random bits. First of all, we may restrict our attention to deterministic functions by arranging that each player $i$ supply, as part of its input, a uniformly random sequence of $\rho(n,m)$ bits, $\vec{r}_i = (r_{i1}, \ldots, r_{i,\rho(n,m)})$. It is easy to see that as long as one player supplies uniformly random bits, it suffices to compute the deterministic function $\check{F}((x_1, \vec{r}_1), \ldots, (x_n, \vec{r}_n))$ defined as $\hat{F}(x_1, \ldots, x_n, R_1, \ldots, R_{\rho(n,m)})$ where $R_j = r_{1j} \oplus r_{2j} \oplus \cdots \oplus r_{nj}$.

Secondly, it suffices to consider functions whose single result is revealed in its entirety to all players. Observe that if each player provides a mask, namely a sequence of random bits $r_i$, it suffices to compute $(y_1 \oplus r_1, \ldots, y_n \oplus r_n)$ and reveal the entire vector to all players. Unless the adversary specifically corrupts a given player $i$, it gains no information about the result $y_i$ from the publicized value $y_i \oplus r_i$.

Thus, without loss of generality our protocol descriptions consider each $F$ to be a family of deterministic functions, with one $m$-bit output or $n$ different $m$-bit outputs as convenient. A general protocol compiler for families in $\mathcal{F}^{det,poly}$ is also one for $\mathcal{F}$.

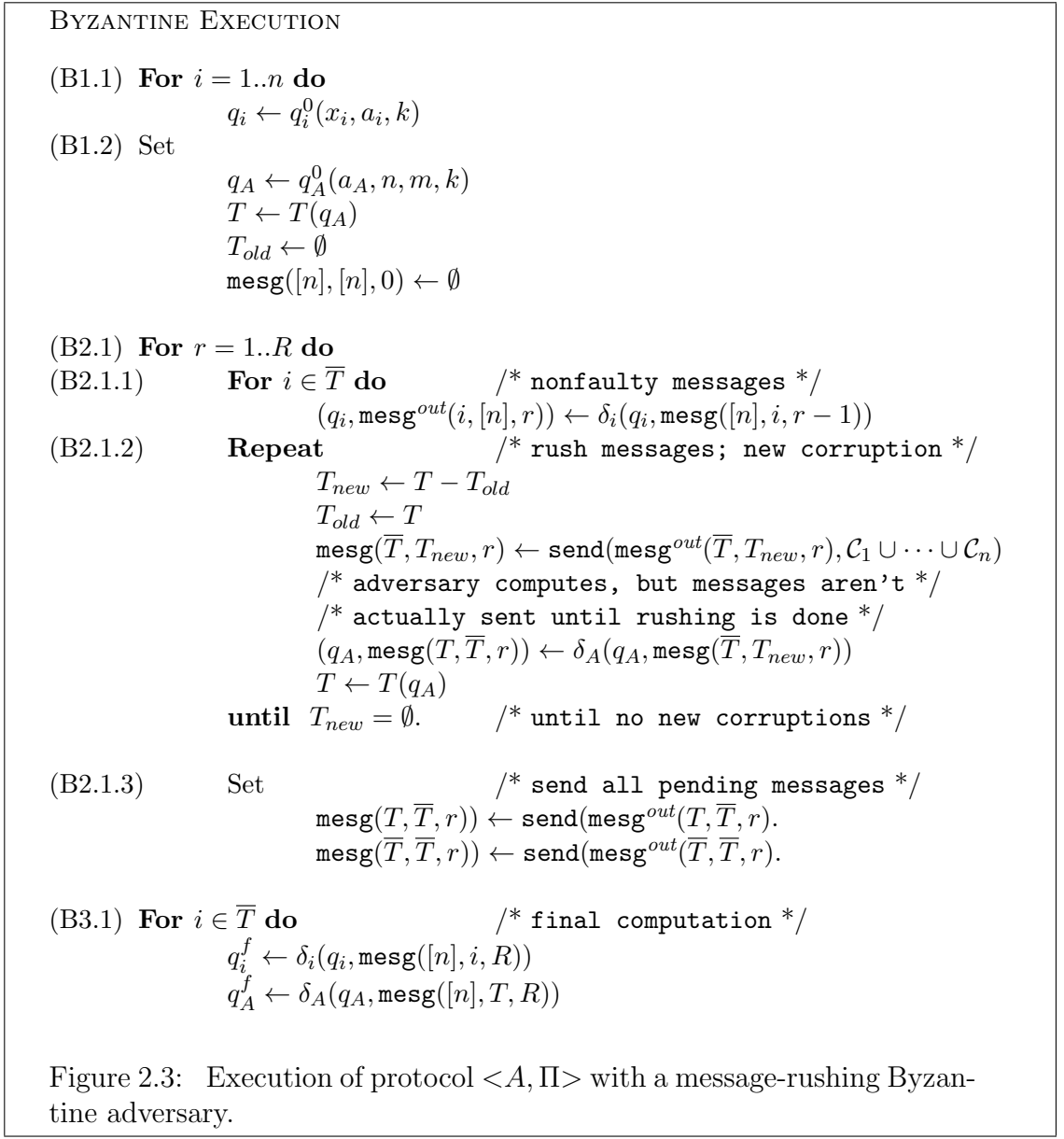

Byzantine Execution

**(B1.1) For** $i = 1..n$ **do**

$$q_i \leftarrow q_i^0(x_i, a_i, k)$$

**(B1.2)** Set

$$q_A \leftarrow q_A^0(a_A, n, m, k)$$
$$T \leftarrow T(q_A)$$
$$T_{old} \leftarrow \emptyset$$
$$\mathtt{mesg}([n], [n], 0) \leftarrow \emptyset$$

**(B2.1) For** $r = 1..R$ **do**

**(B2.1.1)**     **For** $i \in \overline{T}$ **do**      /* nonfaulty messages */

$$(q_i, \mathtt{mesg}^{out}(i, [n], r)) \leftarrow \delta_i(q_i, \mathtt{mesg}([n], i, r-1))$$

**(B2.1.2)**     **Repeat**      /* rush messages; new corruption */

$$T_{new} \leftarrow T - T_{old}$$
$$T_{old} \leftarrow T$$
$$\mathtt{mesg}(\overline{T}, T_{new}, r) \leftarrow \mathtt{send}(\mathtt{mesg}^{out}(\overline{T}, T_{new}, r), \mathcal{C}_1 \cup \cdots \cup \mathcal{C}_n)$$

/* adversary computes, but messages aren't */
/* actually sent until rushing is done */

$$(q_A, \mathtt{mesg}(T, \overline{T}, r)) \leftarrow \delta_A(q_A, \mathtt{mesg}(\overline{T}, T_{new}, r))$$
$$T \leftarrow T(q_A)$$

    **until** $T_{new} = \emptyset$.      /* until no new corruptions */

**(B2.1.3)**     Set      /* send all pending messages */

$$\mathtt{mesg}(T, \overline{T}, r) \leftarrow \mathtt{send}(\mathtt{mesg}^{out}(T, \overline{T}, r).$$
$$\mathtt{mesg}(\overline{T}, \overline{T}, r) \leftarrow \mathtt{send}(\mathtt{mesg}^{out}(\overline{T}, \overline{T}, r).$$

**(B3.1) For** $i \in \overline{T}$ **do**      /* final computation */

$$q_i^f \leftarrow \delta_i(q_i, \mathtt{mesg}([n], i, R))$$
$$q_A^f \leftarrow \delta_A(q_A, \mathtt{mesg}([n], T, R))$$

Figure 2.3: Execution of protocol $<A, \Pi>$ with a message-rushing Byzantine adversary.

# Chapter 3

# Formal Definitions for Reliability and Security

> To think is to forget differences, generalize, make abstractions. In the teeming world of Funes, there were only details, almost immediate in their presence.
>
> — Jorges Luis Borges, *Funes the Memorious*

The most striking problem with current definitions of security and reliability, apart from their scarcity, is the *ad hoc* and incremental nature in which they have developed. Correctness and privacy are naturally the most important goals of a reliable, secure computation. Their formal specifications, however, are more subtle than their intuitive simplicity suggests. Furthermore, even though protocols for distributed system security have been designed with such intuitively simple properties in mind, other equally desirable security properties have since arisen, requiring each protocol to be reconsidered — if not redesigned — in light of these new properties. Definitions of security have thus been clouded not only be unforeseen subtleties in the precise formulation of very simple properties but by the confusing and disharmonious set of overly detailed formalizations tailored differently to each property.

We propose a new and concise definition of a property we call *resilience* that captures at a single blow every desirable property of a distributed protocol. Not only is this new definition simple and broad enough to avoid the pitfalls of previous approaches, it appears to capture *a priori* all the natural properties one might imagine. With such a unifying definition, we need design protocols and prove their resilience only once; new and unforeseen properties are simply new, previously unnoticed aspects of the single property of resilience.

In this chapter, we take a brief look at previous, *ad hoc* approaches to formal definitions and discuss their difficulties and insufficiencies. We then introduce the most important tool of this work: a means to compare the resilience — namely the security and reliability — of two arbitrary protocols. By defining a specific, *ideal* protocol, we provide a standard by which to measure the resilience of *any* protocol. In an ideal protocol, a trusted and incorruptible host receives all the inputs and returns the correct outputs. Though such a situation cannot be guaranteed in reality, its robustness is the absolute standard we seek. Our techniques and standards provide not only a concise way to define and to understand security and reliability but also, as we shall see in Chapter 4, a concise and formal means to consider protocol composition and modularity.

Our point is this: a protocol is not meant to compute $F$ while satisfying a list of desirable properties; rather, *it must achieve the same results, in a rigorous sense (see §3.4), as an ideal protocol in which a trusted host performs the computation.* The ability to compare the results of two protocols is essential to showing such a case is true. This chapter addresses these two new and crucial observations.

## 3.1 Where *ad hoc* Definitions Fail

Privacy and correctness are the most obvious properties required by secure and reliable computations. They are intuitively easy to understand: a protocol reveals no unnecessary information if the view of an adversary can be generated from the bounded information to which it is entitled (inputs and auxiliary inputs of corrupted players). A protocol is correct if the output of each player is in fact the function $F(x_1, \ldots, x_n)$ applied to the inputs.

At closer inspection, subtle problems appear. Definitions of privacy often concentrate too closely on the privacy of the inputs and make little reference to other information that may be leaked. ([Hab88] is a notable exception; [AFK89] and [BF90] introduce a specific measure of information that is hidden or leaked for the related problem of instance-hiding schemes.) It may not be the case that, running one protocol after another, the second protocol preserves the privacy of the information in the first. Each protocol might hide its own inputs but leak all the secret information of another protocol if the definitions are not made carefully. Since the protocols in the literature usually happen to satisfy the stronger requirement that no information except $F(x_1, \ldots, x_n)$ is leaked, this issue is overlooked. In order to compose protocols, however, a careful measure of the information leaked by each subprotocol is necessary.

Even the definition of as obvious an idea as an *input* to a protocol requires great care. Defining inputs and outputs is a far more sensitive task than the established study of single-processor computations would suggest. If a player is *supplied* with an input, what happens if it behaves absolutely properly but as though it received a different input? Does one define correctness with respect to behavior or with respect to some "actual" input? What does "actual" mean in this case?

Often, rather than delve into the subtleties, researchers make particular assumptions (*e.g.*, encryptions of every input are supplied to each party, or every input is "secretly shared") in order to finesse the issue. The notion of *committal* to an input plays an important part: after an initial stage (either all processors are supplied with immutable encryptions of other processors' inputs, or the inputs are validly shared), correctness can be defined with respect to the committed inputs. While such approaches are interesting as particular methods to accomplish security goals, they are too specific to merit approval as general definitions of security.

Definitions for the related problem of Byzantine Agreement do not suffice, since the goal of that problem is simply to agree on a common value, a process that is fairly insensitive to individual variations in a fraction of the input values, as opposed to determining a result that may be extremely sensitive to individual inputs. We must find a broader means to understand and define correctness without resorting to such specific aspects of the process as committal, which, although it turns out to be a necessary implication of protocol resilience, is not the primary focus.

Hindering us in our search for clean definitions is the problem of when to stop looking for new properties. Privacy and correctness were once thought sufficient. Other important properties have since become evident. For example, what if, in a secret ballot, one voter were able to cast a vote opposite to that of another, through some clever and malicious manipulation of messages in an intricate protocol, even that corrupt player were not able to learn the vote that it cast? A two-thirds majority could always be prevented by malicious coalitions. Or, for example, what if the system must generate a random bit upon which some decision must be based? If malicious players can choose their inputs depending on those of reliable players, then using the parity of random bits supplied by each player will fail. The property of *independence of input selection* is immediately seen as desirable, even though it is orthogonal to the properties of correctness and privacy. Beaver ([Bea88], see technical report) gave a broad definition of independence with respect to a broad (*i.e.* nonspecific and general) committal function defined on transcripts of protocol executions, but this early approach is unsatisfying for the reasons noted above. Notice that when players are computationally unbounded, independence of inputs can be neatly formulated via independence of random variables describing inputs. When players and adversaries are computationally bounded, however, subtle and difficult points arise: for example, a faulty player may use the *encrypted value* it sees of another player's input as its own input, in which case the two inputs are information-theoretically quite dependent, but formulating that they are independent in some computational task is intricate and unclean.

Another example, occurring primarily in the domain of protocols tolerating extremely high fault rates, is that of *fairness*. Reliable players must not be prevented from learning their outputs of the protocol if corrupted players are able to learn theirs. That is, even if an adversary is able to bring the whole computation to a halt — which, fortunately, is not an issue in protocols with a faulty minority — then it should not enjoy the fruits of the computation while denying them to reliable participants. In order to maintain parity in the knowledge gained by adversary and by reliable players, methods for gradual disclosure of the results are employed [LMR83, Cle86, Yao86, GHY87]. The two primary definitions for fairness are related to the algorithmic solutions provided by their proponents. Yao, and later Galil, Haber, and Yung, allow the reliable players to run a recovery algorithm based on the program used by the adversary, if a majority of players have become faulty [Yao86, GHY87]. Beaver and Goldwasser propose a stronger definition in which the reliable players do not have access to the program of the adversary, thereby allowing the adversary itself to depend on the programs of the reliable players. Further discussion can be found in Chapter 9, where protocols tolerating a faulty majority are presented, and in §3.6, where we demonstrate how our unified definition implies the current set of desired properties.

Our definitions capture all of these properties in a concise and simple manner by avoiding the *ad hoc*, divide-and-conquer approach. The properties we have discussed are related to each other: they are each aspects of an ideal protocol. The ideal protocol and the notion of relative resilience bind them together.

## 3.2 Relative Resilience: Simulators and Interfaces

The key idea behind our approach is a means to compare one protocol to another in terms of their security and reliability. We call the combination of security and reliability, *resilience*. We measure not only the *information* gained by an adversary during an execution of a protocol, but the *influence* it has on the outputs.

This essential idea, unnoticed, was the source of many of the inherent difficulties in previous approaches: even in an *ideal* protocol, with a trusted host who computes results accurately, an adversary has *some* influence over the outputs. An adversary who is not able to attack the trusted host is nevertheless able to choose inputs that corrupted players will send to the host, thereby having some limited but unavoidable effect on the computation. Thus, we must examine an attack not only with respect to what the adversary learns (ideally, only the inputs and outputs of corrupted players), but with respect to how the adversary affects the computation (ideally, only through its choice of inputs for corrupted players).

Before delving into the nature of the *ideal* protocol that will become our standard, we first investigate how to compare the resilience of two arbitrary protocols. The notion of privacy introduced by Goldwasser, Micali, and Rivest [GMR89] in the realm of zero-knowledge proof systems and later applied to multiparty protocols (*cf.* [GMW87, Hab88]) provides a springboard for the ultimate definition of resilience.

As described in §2.2.4, zero-knowledge proof systems make use of a *simulator* to demonstrate that the conversation seen by a corrupted verifier is in fact generable solely from the theorem, "$x \in L$." We phrase it in a more suggestive manner: the simulator demonstrates that the conversation is generable solely from the information (*i.e.* "$x \in L$") that it would obtain in an ideal situation (given a trusted oracle, *i.e.* prover). In any general interactive protocol execution, we would like to ensure that the adversary can simulate its portion of the history of a protocol using only the information that it is entitled to learn. Galil, Haber, and Yung [GHY89] utilize such a generalization and make use of a *fault oracle* that supplies the bounded information (inputs and outputs of corrupted players) to which the adversary is entitled. A simulator must, with that information, generate an accurate view of the real protocol execution. In computationally bounded models, where the verifier has limited resources, the simulator must also use such bounded resources, in order to demonstrate not just that no extra *information* is leaked but that no results that are *computationally infeasible* to generate are leaked.

A related idea that avoids the computational orientation of simulation says that the distribution on views seen by the adversary must be *independent* of the reliable player's information, or more generally that the distribution depends *only* on the final output and the inputs of the corrupted players. This sort of approach forms the basis for the instance-hiding schemes of [AFK89, BF90]. When the players are not computationally bounded, defining privacy according to independence of variables is often a cleaner approach (see Chapters 8,11,12). In general, if a distribution is independent of certain variables, then it can be simulated without the "knowledge" of the values of those variables (though, of course, there is no *a priori* guarantee that simulating the distribution is efficient). With this in mind, we shall adopt the simulation approach, since it also has the advantage of covering computationally-bounded models.

But such definitions for privacy are not sufficient to analyze the security and reliability of interactive multiparty protocols. Simulation is essentially a passive approach: give the simulator information, and let it create a view for the adversary. The *influence* of an adversary on others is not taken into account.

Let us consider two interactive multiparty protocols, $\alpha$ and $\beta$. Each has an associated class of allowable adversaries, $\mathcal{A}_\alpha$ and $\mathcal{A}_\beta$. To compare the resilience of protocol $\alpha$ against an adversary $A \in \mathcal{A}_\alpha$ to the resilience of protocol $\beta$, we should like to allow $A$ to wreak havoc on protocol $\beta$. Unfortunately, $\alpha$ and $\beta$ may be radically different protocols. One might have many more players than the other, one might disallow certain players from being corrupted, one might be written in C while the other is written in FORTRAN, *etc.* We cannot simply run protocol $\beta$ with adversary $A$.

Instead, we surround $A$ by an *interface*, $\mathcal{S}$. The interface $\mathcal{S}$ creates an environment for $A$ so that $A$ will believe it is participating in protocol $\alpha$. On the other hand, $\mathcal{S}$ itself is allowed to wreak havoc on protocol $\beta$. In other words, the combination $\mathcal{S}(A, \cdot)$ with $A$ built into $\mathcal{S}$ is used itself as the adversary to an execution of protocol $\beta$. (The notation indicates that $\mathcal{S}$ has two communication lines, one of which is used to communicate with $A$, and the other of which is used as its adversarial input/output line in a protocol execution.) The combination $\mathcal{S}(A, \cdot)$ must be a permissible adversary in $\mathcal{A}_\beta$.[1]

Consider, for example, the special case where protocol $\beta$ is the particular and simple protocol in which an incorruptible host sends one message (containing "$x \in L$") to the corruptible player. An adversary to protocol $\beta$ can obtain only this information (and has no influence on the results, in this particular simple case). An interface $\mathcal{S}$ receives this message and must present a proper environment for $A$. In this manner, the interaction corresponds to the special case of zero-knowledge proof systems (see §3.4.2), and the interface is used only as a simulator.

In actuality, simulation is only half of its job. We shall later discuss how the interface covers all aspects of the definition of zero-knowledge; first, let us examine

---

[1] Technically, there must be a machine in $\mathcal{A}_\beta$ with an identical input/output behavior as the single tape machine $\mathcal{S}(A, \cdot)$.

how it captures all aspects of multiparty security, including privacy and correctness.

**Definition 3.1 (Interface)** *An **interface** $\mathcal{S}(\cdot, \cdot)$ is a machine (interactive Turing machine or general player) with two communication lines; the first is called an **environment simulation** line, and the second is called an **adversarial** line. If $\alpha$ admits adversary class $\mathcal{A}_\alpha$ and $\beta$ admits adversary class $\mathcal{A}_\beta$, we say $\mathcal{S}$ is an **interface from** $\alpha$ **to** $\beta$ if for every $A \in \mathcal{A}_\alpha$, $\mathcal{S}(A, \cdot) \in \mathcal{A}_\beta$.*

We denote an interface by $\mathcal{S}(\cdot, \cdot)$. When hooked up to an adversary $A$ (with auxiliary input $a$) which itself is simply an interactive machine having a single communication line, we consider the combination $\mathcal{S}(A(a), \cdot)$ as an adversary of its own right. As a unit, it has a single communication tape, namely the adversarial tape. We treat $\mathcal{S}(A(a), \cdot)$ as an adversary to $\beta$ simply by using the corruption requests and corrupt messages $\mathcal{S}$ writes on its adversarial line and by supplying $\mathcal{S}$ with the requested messages and information held by corrupted players in $\beta$, as would normally happen in the execution of protocol $\beta$ with an adversary.

We should like to formalize the intuitive statement that adversary $A$ can wreak no more havoc on protocol $\alpha$ than it could on protocol $\beta$ — given an interface to translate its requests for corruptions and specifications of corrupted messages. If this is the case, we shall say that $\alpha$ is *as resilient as $\beta$*.

In each protocol, every reliable player ends up in a state according to some distribution. We should like that the adversary's influence on the state of reliable players is the same, whether it attacks $\alpha$ or with help from an interface it attacks $\beta$. The adversary itself ends up in some state when it attacks protocol $\alpha$; the distribution on its final state should be the same whether it attacks $\alpha$ or is fed the environment simulated by $\mathcal{S}$. (Recall that an *attack* of an adversary is just a sequence of strings including requests for states of newly corrupted players, messages to replace those of corrupted players, the states of those players, and messages sent by reliable players to corrupted players. The states and reliable messages are supplied either by the formal execution of protocol $\alpha$ or by the computations of interface $\mathcal{S}$.) This says that the adversary gains no more information in $\alpha$ than it would in protocol $\beta$.

Formally, let us consider the distributions $[A, \alpha](n, m)(\vec{x} \circ \vec{a} \circ a, k)$ and $[A, \mathcal{S}, \beta](n, m)(\vec{x} \circ \vec{a} \circ a, k)$. The former is induced by running protocol $\alpha$ with adversary $A(a)$ and the given inputs and auxiliary inputs. The latter is induced by running protocol $\beta$ with adversary $\mathcal{S}(A(a), \cdot)$ and the same inputs and auxiliary inputs. When $n$, $m$, $\vec{x}$, $\vec{a}$, and $a$ are allowed to vary, we have two protocols that each induce an ensemble, $[A, \alpha]$ and $[A, \mathcal{S}, \beta]$, respectively. These two ensembles describe the final states of the players, reflecting the *influence* of the adversary, and the final state of the adversary, reflecting the *information* gained by the adversary.

**Definition 3.2 (Black-Box Auxiliary-Input Relative Resilience)** *A protocol $\alpha$ is as $(\mathcal{A}_\alpha, \mathcal{A}_\beta)$-resilient as protocol $\beta$, written*

$$\alpha \succeq_{(\mathcal{A}_\alpha, \mathcal{A}_\beta)} \beta,$$

*if there exists an interface $\mathcal{S}$ from $\alpha$ to $\beta$ such that for all adversaries $A \in \mathcal{A}_\alpha$,*

$$[A, \alpha] \asymp [A, \mathcal{S}, \beta]$$

The subscript $(\mathcal{A}_\alpha, \mathcal{A}_\beta)$ is omitted where clear from context. We say the protocols are *exponentially* ($\succeq^e$), *statistically* ($\succeq^s$), or *computationally* ($\overset{c}{\succeq}$) relatively resilient according to how indistinguishable the induced ensembles are. It is clear that the use of auxiliary inputs can be removed simply by omitting them from the definitions, if one wishes to consider strictly uniform computations. To avoid potential confusion, we remark that definition of indistinguishability already takes the inputs and auxiliary inputs into account: for *every* $z$, namely for every value of $n$, $m$, and $\vec{x} \circ \vec{a} \circ a$, the two sequences of distributions induced by $(\vec{x} \circ \vec{a} \circ a, k)$ must approach each other as $k$ gets large. A trivial modification of the definitions removes auxiliary inputs from consideration, which may be useful if one wishes to consider strictly uniform computations with uniform adversaries. Auxiliary inputs, as mentioned, are useful for other reasons, and we include them in our definition.

When $n$ and $m$ are fixed, we may compare two protocols in the same manner: $\alpha(n, m)$ is as resilient as $\beta(n, m)$, written $\alpha(n, m) \unrhd \beta(n, m)$, if there exists an interface $\mathcal{S}$ such that $[A, \alpha](n, m) \approx [A, \mathcal{S}, \beta](n, m)$. The technical distinction is that here we compare ensembles, whereas above we compare families of ensembles.

The study of zero-knowledge often considers a weaker form of simulation in which there need not be a universal simulator $\mathcal{S}$ but rather there need only be a specific simulator $\mathcal{S}_A$ for each adversary $A$. The simulator can depend on the internal structure of the adversary. For completeness, we give the analogous definition for resilience, considering an interface $\mathcal{S}_A$ that acts as an adversary to protocol $\beta$ in the same way $\mathcal{S}(A, \cdot)$ did previously. We shall not consider this definition further.

**Definition 3.3 (Weak Auxiliary-Input Relative Resilience)** *A protocol $\alpha$ is* **weakly as $(\mathcal{A}_\alpha, \mathcal{A}_\beta)$-resilient as** *protocol $\beta$, if for any adversary $A \in \mathcal{A}_\alpha$, there exists an interface $\mathcal{S}_A$ from $\alpha$ to $\beta$ such that*

$$[A, \alpha] \asymp [A, \mathcal{S}, \beta]$$

## 3.3   Ideal Interaction With a Trusted Host

Given a means to compare the resilience of protocols, we come now to the other important half of the approach we propound: *a secure and reliable protocol must achieve the same results as an* ideal *protocol in which a trusted host performs the computation.*

Chapter 2 describes the mechanics of a general protocol, whether generic or attacked by an adversary. A *real* protocol is one having a fault class consisting of arbitrary subsets of up to $t$ players, and having some arbitrary set of communication channels. For example, a real protocol may provide broadcast channels and private channels between each pair of players. The essential point is that no player need be incorruptible.

An *ideal* protocol, on the other hand, provides an absolutely reliable and secure trusted host, even though in the "real" world, it is neither prudent nor efficient to rely upon such an immensely vulnerable, centralized situation. Computation with a trusted host must nevertheless take into account the presence of an adversary. In the ideal case, however, the limited powers of the adversary are more clearly delineated than among the intricacies of a complex protocol, giving a convincing justification for claims of absolute security and reliability.

Figure 3.1 describes the *ideal protocol* for $F$. The participants are the usual $n$ players along with a central, trusted host, player $(n+1)$. Each player has a private communication line to player $(n+1)$, so we need not consider information leaked implicitly through interaction among the players.

---

Ideal-Protocol <ID($F$)>

(I1)    Each player $i$ ($1 \leq i \leq n$) starts with input and auxiliary input $(x_i, a_i)$. Player $(n+1)$, the trusted host, has no input. Each player $i$ sends $x_i$ to player $(n+1)$, for $1 \leq i \leq n$.

(I2)    Player $(n+1)$ sets $x_i^\star = \Lambda$ for any messages not falling in the domain $X_i$ and sets $x_i^\star$ to the message from player $i$ otherwise. It computes the sample $(y_1, \ldots, y_n) \leftarrow F(x_1^\star, \ldots, x_n^\star)$. (Without loss of generality assume that $F$ is defined on input $\Lambda$.) Player $(n+1)$ sends $y_i$ to player $i$ for $1 \leq i \leq n$.

Figure 3.1:    Ideal computation with a trusted host and reliable parties.

---

When an adversary is allowed to attack the ideal protocol, its powers are strictly limited to learning and influencing the inputs of players in the range $\{1, \ldots, n\}$; player $(n+1)$ is immune to corruption. That is, the *ideal-t-fault-class* $\mathcal{T}_{ideal}^t$ consists of all subsets of $\{1, \ldots, n\}$ of size $t$ or less. The *ideal-t-adversary class* $\mathcal{A}_{ideal}^t$ consists of all adversaries that only request coalitions in the ideal $t$-fault class. For the purposes of discussing complexity-based security, in which adversaries and players are assumed to perform probabilistic polynomial-time computations, we may restrict this class to computationally-bounded adversaries. In that case, the adversary must not only request coalitions strictly in the specified fault class, but it must be a polynomial time Turing machine as well.

The execution of the ideal protocol in the presence of a dynamic, Byzantine adversary is described informally in Figure 3.2. The formal specifications of the distributions and histories obtained in this interaction should be clear from the definitions given in Chapter 2. The interactions in the presence of a weaker adversary, such as a passive or static one, should also be clear. Notice that a dynamic adversary may base its choice of players to corrupt on information gained from previously corrupted players in an adaptive fashion. It may also choose to corrupt more players after $F$ has been computed.

The idea of committal to inputs corresponds to the end of round 1: each player has sent its input to the host, and the computation of $F$ has not yet begun. A special protocol that is useful in formally capturing ideas of commitment and privacy is the *ideal vacuous protocol*, $\mathtt{ID}(\emptyset)$, which is the ideal protocol in which each player must supply a 0 input; the trusted host returns a string of $n$ bits, each of which is 0 if the corresponding player supplied a 0 and is 1 otherwise. Essentially, no information besides the identities of the cheating parties is returned.

---

IDEAL-PROTOCOL-WITH-ADVERSARY $<A, \mathtt{ID}(F)>$

(IA1)

- Each player $i$ $(1 \leq i \leq n)$ starts with input and auxiliary input $(x_i, a_i)$. Player $(n+1)$, the trusted host, has no input.

- The adversary computes and chooses an allowable subset $T \subseteq \{1, \ldots, n\}$ of players to "corrupt." (The trusted host is *not* included in the class of allowable faults.) It obtains the inputs $(\vec{x}_T, \vec{a}_T)$ of those players, and nothing else. A dynamic adversary may repeat this step as often as it pleases (modulo resource bounds, and modulo the requirement that new coalitions contain old ones).

- The adversary chooses an alternate set of effective inputs $x_i'$ for the members of coalition $T$, and sends them to the trusted host.

- Every uncorrupted player sends its input $x_i$ to the trusted host.

(IA2)

- The trusted host sets $x_i^\star$ to be $\Lambda$ for each $i$ from which it received a message out of range, and otherwise sets $x_i^\star$ to be the message it received. The trusted host samples $(y_1, \ldots, y_n) \leftarrow F(x_1^\star, \ldots, x_n^\star)$, where $x_i^\star$ is either $x_i$, $\Lambda$, or a corrupted but possible input choice.

- The trusted host sends $y_i$ to player $i$ for each $i$. The adversary thereby receives outputs $\vec{y}_T$.

- The adversary may choose more players to corrupt, obtaining their inputs $(x_i, a_i)$ and their output values $y_i$, with the same restrictions on coalitions as before.

---

Figure 3.2: Ideal computation with a trusted host, dynamic malicious adversary, and reliable parties. This protocol requires two rounds.

## 3.4 Resilience: The Unified Theory

The groundwork is laid for the concise and precise definition of security and fault-tolerance for multiparty computations. Using the concept of *relative resilience* and using the ideal protocol as a standard:

**Definition 3.4 (Resilience)** *A protocol* $\Pi$ *with fault class* $\mathcal{T}$ *is* $\mathcal{T}$**-resilient leaking** $F$ *if*

$$\Pi \ \succeq_{(\mathcal{T}, \mathcal{A}^t_{ideal})} \ \mathtt{ID}(F).$$

We say *exponentially, statistically,* or *computationally* if $\succeq^e$, $\succeq^s$, or $\overset{.}{\succeq}$ holds, respectively.

### 3.4.1 Three Scenarios: A Summary

The diagram in Figure 3.3 illustrates the three scenarios of key importance. The first represents the ideal world, in which a trusted and reliable host is available and ensures that the adversary is truly restricted to gaining only the inputs and outputs of players of its choice. The third represents the real world, in which no player can be trusted but a protocol must be designed to perform the same computations as in the ideal world, correctly and privately. The second and intermediate scenario joins the two, modelling the interaction in an ideal protocol attacked by a real adversary that is assisted by an interface. Allowed to attack the ideal protocol as best it may, the information and influence of the adversary is clearly delineated in this central case. It connects the clear measurements of security and reliability in the ideal case to the less easily understood powers of the adversary in the real world.

### 3.4.2 Zero-Knowledge at One Blow

Resilience captures all aspects of zero-knowledge proof systems, not just privacy, in a concise way. Consider a zero-knowledge proof system as a two-party protocol in which any number of players may be corrupted. The *ideal* protocol provides a trusted host. The prover tells the trusted host, "$x \in L$," for some $x \in L$, or declines to participate by sending some other message. The host simply sends "$x \in L$" or "Cheating" to the verifier, accordingly.

It is clear that the most an adversary can do is to cause the prover detectably to cheat without convincing the verifier, or to gain at most the information that the verifier learns, which in this case amounts only to "$x \in L$." Separately, these possibilities correspond to soundness and to privacy. Resilience unifies them.

The ideal protocol, denoted $\mathtt{ID}(F_L) = \langle P_L, V_L \rangle$, computes the function $F_L$ defined by

$$F_L(p, 0) = \left\{ \begin{array}{ll} \text{``}x \in L\text{''} & \text{if } p = \text{``}x \in L\text{'' and } x \in L \\ \text{``Cheating''} & \text{otherwise} \end{array} \right.$$

**Ideal Protocol**

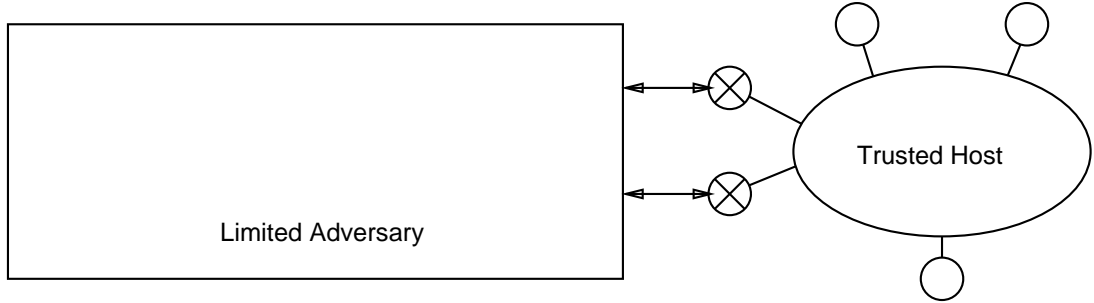

**Protocol with Interface**

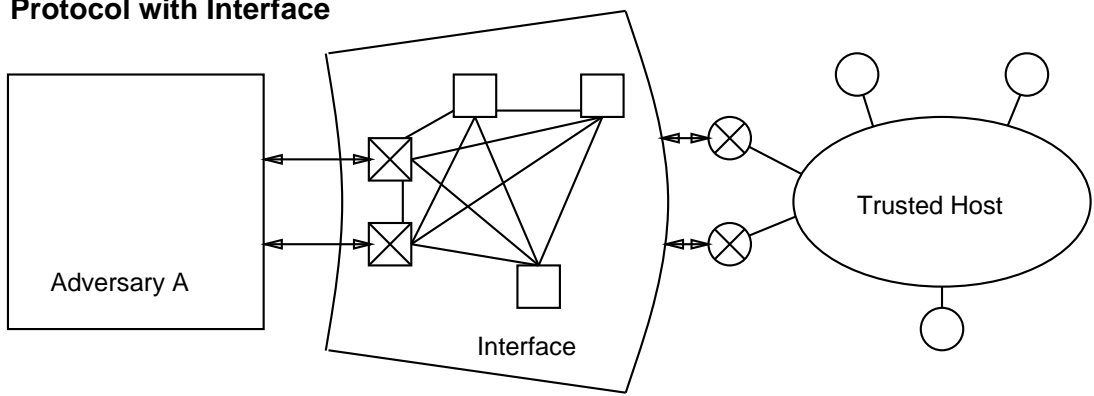

**Real Protocol**

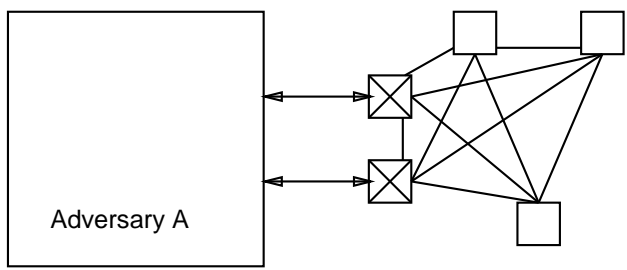

Figure 3.3: Three scenarios: first, an ideal protocol with a trusted host, clearly delineating the limited information and influence of an adversary; second, an adversary attacking a trusted host by way of an interface that creates a simulated environment for it; third, an adversary attacking a real protocol with no trusted parties. Squares and circles indicate players; squares within the interface are simulated players. The **X** marks indicate corrupted players.

**Theorem 3.5** $<P, V>$ *is a (statistical, computational) zero-knowledge proof system for $L$ if and only if it is a (statistically, computationally) 2-resilient protocol for $F_L$.*

**Proof:** If $<P, V>$ is a zero-knowledge proof system for L, then there exists a black-box simulator $S_{PV}$. Construct an interface $\mathcal{S}$ that does the following. Do nothing until adversary $A$ requests a corruption. If $A$ corrupts $V$, request corruption of $V_L$ in protocol $\texttt{ID}(F_L)$, and obtain $V_L$'s output, "$x \in L$." Adversary $A$ may have waited until some later round $r$ to request a corruption, so, by internally simulating $V$ and running the simulator $S_{PV}(x)$ to generate a transcript, create a history of the protocol through round $r$. Note that the output of the simulator is accurate for a proof system with no cheating (*i.e.* only passive corruption) and with $x \in L$. Therefore the prefix is accurate. Now, supply $A$ with the current state of internally simulated $V$, and run $S_{PV}(x)$ to the end, with access to $A$ as the corrupt verifier.[2]

If $A$ corrupts $P$, then $\mathcal{S}$ corrupts $P_L$. The interface $\int$ operates $V$ internally[3] carrying on a correspondence with $A$ as the prover. At the end of the process, the internal $V$ either accepts some $x \in L$ or rejects. If $V$ accepts, $\mathcal{S}$ requires the corrupt $P_L$ to send "$x \in L$" in the ideal protocol; the host will approve and $V_L$ will accept. If $V$ rejects, $\mathcal{S}$ requires the corrupt $P_L$ to send $\Lambda$ in the ideal protocol; the host and $V_L$ reject. In either case the final outputs of the uncorrupted players, $V$ and $V_L$, are identical. The adversary $A$ receives a view of an interaction with an honest verifier.

Finally, if at some point $A$ corrupts both $P$ and $V$, $\mathcal{S}$ stops interacting with $A$, since there are no nonfaulty outputs and no more nonfaulty players to corrupt or to receive messages from.

Now, if $<P, V>$ is statistically 2-resilient, then for some interface $\mathcal{S}$,

$$[P, V] \succeq^s [\mathcal{S}(A), P_L, V_L].$$

Hence for any adversary that corrupts $V$,

$$[P, V]^{Y_A} \succeq^s [\mathcal{S}(A), P_L, V_L]^{Y_A}$$

so $<P, V>$ is statistically $t$-private, and for any adversary that corrupts $P$,

$$[P, V]^{Y_V} \succeq^s [\mathcal{S}(A), P_L, V_L]^{Y_V}$$

so with asymptotically high probability, an honest $V$'s output is 1 when $x \in L$ and 0 otherwise. ∎

---

[2] Weak zero-knowledge, in which the simulator is not restricted to unresettable, black-box use of $V$, corresponds to weak resilience. The interface $\mathcal{S}$ simply awaits the output of such a simulator and then uses it as its own output.

[3] $V$ is polynomial time and its tapes are public.

# 3.5 Fault Oracles

An intermediate approach pursued by Beaver [Bea89b] and independently by Kilian, Micali, and Rogaway [KMR90] uses the idea of a fault-oracle that provides limited information from which to construct a simulated view for an adversary. Fault-oracles are a natural extension of zero-knowledge. Galil, Haber, and Yung [GHY89] proposed a function-oracle for the case of two-party minimum knowledge *decision* proof systems, in which the prover proves either $x \in L$ or $x \notin L$. Because the verifier should receive the result of a *decision* about membership in $L$ in addition to a proof, the simulator that produces its view ought to receive exactly that much information. Thus, the simulator is allowed a single query to an oracle that computes the characteristic function for $L$.

To define multiparty *privacy*, Beaver [Bea89b] and Rogaway [Rog89] extended the function-oracle to a fault-oracle. Some additional requirements are needed, since the powers of an adversary in a protocol are more diverse. Essentially, the simulator having access to $\mathcal{O}$ is allowed to request inputs of up to $t$ players and to substitute its own. Then it receives from $\mathcal{O}$ a single computation of $F$ using the unrevealed inputs of uncorrupted players. In Beaver's model, it is also allowed to request more corruptions (up to a total of $t$, and obtaining inputs and outputs) after the computation, since a dynamic adversary does have the option of doing so after a normal protocol.

As with zero-knowledge, leaked information is bounded by the amount of information a simulator needs to construct an accurate view of a real protocol execution.

A key step, noted independently by Beaver and by Kilian, Micali, and Rogaway, is that the computation of $F$ *induces output values $F_i(x_1, \ldots, x_n)$ for nonfaulty players,* even though these values are not returned. Correctness can thus be defined to hold when these induced outputs match those in the real protocol, for then, since the induced outputs are correct, so must the real ones be.

The definition of security becomes one of matching simulated to real adversarial views, and matching induced to real nonfaulty outputs. It is remarkably more concise and comprehensible than earlier approaches. Unfortunately, the definition is not overly flexible. For example, the concatenation of two secure protocols may correspond to two fault-oracle queries, but to prove the concatenation secure, one either allows only one query or must redefine security and the fault oracle.

The essential step that was not previously observed was that the oracle is far better regarded as a trusted host actually participating in a protocol. This observation paves the way for developing a general concept of protocol comparison. It serves as the basis to our approach. The definition of *relative resilience* is as concise, if not more so, as the definition of a fault oracle and of security with respect to a fault oracle. The fault oracle, even though a key stage in the development of good definitions, is a very limited and inflexible application of a very broad and powerful concept.

## 3.6 Satisfying the *ad hoc* Properties

The important properties of the trusted host are several. First, the trusted host ensures that the inputs are not revealed, apart from information leaked solely from the output of the function itself. Second, it returns a correct set of values based on the inputs it has received. Third, because it waits until all inputs are received (privately) before evaluating the function, it ensures that faulty or maliciously chosen inputs do not depend on the choices made by reliable players. Finally, every player is assured of receiving its respective output, and fairness is achieved.

We claim that the ideal case captures *a priori* all desirable properties that are as yet unconsidered; that is, it declares precisely what needs to be accomplished by a secure protocol and it specifies how malicious influences must be limited. New properties that one might like to consider reflect a deeper understanding of the ideal protocol as opposed to a change in the model and in the idea of security.

We can analyze the *ad hoc* list of individual properties as aspects of our definition of resilience. *Correctness* addresses the accuracy of the results computed by reliable players; *privacy* addresses what results the adversary can compute based on what he sees.

### Definition 3.6 (Correctness and Privacy)

- *A protocol $\alpha$ is $t$-**correct leaking** $F$ if there exists an interface $\mathcal{S}$ such that*

$$[A, \Pi]^{\vec{Y}} \asymp [A, \mathcal{S}, \text{ID}(F)]^{\vec{Y}}.$$

- *Protocol $\Pi$ is $t$-**private leaking** $F$ if there exists an interface $\mathcal{S}$ such that*

$$[A, \Pi]^{Y_A} \asymp [A, \mathcal{S}, \text{ID}(F)]^{Y_A}.$$

*Independence of inputs,* a more tricky property to define formally, is also captured by the ideal case. In the ideal case there is absolutely no interaction among players; the choice of input by a faulty player is completely independent of messages sent by reliable players to the trusted host, since those messages are carried on private channels. Since the adversary cannot capture a reliable player's input or be influenced by it in the ideal case without deciding to corrupt that player, its choice of inputs is the same regardless of the information held by reliable players. Resilience implies that the adversary's behavior and hence its choice of inputs is the same in real and ideal cases, so the faulty inputs are not influenced by the reliable ones.[4]

Our definition implies *fairness,* and in fact an even stronger property: not only do all reliable players obtain their results whenever the faulty players do, they obtain them regardless of the faulty players' behavior.

---

[4]One could certainly make a weaker statement that if the outputs of nonfaulty players have the same distribution in the real and ideal cases, then even if the adversary's choice of *inputs* depends on nonfaulty player's inputs, the final effect amounts to nothing. But because the *adversary's* view, containing the choice of inputs, must be identical in both cases, we need not weaken the claim.

Yao [Yao86] and Beaver and Goldwasser [BG89] (presented in Chapter 9) examine protocols where $t \geq n/2$. In this scenario, it is impossible for reliable players to force a computation to finish. It is impossible to achieve resilience. But it may be possible to ensure correctness and privacy if a computation is not halted. The issue of *fairness* has more meaning here. What if some players should learn the results while others do not?

Whereas perfect fairness is impossible, [BG89] use a weaker property to attain an approximation to fairness. The weaker property states that, if any reliable player fails to obtain an output, then all reliable players detect cheating. With cryptographic assumptions or given a network supporting oblivious transfer, [BG89] showed that misbehaving players can be identified. While this doesn't prevent other faulty players from changing their inputs if the protocol were to be run a second time after detecting some faults, it is a useful outcome (especially in a litigious society). Hence we may extend the output to be a vector of values or a value $(T, \text{CHEATING})$ which states that coalition $T$ was agreed by all reliable players to be cheating.

An approximation to fairness based on the ability to detect cheating is attainable [BG89]. If cheating does not occur, then all players can progress slowly but equally toward knowing their output. Nevertheless, defining this approximation is still tricky, since it is not clear what an *ideal* protocol would achieve. A rough sketch is as follows. A trusted party would accept inputs and compute the function. It would then take several rounds to reveal the results gradually. Intuitively, the odds of each player to know its result $y_i$ after each round must advance in lock-step. A more precise treatment of this concept appears in Chapter 9.

## 3.7   Strong Security: Solving a Subtle Error

A simple yet subtle bug exists in all known multiparty protocols. If the $t$ corrupted players send all their information to a single reliable player $i$, then that player now knows *every* input. For example, protocols based on secret sharing (*cf.* [Sha79, GMW87, BGW88, CCD88]) have the property that the information held by any $t+1$ players determines all the inputs. Therefore if $t$ players give up their information to player $i$, player $i$ knows everything. One way to circumvent this is to say that player $i$ is "reliable" and hence "ignores" improper messages. What reliable person would believe what the adversary says, anyway? An honest-but-curious player (one who behaves in terms of messages but may try to determine additional information) may be tempted to gamble that the adversary is telling the truth in leaking information, and may benefit from improper extra knowledge if the adversary is indeed leaking information.

But this sort of philosophizing is far from formal. If the information arrives at a node, the bottom line is that it should not be there. More formally, the set of messages given to the honest-but-curious player $i$ ought to be simulatable from the knowledge held by player $i$ and the adversary; the inputs of other uncorrupted players should not be compromised, and hence should not be needed to simulate the conversations seen by player $i$.

We may consider this as an example of the following situation: in addition to the malicious adversary $A$, there are one or more passive adversaries $B_1, B_2, \ldots$. A passive adversary $B$, like the Byzantine adversary $A$, is an automaton which requests states of players, but which does not replace messages from those players. Each reliable player is a fixed passive adversary with access only to its own communications. We would like the protocol to be secure simultaneously against $A$ and at least against such passive adversaries.

We address this *formally* in the following manner. Denote the distribution on *all* output strings and views by

$$[A, B, \Pi]^{hist}(n, m)(\vec{x} \circ \vec{a} \circ a \circ a_B, k) \;=\;$$
$$(Y_A, Y_B, Y_1, \ldots, Y_n, \texttt{view}_A, \texttt{view}_B, \texttt{view}_1, \ldots, \texttt{view}_n).$$

Here, $a_B$ is the auxiliary input of the passive adversary $B$.

We extend the interface $\mathcal{S}$ to have a third communication tape, to interact with a (black-box) passive adversary $B$. As before, consider an execution of the *ideal* protocol with $\mathcal{S}$ acting as the Byzantine adversary. We also allow $\mathcal{S}$ to corrupt a second set of players passively, just as B would. The number of additional players that $\mathcal{S}$ can passively compromise is bounded by $t_B$, the bound on the number of players that $B$ could corrupt in a real protocol. (We would like at least to ensure security for $t_B = 1$, corresponding to a single honest-but-curious player.)

Let $[A, B, \mathcal{S}, \texttt{ID}]^{hist}(n, m)(\vec{x} \circ \vec{a} \circ a \circ a_B, k)$ denote the distribution on outputs and views during an ideal protocol execution with $\mathcal{S}$ playing the part of the adversary, allowing $\mathcal{S}$ to request $(t + t_B)$ inputs.

We define the ensemble families $[A, B, \Pi]$ and $[A, B, \mathcal{S}, \texttt{ID}]$ as subsets of the random variables, namely just the $Y$ variables in $[A, B, \Pi]^{hist}$ and $[A, B, \mathcal{S}, \texttt{ID}]^{hist}$ as earlier. Then the more careful notion of security is the following:

**Definition 3.7** *Protocol* $\Pi$ *is* **strongly $t$-resilient leaking** $F$ *if there exists an interface* $\mathcal{S}$ *such that for any Byzantine $t$-adversary $A$, for any passive $1$-adversary $B$,*

$$[A, B, \Pi] \asymp [A, B, \mathcal{S}, \texttt{ID}]$$

We remark that similar modifications apply to the case of *relative* resilience, though we shall not occupy additional space with them. Auxiliary inputs can, as before, be excluded if so desired.

# Chapter 4

# Modular Protocol Design

Secure multiparty protocols are often based on a paradigm introduced by Goldreich *et al* [GMW86, GMW87]: "Share Inputs; Compute New Secrets; Reconstruct Results." The value of reducing a function computation to a circuit evaluation is clear: by constructing subprotocols to add and multiply secretly shared values, one provides the modules for a simple and modular protocol to compute the overall function.

Though the direct application of this approach to circuit evaluation results in inefficient protocols, we shall follow it and modify it to provide a more general and far more efficient modular approach. Rather than decomposing a function into subprotocols centered around each gate of a circuit, we focus on constructing a sequence of intermediate functions whose results can be revealed, unlike the outputs of intermediate gates in a circuit, without compromising security. We propose a more general modular approach.

To support a modular approach, however, the concatenation of protocols and the composition of functions require particular attention in terms of security and reliability. In this chapter we address the particular intuitive and formal issues arising from the pursuit of a modular approach. We prove some intuitive lemmas regarding composition and concatenation, some of which are intuitive results — such as the validity of composing polynomially many protocols — whose statements as folk-theorems are sufficiently imprecise that they do not hold. We give the formal requirements under which they do hold. In the next chapter we present particular protocols; this chapter focuses on technical lemmas and methodology.

## 4.1   Robust and Private Representations

A threshold scheme allows one player, the *dealer,* to distribute a secret value $s$ among the network in such a way that only certain coalitions of players have sufficient information to determine $s$. When a bound $t$ on the number of faulty or curious participants is given, the power to reconstruct $s$ is given to groups of size $t + 1$ or more, but not to any group of size $t$ or less.

There are two properties we should like to satisfy. The first states that there is a means to reconstruct a shared value despite errors. Given a vector $\vec{y}$ of $n$ values, a *t-modification* of $\vec{y}$ is a vector differing in at most $t$ places from $\vec{y}$, namely a vector of Hamming distance $t$ or less from $\vec{y}$.

**Definition 4.1** *A function* $\mathtt{sha} : S \to (\Sigma^\star)^n$ *is a* **t-robust representation** *if there exists a function* $\mathtt{rec}$ *such that, for all* $s \in S$, *for all t-modifications* $\vec{y}\prime$ *of* $\vec{y} = \mathtt{sha}(s)$, *we have* $\mathtt{rec}(\vec{y}\prime) = s$.

Broadcast channels are useful to maintain robustness, in which case the value is not kept private. Secret sharing, which does achieve privacy at the same time, is also useful.

The second property states that the information distributed to the players preserves the privacy of $s$. Recall that the vacuous protocol reveals no information (§3.3).

**Definition 4.2** *A function* $\mathtt{sha}$ *is a* **t-private function** *if there exists a protocol* SHARE *to compute* $\mathtt{sha}$ *that is as t-resilient as the ideal vacuous protocol.*

Any function that produces the same results regardless of its inputs is certainly private, though not necessarily robust. Though such a function is apparently useless at first glance, we shall find important uses. Robustness and privacy together define secret sharing:

**Definition 4.3** *A* **threshold scheme** *with threshold* $t$ *is a pair of protocols* Sha *and* Rec, *where* Sha *computes a t-robust and t-private representation* $\mathtt{sha}$, *and* Rec *computes the function* $\mathtt{rec}$.

## 4.2 Composing Functions

In order to develop modular protocols, let us consider protocols intended to compute some sequence of intermediate functions. For clarity of exposition, first consider only two functions, $F(x_1(1), \ldots, x_n(1))$ and $G(x_1(2), \ldots, x_n(2))$, in sequence. To be precise, the inputs to $G$ may include or be influenced by the inputs and outputs of $F$. We consider $x_1(2)$ to be a function of $y_1(1)$ (which itself specifies $x_1(1)$, without loss of generality) and of $x_1^{new}(2)$, a new portion of the input. The desired application may certainly specify that $x_1^{new}(2)$ is ignored; our protocols do not actually take advantage of new inputs, and we shall soon omit any mention of them for the sake of readability.

There are a few different ways to combine functions that are useful (*cf.* §2.7):

1. $(F, G)$ : the computation of the function $H(x_1, \ldots, x_n)$ that concatenates the outputs of $F$ and $G$ : namely $H_i(x_1, \ldots, x_n) = (F_i(x_1, \ldots, x_n), G_i(x_1, \ldots, x_n))$. The inputs to $G$ do not depend on the outputs of $F$.

2. $(F; G)$ : the *sequential* computation of $G$ with $F$. The inputs to $G$ do not necessarily depend on the outputs of $F$.

3. $\circ\{F, G\}$ or $(F; G \circ F)$ : the *open composition* of $F$ then $G$, using the outputs of $F$ in the computation of $G$.

4. $\diamondsuit\{F, G\}$ or $(G\diamondsuit F)$ : the *hidden composition* of $G$ with $F$, revealing the results of function $G \circ F$ without revealing the results of $F$.

The two of greatest interest are (3) *open composition* and (4) *hidden composition*. The idea of computing on hidden values, using the results of $F$ to compute $G$ without revealing them in the interim, is extremely useful.

A *collection* of function families is denoted $\mathcal{F} = \{F^1, F^2, \ldots\}$ where each $F^i$ is a family $F^i = \{F^{i,n,m}\}$. Let $f(n, m, k)$ be some function from integers to integers.

**Definition 4.4** *The* **closed composition** *of $f(n, m, k)$ functions from $\mathcal{F}$, written $\diamondsuit\mathcal{F}$ or informally $F^{f(n,m,k)}\diamondsuit \cdots \diamondsuit F^1$, is defined as the function family $F = \{F^{n,m}\}$, where each $F^{n,m}$ is:*

$$F^{n,m} = F^{f(n,m,k),n,m} \circ F^{f(n,m,k)-1,n,m} \circ \cdots \circ F^{1,n,m}.$$

**Definition 4.5** *The* **open composition** *of $f(n, m, k)$ functions from $\mathcal{F}$, written $\circ\mathcal{F}$, includes the progressive results in the composition of the $f(n, m, k)$ functions:*

$$F^{n,m \cdot f(n,m,k)} = (F^{1,n,m}, F^{2,n,m} \circ F^{1,n,m}, \ldots, F^{f(n,m,k),n,m} \circ \cdots \circ F^{1,n,m}).$$

*Note the use of $m \cdot f(n, m, k)$ in the indexing; the function has an output that is $f(n, m, k)$ times as large. (We define the functions $F^{n,m \cdot f(n,m,k)+a}$ for $0 < a < f(n, m, k)$ to be identically 0.)*

## 4.3   Concatenating Protocols

We shall speak of protocols, protocol families, and collections of protocol families; to make the presentation more clear, Figure 4.1 outlines the various levels on which we group together and discuss protocols and the events they induce. Recall that a protocol execution on a particular set of inputs $\vec{x}$, $\vec{a}$, and $a$, with particular values of $n$, $m$, and $k$, induces a *distribution* on histories. By varying these parameters, we consider larger and larger conglomerations of protocols (*e.g.* protocols, families of protocols, *etc.*) which induce larger and larger probabilistic conglomerations (*e.g.*, distributions, ensembles, families of ensembles, *etc.*).

### 4.3.1   Concatenating General Protocols

Let $\alpha = \{\alpha(1), \alpha(2), \ldots\}$ be a collection of protocol families, and let $f(n, m, k)$ be some function from $\mathbf{N}$ to $\mathbf{N}$. Let $\{x_i^{new}(r)\}_{i,r\in\mathbf{N}}$ be a collection of inputs that are to be supplied as the execution of the protocol progresses. Let $\{a_i^{new}(r)\}_{i,r\in\mathbf{N}}$ be a collection of auxiliary inputs that are to be supplied as the execution of the protocol progresses. Let $x_i(1) = x_i^{new}(1)$ and $a_i(1) = a_i^{new}(1)$ for all $i$. Let $\{\chi^{n,i,r}\}_{n,i,r\in\mathbf{N}}$ be a collection of probabilistic functions mapping an output and auxiliary input to a new input.

| Fixed Parameters | Protocol Group | History Group |
|---|---|---|
| $i, n, m,$ $k, \vec{x}, \vec{a}, a$ | protocol on particular inputs $\alpha(i)(n,m)(\vec{x} \circ \vec{a} \circ a, k)$ | distribution $[A, \alpha(i)](n,m)(\vec{x} \circ \vec{a} \circ a, k)$ |
| $i, n, m$ | protocol $\alpha(i)(n,m)$ | ensemble $[A, \alpha(i)](n,m)$ |
| $i$ | family of protocols $\alpha(i) = \{\alpha(i)(n,m)\}_{n,m \in \mathbf{N}}$ | family of ensembles $[A, \alpha(i)] = \{[A, \alpha(i)(n,m)]\}_{n,m \in \mathbf{N}}$ |
| | collection of families of protocols $\alpha = \{\alpha(i)\}_{i \in \mathbf{N}}$ | collection of families of ensembles $[A, \alpha] = \{[A, \alpha(i)]\}_{i \in \mathbf{N}}$ |

Figure 4.1: Hierarchy of protocols and associated probability spaces. After the first row, inputs $(\vec{x} \circ \vec{a} \circ a, k)$ are taken to vary over all possible values.

The *sequential concatenation* of protocols in $\alpha$ is the protocol described in Figure 4.2 and is written $\circ \alpha$, or $\circ_{i=1}^{f(n,m,k)} \alpha(i)$, or $(\alpha(1); \alpha(2); \ldots ; \alpha(f(n,m,k)))$. For each $n$, $f(n,m,k)$ protocols are executed in turn. The input of player $i$ to the $r^{th}$ protocol is denoted $x_i(r)$. Before the $r^{th}$ protocol is executed, each player $i$ obtains an external auxiliary input, $a_i^{new}(r)$, which is combined with its view of previous executions to provide the auxiliary input $a_i(r)$ for the $r^{th}$ protocol. Player $i$ chooses an input $x_i(r)$ for the $r^{th}$ protocol based on the output of the previous protocol $y_i(r-1)$ and its new input $x_i^{new}(r)$, according to the function $\chi^{n,i,r}$. This allows the player to base its next input on the results of previous protocols as well as other information it may have. The functions $\chi^{n,i,r}$ are part of the protocol and players are required to use them; the protocol designer is free to specify that players ignore previous outputs or that they use them in some particular way. Note that players corrupted by a Byzantine adversary are certainly allowed to specify inputs chosen differently.

We must include one restriction on the adversary: the set of players it corrupts throughout the entire concatenated protocol must be a subset of some allowable coalition in the fault class. Formally, if the adversary classes are $\mathcal{A}_1, \mathcal{A}_2, \ldots$, then the adversary class for the concatenation of protocols is $\mathcal{A} = \cap \mathcal{A}_r$. In each successive execution, the adversary may be treated as a new adversary receiving the view of the previous execution as its auxiliary input.

## 4.3.2 Concatenating Ideal Protocols

Though the definitions of §4.3.1 demonstrate how ideal protocols are concatenated, the direct concatenation of ideal protocols will not provide quite the level of security and reliability we desire. Operating two protocols $\mathtt{ID}(F)$ and $\mathtt{ID}(G)$ in sequence does not ensure that the inputs to the second protocol are in any way related to the inputs

---

CONCAT-PROTOCOL$(\alpha, f(n, m, k))$

(CP1)  Run protocol $\alpha(1)$ :

$$\mathtt{view}(1) \leftarrow [A, \alpha(1)](n, m)(\vec{x} \circ \vec{a} \circ a, k)$$
$$y_i(1) \leftarrow Y_i(\mathtt{view}_i(1)) \qquad (1 \le i \le n)$$

(CP(r))**For** $r = 2..f(n, m, k)$ **do**

$$a_i(r) \leftarrow (\mathtt{view}_i(r-1), a_i^{new}(r)) \qquad (1 \le i \le n)$$
$$a(r) \leftarrow (\mathtt{view}_A(r-1), a^{new}(r))$$
$$x_i(r) \leftarrow \chi^{n,i,r}(y_i^{r-1}, x_i^{new}(r)) \qquad (1 \le i \le n)$$
$$\mathtt{view}(r) \leftarrow [A, \alpha(r)](n, m)(\vec{x}(r) \circ \vec{a}(r) \circ a(r), k)$$
$$y_i(r) \leftarrow Y_i(\mathtt{view}_i(r)) \qquad (1 \le i \le n)$$

Figure 4.2:  Concatenation of $f(n, m, k)$ protocols from the collection of protocol families $\alpha = \{\alpha(1), \alpha(2), \dots\}$.

or outputs of the first. That is, a corrupt player might obtain $y_i(1)$ as its output of $F$ but instead supply $(y_i(1) + 1, x_i^{new}(2))$ as its input $x_i(2)$ to $G$.

In many circumstances this is quite undesirable. For example, the employees of a company might wish to calculate their average overall salary and the average salary of management. First, they compute the overall average. For the second computation, however, the management may report lower salaries to obtain an advantage in salary negotiations.

As defined in §2.7, let $\mathcal{F} = \{F^i\}_{i \in \mathbf{N}}$ be a possibly infinite collection of families $F^i = \{F^{i,n,m}\}$ of finite functions, and let $f : \mathbf{N}^3 \to \mathbf{N}$. (We normally take $f(n, m, k)$ to be polynomial.)

**Definition 4.6** *The* **open concatenation of** $f$ **ideal protocols** *using* $\{\chi^{n,i,r}\}_{n,i,r \in \mathbf{N}}$ *is the following protocol, denoted by* $\circ \{\mathtt{ID}(\mathcal{F})\} = \circ_i \mathtt{ID}(F^i)$. *Consider* $n + f(n, m, k)$ *players. A* $t$-*fault class is, as usual, the collection of subsets of* $[n]$; *players* $i > n$ *are incorruptible. The protocol requires* $2f(n, m, k)$ *rounds; let* $r$ *range from* $1$ *to* $f(n, m, k)$ :

**(Round 1)** *Each player* $i \in [n]$ *sends* $x_i(1) = x_i$ *to trusted player* $(n + 1)$.

**(Round 2)** *Player* $(n+1)$ *computes the value* $F^1(x_1, \dots, x_n)$ *and returns the results.*

**(Round $2r - 1$)** *Each player* $i \in [n]$ *chooses an input* $x_i(r) \leftarrow \chi^{n,i,r}(y_i(r-1), x_i^{new}(r-1))$ *and sends it to trusted player* $(n + r)$.

**(Round $2r$)** *Player* $(n + r)$ *computes the value* $F^r(x_1, \dots, x_n)$ *and returns* $y_i(r)$ *to each player* $i \in [n]$.

Technically, the objection to simply concatenating protocols directly is the following. Operating ideal protocols in sequence invokes *different* trusted parties, one to compute $F^1$, one to compute $F^2$, and so on. None of the trusted parties share any information or communicate at all.

The ideal *composite* protocol should have *one* trusted party. The protocol is slightly longer: first the trusted host requests $x_1(1), \ldots, x_n(1)$, and returns $F^1(x_1(1), \ldots, x_n(1))$. The host retains the values of the outputs. Since the inputs to the next computation are each some function of the previous output and a new input, the players themselves need only supply the new input, and the host can compute the next round's inputs using $x_i(r) \leftarrow \chi^{n,i,r}(y_i(r-1), x_i^{new}(r))$. This prevents changing the output of previous protocols, while permitting a protocol designer to include new information during the protocol if desired. More specifically:

**Definition 4.7** *The **ideal open composite protocol** for $p$ functions from $\mathcal{F}$, using $\{\chi^{n,i,r}\}_{n,i,r \in \mathbf{N}}$, is the following protocol, denoted $\texttt{ID}(\circ\mathcal{F})$. For each $n$, consider $n+1$ players. Player $(n+1)$ is incorruptible; the $t$-fault class is the collection of subsets of $[n]$. The protocol requires $2f(n, m, k)$ rounds:*

**(Round 1)** *Each player $i \in [n]$ sends $x_i(1)$ to trusted player $(n+1)$.*

**(Round 2)** *Player $(n+1)$ computes the value $F^1(x_1, \ldots, x_n)$ and returns $y_i(1) = F_i^1(x_1, \ldots, x_n)$ to each player $i \in [n]$.*

**(Round $2r-1$)** *Each player $i \in [n]$ sends $x_i^{new}(r)$ to the trusted host.*

**(Round $2r$)** *In round $2r$, player $(n+1)$ (not player $i$) computes $x_i(r) \leftarrow \chi^{n,i,r}(y_i(r-1), x_i^{new}(r))$, the next "input" for player $i$. Player $(n+1)$ then computes the value $F^r(x_1(r), \ldots, x_n(r))$ and returns $y_i(r)$ to each player $i \in [n]$.*

**Definition 4.8** *The **ideal hidden composite protocol**, $\texttt{ID}(\Diamond\mathcal{F})$, is the same as $\texttt{ID}(\circ\mathcal{F})$ except that in intermediate rounds the trusted host always returns the same results as the vacuous protocol (i.e. a vector describing honest and cheating players) instead of $y_i^r$. In the final round, the trusted host does return the final result.*

Our ultimate goal is to achieve the two protocols $\texttt{ID}(\circ\mathcal{F})$ and $\texttt{ID}(\Diamond\mathcal{F})$, since they ensure that information is consistent from one function evaluation to the next. Section 4.4 presents a method to ensure that the concatenation of ideal protocols does in fact achieve the same results as an ideal composite protocol with a single trusted host. Verifiable secret sharing is essential, both for maintaining consistency and for hiding progressive results. For clarity, we shall omit mention of the new inputs $x_i^{new}$ and of the functions $\chi^{n,i,r}$.

## 4.4 The Modular Approach

The key to achieving efficiency is to avoid evaluating a circuit for $F$ directly, and instead to break the computation of $F$ into the computation of several functions $F^1, \ldots, F^{f(n,m,k)}$, each of which reveals nothing about the inputs. Each of these computations is itself broken up into a composition of fundamental operations $G$, such as addition and multiplication, according to the established paradigm. Each input is shared as a secret using a robust representation. Instead of using the fundamental computations, which would reveal information, *robust* and *private* representations `hide`$(G)$ are used to compute pieces of new secrets from the old ones. For example, secrets are added or multiplied, but the inputs and outputs are maintained in shared form. The use of a robust and private representation (secret sharing) allows us to compute *open* compositions of functions, which allows us simply to concatenate sub-protocols. It also provides the glue to ensure that the computations are insensitive to faulty players who fail to supply the output of one subprotocol as the input to the next.

The supporting machinery [BGW88, CCD88] establishes how to compute each intermediate function $F^i$ by reducing it to simple operations on secrets. We shall use the protocols of Ben-Or *et al* (see §5.2) for these simple operations. We achieve vast improvements in efficiency through careful choice of the intermediate functions (see §5.4). In the remainder of this chapter we give formal arguments for the validity of our methods.

Using more formal notation, the general technique is as follows. Function family $F$ is first represented as the open composition $\circ \mathcal{F}$ of some collection of function families, each of which is private. In other words, $F$ is decomposed into a sequence of intermediate functions whose results can be revealed openly without compromising essential information. Then, each family $F^i$ in the collection is itself written as a *hidden* composition of a collection $\mathcal{G}^i$ containing more fundamental functions (such as addition and multiplication). In other words, we should like that each $F^i$ is computed as a composition of fundamental functions $\{G^{ij}\}$, even though the intermediate results of these computations *cannot* be revealed. Finally, to facilitate the hidden composition of these fundamental functions, each fundamental function $G^{ij}$ must have a private and robust representation `hide`$(G^{ij})$. In other words, `hide`$(G^{ij})$ is essentially a function that operates on secret pieces of the inputs to $G^{ij}$ and produces secret pieces of the results of $G^{ij}$, rather than producing the results themselves. Diagrammatically:

$$
\begin{aligned}
F &\rightarrow \circ_i \{F^i\} \\
F^i &\rightarrow \Diamond_j \{G^{ij}\} \\
G^{ij} &\rightarrow \texttt{hide}(G^{ij})
\end{aligned}
$$

## 4.5   Indistinguishability

In §2.1 we defined how two ensembles are indistinguishable, and we defined how two families of ensembles (parametrized by $n$ and $m$) are indistinguishable. In order to examine concatenations of many protocols and in order to facilitate proofs of resilience, we define indistinguishability for *collections* of ensemble families and collections of protocol families.

### 4.5.1 Collections of Families of Ensembles

Let $P = \{\mathcal{P}(i)\}_{i \in \mathbf{N}}$ be a collection of families of ensembles. For concreteness, note that $\mathcal{P}(i)$ is a family of ensembles, $\mathcal{P}(i)(n, m)$ is an ensemble, and $\mathcal{P}(i)(n, m)(z, k)$ is a distribution on strings. Let $Q = \{\mathcal{Q}(i)\}_{i \in \mathbf{N}}$ also be a collection of families of ensembles. In the spirit of mathematical analysis we define uniform indistinguishability, where "uniform" indicates not Turing machines but the flavor of uniform *vs.* pointwise convergence. We require that for any $n$ and $m$, there is a bound $\delta(i, k)$ on the closeness of corresponding ensembles $\mathcal{P}(i)(n, m)$ and $\mathcal{Q}(i)(n, m)$, and that the bound is approached uniformly for all $i$ :

**Definition 4.9** *Two collections of ensemble families $P = \{\mathcal{P}(i)\}$ and $Q = \{\mathcal{Q}(i)\}$ are*

- **pairwise $O(\delta(i, k))$-indistinguishable***if for all $i$, $\mathcal{P}(i) \asymp {}^{O(\delta(i,k))} \mathcal{Q}(i)$.*

- **uniformly pairwise $O(\delta(i, k))$-indistinguishable** *if*

$$(\exists \Delta : \mathbf{N} \to \mathbf{N}) \quad (\forall i) \;\; \mathcal{P}(i) \approx {}^{\Delta(i,k)} \mathcal{Q}(i) \qquad \text{and} \;\; \Delta(k) = O(\delta(k)).$$

- **pairwise $O(\delta(i, k))$-computationally indistinguishable** *if for all $i$, $\mathcal{P}(i) \overset{\cdot}{\asymp} {}^{O(\delta(i,k))} \mathcal{Q}(i)$.*

- **uniformly pairwise $O(\delta(i, k))$-computationally indistinguishable** *if*

$$(\forall n, m)(\forall A \in \mathrm{PPC})(\exists c, k_0)(\forall k \geq k_0)(\forall z)(\forall i)$$

$$|A_{\mathcal{P}(i)(n,m)(z,k)} - A_{\mathcal{Q}(i)(n,m)(z,k)}| \leq c \cdot \delta(i, k).$$

*We write $P \simeq {}^{O(\delta(k))} Q$ for the uniform case and $P \overset{\cdot}{\simeq} {}^{O(\delta(k))} Q$ for the uniform computational case.*

The essential point is that for *uniform* indistinguishability, the same convergence parameter $k_0$ applies to all the families simultaneously. That is, the closeness of the corresponding pairs converges for sufficiently large $k$, but the rate of convergence (*i.e.* the point at which the families are thereafter close) is the same for all pairs at once. Clearly, any two equally-sized *finite* collections of ensemble families are *uniformly* pairwise $O(\delta(i, k))$-indistinguishable if they are $O(\delta(i, k))$-indistinguishable.

We also examine a *single* collection of ensemble families with the property that each ensemble is indistinguishable from its successor. Before, there need be no relationship among families in the same collection, whereas now we require one.

**Definition 4.10** *A collection $P = \{\mathcal{P}(i)\}$ of families of ensembles is*

- **sequentially $O(\delta(k))$-indistinguishable** *if for all $i$, $\mathcal{P}(i) \asymp {}^{O(\delta(k))} \mathcal{P}(i+1)$.*

- **uniformly sequentially $O(\delta(k))$-indistinguishable** *if*

$$(\exists \Delta : \mathbf{N} \to \mathbf{N}) \quad (\forall i) \;\; \mathcal{P}(i) \asymp {}^{\Delta(k)} \mathcal{P}(i+1) \qquad \text{and} \;\; \Delta(k) = O(\delta(k)).$$

- **sequentially** $O(\delta(k))$**-computationally indistinguishable** *if for all $i$,* $\mathcal{P}(i) \stackrel{O(\delta(k))}{\backsimeq} \mathcal{P}(i+1)$.

- **uniformly sequentially** $O(\delta(k))$**-computationally indistinguishable** *if*

$$(\forall n, m)(\forall A \in \mathrm{PPC})(\exists c, k_0)(\forall k \geq k_0)(\forall z)(\forall i)$$

$$|A_{\mathcal{P}(i)(n,m)(z,k)} - A_{\mathcal{P}(i+1)(n,m)(z,k)}| \leq c \cdot \delta(k).$$

Clearly, any *finite* collection of ensemble families is *uniformly $O(\delta(k))$-*indistinguishable if each successive pair of families is $O(\delta(k))$-indistinguishable.

## 4.5.2 Collections of Families of Protocols

The relative resilience of two collections of *protocol* families is defined in a similar fashion as ensemble families: according to the relative resilience of each corresponding pair of protocols.

**Definition 4.11** *Let $\alpha = \{\alpha(i)\}$ and $\beta = \{\beta(i)\}$ be collections of protocol families. Then $\alpha$ is*

- **pairwise** $O(\delta(i,k))$**-resilient as** $\beta$ *if for all $i$,* $\alpha(i) \succeq^{O(\delta(i,k))} \beta(i)$.

- **uniformly pairwise** $O(\delta(i,k))$**-resilient as** $\beta$ *if*

$$(\exists \Delta : \mathbf{N}^2 \to \mathbf{N}) \quad (\forall i) \quad \alpha(i) \succeq^{\Delta(i,k)} \beta(i) \qquad and \; \Delta(i,k) = O(\delta(i,k)).$$

*We write $P \sqsupseteq^{O(\delta(i,k))} Q$ for the uniform case.*

**Definition 4.12** *A collection $\alpha = \{\alpha(i)\}$ of protocol families*

- **sequentially** $O(\delta(k))$**-resilient** *if for all $i$,* $\alpha(i) \succeq^{O(\delta(k))} \alpha(i+1)$.

- **uniformly sequentially** $O(\delta(k))$**-resilient** *if*

$$(\exists \Delta : \mathbf{N} \to \mathbf{N}) \quad (\forall i) \quad \alpha(i) \succeq^{\Delta(k)} \alpha(i+1) \qquad and \; \Delta(k) = O(\delta(k)).$$

The adjectives *perfect, exponential, statistical,* and *computational* apply respectively, as in §2.1 and §4.5.1, *mutatis mutandis.*

# 4.6 Indistinguishable Ensembles: Technical Lemmas

Two indistinguishable ensembles remain indistinguishable when restricted to simple subranges of the samples.

**Lemma 4.13** *(Indistinguishable Substrings) Let $\mathcal{G} = \{G(z,k)\}$ and $\mathcal{H} = \{H(z,k)\}$ be ensembles where each $G(z,k)$ and $H(z,k)$ is a distribution on $\Sigma^{f(|z|,k)}$. Let $a(z,k)$ and $b(z,k)$ be polynomial-time computable indices in the range $1, \ldots, f(|z|,k))$, with $a(z,k) \leq b(z,k)$. Let $G[a,b](z,k)$ be the distribution*

$$\{\sigma \leftarrow G(z,k); \tau \leftarrow \sigma[a(z,k)..b(z,k)] : \tau\}$$

*and define $H[a,b](z,k)$ similarly. Then subranges are indistinguishable:*

$$\mathcal{G} \approx \mathcal{H} \quad \Rightarrow \quad \mathcal{G}[a,b] \approx \mathcal{H}[a,b] \tag{4.1}$$

$$\mathcal{G} \approx^e \mathcal{H} \quad \Rightarrow \quad \mathcal{G}[a,b] \approx^e \mathcal{H}[a,b] \tag{4.2}$$

$$\mathcal{G} \approx^s \mathcal{H} \quad \Rightarrow \quad \mathcal{G}[a,b] \approx^s \mathcal{H}[a,b] \tag{4.3}$$

$$\mathcal{G} \stackrel{.}{\approx} \mathcal{H} \quad \Rightarrow \quad \mathcal{G}[a,b] \stackrel{.}{\approx} \mathcal{H}[a,b] \tag{4.4}$$

**Proof:** Statements (4.1)-(4.3) follow by observing that $\mathcal{G} \approx^{\delta(k)} \mathcal{H} \Rightarrow \mathcal{G}[a,b] \approx^{\delta(k)} \mathcal{H}[a,b]$, as shown by:

$$
\begin{aligned}
|\mathrm{Pr}_{\mathcal{G}[a,b]}[\tau] - \mathrm{Pr}_{\mathcal{H}[a,b]}[\tau]| &= |\sum \mathrm{Pr}_{\mathcal{G}}[\sigma \mid \tau] - \mathrm{Pr}_{\mathcal{H}}[\sigma \mid \tau]| \\
&= |\sum_{\sigma|\sigma[a..b]=\tau} \mathrm{Pr}_{\mathcal{G}}[\sigma] - \sum_{\sigma|\sigma[a..b]=\tau} \mathrm{Pr}_{\mathcal{H}}[\sigma]| \\
&\leq \delta(k),
\end{aligned}
$$

which holds for for sufficiently large $k$, by the indistinguishability of $\mathcal{G}$ and $\mathcal{H}$.

To verify (4.4), which describes computational indistinguishability, let us assume it fails. Then there exists a machine $M$ that distinguishes $G[a,b]$ from $H[a,b]$. Construct machine $M'$ which does the following: on input $\sigma$ of length $n^c$, set $\tau = \sigma[a(n)..b(n)]$, run $M$ on $\tau$, and use its output. Now, for any $c$, the distinguisher $M$ outputs 0 with different probabilities: $|M_{G[a,b]}[\tau] - M_{H[a,b]}[\tau]| \geq n^{-c}$ infinitely often. Clearly, $M'_G[\sigma] = M'_{G[a,b]}[\tau]$ and $M'_H[\sigma] = M'_{H[a,b]}[\tau]$. Therefore for any $c$, $|M'_G[\sigma] - M'_H[\sigma]| \geq n^{-c}$ infinitely often, contradicting the indistinguishability of $\mathcal{G}$ and $\mathcal{H}$. ∎

Observe that replacing one ensemble in a series by an indistinguishable ensemble gives rise to an indistinguishable entire series:

**Lemma 4.14** *Let $\mathcal{P}_1, \mathcal{P}_2^\alpha, \mathcal{P}_2^\beta,$ and $\mathcal{P}_3$ be ensembles. Define the ensembles*

$$
\begin{aligned}
\mathcal{Q}^\alpha(z,k) &= \{z_1 \leftarrow \mathcal{P}_1(z,k); z_2^\alpha \leftarrow \mathcal{P}_2^\alpha(z_1,k); z_3^\alpha \leftarrow \mathcal{P}_3(z_2^\alpha,k) : z_3^\alpha\} \\
\mathcal{Q}^\beta(z,k) &= \{z_1 \leftarrow \mathcal{P}_1(z,k); z_2^\beta \leftarrow \mathcal{P}_2^\beta(z_1,k); z_3^\beta \leftarrow \mathcal{P}_3(z_2^\beta,k) : z_3^\beta\}
\end{aligned}
$$

*If $\mathcal{P}_2^\alpha \approx^{O(\delta(k))} \mathcal{P}_2^\beta$, then $\mathcal{Q}^\alpha \approx^{O(\delta(k))} \mathcal{Q}^\beta$. The convergence parameters are identical.*

**Proof:** Let $k_0$ be the convergence parameter for $\mathcal{P}_2^\alpha$ and $\mathcal{P}_2^\beta$, with associated constant $c_0$. Assume by way of contradiction that $\mathcal{Q}^\alpha \not\approx^{O(\delta(k))} \mathcal{Q}^\beta$. Then there is a $k \geq k_0$ and

a $z$ such that

$$\sum_{z_1,z_2,z_3} |\Pr_{\mathcal{P}_3(z_2,k)}[z_3]\Pr_{\mathcal{P}_2^\alpha(z_1,k)}[z_2]\Pr_{\mathcal{P}_1(z,k)}[z_1]$$
$$-\Pr_{\mathcal{P}_3(z_2,k)}[z_3]\Pr_{\mathcal{P}_2^\beta(z_1,k)}[z_2]\Pr_{\mathcal{P}_1(z,k)}[z_1]| \;\; > \;\; c_0 \cdot \delta(k)$$

Thus there exists a $z_1$ such that

$$\sum_{z_2,z_3} |\,(\Pr_{\mathcal{P}_3(z_2,k)}[z_3]\Pr_{\mathcal{P}_2^\alpha(z_1,k)}[z_2]$$
$$-\Pr_{\mathcal{P}_3(z_2,k)}[z_3]\Pr_{\mathcal{P}_2^\beta(z_1,k)}[z_2])\,| \;\; > \;\; c_0 \cdot \delta(k)$$

It follows that

$$\sum_{z_2} |\Pr_{\mathcal{P}_2^\alpha(z_1,k)}[z_2] - \Pr_{\mathcal{P}_2^\beta(z_1,k)}[z_2]| \;\; =$$

$$\sum_{z_2}\left(\sum_{z_3}\Pr_{\mathcal{P}_3(z_2,k)}[z_3]\cdot|\Pr_{\mathcal{P}_2^\alpha(z_1,k)}[z_2] - \Pr_{\mathcal{P}_2^\beta(z_1,k)}[z_2]|\right) \;\; > \;\; c_0 \cdot \delta(k).$$

Since $k \geq k_0$, this contradicts $\mathcal{P}_2^\alpha \approx^{O(\delta(k))} \mathcal{P}_2^\beta$. ∎

**Lemma 4.15** *Let $\mathcal{P}_1, \mathcal{P}_2^\alpha, \mathcal{P}_2^\beta,$ and $\mathcal{P}_3$ be polynomially generable and let $\mathcal{Q}^\alpha$ and $\mathcal{Q}^\beta$ be as in Lemma 4.14. If $\mathcal{P}_2^\alpha \overset{\cdot}{\approx}^{\delta(k)} \mathcal{P}_2^\beta$ then $\mathcal{Q}^\alpha \overset{\cdot}{\approx}^{\delta(k)} \mathcal{Q}^\beta$.*

**Proof:** Let $k_0$ be the convergence parameter for $\mathcal{P}_2^\alpha$ and $\mathcal{P}_2^\beta$. Assume by way of contradiction that $\mathcal{Q}^\alpha \overset{\cdot}{\not\approx}^{\delta(k)} \mathcal{Q}^\beta$. Then there exists a probabilistic polynomial size distinguisher $A$ such that for some $z$ and infinitely many $k$,

$$|A_{\mathcal{Q}^\alpha(z,k)} - A_{\mathcal{Q}^\beta(z,k)}| > \delta(k).$$

Now, $A_{\mathcal{Q}^\alpha(z,k)}$ is just

$$\sum_{z_1,z_2,z_3}\Pr_{A(z_3)}[1]\Pr_{\mathcal{P}_3(z_2,k)}[z_3]\Pr_{\mathcal{P}_2^\alpha(z_1,k)}[z_2]\Pr_{\mathcal{P}_1(z,k)}[z_1]$$

so there exists a $z_1$ such that

$$\sum_{z_2,z_3} |\,(\Pr_{A(z_3)}[1]\Pr_{\mathcal{P}_3(z_2,k)}[z_3]\Pr_{\mathcal{P}_2^\alpha(z_1,k)}[z_2]$$
$$-\Pr_{A(z_3)}[1]\Pr_{\mathcal{P}_3(z_2,k)}[z_3]\Pr_{\mathcal{P}_2^\beta(z_1,k)}[z_2])\,|$$
$$> \;\; \delta(k).$$

Let $A'$ be the distinguisher that does the following. On input $\sigma$, sample $z_3 \leftarrow \mathcal{P}_3(\sigma,k)$,

and run $A(z_3)$. Return the result of $A$. It follows that

$$| A'_{\mathcal{P}_2^\alpha(z_1,k)} - A'_{\mathcal{P}_2^\beta(z_1,k)} |$$
$$= | \Pr_{A(z_3)}[1] \Pr_{\mathcal{P}_3(z_2,k)}[z_3] \Pr_{\mathcal{P}_2^\alpha(z_1,k)}[z_2]$$
$$- \Pr_{A(z_3)}[1] \Pr_{\mathcal{P}_3(z_2,k)}[z_3] \Pr_{\mathcal{P}_2^\beta(z_1,k)}[z_2] |$$
$$> \delta(k).$$

Since this holds for some $z$ and $z_1$, and for infinitely many $k$, $A'$ actually distinguishes $\mathcal{P}_2^\alpha$ from $\mathcal{P}_2^\beta$, so $\mathcal{P}_2^\alpha \stackrel{\delta(k)}{\not\asymp} \mathcal{P}_2^\beta$. ∎

Given a collection of ensemble families and a function $f(n,m,k)$, we may define a particular ensemble family using the $k^{th}$ distribution from the $f(n,m,k)^{th}$ family.

**Definition 4.16** *The* **diagonal ensemble family** $\mathcal{P}^f$ *of a collection of ensemble families* $P = \{\mathcal{P}(0), \mathcal{P}(1), \ldots\}$ *is, for a given function* $f : \mathbf{N}^3 \to \mathbf{N}$, *defined by the following:*

$$\mathcal{P}^f(n,m)(z,k) = \mathcal{P}(f(n,m,k))(n,m)(z,k).$$

*The* **diagonal protocol family** $\alpha^f$ *for a collection of protocol families* $\alpha = \{\alpha(1), \alpha(2), \ldots\}$ *is the protocol defined by:*

$$\alpha^f(n,m)(\vec{x} \circ \vec{a} \circ a, k) = \alpha(f(n,m))(n,m)(\vec{x} \circ \vec{a} \circ a, k).$$

The first ensemble family in a sequentially indistinguishable collection is indistinguishable from the diagonal family to a certain degree:

**Lemma 4.17** *Let* $P = \{\mathcal{P}(0), \mathcal{P}(1), \ldots\}$ *be a uniformly sequentially $O(\delta(k))$-indistinguishable collection of families of ensembles, and let* $f : \mathbf{N}^3 \to \mathbf{N}$. *Then* $\mathcal{P}(0) \asymp^{O(\delta(k) \cdot f(n,m,k))} \mathcal{P}^f$.

**Proof:** Let $k_0$ be the convergence parameter for $P$. Assume by way of contradiction that $\mathcal{P}(0) \not\asymp^{\delta(k) \cdot f(n,m,k)} \mathcal{P}^f$. Then there are $n$ and $m$ such that $\mathcal{P}(0)(n,m) \not\asymp^{\delta(k) \cdot f(n,m,k)} \mathcal{P}^f(n,m)$. Let $c$ be arbitrary; there exists a $k \geq k_0$ and a $z$ such that

$$|\mathcal{P}(0)(n,m)(z,k) - \mathcal{P}(f(n,m,k))(n,m)(z,k)| > c \cdot \delta(k) f(n,m,k).$$

That is,

$$\left| \sum_{i=0}^{f(n,m,k)-1} \mathcal{P}(i)(n,m)(z,k) - \mathcal{P}(i+1)(n,m)(z,k) \right| > c \cdot \delta(k) f(n,m,k).$$

So for some $i_0$, we have $k \geq k_0$ and

$$|\mathcal{P}(i_0)(n,m)(z,k) - \mathcal{P}(i_0+1)(n,m)(z,k)| > \frac{c \cdot \delta(k) f(n,m,k)}{f(n,m,k)} = c \cdot \delta(k),$$

contradicting the uniform $\delta(k)$-indistinguishability of $P$. ∎

Adapting this proof to computational indistinguishability in a fashion similar to the proofs of of Lemmas 4.14 and 4.15, a similar result holds for *computational* indistinguishability. that

**Lemma 4.18** *Let $P$ be a uniformly sequentially $O(\delta(k))$-computationally indistinguishable collection of families of ensembles. Let each ensemble be polynomially generable. Let $f(n, m, k)$ be a polynomial. Then $\mathcal{P}(0) \overset{.}{\asymp}{}^{O(\delta(k) \cdot f(n,m,k))} \mathcal{P}^f$.*

The following lemma summarizes direct consequences of Lemmas 4.17 and 4.18:

**Lemma 4.19** *Let $P$ be a uniformly sequentially $O(\delta(k))$-indistinguishable collection of families of ensembles. Let $f(n, m, k)$ be a polynomial. Then the following hold according to the indistinguishability of $P$ :*

| | | | | | | |
|---|---|---|---|---|---|---|
| *perfect* | $[(\forall i)$ | $\mathcal{P}(i)$ | $\asymp$ | $\mathcal{P}(i+1)] \Rightarrow$ | $\mathcal{P}(0)$ | $\asymp$ | $\mathcal{P}^f$ |
| *exponential* | $[(\forall i)$ | $\mathcal{P}(i)$ | $\asymp^{O(c^{-k})}$ | $\mathcal{P}(i+1)] \Rightarrow$ | $\mathcal{P}(0)$ | $\asymp^e$ | $\mathcal{P}^f$ |
| *statistical* | $[(\forall i)$ | $\mathcal{P}(i)$ | $\asymp^{O(k^{-c})}$ | $\mathcal{P}(i+1)] \Rightarrow$ | $\mathcal{P}(0)$ | $\asymp^e$ | $\mathcal{P}^f$ |
| *computational* | $[(\forall i)$ | $\mathcal{P}(i)$ | $\overset{.}{\asymp}{}^{O(k^{-c})}$ | $\mathcal{P}(i+1)] \Rightarrow$ | $\mathcal{P}(0)$ | $\overset{.}{\asymp}$ | $\mathcal{P}^f$ |

We also consider the indistinguishability of two diagonal ensemble families taken from two pairwise uniformly indistinguishable collections.

**Lemma 4.20** *Let $P = \{\mathcal{P}(i)\}$ and $Q = \{\mathcal{Q}(i)\}$ be uniformly $O(\delta(i, k))$ pairwise indistinguishable collections of families of ensembles, and let $f : \mathbf{N}^3 \to \mathbf{N}$. Then $\hat{\mathcal{P}}^f \asymp^{O(\delta(f(n,m,k),k))} \hat{\mathcal{Q}}^f$.*

**Proof:** Let $k_0$ be the convergence parameter for $P$. Assume by way of contradiction that $\hat{\mathcal{P}}^f \not\asymp^{O(\delta(i,k) \cdot f(n,m,k))} \hat{\mathcal{Q}}^f$. Then there are $n$ and $m$ such that $\mathcal{P}(f(n, m, k))(n, m) \not\asymp^{O(\delta(i,k) \cdot f(n,m,k))} \mathcal{Q}(f(n, m, k))(n, m)$. Hence for an arbitrary $c$, there is a $k \geq k_0$ and a $z$ such that

$$|\mathcal{P}(f(n, m, k))(n, m)(z, k) - \mathcal{Q}(f(n, m, k))(n, m)(z, k)| > c \cdot \delta(f(n, m, k), k)$$

contradicting the *uniform* pairwise $O(\delta(i, k))$ indistinguishability of $P$ and $Q$. ∎

As before, the proof is adaptable to *computational* indistinguishability, giving:

**Lemma 4.21** *Let $P = \{\mathcal{P}(i)\}$ and $Q = \{\mathcal{Q}(i)\}$ be uniformly $O(\delta(i, k))$ pairwise computationally indistinguishable collections of families of ensembles, and let $f$ be a polynomial. Then $\hat{\mathcal{P}}^f \overset{.}{\asymp}{}^{O(\delta(f(n,m,k),k))} \hat{\mathcal{Q}}^f$.*

## 4.7 Formal Proofs

Generally, an interface creates a *simulated environment* for the adversary. Consider the collection of random variables describing the views of all the players generated during a run of a given protocol:

$$\{\mathtt{view}_1^1, \ldots, \mathtt{view}_n^1, \mathtt{view}_A^1; \mathtt{view}_1^2, \ldots, \mathtt{view}_n^2, \mathtt{view}_A^2; \ldots; \mathtt{view}_1^R, \ldots, \mathtt{view}_n^R, \mathtt{view}_A^R\}.$$

With static adversaries, the interface need only keep track of a fixed subset of these views. Interfacing becomes more difficult when dynamic adversaries are considered: even though the variables are dependent on one another, the interface will not be able to specify in advance all of the necessary views. The interface must be able to answer two types of queries from an adversary: requests for rushed messages, and requests for the view of a newly corrupted player. It computes these incrementally; as the interaction between $\mathcal{S}$ and $A$ progresses, $\mathcal{S}$ fills in some of the views (corresponding to newly corrupted players) as necessary. For instance, if player $i$ is corrupted at round $r$, the interface must sample the values of the variables $\mathtt{view}_i^1, \mathtt{view}_i^2, \ldots, \mathtt{view}_i^r$, even though it had not previously "filled in" these values. To do this, it must use appropriately chosen conditional distributions.

When a protocol is a composition of several subprotocols, the interface $\mathcal{S}$ actually runs sub-simulations $\mathcal{S}_j$ for each of the intermediate executions. The simulation is not quite as simple as running sub-simulations, however: the view an adversary gains when corrupting a new player $i$ in the third subprotocol, say, includes not only the original auxiliary input $a_i$ — which is all that $\mathcal{S}$ would obtain if it requests the corruption of player player $i$ in the ideal protocol — but also the view $\mathtt{view}_i(1) \circ \mathtt{view}_i(2)$ of player $i$ in the earlier two subprotocols. In particular, in order to be able to run interface $\mathcal{S}_{j+1}$, $\mathcal{S}$ must be able to generate auxiliary input $a_i(j)$, the auxiliary input of player $i$ in the $j^{th}$ subprotocol, when $\mathcal{S}_{j+1}$ requests it.

Thus, before we are able to discuss the resilience of concatenating resilient protocols, we must first consider (1) how to obtain an accurate final set of views by sampling a subset of the random variables in some order, and (2) how to ensure that a sub-simulation is not just resilient but resilient enough to provide views even after the sub-simulation has completed.

### 4.7.1 Random Variables and Tableaus

The description of a general protocol execution in the presence of an adversary (Chapter 2, §2.5.2) and the particular specification of the protocol describe how to sample the collection of random variables parametrized by $\mathtt{Vnames} \times [R]$ by sampling some of them and later applying probabilistic functions to the results in a specified order. We are concerned on the one hand with the distributions $\mathtt{V}(y_1, R), \mathtt{V}(y_2, R), \ldots, \mathtt{V}(y_n, R), \mathtt{V}(y_A, R)$, which make up the ensemble $[A, \Pi](\vec{x} \circ \vec{a} \circ a, k)$ generated by running protocol $\Pi(n, m, k)(\vec{x}, \vec{a}, A(a))$. On the other hand are the distributions $\mathtt{V}_{ideal}(y_1, R), \mathtt{V}_{ideal}(y_2, R), \ldots, \mathtt{V}_{ideal}(y_n, R), \mathtt{V}_{ideal}(y_A, R)$, which make up the ensemble $[A, \mathtt{ID}](\vec{x} \circ \vec{a} \circ a, k)$ generated by running protocol $\mathtt{ID}(F)(n, m, k)(\vec{x}, \vec{a}, S(A(a), \cdot))$. We must eventually show that the interaction of interface with black-box adversary and ideal protocol produces the same ensembles, even though the process of sampling various distributions $(\{\mathtt{Vnames}_{ideal}(v, r)\})$ lead-ing up to these final ones is different than in the real protocol (*i.e.*, local variables of reliable players are *never* assigned, and distributions are sampled in a different order).

To illustrate the principle behind our arguments, let us consider two experiments. Let $X_0$ be a distribution on some set $S$, and let $\chi_1$ and $\chi_2$ be functions on $S$ such that $x_0 = (\chi_1(x_0), \chi_2(x_0))$ for all $x_0 \in S$. Let $f_3$ and $f_4$ be probabilistic functions.

The first experiment generates the following distribution:

$$P = \{x_0 \leftarrow X_0; x_1 \leftarrow \chi_1(x_0); x_2 \leftarrow \chi_1(x_0); x_3 \leftarrow f_3(x_2); x_4 \leftarrow f_4(x_1, x_3) : x_4\}.$$

At the point at which $x_3$ is computed, the value of $x_1$ does not matter; in a sense, though it has been specified, it is a hidden variable. It comes into play later, when $x_4$ is computed. Intuitively, this experiment is analogous to the following sequence in a real protocol: a reliable player computes its transition function, specifying hidden information (new state and messages to reliable players) and compromised information (messages known to the adversary). Later, the adversary decides to corrupt the player, and obtains the values of the hidden variables, which it uses in later computations.

Now define the following distributions:

$$X_1 = \{x_0 \leftarrow X_0 : \chi_1(x_0)\}$$

$$X_2 = \{x_0 \leftarrow X_0 : \chi_2(x_0)\}$$

$$(X_1 \mid x_2) = \{x_0 \leftarrow X_0 : \chi_1(x_0) \mid \chi_2(x_0) = x_2\}$$

The second experiment is the following:

$$Q = \{x_2 \leftarrow X_2; x_3 \leftarrow f_3(x_2); x_1 \leftarrow (X_1 \mid x_2); x_4 \leftarrow f_4(x_1, x_3) : x_4\}.$$

In a sense, $x_1$ is not computed until after $x_3$ has been computed. This is analogous to the operation of an interface: first the interface computes the variable $X_2$ that is initially "known" to the adversary, and it does not attempt to specify the variable $X_1$. Later, the adversary corrupts a new player, and requests the value of $x_0$ (that is, of $x_1$ and $x_2$). The interface must now sample $X_1$ given the information it has already committed to the adversary, namely given $x_2$.

It is not hard to see that distributions $P$ and $Q$ are identical. The essential principle to note is that the *order* of sampling the distributions is irrelevant as long as the appropriate conditional distributions are used.

The task of proving privacy reduces to showing that the interface can gradually fill in the necessary portions of a simulated tableau by sampling the appropriate conditional distributions $(X_1 \mid x_2)$ *given* the random variables $(x_2)$ which it has already sampled and committed to $A$, *by using* the information it has at hand (inputs obtained by corruptions in the *ideal* protocol).

### 4.7.2 Post-Protocol Corruption

For proofs involving dynamic adversaries, simulating the view of a newly corrupted player is essential. Certainly, during the run of a protocol, an interface must be able to compute such a view accurately. In the case of the sequential concatenation of protocols, however, the view of a reliable player includes its view during previous protocols. The task becomes more difficult; despite the fact that the simulation of the earlier subprotocol has already occurred, the earlier view must be generated so that it can be included in the overall view of the newly corrupted player. We therefore consider an additional requirement on the interface, to facilitate proofs of resilience. This requirement is necessarily satisfied in the case of *perfect* and *exponential* resilience, in the case of *static* adversaries, or trivially in *memoryless* protocols.

Consider an execution of a protocol $\beta$ with $\mathcal{S}(A, \cdot)$, giving distribution $[A, \texttt{ID}](\vec{x} \circ \vec{a} \circ a, k)$. At the conclusion of the protocol, $\mathcal{S}(A(a), \cdot)$ has corrupted some coalition $T$. It may be the case that $|T| < t$, in which case $\mathcal{S}(A(a), \cdot)$ would have been permitted $(t - |T|)$ more corruptions. The interface $\mathcal{S}$ can continue to run, since it simply receives requests for new corruptions. There is, in general, no guarantee that if it is supplied with *additional* requests after $A(a)$ has halted, the views $\mathcal{S}$ returns are accurate in any fashion. Post-protocol corruptibility ensures that the interface does have this power, namely that even after $A(a)$ has halted, producing its output $Y_A$, the interface can accurately output $\texttt{view}_i$ for other players.

Let $T'$ be a set of players. When $\mathcal{S}$ continues to run after $A(a)$ has halted and is supplied in turn with each $i \in T'$ written to its input tape as a request for a new corruption, it returns the views $\{\texttt{view}_i \mid i \in T'\}$, inducing the ensemble of distributions

$$\beta^{\texttt{view}_{T'}}(\vec{x} \circ \vec{a} \circ a, k).$$

**Definition 4.22** *An interface $\mathcal{S}(\cdot, \cdot)$ is **post-protocol compatible** with respect to protocols $\alpha$ and $\beta$ if for any adversary $A$, for any execution of $<A, \mathcal{S}, \beta>$, for any $(|T| - t)$-coalition $T' \subseteq [n] - T$ where $T$ is the final coalition selected by adversary $A(a)$, the following ensembles are indistinguishable:*

$$\begin{aligned}
\alpha(\vec{x} \circ \vec{a} \circ a, k) &\approx \beta(\vec{x} \circ \vec{a} \circ a, k) \\
\alpha^{\texttt{view}_{T'}}(\vec{x} \circ \vec{a} \circ a, k) &\approx \beta^{\texttt{view}_{T'}}(\vec{x} \circ \vec{a} \circ a, k)
\end{aligned}$$

**Definition 4.23** *A protocol $\alpha$ is **post-protocol corruptible** with respect to $\beta$ if it admits a post-protocol compatible interface $\mathcal{S}$.*

Memoryless protocols are trivially post-protocol corruptible, since the views are erased. Protocols secure against static adversaries are also post-protocol corruptible, because the set of corrupted players is initially maximal so that $T' = \emptyset$ always.

A protocol $\alpha$ that is perfectly as resilient as protocol $\beta$ (*e.g.* a real protocol $\Pi$ that is perfectly resilient) against dynamic adversaries is of necessity post-protocol corruptible by the following argument. For any adversary $A$, let $A'$ be the adversary that operates $A$ until the final round and then does the following. If the coalition $T$ has size less than $t$, then $A'$ chooses $T' = \{\sigma_1, \ldots, \sigma_\tau\}$ uniformly at random from all subsets of $[n] - T$ of size $(t - |T|)$ or less. Then $A'$ requests the corruption of the players in $T'$. If follows that for all $T'$, $\texttt{view}_{\sigma_1}, \ldots, \texttt{view}_{\sigma_\tau}$, and $\texttt{view}_A$, the resilience of $\alpha$ against $A'$ implies that

$$\Pr{}_\alpha\left[T', \texttt{view}_{T'}, \texttt{view}_A\right] = \Pr{}_\beta\left[T', \texttt{view}_{T'}, \texttt{view}_A\right]$$
$$\Pr{}_\alpha\left[T'\right]\Pr{}_\alpha\left[\texttt{view}_{T'} \mid \texttt{view}_A\right]\Pr{}_\alpha\left[\texttt{view}_A\right] = \Pr{}_\beta\left[T'\right]\Pr{}_\beta\left[\texttt{view}_{T'} \mid \texttt{view}_A\right]\Pr{}_\beta\left[\texttt{view}_A\right]$$

Summing over all possible views of $A$,

$$\sum_{\texttt{view}_A}\Pr{}_\alpha\left[\texttt{view}_{T'} \mid \texttt{view}_A\right]\Pr{}_\alpha\left[\texttt{view}_A\right] = \sum_{\texttt{view}_A}\Pr{}_\beta\left[\texttt{view}_{T'} \mid \texttt{view}_A\right]\Pr{}_\beta\left[\texttt{view}_A\right]$$
$$\Pr{}_\alpha\left[\texttt{view}_{T'}\right] = \Pr{}_\beta\left[\texttt{view}_{T'}\right]$$

Similar arguments show that any protocol that is exponentially or statistically resilient against a dynamic adversary also must be post-protocol corruptible. If we require that the convergence parameter is bounded by a polynomial in $n$ and $m$, *i.e.* that security is achieved when $k$ is of size polynomial in $n$ and $m$, then any protocol that is exponentially resilient must be post-protocol corruptible.

## 4.8   Relatively Resilient Protocols and Concatenation

We observe that in a sequence of protocols, each as resilient as its successor (to within $\delta(k)$), the first protocol is as resilient (to within $\delta(k) \cdot f(n, m, k)$) as the particular protocol $\alpha^f$ constructed by using the $f(n, m, k)^{th}$ protocol from each family.

**Lemma 4.24** *Let $\alpha = \{\alpha(0), \alpha(1), \ldots\}$ be a uniformly sequentially $O(\delta(k))$-relatively resilient collection of protocol families. Let $f : \mathbf{N}^3 \to \mathbf{N}$. Then $\alpha(0) \succeq^{O(\delta(k)f(n,m,k))} \alpha^f$.*

**Proof:** Let $A$ be an adversary for protocol $\alpha(0)$ and let $\mathcal{S}_0, \mathcal{S}_1, \ldots$ be the interfaces postulated by the sequential resilience of the collections. Let $\mathcal{S}^j(A, a)$ denote the nested interface $\mathcal{S}_j(\cdot, \mathcal{S}_{j-1}(\cdot, \cdots(\cdot, \mathcal{S}_0(A, \cdot))))$. Let $\mathcal{P}(j)(n, m)(\vec{x} \circ \vec{a} \circ a, k)$ be the ensemble generated by running $\alpha(j)(n, m, k)(\vec{x}, \vec{a}, A(a))$. Because $\alpha$ is uniformly sequentially $O(\delta(k))$ resilient, the collection $P = \{\mathcal{P}(0), \mathcal{P}(1), \ldots\}$ is uniformly sequentially $O(\delta(k))$ indistinguishable. Construct an interface $\mathcal{S}$ that runs $\mathcal{S}^k$ when given security parameter $k$, and note that $\mathcal{P}^f(n, m)(\vec{x} \circ \vec{a} \circ a, k)$ is the ensemble generated by running protocol $\alpha^f(n, m, k)(\vec{x}, \vec{a}, \mathcal{S}^j(A, a))$. Lemma 4.17 implies:

$$\mathcal{P}(0) \asymp^{\delta(k) \cdot f(n,m,k)} \mathcal{P}^f.$$

The corresponding result holds for computational resilience:

**Lemma 4.25** *Let $\alpha = \{\alpha(0), \alpha(1), \ldots\}$ be a uniformly sequentially $O(\delta(k))$ computationally relatively resilient collection of protocol families. Let $f(n, m, k)$ be a polynomial. Then, computationally, $\alpha(0) \succeq {}^{O(\delta(k)f(n,m,k))} \alpha^f$.*

The transitivity of $\succeq$ falls out as an immediate corollary, recalling that a *finite* sequentially resilient collection of protocol families is *uniformly* sequentially resilient.

**Lemma 4.26** *(Transitivity) Let $P_1, P_2$, and $P_3$ be protocol families. If $P_1 \succeq P_2$ and $P_2 \succeq P_3$, then $P_1 \succeq P_3$. Furthermore, if $P_1 \succeq {}^{O(\delta(k))} P_2$ and $P_2 \succeq {}^{O(\delta(k))} P_3$, then $P_1 \succeq {}^{O(\delta(k))} P_3$.*

Let $\alpha = \{\alpha(i)\}$ and $\beta = \{\beta(i)\}$ be two collections of protocol families. Consider the $j$-fold concatenation of protocols from each: $\hat{\alpha}(j) = \circ_{i=1}^{j} \alpha(i)$ and $\hat{\beta}(j) = \circ_{i=1}^{j} \beta(i)$.

**Lemma 4.27** *(Protocol Concatenation) Fix $j$. If $\alpha \sqsupseteq {}^{O(\delta(k))} \beta$ with post-protocol compatible interfaces and with convergence parameter $k_0$, then $\hat{\alpha}(j) \succeq {}^{O(j \cdot \delta(k))} \hat{\beta}(j)$ with convergence parameter $k_0$.*

**Proof:** We gradually replace $\alpha$'s by $\beta$'s. Define the hybrid protocol family $H = \{H(0), H(1), \ldots\}$ by:

$$
\begin{aligned}
h(i, l)(n, m) &= \begin{cases} \beta(i)(n, m) & i < l \\ \alpha(i)(n, m) & i \geq l \end{cases} \\
H(l)(n, m) &= \circ_{i=1}^{l} h(i, l)(n, m) \\
&= h(l, l)(n, m) \circ h(l-1, l)(n, m) \circ \cdots \circ h(1, l)(n, m)
\end{aligned}
$$

It suffices to show that $H$ is uniformly sequentially $O(\delta(k))$ relatively resilient with convergence parameter $k_0$, since Lemma 4.24 with $f(n, m, k) = j$ implies that

$$
\hat{\alpha}(j) = H(0) \succeq {}^{O(\delta(k) \cdot f(n,m,k))} H^f = \hat{\beta}(j).
$$

Assume otherwise; then for some $n, m$, and $l$,

$$
H(l)(n, m) \not\succeq {}^{O(\delta(k))} H(l+1)(n, m).
$$

In particular, for all $c$ there are $\vec{x}, \vec{a}, A, a$, and $k \geq k_0$ such that

$$
|H(l)(n, m)(\vec{x} \circ \vec{a} \circ a, k) - H(l+1)(n, m)(\vec{x} \circ \vec{a} \circ a, k)| > c \cdot \delta(k) \tag{4.5}
$$

Let $\mathcal{S}_1, \ldots, \mathcal{S}_j$ be the post-protocol compatible interfaces for the $j$ subprotocols. Because the interfaces are post-protocol corruptible, we may treat the sequential execution of the subprotocols in $H(l)$ or in $H(l+1)$ as a sequence of three probabilistic experiments, $(\mathcal{P}_1, \mathcal{P}_2^\alpha, \mathcal{P}_3)$, or $(\mathcal{P}_1, \mathcal{P}_2^\beta, \mathcal{P}_3)$, where:

$$
\begin{aligned}
\mathcal{P}_1 &= \alpha(l-1)(n, m)(\vec{x}(l-1) \circ \vec{a}(l-1) \circ a(l-1), k) \circ \cdots \\
&\quad \circ \alpha(1)(n, m)(\vec{x} \circ \vec{a} \circ a, k)
\end{aligned}
$$

$$\begin{aligned}
\mathcal{P}_2^\alpha &= \alpha(l)(n,m)(\vec{x}(2) \circ \vec{a}(2) \circ a(2), k) \\
\mathcal{P}_2^\beta &= \beta(l)(n,m)(\vec{x}(l) \circ \vec{a}(l) \circ a(l), k) \\
\mathcal{P}_3 &= \alpha(j)(n,m)(\vec{x}(j) \circ \vec{a}(j) \circ a(j), k) \circ \cdots \\
&\quad \circ \alpha(l+1)(n,m)(\vec{x}(l+1) \circ \vec{a}(l+1) \circ a(l+1), k)
\end{aligned}$$

By assumption, $\mathcal{P}_2^\alpha \approx^{O(\delta(k))} \mathcal{P}_2^\beta$ with convergence parameter $k_0$. By Lemma 4.14, $\mathcal{Q}^\alpha \approx^{O(\delta(k))} \mathcal{Q}^\beta$ with parameter $k_0$. But $\mathcal{Q}^\alpha = H(l)(n,m)(\vec{x} \circ \vec{a} \circ a, k)$ and $\mathcal{Q}^\beta = H(l+1)(n,m)(\vec{x} \circ \vec{a} \circ a, k)$. This contradicts (4.5). ∎

We are now ready to measure the resilience achieved by concatenating a growing number of subprotocols.

**Theorem 4.28** *(Protocol Concatenation) Let $f : \mathbf{N}^3 \to \mathbf{N}$. If $\alpha \sqsupseteq^{\delta(i,k)} \beta$ with post-protocol compatible interfaces and with convergence parameter $k_0$, then concatenating $f(n,m,k)$ protocols from each satisfies:*

$$\circ_{i=1}^f \alpha(i) \succeq^{O(\delta(f(n,m,k),k) \cdot f(n,m,k))} \circ_{i=1}^f \beta(i).$$

*In particular,*

$$\begin{array}{llllll}
\alpha & \sqsupseteq & \beta & \Rightarrow & \circ\alpha & \succeq & \circ\beta & & (1) \\
\alpha & \sqsupseteq^e & \beta & \Rightarrow & \circ\alpha & \succeq^e & \circ\beta & (\textit{if } f(n,m,k) = O(k^c)) & (2) \\
\alpha & \sqsupseteq^s & \beta & \Rightarrow & \circ\alpha & \succeq^s & \circ\beta & (\textit{if } f(n,m,k) = O(k^c)) & (3) \\
\alpha & \sqsupseteq^{\cdot} & \beta & \Rightarrow & \circ\alpha & \succeq^{\cdot} & \circ\beta & (\textit{if } f(n,m,k) = O(k^c)) & (4)
\end{array}$$

**Proof:** We define two collections of protocol families as follows. The first is $\mu = \{\mu(1), \mu(2), \ldots\}$, the concatenation of linearly many protocols from $\alpha$ :

$$\mu(j)(n,m,k) = \circ_{i=1}^j \alpha(i)(n,m).$$

The second is defined similarly: $\nu = \{\nu(1), \nu(2), \ldots\}$, where

$$\nu(j)(n,m,k) = \circ_{i=1}^j \beta(i)(n,m).$$

Assume $\alpha \sqsupseteq^{O(\delta(i,k))} \beta$ pairwise uniformly with convergence parameter $k_0$. By Lemma 4.27, $\alpha \sqsupseteq^{\delta(i,k)} \beta$ implies that, for each $i$, $\mu(i) \sqsupseteq^{i\delta(i,k)} \nu(i)$ with convergence parameter $k_0$. It follows that $\mu \sqsupseteq^{i\delta(i,k)} \nu$, pairwise uniformly with convergence parameter $k_0$. By Lemma 4.20, the diagonal protocol families $\hat{\mu}^f$ and $\hat{\nu}^f$ satisfy

$$\hat{\mu}^f \succeq^{O(f(n,m,k) \cdot \delta(f(n,m,k),k))} \hat{\nu}^f.$$

But

$$\begin{aligned}
\hat{\mu}^f &= \circ_{i=1}^{f(n,m,k)} \alpha(i) \\
\hat{\nu}^f &= \circ_{i=1}^{f(n,m,k)} \beta(i)
\end{aligned}$$

It follows that

$$\circ_{i=1}^{f(n,m,k)}\alpha(i) \succeq {}^{O(f(n,m,k)\cdot\delta(f(n,m,k),k))} \circ_{i=1}^{f(n,m,k)} \alpha(i)$$

Implication (1) follows directly; implications (2), (3), and (4) are straightforward. For the proof of (4), we invoke the computational versions of the corresponding supporting lemmas. ∎

If intermediate functions are private, then computing their open composition is as secure as computing their closed composition:

**Lemma 4.29** *(Open Function Composition) Let $f : \mathbf{N}^3 \to \mathbf{N}$. Let $\mathcal{F} = \{F^i\}$ be a collection of function families where $F^{1,n,m}, \ldots, F^{f(n,m,k)-1,n,m}$ are private. Let $F = \circ \mathcal{F}$. Then the ideal open composite protocol $\mathtt{ID}(\circ\mathcal{F})$ for $\mathcal{F}$ is as resilient as the ideal protocol $[A, \mathtt{ID}]$ for $F$:*

$$\mathtt{ID}(\circ\mathcal{F}) \succeq \mathtt{ID}(\Diamond\mathcal{F}) \succeq [A, \mathtt{ID}].$$

*If $f(n,m,k)$ is polynomial, then this holds for exponential, statistical, and computational relative resilience.*

**Proof:** Since each $F^{1,n,m}, \ldots, F^{f(n,m,k)-1,n,m}$ is *private*, each is as resilient as the vacuous protocol. So for each intermediate function there exists a protocol $\mathtt{ID}(F)(F^j)$ and an interface $\mathcal{S}_j$ showing $\mathtt{ID}(F)(F^j)$ is as resilient as the ideal vacuous protocol. Recall that the ideal open composite protocol runs for $f(n,m,k)$ rounds, accepting an initial set of inputs $x_1^1, \ldots, x_n^1$, returning the value of $F^r \circ \cdots \circ F^1(x_1^1, \ldots, x_n^1)$ at each round $r$. The ideal hidden composite protocol returns only the vectors of identities of "cheating" players (players not supplying a valid input) and returns the final result.

The interface $\mathcal{S}$ uses each $\mathcal{S}_j$ to create the needed intermediate function values $F_j \circ \cdots \circ F_1$ that are supplied to corrupted players. That is, at the first round, $\mathcal{S}$ collects requests for corruptions from $A$, supplies them to $\mathcal{S}_1$, collects requests for corruptions from $\mathcal{S}_1$, requests those corruptions itself, and returns the results along the same paths. When $A$ and $\mathcal{S}_1$ are done requesting corruptions, $\mathcal{S}_1$ supplies messages from corrupted players (either 0 or 1 depending on if the player decides to cheat detectably), $\mathcal{S}$ sends them in the hidden protocol, and $\mathcal{S}$ obtains the vacuous output from the trusted host in round 1 of the hidden protocol. Interface $\mathcal{S}$ then supplies the vacuous output to $\mathcal{S}_1$, who may request more corruptions but eventually computes the values of $F_1$ to be sent to corrupted players. When $\mathcal{S}_1$ is finished, $\mathcal{S}$ runs $\mathcal{S}_2$ as before; if $\mathcal{S}_2$ corrupts a player, $\mathcal{S}$ must make a post-protocol corruption request to $\mathcal{S}_1$ to produce the view of that player during round 1. When $\mathcal{S}_2$ is finished corrupting players, $\mathcal{S}$ requests the vacuous output of round 2 of the hidden protocol and supplies it to $\mathcal{S}_2$.

This continues for $f(n,m,k) - 1$ steps. At the last step, $\mathcal{S}$ collects requests for corruptions directly from $A$ and supplies responses as before. When $A$ requests the $f(n,m,k)^{th}$ output from the open protocol, namely the value of $F^{f(n,m,k)} \circ \cdots \circ F^1$, $\mathcal{S}$ requests it in the hidden protocol and returns the result.

The second statement, $\mathtt{ID}(\Diamond\mathcal{F}) \succeq [A, \mathtt{ID}]$, is easy to see by noting that an interface itself can compute the vacuous messages returned to $A$ for each round up to the penultimate one, based on the messages sent to $\mathcal{S}$ by $A$; its only interaction in protocol $[A, \mathtt{ID}]$ is to request original inputs of corrupted players. In the final round, $\mathcal{S}$ obtains the results for corrupted players in $[A, \mathtt{ID}]$ and returns them to $A$. ∎

Concatenating ideal protocols for robust functions is as resilient as a single ideal composite protocol that computes their open composition — this is the motivation for robustness:

**Lemma 4.30** *(Ideal Protocol Concatenation) Let $f(n, m, k)$ be a polynomial. Let $\mathcal{F} = \{F^i\}$ be a collection of function families where $F^{1,n,m}, \ldots, F^{f(n,m,k)-1,n,m}$ are robust. Then the concatenation of ideal protocols for $\mathcal{F}$ is as resilient as the ideal open composite protocol for $\mathcal{F}$:*

$$\circ\{\mathtt{ID}(\mathcal{F})\} \succeq \mathtt{ID}(\circ\mathcal{F})$$

**Proof:** Protocol $\mathtt{ID}(F)(\circ\mathcal{F})$ lasts $f(n, m, k)$ rounds. For the first $f(n, m, k) - 1$ rounds, whenever adversary $A$ requests the corruption of player $i$, interface $\mathcal{S}$ corrupts player $i$ in protocol $\mathtt{ID}(F)(\Diamond\mathcal{F})$ and returns the original input, along with the list of outputs generated thus far. The output of each round is, as in the vacuous protocol, a list of $n$ bits which are 0 for every uncorrupted player and are 0 or 1 for corrupted players depending on whether the adversary forced them to send a 0 message or not. (Note that the vacuous output is the same for all players; $\mathcal{S}$ simply records what it has generated at each round of the simulation in order to include it in the information of later corruptions.)

At round $f(n, m, k)$, however, when the adversary generates its last set of messages for players in the $\mathtt{ID}(F)(\circ\mathcal{F})$ protocol, the interface $\mathcal{S}$ sends the vector of messages it received for corrupted players in the first round of the $\mathtt{ID}(F)(\circ\mathcal{F})$ protocol: these are the inputs upon which the computation of $F^{f(n,m,k)}$ is based. Interface $\mathcal{S}$ receives a message from the trusted host in protocol $\mathtt{ID}(F)(\Diamond\mathcal{F})$ that contains the outputs for corrupted players, and it relays this message to $A$. That the final messages are identically distributed in $\mathtt{ID}(F)(\circ\mathcal{F})$ and $\mathtt{ID}(F)(\Diamond\mathcal{F})$ is ensured by the robustness of each intermediate result. ∎

Let $\mathtt{Sha}$ be a protocol that implements a robust and private representation $\mathtt{sha}$, and let $\mathtt{Rec}$ be a protocol that reconstructs the represented value according to the $\mathtt{rec}$ function. The robust and secret version of an arbitrary function $H$ is the function:

$$\mathtt{hide}(H) = \mathtt{sha} \circ H \circ \mathtt{rec}$$

We have,

$$\mathtt{rec} \circ \mathtt{hide}(H) \circ \mathtt{sha} = H$$

or more generally, for a collection of families $\{H^j\}$,

$$\mathtt{rec} \circ (\circ_j \mathtt{hide}(H^j)) \circ \mathtt{sha} = \circ_j H^j \tag{4.6}$$

The protocols in [GMW87, BGW88, CCD88], for example, are of the form

$$\texttt{Rec} \circ (\circ_j \Pi(\texttt{hide}(G^j))) \circ \texttt{Sha},$$

where each $G^j$ is an arithmetic or Boolean function (a set of gates on one level of a circuit). In particular, the players share the inputs, evaluate functions $G^1, \ldots, G^{f(n,m,k)}$ on them, and reveal only the final result. Our protocols will be of a somewhat more general form, computing *several* intermediate functions whose values *are* revealed on the path to computing $F$ :

$$\circ_i(\texttt{Rec} \circ (\circ_j \Pi(\texttt{hide}(G^{ij}))) \circ \texttt{Sha}).$$

The following theorem demonstrates the resilience of our modular approach:

**Theorem 4.31** *(Modular Protocol Construction) Let $p(n, m, k) = O((nmk)^c)$ and $q(n, m, k) = O((nmk)^c)$ for some c. Let $\{\mathcal{G}^i\}$ be a set of collections of function families, where $\mathcal{G}^i = \{G^{i,j}\}$, and $G^{i,j} = \{G^{i,j,n,m}\}$. Let $F^i = \circ \mathcal{G}^i$ be the composition of $p(n, m, k)$ functions from $\mathcal{G}^i$. Let $\mathcal{F} = \{F^i\}$, and let $F = \circ \mathcal{F}$ be the composition of $q(n, m, k)$ functions from $\mathcal{F}$.*

*If each $F^{1,n,m}, \ldots, F^{q(n,m,k)-1,n,m}$ is private and robust, and if there exist t-resilient protocols $\Pi(\texttt{hide}(G^{ij}))$ for each $i, j$, then:*

$$\circ_{i=1}^q(\texttt{Rec} \circ (\circ_{j=1}^p \Pi(\texttt{hide}(G^{ij}))) \circ \texttt{Sha}) \succeq \texttt{ID}(F)$$

*In other words, the concatenation of the t-resilient protocols as described above is a t-resilient protocol for F. This is also true of exponential, statistical, and computational resilience.*

**Proof:** First let us examine the resilience of computing each $F^i$ by concatenating the ideal protocols for the $G^{ij}$ functions:

$$
\begin{aligned}
\texttt{Rec} \circ [\circ_j \texttt{ID}(\texttt{hide}(G^{ij}))] \circ \texttt{Sha} \quad &\succeq \quad \texttt{Rec} \circ \texttt{ID}(\circ_j \texttt{hide}(G^{ij})) \circ \texttt{Sha} \quad &(4.30, 4.27)\\
&\succeq \quad \texttt{Rec} \circ \texttt{ID}(\Diamond_j \texttt{hide}(G^{ij})) \circ \texttt{Sha} \quad &(4.29, 4.27)\\
&\succeq \quad \texttt{ID}(\texttt{rec}\Diamond[\Diamond_j \texttt{hide}(G^{ij})]\Diamond\texttt{sha}) \quad &(4.29, 4.30, \ 4.27)\\
&\succeq \quad \texttt{ID}(\Diamond_j G^{ij}) \quad &(4.6)\\
&\succeq \quad \texttt{ID}(F^i) \quad &(4.29)
\end{aligned}
$$

At this stage, we have demonstrated enough to support the methods of [BGW88, CCD88]. We wish to concatenate protocols that *do* reveal (restricted) intermediate values:

$$
\begin{aligned}
\circ_i(\texttt{Rec} \circ [\circ_j \texttt{ID}(\texttt{hide}(G^{ij}))] \circ \texttt{Sha}) \quad &\succeq \quad \circ_i \texttt{ID}(F^i) \quad &(4.31)\\
&\succeq \quad \texttt{ID}(\Diamond_i F^i) \quad &(4.30, 4.29)\\
&\succeq \quad \texttt{ID}(F) \quad &(4.29)
\end{aligned}
$$

The theorem stipulates the existence of resilient protocols $\Pi(\texttt{hide}(G^{ij}))$ for each $i, j$. That is,

$$\Pi(\texttt{hide}(G^{ij})) \succeq \texttt{ID}(\texttt{hide}(G^{ij}))$$

Hence

$$
\begin{aligned}
\circ_i(\texttt{Rec} \circ [\circ_j \Pi(\texttt{hide}(G^{ij}))] \circ \texttt{Sha}) \quad &\succeq \quad \circ_i(\texttt{Rec} \circ [\circ_j \texttt{ID}(\texttt{hide}(G^{ij}))] \circ \texttt{Sha}) \\
&\succeq \quad \texttt{ID}(F)
\end{aligned}
$$

∎

# Chapter 5

# Efficient, Unconditionally Secure Multiparty Protocols

> They developed a tendency to shirk every movement that didn't seem absolutely necessary or called for efforts that seemed too great to be worthwhile. Thus these men were led to break, oftener and oftener, the rules of hygiene they themselves had instituted, to omit some of the numerous disinfections they should have practiced, and sometimes to visit the homes of people suffering from pneumonic plague without taking steps to safeguard themselves against infection, because they had been notified only at the last moment and could not be bothered with returning to a sanitary service station, sometimes a considerable distance away, to have the necessary injections. There lay the real danger; for the energy they devoted to fighting the disease made them all the more liable to it.
>
> — Albert Camus, *The Plague*

Ben-Or, Goldwasser, Wigderson, and Chaum, Crépeau, and Damgaard [BGW88, CCD88] show how, given private communication channels and a broadcast channel,[1] a network can evaluate a circuit $C_F$ for a function $F(x_1, \ldots, x_n)$ in the presence of $t$ Byzantine faults, as long as $t < \frac{n}{3}$. No complexity-theoretic assumptions are necessary.

Their construction requires a number of rounds of interaction proportional to the depth of $C_F$ and messages which grow with its size. In fact, since most early protocols for securely computing a function $F$ were based on evaluating a circuit for $F$ gate by gate, this led many researchers to conjecture that the circuit depth of $F$ was a lower bound on the number of rounds of interaction, and the circuit size a lower bound on the message size.

We prove to the contrary that the communication complexity of securely computing $F$ need not be related to the computational (circuit) complexity of $F$. First, we show that the number of rounds of interaction can be reduced by a logarithmic factor while maintaining small message sizes. The number of rounds required for any function in the broad class $NC^1$ is constant. In fact, the number of rounds for *any* function is reducible to a constant, though at the price of larger messages (see §5.5) or lower fault tolerance (see Chapter 14).

---

[1]The requirement of a broadcast channel can be removed at the cost of a constant factor increase in the number of rounds, using a Byzantine Agreement protocol of Feldman and Micali [FM88]; the communications during the protocol are independent of all values except the broadcast value, so that the protocol maintains privacy and resilience.

Our protocols are simple to implement: they employ matrix multiplication, polynomial evaluation, and polynomial interpolation as their most complicated subroutines. All of the local computations are fast.

The main result of this chapter is a protocol to evaluate a function in a constant number of rounds using a technique that avoids gate-by-gate simulation. (see §5.3–§5.3.5) Before presenting our results, we must first describe the threshold schemes that underlie most of the protocols in this dissertation. By recombining values that have been distributed according to the threshold scheme, new, robustly shared results are constructed. We give a method, similar to yet more efficient than that of [BGW88, CCD88], for evaluating the AND, OR, and NOT gates of a circuit, representing the results using the threshold scheme. Because proofs and formal specifications of the protocols in [BGW88, CCD88] have not appeared, we must provide proofs of the underlying methods as well. We present a general paradigm for modular protocol design and prove it robust.

**Assumptions made in this chapter.** The network is complete, with private lines, $n$ processors, and at most $t < \frac{n}{3}$ Byzantine faults, chosen dynamically. The protocols are information-theoretically secure, or in other words, the results are perfectly resilient.

## 5.1 Threshold Schemes

A particularly efficient and simple threshold scheme is presented by Shamir [Sha79]; we refer to it as *secret sharing*. Fix a finite field $E$ of size greater than $n$, the number of players. For example, the set $\mathbf{Z}_p$ of integers taken modulo $p$ will do, where $p$ is a prime number in the range $n < p < 2n$. Arithmetic over $\mathbf{Z}_p$ is extremely easy to compute. Another convenient field to use, though slightly less intuitive to implement, is $GF(2^n)$, which has the often useful property that $x + x = 0$ for any $x$.

In any case, the dealer *shares* $s \in E$ by the method displayed in Figure 5.1. The dealer chooses $t$ random coefficients $a_t, \ldots, a_1$ in $E$ and creates a polynomial $p(n)$ of degree $t$ using these coefficients, with $p(0) = s$. Then, for fixed, public, nonzero evaluation points $\alpha_1, \ldots, \alpha_n$, he constructs the "piece" $\text{PIECE}_i(s)$ by evaluating $p(\alpha_i)$. Finally, he sends $\text{PIECE}_i(s)$ to player $i$, privately. Distributions $\mathit{UPoly}\,(t, s)$ and $\mathit{UPieces}\,(n, t, s)$, discussed in §2.1, describe the distributions on polynomials and vectors of pieces so generated.

Any $t+1$ players can determine $s$ easily, since $t+1$ points determine a polynomial of degree $t$. By interpolating the polynomial of degree $t$ passing through their points, a sufficiently large collection of players can recover $p(0) = s$. As long as the number of omitted pieces, which is bounded by $t$, leaves $t+1$ valid pieces remaining, the

---

Sʜᴀʀᴇ $(s)$

- Fix field $E$, $|E| > n$, and elements $\alpha_1, \ldots, \alpha_n \neq 0$.

- Dealer selects $a_t, \ldots, a_1 \in E$ at random.

- Dealer sets $p(u) = a_t u^t + \cdots + a_1 u + s$.

- Dealer computes ᴘɪᴇᴄᴇ$_i(s) \leftarrow p(\alpha_i)$.

- Dealer sends ᴘɪᴇᴄᴇ$_i(s)$ to player $i$.

Figure 5.1: Protocol for dealer to secretly share value $s \in E$.

---

reconstruction is possible; hence $n - t \geq t + 1$ must be satisfied, or equivalently $2t < n$. Thus at the appropriate time specified by the protocol, it is easy to reconstruct the secret from the information held by reliable players.

---

Rᴇᴄᴏɴsᴛʀᴜᴄᴛ-Sᴇᴄʀᴇᴛ $($ᴘɪᴇᴄᴇ$_{i_1}(s), \ldots,$ ᴘɪᴇᴄᴇ$_{i_{t+1}}(s))$

- Each player $i$ sends ᴘɪᴇᴄᴇ$_i(s)$ to player $j$.

- Player $j$ selects $t + 1$ of the pieces he collects. Player $j$ interpolates the polynomial $p(u)$ passing through $(i_1,$ ᴘɪᴇᴄᴇ$_{i_1}(s)), \ldots, (i_{t+1},$ ᴘɪᴇᴄᴇ$_{i_{t+1}}(s))$.

- Player $j$ computes $s = p(0)$.

Figure 5.2: Protocol to reveal secret $s$ to player $j$.

---

Intuitively speaking, any $t$ or fewer evaluation points of a $t^{th}$-degree polynomial do not determine that polynomial. Thus, an adversary who gains the information held by a coalition of $t$ or fewer players learns nothing; it obtains a uniformly distributed vector of values, regardless of the value of $s$.

**Lemma 5.1** *(Shamir 1979) Let $E$ be a field and let $1 \leq t < |E|$. Then for any $s \in E$, for any $t' \leq t$, and for any set $T = \{\alpha_1, \ldots, \alpha_{t'}\} \subset E - \{0\}$ of size $t'$,*

$$\{\vec{a} \leftarrow \mathtt{uniform}(E^t) : (p_{s,\vec{a}}(i_1), \ldots, p_{s,\vec{a}}(i_{t'}))\} = \mathtt{uniform}(E^{t'})$$

*where $p_{s,\vec{a}}(u) = s + \sum_{j=1}^{t} a_j u^j$. In particular, for any $n < |E|$, $\alpha_1, \ldots, \alpha_n, s \in E$, and $T \subseteq [n]$ with $|T| \leq t$,*

$$\{\vec{p} \leftarrow \mathtt{UPieces}(n, t, s) : \vec{p}_T\} = \mathtt{uniform}(E^{|T|})$$

Given some pieces of a secret, new pieces remain uniformly random, up to a total of $t$ :

**Lemma 5.2** *Let $E$ be a field and let $1 \leq t < n < |E|$. Then for any $s \in E$, for any disjoint $T, T' \subseteq [n]$ satisfying $|T \cup T'| \leq t$, and for any $\vec{q} \in E^n$,*

$$\{\vec{p} \leftarrow \mathtt{UPieces}(n, t, s) : \vec{p}_{T'} \mid \vec{p}_T = \vec{q}_T\} = \mathtt{uniform}(E^{|T'|})$$

**Proof:** Elementary linear algebra. ▌

**Lemma 5.3** *For $3t < n$, in a completely connected network of $n$ processors with private channels, the method for secret sharing presented above is $t$-resilient against dynamic, passive adversaries.*

**Proof:** In the ideal protocol, the dealer sends $n$ pieces to the trusted host, who then distributes them. Before the adversary corrupts the dealer, the interface need only generate random field elements in response to corruption requests, which it does according to Lemma 5.2 by generating uniformly random field elements. If the adversary corrupts the dealer, then the interface first chooses $t - |T|$ remaining pieces uniformly at random as directed by Lemma 5.2, and corrupts the dealer in the ideal protocol, obtaining $s$. Given $s$ and $t$ pieces, the polynomial is uniquely determined and $\mathcal{S}$ simply solvers for the remaining pieces. It constructs a view of the dealer consisting of $s$, the coefficients, and the entire list of pieces, and provides the list to the adversary. Subsequent requests for corruptions are computed by evaluating the polynomial. ▌

Polynomial interpolation is extremely fast and efficient, especially at fixed interpolation points. One way to perform the interpolation of $p(x)$ from values $p(1), \ldots, p(n)$ is through the LaGrange interpolation formula:

$$p(x) \;=\; \sum_{1}^{n} L_i(x) p(i) \tag{5.1}$$

where the LaGrange polynomial $L_i(x)$ is defined by

$$L_i(x) \;=\; \prod_{j=1..n; j \neq i} \frac{x - j}{i - j}.$$

Each $p(i)$ is a constant in Equation 5.1, so that $p(x)$ is the weighted sum of polynomials $L_i(x)$. This sum is easily and efficiently computable. (Note that even if the degree of $p(x)$ is $t < n - 1$, the polynomial thus interpolated will be identical to $p(x)$; the terms will cancel properly to ensure the final polynomial is of degree $t$.)

The coefficients of the polynomials $L_i(x)$ are easy to compute as well, and need only be computed once. An alternative means to compute them is to compute the inverse of the Vandermonde matrix $M$ defined as $M_{ij} = (\alpha_i)^{j-1}$.

A particularly suitable set of evaluation points is given by $\alpha_i = \omega^{i-1}$, so that $M_{ij} = \omega^{(i-1)(j-1)}$, where $\omega$ is an $n^{th}$ root of unity over the field $E$ used for secret sharing. The evaluation of the polynomial $p(u)$ becomes a *fast Fourier transform*:

$$M\vec{a} = \begin{bmatrix} 1 & 1 & 1 & \cdots & 1 \\ 1 & \omega & \omega^2 & \cdots & \omega^{n-1} \\ 1 & \omega^2 & \omega^4 & \cdots & \omega^{2(n-1)} \\ \vdots & & & & \vdots \\ 1 & \omega^{n-1} & \omega^{(n-1)2} & \cdots & \omega^{(n-1)(n-1)} \end{bmatrix} \cdot \begin{bmatrix} s \\ a_1 \\ a_2 \\ \vdots \\ a_t \\ 0 \\ \vdots \\ 0 \end{bmatrix} = \begin{bmatrix} p(\omega^0) \\ p(\omega^1) \\ p(\omega^2) \\ \vdots \\ p(\omega^{t-1}) \\ p(\omega^t) \\ \vdots \\ p(\omega^{n-1}) \end{bmatrix} = \vec{p}$$

so that the inverse of M gives the desired weights:

$$\vec{a} = M^{-1}\vec{p}.$$

As observed crucially by Ben-Or *et al* [BGW88], the use of these particular evaluation points supports the use of error-correcting codes, which will be essential to defend against malicious errors.

Regardless of the choice of evaluation points, the weights can be precomputed at the cost of a matrix inversion. Using roots of unity as evaluation points, evaluating $p$ and interpolating $p$ is even faster ($O(n\log n)$ time *vs.* $O(n^3)$) since the computations become fast Fourier transforms. Using the precomputed weights in the online interpolation of $p(u)$ costs a matrix multiplication.

### 5.1.1 Verifiability

The astute reader will note that a single error in any of the values will throw off the interpolation and produce an incorrect value for $s$. Certainly, $t+1$ reliable players will have sufficient information to reconstruct $s$ at the appropriate time (the "appropriate" time is specified by their programs, which, since they are reliable, is specified by the protocol designer), but they must be able to distinguish correct pieces from incorrect ones. Furthermore, they ought to be able to tell if the dealer distributed a valid set of pieces in the first place.

This problem is known as *verifiable* secret sharing [CGMA85, BGW88, RB89]. Cryptographic solutions often involve ideas such as digitally signing the pieces to ensure that they aren't changed and zero-knowledge proofs to ensure that they do interpolate to a proper polynomial of degree $t$. In this chapter, however, we are concerned with unconditionally secure protocols: we make no complexity-theoretic assumptions or restrictions, and cannot rely on such cryptographic approaches.

For the range $t < \frac{n}{3}$, Ben-Or *et al* present a technique to ensure verifiability with unconditional security. In Chapter 6 we give a method for verifiable secret sharing for $t < \frac{n}{2}$, which is unconditionally secure with high probability.

Ben-Or *et al* base their methods on BCH codes, using particular values $\alpha_1 = \omega^0, \alpha_2 = \omega^1, \ldots, \alpha_n = \omega^{n-1}$ as the evaluation points for $p(u)$, where $\omega$ is a primitive $n^{th}$ root of unity in the field $E$ used for secret sharing. We shall not list the details of error correction here, instead referring the interested reader to [BGW88, PW72]. The essential property to note is that there is an error-correcting method to interpolate polynomials evaluated at roots of unity that tolerates omissions and changes in up to a third of the values.

For completeness and for the purposes of proving resilience (a proof has not appeared), we describe their method for verifiability using BCH codes. The dealer chooses a bivariate polynomial $p(u, v)$ of degree $t$ in $u$ and $v$, subject to $p(0, 0) = s$, and sends the polynomials $p_i(u) = p(u, \omega^{i-1})$ and $q_i(v) = p(\omega^{i-1}, v)$ to player $i$. The original piece $\text{PIECE}_i(s)$ corresponds simply to $p_i(0)$; the other information is for verification. Then each player $i$ sends $p_i(\omega^{j-1})$ to player $j$. Player $j$ can check for himself that the values $p_1(\omega^{j-1}), \dots, p_n(\omega^{j-1})$ match his values for $q_j(\omega^0), \dots, q_j(\omega^{n-1})$; if not, he requests that the dealer broadcast the true values of $p(i, j)$ at the discrepancies.

Now, if any player detects more than $t$ errors or had to correct one of his values, then he *impeaches* the dealer. An impeachment is a broadcast request by player $i$ to reveal $p_i(u)$ and $q_i(v)$. If a player finds a contradiction with the polynomials broadcast by the dealer, he too impeaches the dealer. Finally, any player seeing $t + 1$ or more impeachments decides that the dealer is faulty, and otherwise accepts the secret, since at least $n - t \geq t + 1$ reliable players have accepted their values and therefore determine a secret uniquely. Figure 5.19 lists the steps involved. The following theorem is proved in §5.6.2.

**Theorem 5.4** *[after Ben-Or, Goldwasser, Wigderson 1988] For $3t < n$, in a completely connected network of $n$ processors with private channels, protocol* VSS *is $t$-private and correct against Byzantine adversaries.*

## 5.2 Fundamentals: Computing With Secrets

Let $C_F$ be a boolean circuit for $F$ over the gates $\{\wedge, \vee, \neg\}$, having inputs $x_1, \dots, x_n$. Such a circuit is represented as an *arithmetic* circuit over field $E$ using the standard mapping $\phi$ taking $x_i^\phi \mapsto x_i, (x \wedge y)^\phi \mapsto xy, (x \vee y)^\phi \mapsto x + y - xy, (\neg x)^\phi \mapsto 1 - x$.

The three-stage paradigm, "Share Inputs; Compute New Secrets; Reconstruct Results," forms the basis for constructing multiparty protocols to compute a function $F(x_1, \dots, x_n)$. The intermediate stage is the backbone of secure multiparty protocols: a collection of subprotocols to add and to multiply secrets, creating new secrets from old, and thus to evaluate $C_F$ gate by gate, secretly. In this section we review and improve techniques of [BGW88] for evaluating $F$ via $C_F$. Formally speaking, for now we consider computing $F$ directly without dividing it into several intermediate functions $F^i$. Our more general techniques will be introduced in §5.3 and §5.4.

### 5.2.1 Linear Combinations

Given secretly shared values $X_1$ and $X_2$, a protocol to create a new secretly shared $Y$ whose value is their sum is easy to construct (see Figure 5.3). Each player computes his new piece as $\text{PIECE}_i(Y) = \text{PIECE}_i(X_1 + X_2) \leftarrow \text{PIECE}_i(X_1) + \text{PIECE}_i(X_2)$. Behind the scenes, if $X_1$ and $X_2$ are shared using polynomials $p_{X_1}(u)$ and $p_{X_2}(u)$, then the polynomial defined by $\text{PIECE}_1(Y), \ldots, \text{PIECE}_n(Y)$ is $p_Y(u) = p_{X_1}(u) + p_{X_2}(u)$, a polynomial of degree $t$ and with uniformly random coefficients subject to $p_Y(0) = Y = X_1 + X_2$. This protocol requires no communication. Noting that $(c \cdot p)(X) = c \cdot p(X)$ is uniformly distributed and of degree $t$ with free term $c \cdot p(0) = cX$, it easily generalizes to multiplying secrets by fixed constants and hence to linear combinations of secrets. To effect a constant multiplication, each player multiplies his piece by that constant, since $(c \cdot p)(\alpha_i) = c \cdot p(\alpha_i)$.

---

LINEAR-COMBINE-ONE

**For $i = 1..n$ do in parallel**

    Player $i$ sets

        $\text{PIECE}_i(Y) \leftarrow c_0 + c_1 \cdot \text{PIECE}_i(X_1) + \cdots + c_N \cdot \text{PIECE}_i(X_N)$.

Figure 5.3: Protocol to create a new secret $Y = c_0 + c_1 X_1 + \cdots + c_N X_N$, where $c_0, \ldots, c_N$ are fixed constants and $X_1, \ldots, X_N$ are secret.

---

The protocol to compute several linear combinations simultaneously is simple: repeat the single linear combination protocol in parallel. No interaction is needed. Figure 5.4 describes the extension.

---

LINEAR-COMBINE $(\{c_{jl}\}, \{X_{jl}\})$

**For $i = 1..n$ do in parallel**

    **For $j = 1..M$ do**

        Player $i$ sets

        $\text{PIECE}_i(Y_j) \leftarrow c_{j0} + c_{j1} \cdot \text{PIECE}_i(X_{j1}) + \cdots + c_{jN} \cdot \text{PIECE}_i(X_{jN})$.

Figure 5.4: Protocol to create new secrets $Y_1, \ldots, Y_M$ whose values are $Y_j = c_{j0} + c_{j1} X_{j1} + \cdots + c_{jN} X_{jN}$. The $\{c_{jl}\}$ values are fixed ("public") constants and the the $\{X_{jl}\}$ are secrets.

---

**Lemma 5.5** *Protocol* LINEAR-COMBINE *is a $t$-resilient protocol to compute a robust and private representation* `hide`$(G)$ *where*

$$G(\{c_{jl}\}, \{X_{jl}\}) = (c_{10} + \sum_1^N X_{1l}, \ldots, c_{M0} + \sum_1^N X_{Ml}).$$

**Proof:** An interface for the protocol is trivial, since no messages need be generated. The interface need only request the inputs and auxiliary inputs of the players that its black-box adversary $A$ requests to corrupt, and supply the values to $A$. Post-protocol corruption is accomplished by the same process. It is easy to see that the results of the protocol are the correct secretly shared values. ∎

## 5.2.2  Multiplying Secrets

The protocol for adding secrets does not immediately generalize to multiplication, since the product $p_Y(u) = p_{X_1}(u)p_{X_2}(u)$ has degree $2t$. If the degree of the representation of the secrets is allowed to grow, there will quickly by an insufficient number of pieces to reconstruct the result.

Ben-Or *et al* and Chaum *et al* present the first methods to reduce the degree of the polynomial represented by having each player compute $\mathrm{PIECE}_i(X_1) \cdot \mathrm{PIECE}_i(X_2)$. They reduce the problem of degree reduction to a linear combination of secrets using matrix operations. Before the degree reduction occurs, a random polynomial of degree $2t$ (and zero free term) is added in order to ensure a uniform distribution on the resulting coefficients.

We present an alternative and independently discovered approach to degree reduction and a faster method for ensuring uniform randomness of the resulting coefficients. These minor optimizations are not the main thrust of this chapter but are included because they also have certain conceptual advantages.

We have seen in Equation 5.1 how a polynomial can be interpolated using a weighted sum of LaGrange polynomials. Let $\overline{p}(u)$ denote the result of truncating all terms of polynomial $p(u)$ having degree higher than $t$; that is, $\overline{p}(u) = p(u) \bmod x^{t+1}$. Then we have

$$\overline{p}(x) \;\; = \;\; \sum_1^n \overline{L}_i(x)p(i). \tag{5.2}$$

Clearly, then, we can express the value of $\overline{p}(u)$ at the point $\alpha_j$ as a weighted sum of the values $p(1), \ldots, p(n)$ :

$$\overline{p}(\alpha_j) \;\; = \;\; \sum_1^n \overline{L}_i(\alpha_j)p(i). \tag{5.3}$$

The weights, $\overline{L}_1(\alpha_j), \ldots, \overline{L}_n(\alpha_j)$, are easily precomputed. Furthermore, $\overline{p}(0) = p(0)$.

If the values $p(1), \ldots, p(n)$ are themselves shared as secrets, then Equation 5.3 indicates how to express each value $\overline{p}(\alpha_j)$ as a linear combination of secrets. Thus, $\overline{p}(u)$ could be used as a secret-sharing polynomial to represent the secret $p(u)$, since it has degree $t$. Unfortunately, the coefficients of $\overline{p}(u)$ are not uniformly random unless the original coefficients of $p(u)$ are.

To randomize the coefficients, it suffices to add a random polynomial of degree $t+1$ and free term 0 to $p(u)$ before truncation. This observation differs from [BGW88], who use $t$ different polynomials to construct a polynomial of degree $2t$ in order to randomize *all* high-order coefficients. However, it is necessary only to randomize coefficients of degree $t$ or less, since the others will be truncated. Their protocol hence requires a $t$-fold increase in the expense of this subprotocol.

If $r(u)$ is completely random and of degree $t$, then $u \cdot r(u)$ is of degree $t+1$ with free term 0 as desired. To create a random polynomial $r(u)$ (whose free term is random), each player $i$ shares a secret $R_i$ using a polynomial $r_i(u)$. Their sum, $r(u) = \sum_i r(u)$, is what we need; each player $j$ holds the value $r(j) = \sum_i r(j) = \sum_i \text{PIECE}_j(R_i)$. The truncation protocol is given in Figure 5.5.

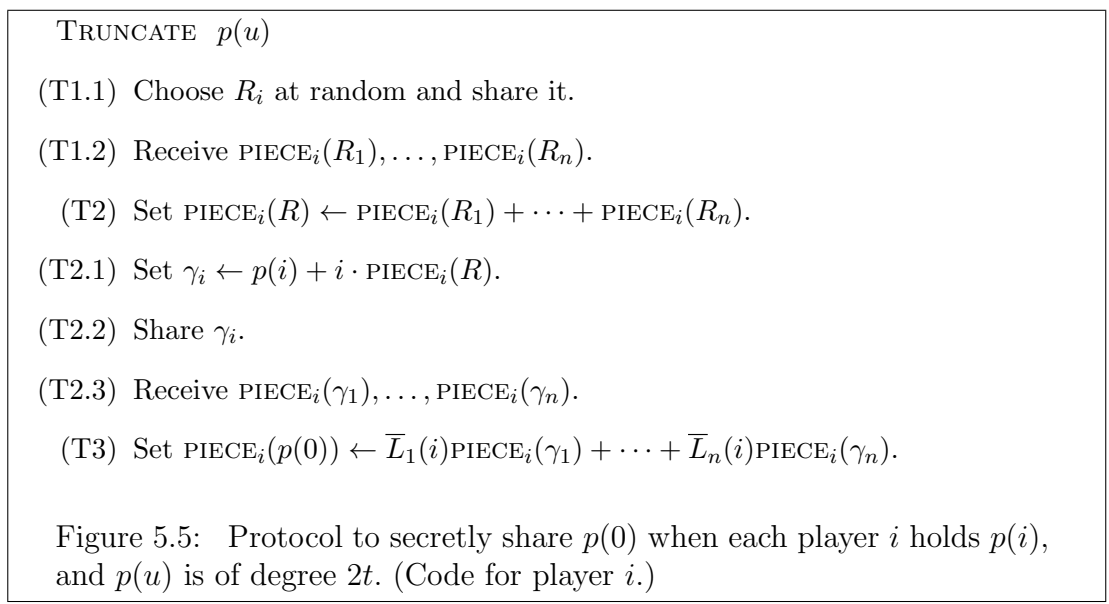

> TRUNCATE  $p(u)$
>
> (T1.1)  Choose $R_i$ at random and share it.
>
> (T1.2)  Receive $\text{PIECE}_i(R_1), \ldots, \text{PIECE}_i(R_n)$.
>
> (T2)  Set $\text{PIECE}_i(R) \leftarrow \text{PIECE}_i(R_1) + \cdots + \text{PIECE}_i(R_n)$.
>
> (T2.1)  Set $\gamma_i \leftarrow p(i) + i \cdot \text{PIECE}_i(R)$.
>
> (T2.2)  Share $\gamma_i$.
>
> (T2.3)  Receive $\text{PIECE}_i(\gamma_1), \ldots, \text{PIECE}_i(\gamma_n)$.
>
> (T3)  Set $\text{PIECE}_i(p(0)) \leftarrow \overline{L}_1(i)\text{PIECE}_i(\gamma_1) + \cdots + \overline{L}_n(i)\text{PIECE}_i(\gamma_n)$.

Figure 5.5:  Protocol to secretly share $p(0)$ when each player $i$ holds $p(i)$, and $p(u)$ is of degree $2t$. (Code for player $i$.)

Given a protocol to truncate a high-degree polynomial secretly, the protocol to compute the product of two secrets can be stated concisely (see Figure 5.7). Simply: each player computes the product of his pieces and participates in a truncation protocol, which involves secretly sharing $\text{PIECE}_i(X_1)\text{PIECE}_i(X_2)$ itself.

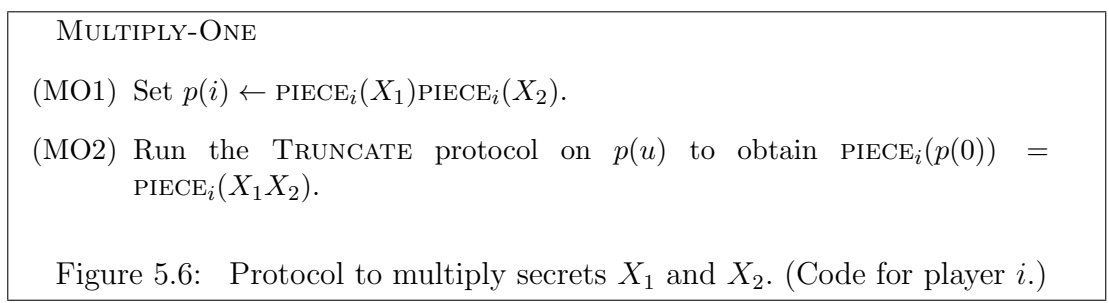

> MULTIPLY-ONE
>
> (MO1)  Set $p(i) \leftarrow \text{PIECE}_i(X_1)\text{PIECE}_i(X_2)$.
>
> (MO2)  Run the TRUNCATE protocol on $p(u)$ to obtain $\text{PIECE}_i(p(0)) = \text{PIECE}_i(X_1X_2)$.

Figure 5.6:  Protocol to multiply secrets $X_1$ and $X_2$. (Code for player $i$.)

## Byzantine Faults

As before, we must be careful to check for misbehavior. The protocols as listed are secure against a passive adversary but are not reliable in the presence of Byzantine

---

MULTIPLY

**For $j = 1..M$ do in parallel**
      Execute protocol MULTIPLY-ONE to compute $Y_j = X_{j1} \cdot X_{j2}$.

Figure 5.7: Protocol to create new secrets $Y_1, \ldots, Y_M$ whose values are $Y_j = X_{j1} \cdot X_{j2}$. The $\{X_{j1}, X_{j2}\}_{j \in [M]}$ are secrets. (Code for player $i$.)

---

faults.

Since the error-correcting codes (polynomials evaluated at roots of unity) are additive, the addition of verified secrets need not be checked. Multiplication, however, requires the resharing of pieces; it must be checked that player $i$ correctly shares $p(i) = \text{PIECE}_i(X_1)\text{PIECE}_i(X_2)$. In this section we review the method of [BGW88]; in Chapter 6, we present a new method to achieve this goal that tolerates a much larger number of faults.

**The ABC problem.** Alice knows the values of secrets $a$ and $b$. Alice must share a new secret $c$ and prove to the other players that the secret value of $c$ is indeed $ab$. If Alice is honest, no information about $a$, $b$, or $c$ should be revealed.

A solution to this problem is detailed in Figure 5.8. Notice that if $f(0)g(0)$ differs from $h(0)$, then it is impossible for Alice to select polynomials of degree $t$ to make $h(x)$ of degree $t$. The use of random coefficients ensures that if Alice is nonfaulty, then an interface can easily generate messages from Alice to faulty players, because these messages will contain uniformly random field elements, according to Lemma 5.2.

The protocol specifies twice that Alice participate in a VSS protocol using particular polynomials or pieces, in order that the system may verify that Alice uses polynomials of degree $t$. In other words, the subprotocols correspond to executing VSS but requiring that Alice send out $(p_i(u), q_i(v))$ values in the first round that match the given polynomials. The recipients of these messages check in the first round whether Alice has sent them consistent values, and if not, they consider Alice to have sent them nothing, and continue the VSS protocol exactly as specified from there.

**Lemma 5.6** *Protocol* MULTIPLY *is a t-resilient protocol to compute a robust and private representation* $\texttt{hide}(G)$ *of the produce of $M$ secrets, where:*

$$G(X_{11}, X_{21}, \ldots, X_{M1}; X_{12}, X_{22}, \ldots, X_{M2}) = (X_{11} \cdot X_{12}, \ldots, X_{M1} \cdot X_{M2}).$$

**Proof:** Step (MO1) of MULTIPLY-ONE is noninteractive and hence trivially $t$-resilient. The TRUNCATE protocol is $t$-resilient because it is the concatenation of $t$-resilient protocols (VSS and LINEAR-COMBINE ). Protocol MULTIPLY is a concatenation of two $t$-resilient subprotocols, namely step (MO1) and the TRUNCATE protocol, so by Theorem 4.31 it is itself $t$-resilient. ∎

Prove-Product **-bgw**$(s)$

(A1)  Alice:  **For** $i = 1..t$ **for** $j = 0..t - 1$ **do**
$$r[i, j] \leftarrow \textbf{\textit{uniform}}(E)$$
**For** $i = 1..t$ **do**
$$h_i(x) = \sum_{j=0}^{t-1} r[i, j] x^j +$$
$$x^t \cdot (c_{t+i} - \sum_{j=1}^{t-1} r[t + 1 - j, j])$$
Alice shares $c_t, \ldots, c_{2t}$ using these polynomials.

(A2)  $(1 \leq i \leq n)$  $h(\alpha_i) = f(\alpha_i) g(\alpha_i) - \sum_{i=1}^{t} (\alpha_i)^t h_i(\alpha_i)$.

(A3)  Alice secretly shares $ab$ using these as $p_i(0)$ values.

Figure 5.8:  Protocol for Alice to verifiably share secret $c = ab$.

## 5.2.3  Disqualification and Fault Recovery

Rather than burden the descriptions of the protocols to come, we implicitly require that, following an impeachment, namely a broadcast message from one player $i$ stating that some player $j$ is faulty, each player checks to see if the impeached player should be disqualified. We consider a processor *disqualified* if $t + 1$ or more players have broadcast impeachments of it. The meaning of "checks to see" is determined by the specifications of the particular protocol. If the player is disqualified, a recovery procedure may be necessary. In this chapter, when the number of faults is less than a third, no recovery procedure is necessary (though of course, misbehavior is eliminated by eliminating the messages of faulty players identified by error-correction) except in the Input stage.

If a fault occurs during the Input stage, namely when a player ought to share its input, the faulty player's secret is replaced by a default value selected (for full generality) according to some samplable distribution: in addition to sharing the input $x_i$, each player $i$ shares a set of random bits which are used to construct a default input for player $j$ should it fail. Let $T$ be the set of players disqualified when sharing the inputs. Then it is easy to describe a set of circuits $\{C_j\}_{j \in T}$ which take as inputs the random bits shared by players $i \notin T$, compute their exclusive or, and output default values for each $x_j$. Rather than evaluating $C_F$ directly, the protocol calls for an evaluation of the circuit $C'_F$ which is the circuit $C_F$ with its $i^{th}$ input set to either the output of $C_i$ or to the value $x_i$ shared by player $i$; the circuit $C'_F$ selects one or the other by considering the sum of extra inputs from each player specifying whether the player impeaches player $i$. If the sum is large enough, the circuit selects the default input, and otherwise selects $x_i$.

We shall normally consider these conditions implicitly in our descriptions of the protocols, for otherwise they would be tremendously difficult to read.

### 5.2.4 Putting It Together

Given subprotocols to add and to multiply secrets, the specification of a protocol to evaluate an arbitrary arithmetic circuit $C_F$ is not hard to describe. Let $E$ be a fixed finite field. A basis for a circuit is a set of functions which the gates may compute. An arithmetic gate is an element of the set $\{\times\} \cup \{(+, a_0, a_1, \ldots, a_m)\}_{a_0, \ldots, a_m \in E}$ where the latter sort of gate is a *linear combination*, producing $a_0 + a_1 x_1 + \cdots + a_m x_m$ on inputs $x_1, \ldots, x_m$.

For ease of description, we shall consider circuits as arrays of gates; this allows us to specify circuits with multiple outputs or to describe straight-line programs (circuits of width 1) as we desire. It also allows us to specify in a natural way that the outputs of particular gates at the final level are given to distinct players. Specifically, a circuit is a $(d+1) \times w$ array of elements of the form $(g_{ij}, \mathtt{in}(g_{ij}))$, where $\mathtt{in}(g_{ij})$ is a set of pairs $(a, b)$ describing the indices of gates whose output leads to $g_{ij}$. We define $\mathtt{in}_k(g_{ij})$ to be $(a, b)$ if $g_{ab}$ is the $k^{th}$ earliest gate (in row-major order) leading into $g_{ij}$. Each input $g_{ab}$ to gate $g_{ij}$ must satisfy $a < i$. The $0^{th}$ layer represents the inputs $x_1, \ldots, x_n$ to the circuit. Let $\mathtt{out}(i)$ denote the set of gates $g_{dj}$ whose output represents the value $F_i$ that player $i$ should learn.

Without loss of generality we assume that even-numbered levels contain only linear combination gates and odd-numbered levels contain only multiplication gates. The depth of the circuit is doubled at most. Denote the coefficients of the additive gate $g_{ij}$ by $a_{i,j,0}, a_{i,j,1}, \ldots, a_{i,j,|\mathtt{in}_{g_{ij}}|}$.

A circuit with *bounded fan-in* satisfies $|\mathtt{in}(g_{ij})| < c$ for some constant $c$ and for all $i, j$. A circuit with *bounded multiplicative fan-in* satisfies $|\mathtt{in}(g_{ij})| < c$ for some constant $c$ and for all $i, j$ such that $g_{ij} = \times$. A $\times$-fanin-2 circuit has multiplicative fan-in of 2.

The class $NC^1$ consists of functions with unbounded fan-in polynomial-size logarithmic-depth circuit families, over the basis $\{\mathrm{AND}, \mathrm{OR}, \mathrm{NOT}\}$. The class $ANC^1$ is described by polynomial-size logarithmic-depth circuit families, over the arithmetic basis $\{(+, a_0, a_1, \ldots, a_m), \times\}$. We consider a circuit *efficient* if it is of polynomial size and polynomial depth.

**Theorem 5.7** *(After [BGW88]) Let $\{F^{n,m}\}$ be a family of functions described by a $\times$-fanin-2 circuit family $C_{F^{n,m}}$. Then for $3t < n$, there exists a $t$-**resilient** protocol leaking $F^{n,m}$. This protocol requires a polynomial number of bits (in the size of $C_{F^{n,m}}$) and $O(\mathtt{depth}(C_{F^{n,m}}))$ rounds of interaction.*

**Proof:** The protocol (see Figure 5.9) consists of secretly sharing all the inputs, then evaluating $C_{F^{n,m}}$ layer by layer, maintaining the gate outputs as secrets. At each layer, all gates are evaluated in parallel. Finally, for $1 \leq i \leq n$, each output secret in $\mathtt{out}(i)$ is reconstructed for player $i$.

By lemmas 5.5 and 5.6 and Theorem 4.31, the overall protocol is $t$-resilient, since it is a concatenation of perfectly $t$-resilient protocols for robust and private functions.

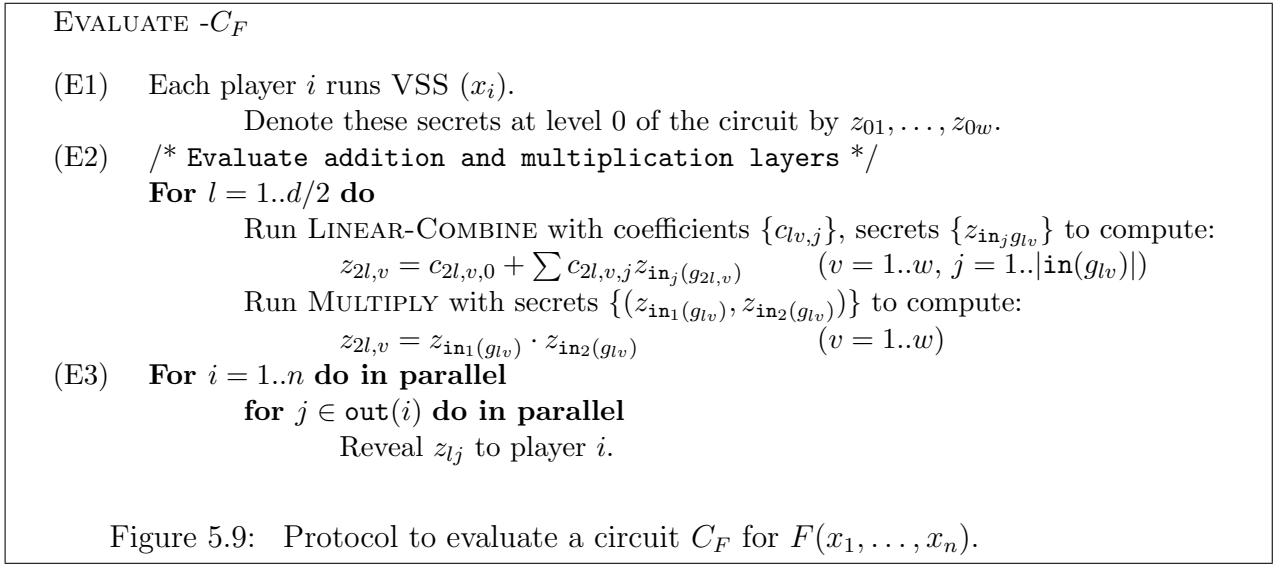

Figure 5.9: Protocol to evaluate a circuit $C_F$ for $F(x_1, \ldots, x_n)$.

## 5.2.5 Some Applications

Protocols for problems such as taking a secret ballot are now relatively easy to describe. Chapter 15 also describes some simple protocols for useful problems, though those protocols have been optimized for efficiency using various tricks not evident in the preceding analysis. Let us examine how, at this stage, secure multiparty protocols may be constructed from simple gate operations.

**Example 5.8 Secret Ballot.**

A *secret ballot* is a private and reliable distributed computation of the sum of 0/1 inputs. Let us give a protocol for taking a secret ballot which is private, correct, ensures that the voters cast votes independently, and reveals the result to everyone, even when up to a third of the voters may be untrustworthy.

The protocol is straightforward, given the secret linear-combination and multiplication protocols. Each voter casts his vote $X_i$ by secretly sharing it. To ensure that each voter cast at most one vote, they secretly compute $Z_i = X_i(X_i - 1)$ for each $i$, and reconstruct the result for everyone. Any voter whose value is nonzero is disqualified. Let $J$ be the set of voters who are not disqualified. Then, together, the voters run the LINEAR-COMBINE protocol to compute $Y = \sum_{i \in J} X_i$. Finally, they reconstruct the result $Y$ for everyone.

It is easy to see that revealing $Z_i$ gives no information about the vote cast by reliable voter $i$. That is, the result for reliable players is *always* 0, and hence a private and robust function. Furthermore, this protocol is fast and efficient, requiring one multiplication and two additions, which take a small constant number of rounds.

**Example 5.9 Unanimous Vote.**

A slightly more complicated goal is to decide if an entire committee votes unanimously or not, without revealing the tally if the committee is not in agreement. As in Example 5.8, the voters secretly compute $Y$, the tally of the valid 0/1 votes. Next, they compute the function $f(y) = 1 - (n - y)^{p-1}$, where $p$ is the size of the field being used for sharing. Notice that $f(n) = 1$ whereas $f(y) = 0$ for $y \neq n$. Finally they reveal $f(y)$ to everyone.

An algebraic circuit of depth $\log p$ suffices to compute $f(y)$, noting that the multiplicative gates are restricted to fan-in 2. If we take $p$ to be on the order of $n$, then the depth is roughly $O(\log n)$. The number of rounds required to perform the unanimous ballot is $O(\log n)$, and the number of bits is polynomial in $n$. This is a prime example of a protocol whose running time is made efficient through our methods for constant-rounds protocols.

---

UNANIMOUS-VOTE

- Each voter $i$ shares $X_i$.

- Compute and reveal $Z_i = X_i(X_i - 1)$ for $1 \leq i \leq n$. Let $J$ be the set of $i$ for which the result is 0.

- Compute $Y = \sum_{i \in J} X_i$.

- Compute $V_1 = Y - n$ and $W_1 = Y - n$.

- **For** $j = 1..\lfloor \log p \rfloor$ **do**

    - Compute $W_{i+1} = W_i^2$.
    - If the $i^{th}$ bit of the binary expansion of $p - 1$ is 1, compute $V_{i+1} = V_i W_{i+1}$, else let $V_{i+1} = V_i$.

- Reconstruct $V_{\log p}$. If 0, output "unanimous," else output "disagreed."

Figure 5.10: Protocol to decide if secret ballot is unanimous without revealing the tally. (Network protocol.)

---

## 5.3 Tools for Efficient Protocols

In this section we present a collection of useful and important subprotocols. The technique for unbounded fan-in multiplication presented in §5.3.5 is the critical tool for developing protocols that require a constant number of rounds as opposed to a number of rounds proportional to the depth of a circuit for $F$.

### 5.3.1 Random Secret Values

We often have need of secretly shared random field elements whose value no player knows. That is, we desire a secret whose value is uniformly random and independent of the information of any $t$ or fewer players.

---

RAND-SECRET ($R$)

- Each player $i$ chooses $R_i$ uniformly at random from the field $E$ and runs VSS to share it. The default value is 0 for misbehaving players.

- Run LINEAR-COMBINE to compute $R = R_1 + \ldots + R_n$.

Figure 5.11: Protocol to create random secret field element $R$.

---

Random secret bits are also useful. They are generated simply by computing the parity of 0/1-valued secrets shared by each player. Over a field of characteristic 2, one computes the parity by computing simply the sum, requiring no interaction beyond the secret sharing. Over other fields, a logarithmic-depth circuit is needed (though 5.16 shows how to perform this in constant rounds). Before computing the parity, each shared bit is verified to be 0 or 1 by computing $b_i(b_i - 1)$, which is 0 iff $b_i$ is 0 or 1, and which is a private function since its value is independent of $b_i$.

---

RAND-BIT ($b$)

- Each player $i$ sets $b_i \leftarrow \boldsymbol{uniform}(\{0, 1\})$ and shares it.

- Run LINEAR-COMBINE and MULTIPLY to compute $c_i = b_i(b_i - 1)$.

- Run LINEAR-COMBINE to compute $b = \sum_{c_i = 0} b_i$.

Figure 5.12: Protocol to create random secret field element $b$ in a field of characteristic 2.

---

### 5.3.2 Groups and Inverses

The best known algebraic circuit to compute multiplicative inverses in a field is fairly deep. It turns out, surprisingly, that the number of rounds required to secretly compute the multiplicative inverse of a secret is the same as that needed to multiply two elements, despite the large number of rounds that would be required by simulating an arithmetic circuit directly. The power of this result will be evident in the next section, where we shall use it to construct a protocol to invert matrices, which in turn will support a constant-round protocol for any $NC^1$ circuit.

**Theorem 5.10 (Constant Rounds for Group Inverses)** *Let $G$ be a group, and let $X \in G$ be secretly shared. Let $\Pi_M$ be a protocol to multiply two elements in $G$ that runs in $T_M$ rounds and uses $C_M$ bits; let $\Pi_R$ be a protocol to generate a random element in $G$ that runs in $T_R$ rounds and uses $C_R$ bits. Let $T = max(T_M, T_R)$ and $C = max(C_M, C_R, C_{share})$, where $C_{share}$ is the number of bits used to secretly share a group element.*

*Then there is a protocol to secretly compute $X^{-1}$ using $O(T)$ rounds and $O(C)$ bits. Specifically, if multiplication of two elements requires constant rounds then inversion requires only constant rounds.*

**Proof:** The protocol is exhibited in Figure 5.13. Its security rests on the elementary observation that, for any $X$, the distribution induced by multiplying by a uniformly random field element is uniform over the group. In other words, the intermediate function $F^1(X) = (V, U)$ is private and robust, since the public results ($V$) have the same distribution regardless of $X$, and the other results are secretly shared; so the computation of $F^1$ is as easy to simulate as the ideal vacuous protocol. The function $F^2$ is such that $F^2(V, U) = V^{-1}U = X^{-1}$ and $F^2$ admits a resilient protocol $\Pi_M$. Thus $F = F^2 \Diamond F^1$ and Theorem 4.31 applies.

---

INVERT $(X)$

- Run $\Pi_R$ to secretly generate a random secret element $U \in G$.

- Run $\Pi_M$ to secretly compute $V \leftarrow UX$.

- Reconstruct $V$ for every player.

- Each player $i$ computes $V^{-1}$ individually. It can now be treated as a fixed public constant.

- Secretly compute $Y \leftarrow V^{-1}U$.

Figure 5.13: Protocol to compute secret inverse of a secret group element.

---

### 5.3.3 Secret Matrices and Matrix Inversion

A *secret $n \times n$ matrix* is a collection of $n^2$ secrets, each representing an element of the matrix. Thus each player holds $n^2$ pieces, one of each entry. Figure 5.14 shows how to multiply matrices secretly.

Each entry of the inverse of a $3 \times 3$ matrix $M$ is easily expressed as a small, constant-depth circuit applied to the nine entries of $M$. By Theorem 5.7, we have the following:

---

MATRIX-MULTIPLY $(A, B, C)$

- **For** $i = 1..\alpha$ **do in parallel** **for** $j = 1..\gamma$ **do in parallel** run LINEAR-COMBINE to compute

$$C_{ij} = \sum_{k=1}^{\beta} A_{ik} B_{kj}.$$

Figure 5.14: Protocol to secretly multiply secret matrices $A$ and $B$ of dimensions $\alpha \times \beta$ and $\beta \times \gamma$, respectively.

---

**Corollary 5.11** *There exists a protocol to invert a full-rank secret $3 \times 3$ matrix using a polynomial number of bits and a fixed constant number of rounds.*

Given a constant-rounds protocol $\Pi_R$ to generate a random $n \times n$ secret matrix of full rank, Theorem 5.10 implies the following result:

**Corollary 5.12** *There exists a protocol to invert a full-rank secret $n \times n$ matrix using a polynomial number of bits and a constant expected number of rounds.*

**Proof:** Consider the group $G$ of full rank matrices under multiplication. To generate random $n \times n$ secret matrices having full rank, it suffices to generate a pair $(R, S)$ of uniformly random secret matrices such that $RS$ has full rank. See Figure 5.15. Because $S$ is therefore uniformly random of full rank, the distribution on $U = RS$ is uniformly random over full-rank matrices regardless of the value of $R$. Hence the desired function (generate a full rank random matrix) can be written as the composition of two private and robust functions, the first computing $U$ and the second computing the full rank matrix. Because the probability of generating a uniformly random matrix of full rank is constant, the expected number of rounds is constant. Clearly, a single repetition suffices with high probability if several ($k$) of these pairs are generated and tested in parallel, and the first qualifying pair is used. Because the subprotocol to generate random group elements uses expected constant rounds, the overall protocol is *expected* constant rounds. This differs from the *fixed*-size matrix inverse problem, whose solution requires *fixed* constant rounds.

∎

## 5.3.4   Random Inverse Pairs and Large Fields

The following theorem describes another useful tool for a computation for which no shallow circuit is known: inverting secret *field* elements. In order to avoid revealing a 0-valued secret because of a failed attempt to invert it, we extend the multiplicative inverse so that $0^{-1} = 0$.

---

FULL-RANK-RANDOM-MATRIX $(X)$

(RM1) Generate uniformly random secret matrices $R, S$, using the RAND-SECRET protocol to create each entry.

(RM2) Secretly compute $U = RS$ and reveal the result. If $U$ has full rank, use $R$. Otherwise go to step 1.

Figure 5.15: Protocol to generate random secret matrix of full rank.

---

**Theorem 5.13 (Constant Rounds for Multiplicative Field Inverses)** *Let $E$ be a field, and let $X \in E$ be secretly shared. Let $\Pi_M$ be a protocol to multiply or add two elements in $E$ that runs in $T_M$ rounds and uses $C_M$ bits; let $\Pi_R$ be a protocol to generate a random element in $E$ that runs in $T_R$ rounds and uses $C_R$ bits. Let $T = max(T_M, T_R)$ and $C = max(C_M, C_R, C_{share}, |E|^4)$, where $C_{share}$) is the number of bits used to secretly share a group element.*

*Then there is a protocol to secretly compute $X^{-1}$ using $O(T)$ rounds and $O(C)$ bits. Specifically, if multiplication of two elements requires constant rounds then inversion requires only constant rounds.*

**Proof:** The protocol is exhibited in Figure 5.16. The distribution on $U_i$ and $V_i$ is clearly independent of $X$. Notice that the distribution on $R_i$, given $U_i = V_i = 0$, is uniform over *all* field elements, *including 0*. Hence $X - R_i$ is uniform over all field elements regardless of $X$, and the intermediate results have the same, uniform distribution for any $X$. It is not hard to see that pairs of extended inverses are the only pairs to give zeros in both tests in step (FI2). Thus, the inverse function $F$ is written as the composition of private and robust intermediate functions, and Theorem 4.31 applies. Choosing $|E|^4$ random pairs ensures that with probability at least $1 - 2^{-|E|}$ all such pairs will appear and that step (FI5) will succeed.

∎

## 5.3.5   Iterated Multiplication

The final set of tools provide the crucial support for reducing the number of rounds of interaction. We show the surprising result that large fan-in multiplication of group elements can be performed in constant rounds. A naive approach would require $\log N$ rounds, where $N$ is the number of elements to multiply. Specifically, we show:

**Theorem 5.14 (Constant Rounds for Iterated Multiplication)** *Let $G$ be a group, and let $X_1, \ldots, X_N \in G$ be secretly shared. Let $\Pi_M$ be a protocol to multiply two elements in $G$ that runs in $T_M$ rounds and uses $C_M$ bits; let $\Pi_R$ be a protocol to generate a random element in $G$ that runs in $T_R$ rounds and uses $C_R$ bits. Let $T = max(T_M, T_R)$ and $C = N \cdot max(C_M, C_R, C_{share})$, where $C_{share}$ is the number of bits used to secretly share a group element.*

---

FIELD-INVERT $(X)$ { Invert a secret field element.}

(FI1)  Generate $|E|^4$ random secret pairs $(R_i, S_i)$.

(FI2)  For each $1 \leq i \leq |E|^4$, secretly compute

$$\begin{aligned} U_i &= R_i(1 - R_i S_i), \\ V_i &= S_i(1 - R_i S_i). \end{aligned}$$

(FI3)  Reveal all the $U_i$ and $V_i$.

(FI4)  For each $i$ such that $U_i = V_i = 0$, compute and reveal $X - R_i$.

(FI5)  Call $Y$ the first secret $S_i$ such that
       $X - R_i = 0$. If none exists, go to step 1.

Figure 5.16:   Protocol to compute secret multiplicative inverse of a secret in a field $E$.

---

*Then there is a protocol to secretly compute $Y = X_1 \cdots X_N$ using $O(T)$ rounds and $O(C)$ bits. Specifically, if multiplication of two elements requires constant rounds then multiplication of $N$ elements requires only constant rounds.*

**Proof:** Figure 5.17 describes the protocol. The following demonstrates that $Y$ is the desired product:

$$\begin{aligned} Y &= R_0 S R_N^{-1} \\ &= R_0 R_0^{-1} X_1 R_1 R_1^{-1} X_2 \cdots X_j R_N R_N^{-1} \\ &= X_1 \cdots X_N \end{aligned}$$

Since each $R_j$ is generated uniformly at random, and since each $R_j$ and each $X_j$ is invertible, revealing $S_1, \ldots, S_N$ gives no information about $X_1, \ldots, X_N$. That is, the list of elements $(S_1, \ldots, S_N)$ is distributed uniformly at random, given by the following easy lemma:

**Lemma 5.15** *Let $G$ be a group. For any $X_1, \ldots, X_N \in G$,*

$$\begin{aligned} \texttt{uniform}(G^N) = \quad &\{(R_0, \ldots, R_N) \leftarrow \texttt{uniform}(G^{N+1})); \\ &S_1 \leftarrow R_0 X_1 R_1^{-1}; \ldots; S_N \leftarrow R_{N-1} X_N R_N^{-1} : (S_1, S_2, \ldots, S_N)\} \end{aligned}$$

Formally speaking, protocol ITERATED-MULTIPLY computes a four robust and private intermediate functions. The first function $F^1$ supplies each player with pieces of random group elements; the second, $F^2$, provides pieces of their inverses; the third, $F^3$, generates a public (hence robust) uniformly distributed (hence private) vector of

---

ITERATED-MULTIPLY $(X_1, \ldots, X_N)$

(IM1)  Generate $N+1$ secret uniformly random group elements $R_0, \ldots, R_N \in G$.

(IM2)  Secretly compute the inverses, $R_0^{-1}, \ldots, R_N^{-1}$.

(IM3)  For $j = 1, \ldots, N$, simultaneously compute the following new secrets:

$$S_j = R_{j-1}^{-1} X_j R_j.$$

(IM4)  Reveal all the secret elements $S_j$.

(IM5)  Each player privately computes

$$S = S_1 \cdots S_N.$$

(IM6)  Secretly compute $Y = R_0 S R_N^{-1}$.

Figure 5.17:   Protocol to compute the product of several elements in a field $E$.

---

values $(S_1, \ldots, S_N)$; and the fourth provides pieces of $Y$. Theorem 4.31 implies the resilience of ITERATED-MULTIPLY .

Applying Theorem 5.10, we need only $O(C)$ rounds to invert each secret $R$ in (IM2). Steps (IM3) and (IM6) use constant depth circuits and are easily performed with a constant number of rounds of interaction. Step (IM4) requires a round of interaction and step (IM5) requires none.   ∎

## 5.4   The Power of Iterated Multiplication

In this section we prove the crucial and surprising result that is the focus of this chapter. The result depends critically on the ability to perform iterated multiplication of secret $3 \times 3$ matrices; Theorem 5.14 of the previous section is key. Let $ANC^1$ denote the class of functions which can be written as polynomial size algebraic formulas, or logarithmic depth circuits, over a fixed finite field $E$. (The abbreviation derives from "Algebraic NC.") The standard approach of simulating a circuit suggests that the number of rounds required to evaluate $F \in ANC^1$ grows unboundedly, and it was conjectured that a logarithmic lower bound on the number of rounds would hold. We prove to the contrary that a fixed number of rounds suffice.

**Theorem 5.16** *For $3t < n$ and any function $F \in ANC^1$ (or $F \in NC^1$), there exists a $t$-resilient protocol to compute $F$ in a constant number of rounds, with polynomial message sizes.*

**Proof:** In [Bar86] Barrington showed that $NC^1$ is equivalent to multiplying polynomially-many permutations of 5 elements. Ben-Or and Cleve [BC88] generalized this to show that computing polynomial-size algebraic formulas (complete for $ANC^1$) over a field $E$ is equivalent to multiplying polynomially-many $3 \times 3$ matrices over that field. Our method for computing $ANC^1$ was facilitated by a table-chaining technique suggested by M. Rabin [Rab88].

Let $F$ be representable by an algebraic formula $\mathcal{F}$ of depth $d$ having variables $X_1, \ldots, X_n$. By [BC88], there is a sequence of $3 \times 3$ matrices $M_1, \ldots, M_{p(d)}$ whose product $M[\mathcal{F}]$ contains $F(X_1, \ldots, X_n)$ in the upper right entry $(1, 3)$. Here, $p(d) = O(4^d)$.

For completeness we describe the sequence of matrices corresponding to $\mathcal{F}$. Let

$$
J_1 = \begin{bmatrix} 0 & 1 & 0 \\ \text{-1} & 0 & 0 \\ 0 & 0 & 1 \end{bmatrix} \quad
J_2 = \begin{bmatrix} 0 & 0 & \text{-1} \\ 1 & 0 & 0 \\ 0 & 1 & 0 \end{bmatrix} \quad
J_5 = \begin{bmatrix} \text{-1} & 0 & 0 \\ 0 & 0 & 1 \\ 0 & 1 & 0 \end{bmatrix}
$$

$$
J_3 = \begin{bmatrix} 0 & 1 & 0 \\ 0 & 0 & 1 \\ 1 & 0 & 0 \end{bmatrix} \quad
J_4 = \begin{bmatrix} 0 & 0 & 1 \\ \text{-1} & 0 & 0 \\ 0 & 1 & 0 \end{bmatrix}
$$

Furthermore, let $M[f] = \begin{pmatrix} 1 & 0 & f \\ 0 & 1 & 0 \\ 0 & 0 & 1 \end{pmatrix}$. Constants and variables are represented by $M[c]$ and $M[x_i]$. The product $M[\mathcal{F}]$ is derived from the following observations:

$$
\begin{aligned}
M[f + g] &= M[f] \cdot M[g] \\
M[f \cdot g] &= J_1 \cdot M[g] J_2 \cdot M[f] \cdot J_3 \cdot M[g] J_4 \cdot M[f] \cdot J_5.
\end{aligned}
$$

The protocol is simple: each $M[x_i]$ is shared by player $i$, and the network multiplies all the matrices together. Notice that the matrices are all full-rank and therefore form a group under multiplication. According to Theorem 5.14 and Corollary 5.11, the secret product $M[\mathcal{F}]$ takes a constant number of rounds to compute. The result is the secret $M[\mathcal{F}](1, 3)$ and is revealed or left secret for use in further protocols.

Figure 5.18 gives more details. For the sake of efficiency, the constant matrices are collapsed at the start: let $H = \{M_i \mid M_i$ contains a variable$\}$, let $q = |H|$, let $h(i)$ be the index of the $i^{th}$ member of $H$, let $G(i) = \{M_j \mid h(i) < j < h(i+1)\}$ be the set of constant matrices to the right of $M_{h(i)}$, and let $G(0)$ be the set of matrices to the left of $M_{h(1)}$. Define $N_1 = [\prod G(0)]H(1)[\prod G(1)]$ and $N_j = H(j)[\prod G(j)]$.

## 5.4.1   Reducing Rounds for Polynomial Size Circuits

The results of the previous section easily show how to reduce the number of rounds for circuit-based protocols.

---

EVALUATE-NC $^1(F)$

- Each player $i$ shares $x_i$.

- Secretly compute each $N_1, \ldots, N_q$ using LINEAR-COMBINE , noninteractively.

- Run ITERATED-MULTIPLY $(N_1, \ldots, N_q)$ to obtain secret matrix $M[\mathcal{F}]$.

- The result is the top right secret of $M[\mathcal{F}]$, *i.e.* $M[\mathcal{F}](1, 3)$.

Figure 5.18: Protocol to evaluate $NC^1$ or $ANC^1$ circuit in constant rounds and polynomial message sizes.

---

**Corollary 5.17** *For $3t < n$ and any function family $F$ described by a polynomial-size circuit family $C_F$, there exists a t-resilient protocol to compute $F$ using $O(\mathtt{depth}(C_F)/(\log nm))$ rounds, with messages of size polynomial in $n$ and $m$.*

**Proof:** Define the slice function $F^i$ to be the set of outputs at the $(i \cdot \log nm)^{th}$ level of circuit $C_F$. Clearly, $F$ can be written as the product of $\mathtt{depth}C_F/(\log nm)$ of these functions, and each function is in $NC^1$. By Theorem 5.16, there is a set of protocols to evaluate each slice function; by eliminating the final step from each protocol, we obtain a protocol that produces secretly shared values rather than revealing the outputs of that level. By Theorem 4.31, the concatenation is $t$-resilient. The number of rounds required by the concatenated protocol is clearly $\log nm$ times the number of rounds to compute a slice function, which is constant. ∎

## 5.4.2 Determinants in Constant Rounds

In fact, Theorem 5.16 gives a stronger result using iterated multiplication of $n \times n$ matrices. By Theorem 5.14 and Corollary 5.12, there is an *expected* constant-rounds protocol to compute the product of a polynomial number of matrices. Cook [Coo85] and Berkowitz [Ber84] show that the iterated product of integer $n \times n$ matrices is complete for $DET^*$, the class of all problems that are $NC^1$ reducible to DET, namely those that are $NC^1$ reducible to computing the determinant of an $n \times n$ matrix.

**Corollary 5.18** *For $3t < n$ and any function $F \in DET^*$, there exists a t-resilient protocol to compute $F$ in a* constant expected *number of rounds, with polynomial message sizes.*

## 5.5 Any Function In Constant Rounds

In fact, at the expense of a possible exponential blowup in message size, it is certainly possible to achieve secure protocols in constant rounds for $3t < n$. The idea is based on representing a function $F(x_1, \ldots, x_n)$ as a weighted sum whose addenda are computable in constant rounds. The message size depends on the number of addenda, which itself depends on $F$. Ignoring the number of addenda, each of which can be computed in parallel and then added non-interactively, the protocol requires constant rounds.

**Lemma 5.19** *Any function $F : E^n \to E$ has a canonical representation as a function $c_F$ such that on $\{0,1\}^n$, $c_F(x_1, \ldots, x_n) = F(x_1, \ldots, x_n)$, in the following manner:*

$$c_F(x_1, \ldots, x_n) = \sum_{(\epsilon_1, \ldots, \epsilon_n) \in \{0,1\}^n} F(\epsilon_1, \ldots, \epsilon_n) \cdot \delta((\epsilon_1, \ldots, \epsilon_n), (x_1, \ldots, x_n))$$

*where $\delta((\epsilon_1, \ldots, \epsilon_n), (x_1, \ldots, x_n)) = 1$ iff each $\epsilon_i = x_i$, and otherwise is 0.*

**Theorem 5.20** *For $3t < n$ and any function $F$, there exists a $t$-resilient protocol to compute $F$ in a constant number of rounds. The message sizes may grow exponentially, depending on the nature of $F$.*

**Proof:** The protocol is trivial, given Lemma 5.19 and a means to compute $\delta$ in constant rounds: in parallel, secretly compute $\delta$ for all $(\epsilon_1, \ldots, \epsilon_n)$ values such that $F(\epsilon_1, \ldots, \epsilon_n) \neq 0$, and then compute the secret linear combination of the results using the publicly-known weights $F(\epsilon_1, \ldots, \epsilon_n)$ as specified.

Define the normalization of $x$ to be $||x|| = 1$ iff $x \neq 0$, and $||0|| = 0$. Then we have

$$\delta((\epsilon_1, \ldots, \epsilon_n), (x_1, \ldots, x_n)) = 1 - ||\textstyle\sum_{i=1}^{n} ||\epsilon_i - x_i|| \; ||$$

But Theorem 5.13 states that there is a constant-rounds protocol for computing the extended multiplicative inverse of a secret. Clearly, $||x|| = x \cdot x^{-1}$, for this extended inverse. Then the protocol to compute $\delta$ is simple: normalize each difference by calling the Field-Invert protocol on $\epsilon_i - x_i$; sum the results non-interactively; normalize the sum; and subtract the result from 1, non-interactively. ∎

An alternative but equivalent formulation arises from the following observation:

**Lemma 5.21** *Any function $F : E^n \to E$ has a canonical representation as a polynomial $c_F$ of degree $n$ such that on $\{0,1\}^n$, $c_F(x_1, \ldots, x_n) = F(x_1, \ldots, x_n)$. In particular, any function $F : \{0,1\}^n \to \{0,1\}$ has such a representation as a polynomial of degree $n$ over $E$.*

**Proof:** Follows from simple algebra and the following definition:

$$c_F(x_1, \ldots, x_n) = \sum_{(\epsilon_1, \ldots, \epsilon_n) \in \{0,1\}^n} F(\epsilon_1, \ldots, \epsilon_n) \cdot \prod_{i=1}^{n} (1 - (x_i - \epsilon_i)(-1)^{\epsilon_i})$$

∎

# 5.6 Formal Proofs

Describing an interface is usually far more tedious and difficult than giving a convincing argument that the information held by an adversary is independent of the inputs of reliable players. It should be remarked that the protocol designer need not concern himself with the details of the interface; given a protocol compiler and a list of subprotocols that can be concatenated, the actual implementation of a protocol has nothing to do with the specification of a interface for the protocol. Thus the only purpose of the following specifications are to prove the resilience of the protocol and are irrelevant to the implementation.

## 5.6.1 Data Structures and Local Variables

In order to examine the progressive states of each player and in order to implement any of the protocols, we need to list explicitly the local variables that make up each player's state. The value of each variable is a string representing a bit, a field element, or a message; if unassigned, the value is $\Lambda$. The local variables listed below are certainly redundant to some degree:

- $x_i$, its input.

- $a_i$, its auxiliary input.

- $y_i$, its auxiliary input.

- $\texttt{rand\_field}_i = (\rho_{i1}, \rho_{i2}, \ldots)$, an array of generic values, some of which may represent secretly shared values (*e.g.* wire values), other public information (*e.g.* publicly known constants and other information), and other information necessary to the particular protocol (*e.g.* information used to verify pieces).

- $\texttt{rand\_bits}_i = (b_{i1}, b_{i2}, \ldots)$, an array of random bits.

- $\vec{s} = (s_1, s_2, \ldots)$, an array of known values, some of which may represent secrets and others of which may represent information necessary to the protocol (such as additional information used to verify pieces).

- $\textsc{pieces}_i = (\textsc{piece}_1(s_1), \ldots, \textsc{piece}_n(s_1); \textsc{piece}_1(s_2), \ldots, \textsc{piece}_n(s_2); \ldots)$, an array of the pieces of all the secrets. Player $i$ will know either one of the pieces or all of the pieces of each given secret.

- $\textsc{disqual}_i = (d_1, d_2, \ldots, d_n)$, an array of disqualifications. $\textsc{disqual}_i[j] = 1$ if player $i$ believes player $j$ to be corrupt; otherwise it is 0.

- $\textsc{glob\_disqual}_i = (g_1, g_2, \ldots, g_n)$, an array of global disqualifications. $\textsc{glob\_disqual}_i[j] = 1$ if player $i$ believes all reliable players have disqualified player $j$. (This should have the same value among all reliable players.)

- $\mathtt{mesg}([n], i, 1..R)$, messages from other players.

- $\mathtt{mesg}(i, [n], 1..R)$, messages to other players.

A state $q_i$ is specified by the vector

$$(x_i, a_i, \mathtt{rand\_field}_i, \mathtt{rand\_bits}_i, \vec{s}_i, \text{PIECES}_i, \text{DISQUAL}_i, \text{GLOB\_DISQUAL}_i,$$
$$\mathtt{mesg}([n], i, 1..R), \mathtt{mesg}(i, [n], 1..R)).$$

As in Chapter 2, §2.5.2, we use these as labels to parametrize distributions in more detail. That is, instead of considering simply the distribution $\mathtt{V}(q_i, r)$ on the state of player $i$ at round $r$, we consider the distribution $\mathtt{V}(\text{PIECE}_i(s_1), r)$ on the value of player $i$'s local variable $\text{PIECE}_i(s_1)$ at round $r$, the distribution $\mathtt{V}(\text{PIECE}_i(s_2), r)$ on the value of player $i$'s local variable $\text{PIECE}_i(s_2)$ at round $r$, and so on. The collection of random variables is parametrized by $r$ and the set $\mathtt{Vnames'}$ of variable names obtained from $\mathtt{Vnames}$ by replacing each $q_i$ in $\mathtt{Vnames}$ by the individual labels listed above.

Our goal is to specify the distributions $\mathtt{V}$ describing variables for VSS and the distributions $\mathtt{V^S}$ describing variables for VSS induced by $\mathcal{S}$ in the ID(SHARE) protocol and to show that they are equal. Some are not sampled by $\mathcal{S}$, and we denote unsampled variable values by $\mathtt{V^S}(v, r) = \lambda$. Implicit in the operation of an interface is the specification that after each round of interaction with $A$, $\mathcal{S}$ sets $\mathtt{V^S}(v, r+1) = \mathtt{V^S}(v, r)$ for each $v$ such that $\mathtt{V^S}(v, r) \neq \lambda$.

In each round we describe corruptions and then rushed messages from $\overline{T}$. After each round of VSS, $\mathcal{S}$ records messages from $A$ in $\mathtt{V^S}(\mathtt{mesg}(T, \overline{T}, r), r)$ accordingly. We give arguments along the way that $\mathtt{V}(v, r) = \mathtt{V^S}(v, r)$ for sampled variables.

## 5.6.2 Proofs of Resilience

### Verifiable Secret Sharing

**Proof of Theorem 5.7:** We would like to show that VSS $\succeq$ ID(SHARE) $\succeq$ ID($\emptyset$). First we construct an interface $\mathcal{S}$ from VSS to ID(SHARE). For clarity we refer to player $i$ in VSS and to player $i_{id}$ in ID(SHARE).

The ID(SHARE) protocol allows the dealer, $D$, to supply pieces to the host, who distributes them if they are properly interpolatable, but otherwise sends CHEATING to all players. The interface runs most of VSS with $A$ before finishing the first round of ID(SHARE).

**Remark.** The intuition that $A$ gains no information by complaining and forcing $D$ to reveal pieces because $A$ knows them already is formalized by $\mathcal{S}$ having at a given stage in VSS set the random variable $\mathtt{V^S}(x, r)$ to some value that either has come from $A$ or has been generated by $S$ for $A$ in response to a corruption request.

If $A$ corrupts $D$ before it supplies secret $s$, $\mathcal{S}$ requests $D_{id}$ be corrupted and supplies $A$ with $s$. If $A$ requests $i$ be corrupted, $\mathcal{S}$ corrupts $i_{id}$ and returns the auxiliary input.

(V1) If $D$ is corrupt, $\mathcal{S}$ does nothing but record the outgoing message of $A$ as $\mathtt{V^S}(\mathtt{mesg}(D, [n], 1), 1)$. Otherwise, $\mathcal{S}$ chooses $\mathtt{V^S}(\mathrm{PIECE}_i(s), 1) \leftarrow \boldsymbol{uniform}(E)$ for all $i \in T$, constructs

$$\{p_i(u) \leftarrow \boldsymbol{UPoly}(n, t, \mathrm{PIECE}_i(s))\}$$

and

$$\{q_i(v) \leftarrow \boldsymbol{UPoly}(n, t, 0) \mid (\forall j \in T) q_i(\omega^{j-1}) = p_j(\omega^{i-1})\},$$

and delivers these messages to $A$. By Lemma 5.2, the conditional distributions satisfy

$$\mathtt{V^S}(\mathtt{mesg}(D, i, 1; (p_i(u), q_i(v))), 1) = \mathtt{V}(\mathtt{mesg}(D_{id}, n+1, 1; (p_i(u), q_i(v))), 1).$$

(V2) If $A$ newly corrupts $D$, $\mathcal{S}$ corrupts $D_{id}$, sets

$$\{g(u) \leftarrow \boldsymbol{UPoly}(n, t, s) \mid (\forall i \in T) g(\omega^{i-1}) = \mathrm{PIECE}_i(s)\},$$

sets for all $i \notin T$

$$\{p_i(u) \leftarrow \boldsymbol{UPoly}(n, t, \mathrm{PIECE}_i(s)) \mid (\forall j \in T) p_i(\omega^{j-1}) = q_j(\omega^{i-1})\},$$

sets

$$\{h(v) \leftarrow \boldsymbol{UPoly}(n, t, s) \mid (\forall i \in T) h(\omega^{i-1}) = q_i(0)\},$$

sets for all $i \notin T$

$$\{q_i(v) \leftarrow \boldsymbol{UPoly}(n, t, h(\omega^{i-1}) \mid (\forall j \in T) p_i(\omega^{j-1}) = q_j(\omega^{i-1})\},$$

and sets $\mathtt{V^S}(\mathtt{mesg}(D, i, 1), 1) = (p_i(u), q_i(v))$ for all $i \in T$. Interface $\mathcal{S}$ sends $A$ the view of $D$ using $x_D$, $a_D$, $\mathtt{V^S}(\mathtt{mesg}(D, T, 1), 1)$. By Lemma 5.2 each of these distributions satisfies $\mathtt{V^S}(x, 1) = \mathtt{V}(x, 1)$ and is polynomial-time computable, being uniform over an easily computed set of solutions determined by the conditions listed above. This computation in effect induces the variable $\mathtt{V^S}(p(u, v), 1) = \mathtt{V}(p(u, v), 1)$.

If $A$ newly corrupts player $i$ we must consider whether $D$ is yet corrupted. If $D$ is corrupt, then all information held by $i$ is known, in particular, $\mathtt{V^S}(\mathtt{mesg}(D, i, 1), 1)$. The interface corrupts $i_{id}$ to obtain $a_{id}$ (note that $x_{id}$ is nil) and constructs $\mathtt{view}_i^1$ from these values. It then provides $A$ with the view. If $D$ is not corrupt, then $\mathcal{S}$ must construct the $(p_i(u), q_i(v))$ message that $D$ sent to $i$ in round (V1).

To generate rushed messages from nonfaulty players, if $D$ is nonfaulty then $\mathcal{S}$ sets $\mathtt{V^S}(\mathtt{mesg}^{broad}(i, [n], 2), 2) = 0$ for all $i \notin T$, since a nonfaulty player will accept the proper (but private) message from a nonfaulty dealer. If $D$ is faulty, then its messages to nonfaulty players are determined by $\mathtt{V^S}(\mathtt{mesg}(D, [n], 1), 1)$, and $\mathcal{S}$ sets $\mathtt{V^S}(\mathtt{mesg}^{broad}(i, [n], 2), 2)$ to be 0 if the message to $i$ describes polynomials of degree $t$, and $\mathcal{S}$ sets the variable to 1 otherwise. Clearly this is the same probabilistic computation $\delta_i$ that each nonfaulty player applies in (V2).

(V3) Now, if $A$ newly corrupts $i$ while $D$ is nonfaulty, $\mathcal{S}$ generates $p_i(u)$ and $q_i(v)$ as in (V1), and returns them. If $D$ is already corrupt, $\mathcal{S}$ uses $\mathtt{V^S}(\mathtt{mesg}(D, T, 1), 1)$ to determine $p_i(u)$ and $q_i(u)$.

To generate messages from nonfaulty players, $\mathcal{S}$ does the following. For each $i \notin T$ and $j \in T$, the $p_i(w^{j-1})$ values sent by nonfaulty players are determined already by the values sent to corrupt players, so $\mathcal{S}$ sets $\mathtt{V}^{\mathtt{S}}(\mathtt{mesg}(i, j, 2; p_i(\omega^{j-1})), 2) = \mathtt{V}^{\mathtt{S}}(\mathtt{mesg}(D, j, 1; p_i(\omega^{j-1})), 1)$.

(V4) If $A$ newly corrupts $D$, corrupt $D_{id}$ and construct $\mathtt{V}^{\mathtt{S}}(\mathtt{mesg}(D, T, 1..2), 1..2)$ as before. If $A$ newly corrupts $i$, $\mathcal{S}$ creates an earlier view as in (V3) and must then construct incoming messages about $p_k(\omega^{i-1})$ in round (V3). For each $k \in T$, check if $k$ sends $i$ a proper $p_k(\omega^{i-1})$ according to whether $\mathtt{V}^{\mathtt{S}}(q_i(\omega^{k-1}), 1) = \mathtt{V}^{\mathtt{S}}(\mathtt{mesg}(i, k, 2; q_i(\omega^{k-1})), 2)$. The former is fixed by previous computations of $\mathcal{S}$. Set $L(i, k)$ accordingly. If $D$ is nonfaulty then set $L(i, k) = 0$ for all $k \,\,/inT$, because all nonfaulty players send what $D$ sent them. If $D$ is faulty but $k \,\,/inT$, use $\mathtt{V}^{\mathtt{S}}(\mathtt{mesg}(D, k, 1), 1)$ and $\mathtt{V}^{\mathtt{S}}(\mathtt{mesg}(D, i, 1), 1)$ to determine whether $i$ complains. This determines $\mathtt{V}^{\mathtt{S}}(L(i, k), 2)$ for all $k$, and $\mathcal{S}$ supplies it along with $a_i$ (obtained by corrupting $i_{id}$) to $A$.

To generate the broadcast messages from nonfaulty players, $\mathcal{S}$ performs the same computation as the new corruption of $i$ just described. It uses the vector $L(i, \cdot)$ as the broadcast value from $i \notin T$.

(V5) The view of a newly corrupt dealer includes that generated according to (V4) along with the broadcast messages $\mathtt{V}^{\mathtt{S}}(\mathtt{mesg}^{broad}(T, [n], 4), 4)$ and $\mathtt{V}^{\mathtt{S}}(\mathtt{mesg}^{broad}(\overline{T}, [n], 4), 4)$ generated by $A$ and $\mathcal{S}$ in round (V4). The view of a newly corrupted $i$ is generated as in (V4), also attaching the list of broadcast messages.

If $D$ is nonfaulty, all disputes involve at least one faulty player $i$, so $\mathcal{S}$ has already specified in $\mathtt{V}^{\mathtt{S}}(\mu(D, i, 1; p_i(\omega^{i-1})), 1)$ the correct value. So $\mathcal{S}$ sets $\mathtt{V}^{\mathtt{S}}(\mathtt{mesg}(D, [n], 4; p(i, j)), 4) = \mathtt{V}^{\mathtt{S}}(\mu(D, i, 1; p_i(\omega^{i-1})), 1)$. If $D$ is faulty, no messages from nonfaulty players need be generated.

(V6) New corruptions of $D$ or player $i$ are treated as in (V5), and $\mathcal{S}$ adds on the messages broadcast by $D$ in (V5), whether generated by $\mathcal{S}$ or by $A$.

If $D$ is nonfaulty, every player $i \notin T$ does not impeach $D$, so $\mathtt{V}^{\mathtt{S}}(\mathtt{mesg}^{broad}(i, [n], 6; M(i)), 6) = 0$. Otherwise it is easy for $\mathcal{S}$ to compute whether player $i \in T$ impeaches $D$ from the values of $\mathtt{V}^{\mathtt{S}}(\mu(D, i, 1), 1)$ and $\mathtt{V}^{\mathtt{S}}(\mu(D, [n], 4), 4)$, since these previous messages from $D$ to $i$ have already been recorded (if not generated) by $\mathcal{S}$, and since $\mathtt{V}^{\mathtt{S}}(\mu(D, i, 1), 1) = \mathtt{V}(\mu(D, i, 1), 1)$ and $\mathtt{V}^{\mathtt{S}}(\mu(D, [n], 4), 4) = \mathtt{V}(\mu(D, [n], 4), 4)$.

(V7) New corruptions of $D$ or $i$ are treated as in (V6), adding the broadcast impeachments from (V6) on the end of the views.

If $D$ is nonfaulty, impeachments come from faulty players only, so $D$ will broadcast $(p_i(u), q_i(v))$ values that $\mathcal{S}$ has already generated:

$$\begin{aligned} \mathtt{V}^{\mathtt{S}}(\mathtt{mesg}^{broad}(D, [n], 7; M'(i)), 7) &= \mathtt{V}^{\mathtt{S}}(\mathtt{mesg}(D, i, 1), 7) \\ &= \mathtt{V}(\mathtt{mesg}(D, i, 1), 7) \\ &= \mathtt{V}(\mathtt{mesg}^{broad}(D, [n], 7; M'(i)), 7) \end{aligned}$$

For $i \notin T$, $D$ broadcasts $M'(i) = 0$, of course. If $D$ is faulty, $\mathcal{S}$ simply records

$\mathtt{V^S}(\mathtt{mesg}^{broad}(D, [n], 7), 7)$ as generated by $A$.

(V8) New corruptions are as in (V7), with the broadcast values $\mathtt{V^S}(\mathtt{mesg}^{broad}(D, [n], 7), 7)$ concatenated at the end.

If player $i$ is nonfaulty, $\mathcal{S}$ checks whether $\mathtt{V^S}(\mathtt{mesg}^{broad}(D, [n], 6; M(i)), 6) = 0$. If so, it sets $\mathtt{V^S}(\mathtt{mesg}^{broad}(i, [n], 8), 8) = 0$. Otherwise, it checks whether, for all $j$ for which $D$ broadcasts $(p_j(u), q_j(v))$, the values agree with $i$'s value $p_j(\omega^{i-1})$ as derived accordingly from $\mathtt{V^S}(\mathtt{mesg}(D, i, 1), 8)$ or $\mathtt{V^S}(\mathtt{mesg}^{broad}(D, [n], 5), 8)$ or $\mathtt{V^S}(\mathtt{mesg}^{broad}(D, [n], 7), 8)$, the choice depending on whether $i$ has complained before. If so, $\mathcal{S}$ sets $\mathtt{V^S}(\mathtt{mesg}^{broad}(i, [n], 8), 8) = 0$, and otherwise sets it to 1.

Finally, if $\mathtt{V^S}(\textsc{glob\_disqual}_i(D), 8) = 1$ for any $i \notin T$ ($\mathcal{S}$ calculates this easily from $\mathtt{V^S}^{broad}(\overline{T}, [n], 8), 8)$, then $\mathcal{S}$ requests that the corrupted dealer $D_{id}$ send $\Lambda$ in the ideal protocol. The trusted host will supply each player with the output, $(0, \mathit{reject})$. Otherwise, if the dealer $D$ is corrupted, $\mathcal{S}$ requests that $D_{id}$ send the pieces that were finally accepted by nonfaulty players, and fills out the list with values for corrupted players $i_{id}$ that provide a polynomial of degree $t$. These corrupted players will receive those values but their outputs are considered to be $\Lambda$, as described in Chapter 2. The nonfaulty players in $\textsc{ID}(\textsc{Share})$ will output $\mathtt{V}(Y_i, 8) = (\textsc{piece}_i(s), \mathit{accept}) = (p_i(0), \mathit{accept}) = \mathtt{V^S}(Y_i, 8)$.

This concludes the description of $\mathcal{S}$. We claim that for every $r$ and every $v$ such that $\mathtt{V^S}(v, r) \neq \lambda$, $\mathtt{V^S}(v, r) = \mathtt{V}(v, r)$. At each step, $\mathcal{S}$ samples new variables $\mathtt{V^S}(v, r)$ as a probabilistic function (sometimes even deterministic, as when reporting values already broadcast) $f$ of earlier samples. Each time, $\mathcal{S}$ uses either:

- a previous output of $A$;

- uniform distributions based on Lemma 5.2;

- direct computation of a nonfaulty player on values already broadcast or known by virtue of correct behavior of other nonfaulty players (*e.g.* nonfaulty players do not impeach other nonfaulty players).

With the aid of Lemma 5.2 and broadcast channel properties, it holds that

$$\mathtt{V^S}(v, r+1) = f(\mathtt{V^S}(v_1, r), \dots, \mathtt{V^S}(v_K, r)) = f(\mathtt{V}(v_1, r), \dots, \mathtt{V^S}(v_K, r)) = \mathtt{V}(v, r+1)$$

for some $K$. In particular, $\mathtt{V}(\mathcal{O}(q_i), 8) = \mathtt{V^S}(\mathcal{O}(q_i), 8)$ for all $i$ and $\mathtt{V}(q_A, 8) = \mathtt{V^S}(q_A, 8)$, hence $[A, \alpha](\vec{x} \circ \vec{a} \circ a, k) = [A, \mathcal{S}, \beta](\vec{x} \circ \vec{a} \circ a, k)$. ∎

VSS $(s)$

(V1) Fix $E = E(m, n)$ and $\omega$.

    $(1 \leq i \leq n)$

    $D:$        $\{p(u, v) \leftarrow \textit{\textbf{uniform}}(p \in E[u, v] \mid p(0, 0) = s, p(u, v) = \sum_1^t \sum_1^t p_{ij} u^i v^j\}$

    $D \rightarrow i:$   $(p_i(u), q_i(v)) = (p(u, \omega^{i-1}), q(\omega^{i-1}, v))$

(V2) $(1 \leq i \leq n)$

    $i \rightarrow [n]:$   **if** degree $t$ **then** 0 **else** 1.

(V3) $(1 \leq i, j \leq n)$

    $i \rightarrow j:$   $p_i(\omega^{j-1})$

(V4) $(1 \leq i, j \leq n)$

    $i:$       $L(i, j) = \begin{cases} 0 & \texttt{mesg}(j, i, 3; p_j(\omega^{i-1})) = \texttt{mesg}(j, i, 3; q_i(\omega^{j-1})) \\ 1 & \text{otherwise} \end{cases}$

    $i \rightarrow [n]:$   $(L(i, 1), \ldots, L(i, n))$

(V5) $(1 \leq i \leq n)$

    $D:$       $L'(i) = \begin{cases} 0 & \texttt{mesg}(i, [n], 4; L(i, j)) = 0 \\ p_i(\omega^{j-1}) & \text{otherwise} \end{cases}$

    $D \rightarrow [n]:$   $(L'(1), \ldots, L'(n))$

(V6) $(1 \leq i \leq n)$

    $i:$       **if** $(\exists j)$ $\texttt{mesg}^{broad}(i, [n], 4; L(i, j)) = 1$

            or $\texttt{mesg}^{broad}(j, [n], 4; L(j, i)) = 1$,

            and $\texttt{mesg}^{broad}(D, [n], 5; p_i(\omega^{j-i})) \neq$

                            $\texttt{mesg}(D, i, 1; q_i(\omega^{i-1}))$

            **then** $M(i) = 1$ **else** $M(i) = 0$.

    $i \rightarrow [n]:$   $M(i)$

(V7) $(1 \leq i \leq n)$

    $D:$       $M'(i) = \begin{cases} 0 & \texttt{mesg}(i, [n], 6; M(i)) = 0 \\ (p_i(u), q_i(v)) & \text{otherwise} \end{cases}$

    $D \rightarrow [n]:$   $(M'(1), \ldots, M'(n))$

(V8) $(1 \leq i \leq n)$

    $i:$       **if** $\texttt{mesg}^{broad}(i, [n], 6; M(i)) = 0$ or

            $\texttt{mesg}^{broad}(D, [n], 7; (p_i(u), q_i(v)))$ is consistent

            **then** $\text{DISQUAL}_i(D) = 0$ **else** $\text{DISQUAL}_i(D) = 1$.

    $i \rightarrow [n]:$   $\text{DISQUAL}_i(D)$

    $(1 \leq i \leq n)$

    $i:$       **if** $t + 1$ players disqualified $D$,

            **then** set $\text{GLOB\_DISQUAL}_i(D) = 1$.

            **if** $\text{GLOB\_DISQUAL}_i(D) = 1$

            **then** $Y_i = (0, \textit{\textbf{reject}})$ **else** $Y_i = (p_i(0), \textit{\textbf{accept}})$.

Figure 5.19: Protocol for dealer to verifiably share secret $s$. See text for more details.

# Chapter 6

# Tolerating a Minority of Faulty Processors

> Democracy is the recurrent suspicion that more than half of the people are right more than half of the time.
>
> — E. B. White, *The Wild Flag*

The methods of [BGW88, CCD88] and even the far more efficient methods we have presented in Chapter 5 allow faults in at most a third of the network ($3t < n$). Perfect privacy is not achievable with higher fault tolerance for even some simple functions like AND [BGW88] (but see Chapter 8 for a characterization of functions computable with perfect privacy at high, passive fault-tolerance levels).

The natural question to ask, though, is whether higher fault tolerance is possible if the assurances of security and reliability need not be absolute. We show that, allowing for a negligible chance of error, $2t < n$ can be achieved. That is, as long as only a minority of the processors are faulty, we can construct protocols to compute any function reliably and securely. For larger numbers of faults, it becomes impossible for the players even to share a secret, which does not immediately imply a negative result but gives a strong intuition as to why higher fault-tolerance is generally impossible without making other assumptions about the network model.

Notice that the major problems with extending the methods of [BGW88, CCD88] to $2t < n$ are that verifiable secret sharing fails and that the specific techniques for the ABC Problem fail.

Rabin [RT88] made initial and significant progress in the direction of improving the fault bounds from $3t < n$ to $2t < n$ by demonstrating a method for verifiable secret sharing for $2t < n$, using a broadcast network. Earlier methods for verifiable secret sharing with a faulty minority required cryptographic assumptions [CGMA85]. The extension to performing *computations* for $2t < n$, however, remained open until this work.

We give a new and efficient method to solve the ABC problem when secret addition is possible, for $2t < n$. Our techniques utilize verifiable secret sharing for $2t < n$, and allow the field used for secret sharing to be of exponential size. Ben-Or [RB89] and Kilian [Kil88b] have independently developed methods to tolerate a faulty minority, based on standard ideas of boolean circuit simulation. Their methods require that the field used for secret sharing be of polynomial size, which restricts the efficiency of their solutions.

**Theorem 6.1** *There exists a protocol* EVALUATE $^{(2t<n)}$ *that is t-resilient against against dynamic Byzantine adversaries for* $2t \leq n$.

Because we can simulate large-field arithmetic operations quickly and directly while alternative methods use bit-simulations or small-field arithmetic, our protocols have the practical advantage of using far fewer rounds of communication for many natural functions. Joined with the techniques of Chapter 5, our techniques use tremendously fewer rounds of interaction. The bit-simulation techniques of [RB89, Kil88b], on the other hand, are incompatible with reducing rounds through the arithmetic circuit reductions of Chapter 5 since they require the added cost of simulating arithmetic operations by bitwise Boolean operations.

As before, our methods are secure against adversaries with unbounded resources, while at the same time requiring only polynomial time to execute.

We shall first describe a modification of the solution in [RT88] for verifiable secret sharing, show how to add secrets, and then give our solution to the ABC problem when $2t < n$. The framework of [BGW88, CCD88] can then be applied, using these more resilient subprotocols to support the share-secrets-create-secrets paradigm.

**Assumptions made in this chapter.** The network is complete, with private lines, broadcast lines, $n$ processors, and at most $t < \frac{n}{2}$ Byzantine faults, chosen dynamically. The protocols are information-theoretically secure with high probability, or in other words, the results are statistically resilient. No unproven complexity-theoretic assumptions are made.

## 6.1 Verifiable Time-Release Messages

There is a long history of solutions for VSS under various unproven assumptions. We wish to avoid unproven cryptographic assumptions and to tolerate $2t < n$. We shall utilize a method for VSS for $2t < n$ very similar to that given by [RT88], but somewhat more convenient and efficient. Briefly, it uses Shamir's method for sharing a secret, and requires that each piece be reshared using a weak form of sharing. The weaker form of sharing includes information called "check vectors," which allow verification of the pieces. Our modification of the protocol is a new method for constructing check vectors. For completeness, because formal proofs of [RT88, RB89] have not appeared, and because we use different subroutines, we shall present the essential details of [RT88, RB89] but we refer the reader to that work for a deeper discussion.

The verification property of Rabin's scheme relies on a subprotocol for what we call Verifiable Time-Release Messages.

We consider three players, a sender S, a receiver R, and an intermediary I. The sender would like to give a secret bit to the intermediary, who will pass it on at a later time to the receiver. The receiver must be able to detect any tampering on the part of the intermediary. At the same time, the intermediary should know whether the information given him will in fact satisfy the receiver at the appropriate time. He must therefore be able to check the behavior of the sender during the initial part of the protocol.

Rabin [RT88] provides an elegant way to satisfy these properties in a manner similar to secret sharing with threshold 1. Her idea involves generating nonzero random numbers $u$ and $v$ mod $p$ and setting $w = b + uv$. The receiver gets $(v, w)$ while $I$ gets $(b, u)$. When $I$ passes the value on to R later, he sends $(b, u)$, and R checks that $w = b + uv$. In order to convince R of an incorrect value, the intermediary must change $u$ correctly, which he cannot do with non-negligible probability. Checking that the sender is correct uses a cut-and-choose idea, which we shall use below.

We present a simple alternative with the nice property that no multiplications are necessary. In terms of polynomial-time algorithms, this makes no difference, but in terms of actual implementations, multiplications over a finite field are considerably more expensive than additions. Our approach is derived from work on a different problem with Feigenbaum and Shoup [BFS90].

If S and R were to share a one-time pad (a private sequence of random bits), then how may S send a secret bit to R via another player I? Player I might change some of the bits. Our simple solution uses $3k$ bits of the one-time pad and ensures that tampering is detected with probability $1 - 2^{-k}$. See Figure 6.1. Intuitively, for I to change the bit $b$ without detection, it must guess the exact sequence of bits $\alpha(1), \ldots, \alpha(k)$, which it can do with probability at most $2^{-k}$.

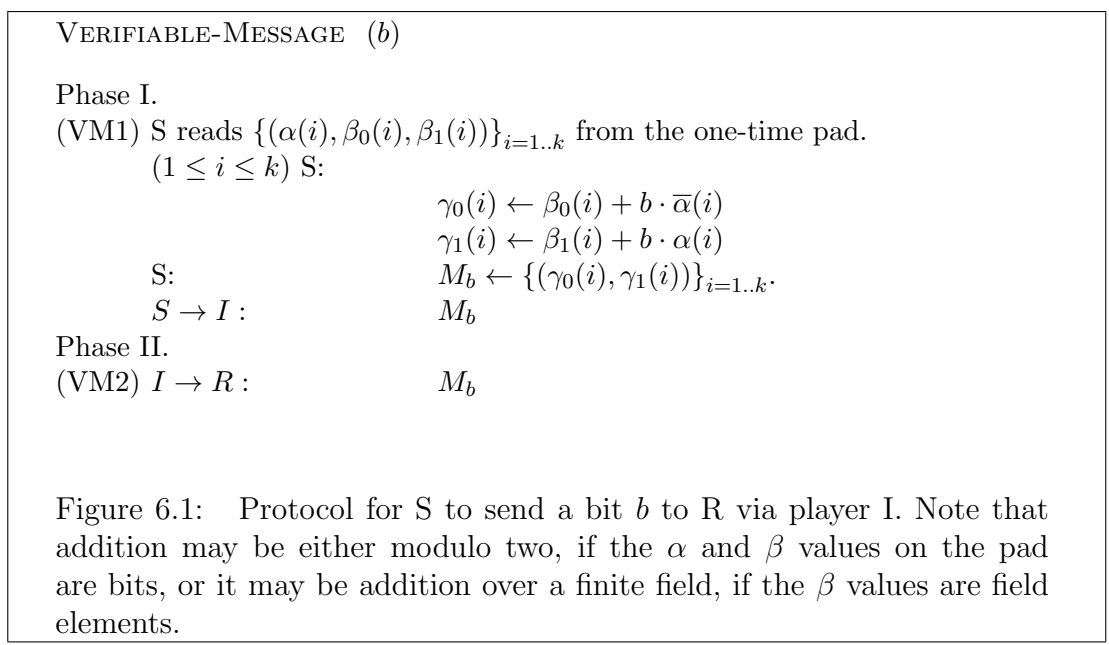

Figure 6.1: Protocol for S to send a bit $b$ to R via player I. Note that addition may be either modulo two, if the $\alpha$ and $\beta$ values on the pad are bits, or it may be addition over a finite field, if the $\beta$ values are field elements.

We use this verifiable one-time pad to construct a *verifiable time-release scheme* as follows. Roughly speaking, the sender sends the one-time pad privately to R, and sends the message $M_b$ to I, who holds onto it until the time it should be released. The one-time pad serves as a *check vector*, *i.e.* a sequence of bits used to check the accuracy of the secret held by I. The property that the intermediary is convinced that R will accept the message later on must also be satisfied, and we now turn our attention to that problem.

As in [RT88], we use a cut-and-choose method. Instead of generating $k$ $(\alpha, \beta_0, \beta_1)$ triples, generate $2k$ of them. Half of these will be revealed to I by R so that I can check for misbehavior. In order for S to cheat undetected, it must misbehave on exactly that half of the triples that R does *not* immediately send to I. Notice that $I$ learns $b$ upon receiving a single $(\alpha, \beta_0, \beta_1)$ triple. To keep $b$ secret from $I$, the sender simply splits $b$ into two random bits as $b = b_I \oplus b_R$, and gives $b_I$ to I and $b_R$ to R.

The protocol is detailed in Figure 6.2. If $M$ is a message that one party is supposed to send to another, we denote by $M'$ the message it actually sends.

Intuitively, if S attempts to cheat, it must guess in advance the exact set of indices $i(1), \ldots, i(k)$ that R selects at random, and it must behave properly on that list but must cheat on every other row of the table. The chances of this are certainly at most $2^{-k}$.

**Theorem 6.2** *Protocol* Verifiable-Time-Release *is exponentially 2-resilient against Byzantine adversaries.*

The ideal protocol ID(VTR) is as follows. In Phase I, the sender $S$ sends two bits, $b_I$ and $b_R$, to the host. The host sends $b_i$ to $I$ and $b_R$ to $R$ if it indeed received two bits; otherwise it sends Cheating. Then, $I$ and $R$ send a 0 to the host to indicate they wish to participate, and the host passes on either 0 or Cheating to $I$ and $R$ to indicate whether they chose to participate. In Phase II, $I$ again sends 0 to indicate it wishes to reveal the value, and the host sends $b_i$ to $R$, or it sends Cheating if $I$ supplied a nonzero value.

In the VSS method of [RT88], all players act both as intermediaries and recipients for pieces of the secret $u$. Each piece $\text{piece}_i(u)$ of $u$ is given to an intermediary $i$, and check vectors $s_j(\text{piece}_i(u))$ are sent out to every other processor $j$. At reconstruction time, the intermediaries $i$ broadcast their check vectors $\vec{s}(\text{piece}_i(u))$, which include $\text{piece}_i(u)$. Each player, acting as a recipient, reconstructs $u$ using the pieces he concludes are accurate.

**Theorem 6.3** *(after T. Rabin [RT88]) There exists a VSS protocol for $2t < n$ that is $t$-resilient.*

## 6.2 Addition of Secrets

It is possible though not trivial to extend the VSS method given above to a method for linearly combining secrets. Say that $u$ and $v$ were shared using polynomials $f(x)$ and $g(x)$, respectively. Let $h(x) = af(x) + bg(x) + c$; then $h(0) = af(0) + bg(0) + c = w$. Sums of the pieces are pieces of the sum; Each player $i$ uses

$$\text{PIECE}_i(w) = a \cdot \text{PIECE}_i(u) + b \cdot \text{PIECE}_i(v) + c$$

as his piece of $w$.

In order that $w$ be a verifiable secret according to the VSS protocol, each $\text{PIECE}_j(w)$ must be reshared in a weak fashion. This involves the use of check vectors for the pieces of $h(j)$. Each player $i$ holds pieces of every other player's pieces, $\text{PIECE}_i(\text{PIECE}_j(u))$ and $\text{PIECE}_i(\text{PIECE}_j(v))$. He sets

$$\text{PIECE}_i(\text{PIECE}_j(w)) = a \cdot \text{PIECE}_i(\text{PIECE}_j(u)) + b \cdot \text{PIECE}_i(\text{PIECE}_j(v)) + c.$$

The only remaining part of the VSS structure is the set of check vectors for the pieces $\text{PIECE}_j(\text{PIECE}_i(w))$. Since player $i$ knows $\text{PIECE}_i(w)$, he creates and distributes new check vectors for the pieces of this value. The other players check the correctness of his vectors using the protocol specified in the VSS scheme [RT88].

The LINEAR-COMBINE protocol is resilient because each of its steps is either a non-interactive computation of a robust and private representation of some secret value, or an application of a subprotocol for VSS that is itself resilient. It is clear that players who deviate from the protocol cannot create convincingly false check vectors with probability exceeding some $\frac{1}{2^{k_1}}$, as determined by the security parameter $k_1$ of the VSS protocol. In this case, an interface would fail to be able to provide a correct argument, since the adversary would learn information to which it is not entitled, and the interface would need to corrupt additional players to obtain that information. The net weight of these events is exponentially small, however. In fact, the only undetectable misbehavior is the possible choice of acceptable check vectors according to an inappropriate distribution. This is, however, easily dealt with by an interface, which needs merely record what the *adversary* specifies; the interface need not itself generate these distributions. Note that nonfaulty players' behavior is independent of how the faulty players generate check vectors, even *acceptable* ones.

## 6.3 Multiplication of Secrets

The protocol for multiplication of secrets is more complicated; the key new idea is a method for giving a proof that the product of two secrets is a third. In fact, the ability to prove products of secrets is the basis for achieving high fault-tolerance in all of the results of this paper.

The protocol we shall present relies on a protocol for addition of secrets.

Our solution follows a few brief steps (cf. [BGW88]). As before, let $u$ and $v$ be shared using $f(x)$ and $g(x)$. Each player $i$ secretly shares the value $f(i)g(i)$ and "proves" that he has in fact shared this value (see §6.3.1). If his proof fails, he is disqualified (see §5.2.3). From the collection of secret products determines the polynomial $h(x) = f(x)g(x)$, of degree $2t$ and free term $uv$. Using a protocol to add a random polynomial of degree $t + 1$ and free term 0 and then to truncate the polynomial to degree $t$ (see §5.2.2), each player $i$ is supplied with the value $h(i)$ for the resulting polynomial $h(x)$ of degree $t$ and free term $uv$. Figure 6.4 describes the MULTIPLY protocol.

## 6.3.1   Verifiable Multiplication

In order to accomplish the resilient verifiable sharing of $f(i)g(i)$, we must provide a solution to the ABC problem when $2t < n$. In fact, we describe a new method that shows that, if secret addition is possible, then multiplication is possible, for $2t < n$.

Given a protocol for the ABC problem, Alice will be able to prove to the network that the new secret which she shares is indeed $f(i)g(i)$.

**Lemma 6.4** (*ABC Lemma.*) *If there exists a t-resilient protocol for linear combinations of secrets, then there exists a t-resilient protocol to solve the ABC problem.*

The protocol is exhibited in Figure 6.5. First, an overview: In the first phase, Alice shares several triples of secrets $(\mathcal{R}, \mathcal{S}, \mathcal{D})$ satisfying a simple equation (of the form $\mathcal{D} = (a + \mathcal{R})(b + \mathcal{S})$), which will be used to ensure that Alice does not misbehave. In the second phase, the players select and reveal combinations of some of these triples in order to confirm that every triple satisfies the simple equation. Finally, each unrevealed triple of secrets gives rise to a simple linear combination of secrets that should equal the desired product $ab$. The third phase checks that the linear combinations are consistent.

Let $k_0$ denote a security parameter; the chance of incorrectness will be bounded by $\frac{1}{2^{k_0}}$. For simplicity we take $k_0 > n$ and $k_0$ a power of two.

Now let us consider the PROVE-PRODUCT protocol in more depth. If Alice is honest, all of the sums $(a + r_j)$ and $(b + s_j)$ are independent of $a$ and $b$, since $r_j$ and $s_j$ are uniformly random field elements. The products $d_j$ are also independent of $a$ and $b$. The choice of $k_0$ indices $j_1, \ldots, j_{k_0}$ to check is independent of $a$ and $b$, given that any one player in the system is honest and shares uniformly random numbers. The partial views obtained by faulty players during the addition protocols are independent of the good players' inputs. Finally, since Alice is honest, $d_j = (a + r_j)(b + s_j) = c + as_j + br_j + rs$, so $c_j = 0$ always. Therefore, the set of messages are $t$-wise independent of $a$ and $b$, and messages from nonfaulty to faulty players are easily generated accurately by an interface.

We would like to show that the chance that $c \neq ab$ without Alice's being detected is smaller than $\frac{1}{k_0}$. We shall show that she must behave properly on exactly the indices in $Y$ and she must misbehave on all the others. Let $X$ be the set of indices $j$ for which Alice shares $d_j$ correctly, that is, for which Alice shares $d_j$ having the value $(a + r_j)(b + s_j)$. The set $Y$ of indices chosen by the system must be a subset of $X$, or else Alice's misbehavior is detected for some $j \in Y \setminus X$. The remaining indices $j \notin Y$ must all satisfy $c_j = c - d_j + as_j + br_j + rs = 0$ or else Alice is caught. If $c \neq ab$ (Alice is cheating), none of the indices $j \notin Y$ is in $X$. Hence $X = Y$, and since $Y$ is chosen randomly after Alice has shared all her secrets, the probability that Alice can cheat without being detected is no more than $\frac{1}{2^{k_0}}$. An interface fails to produce a correct output an exponentially small amount of the time, hence the overall distributions are exponentially indistinguishable.

## 6.4 The Protocol Compiler for Faulty Minority

At this point we have described all the tools necessary to create a multiparty protocol to evaluate any circuit $C_F$ for $F(x_1, \ldots, x_n)$ privately, tolerating $t < \frac{n}{2}$ faults. The protocol is the same as that of protocol EVALUATE (Figure §5.9), with the new LINEAR-COMBINE and MULTIPLY protocols of this chapter substituted for the ones called by that specification. The disqualification procedure is different, as well: after each cycle of evaluating gates, a recovery procedure is initiated if faults have been detected.

Let $\tau$ be the number of faults. The $(n - \tau, t - \tau)$ recovery protocol uses a stripped-down version of the TRUNCATE protocol to reduce the degrees of each polynomial being used by $\tau$. The remaining players in fact reconstruct each *piece* of the various secrets held by the $\tau$ faulty players, using straightforward linear combinations of existing pieces, as seen by the following argument. Without loss of generality take the set of reliable players to be $1, \ldots, t + 1$. Then the LaGrange interpolation gives

$$p(j) = \sum_1^{t+1} L_i(j) p(i),$$

so that the piece $\text{PIECE}_j(s) = p(j)$ is easily computed as a weighted sum of secrets. That is, each player verifiably reshares his pieces and then the system computes the given linear combinations. If more faults occur then the process is restarted. Notice that the players *do not* reconstruct all the information held by the faulty players; instead, they reconstruct *pieces* of secrets held by the faulty processor. This bears no relation to the value of that faulty player's input; such information is not revealed.

Now, each reliable player has learned the pieces held by disqualified players. It now adjusts the value of its own pieces through a simple computation in order to obtain pieces of the original secrets, but shared with polynomials of degree $t - \tau$. For concreteness, assume only one player has been disqualified. Say that secret $s$ is shared via polynomial $p(u) = \sum_0^t a_t u^t$. Player $i$ has $\text{PIECE}_i(s) = p(\alpha_i)$. The newly publicized piece of the disqualified player is $p(\alpha_j) = \beta$. Let $m_1(u) = \prod_{i \neq j}(u - \alpha_i)$, $m_2(u) = (u - \alpha_j)$, and $M(u) = m_1(u) m_2(u)$. Define $g(u) = f(u) \bmod m_1(u)$; then by the Chinese Remainder Theorem,

$$
\begin{aligned}
M_1(u) &= [m_2(u)^{-1} (\bmod\ m_1(u))] \cdot m_2(u) \\
M_2(u) &= [m_1(u)^{-1} (\bmod\ m_2(u))] \cdot m_1(u) \\
f(u) &= M_1(u) g(u) + M_2(u) \beta \qquad (\bmod\ M(u))
\end{aligned}
$$

Then player $i$ computes its new piece of $s$ as:

$$g(\alpha_i) = [f(\alpha_i) - M_2(\alpha_i) \cdot \beta]/M_1(\alpha_i).$$

When using or revealing this value, the verification information is easily adjusted since it verifies the $f(\alpha_i)$ value, which is itself easy to recover from the $g(\alpha_i)$ value since $M_1$, $M_2$, $\alpha_i$, and $\beta$ are publicly known.

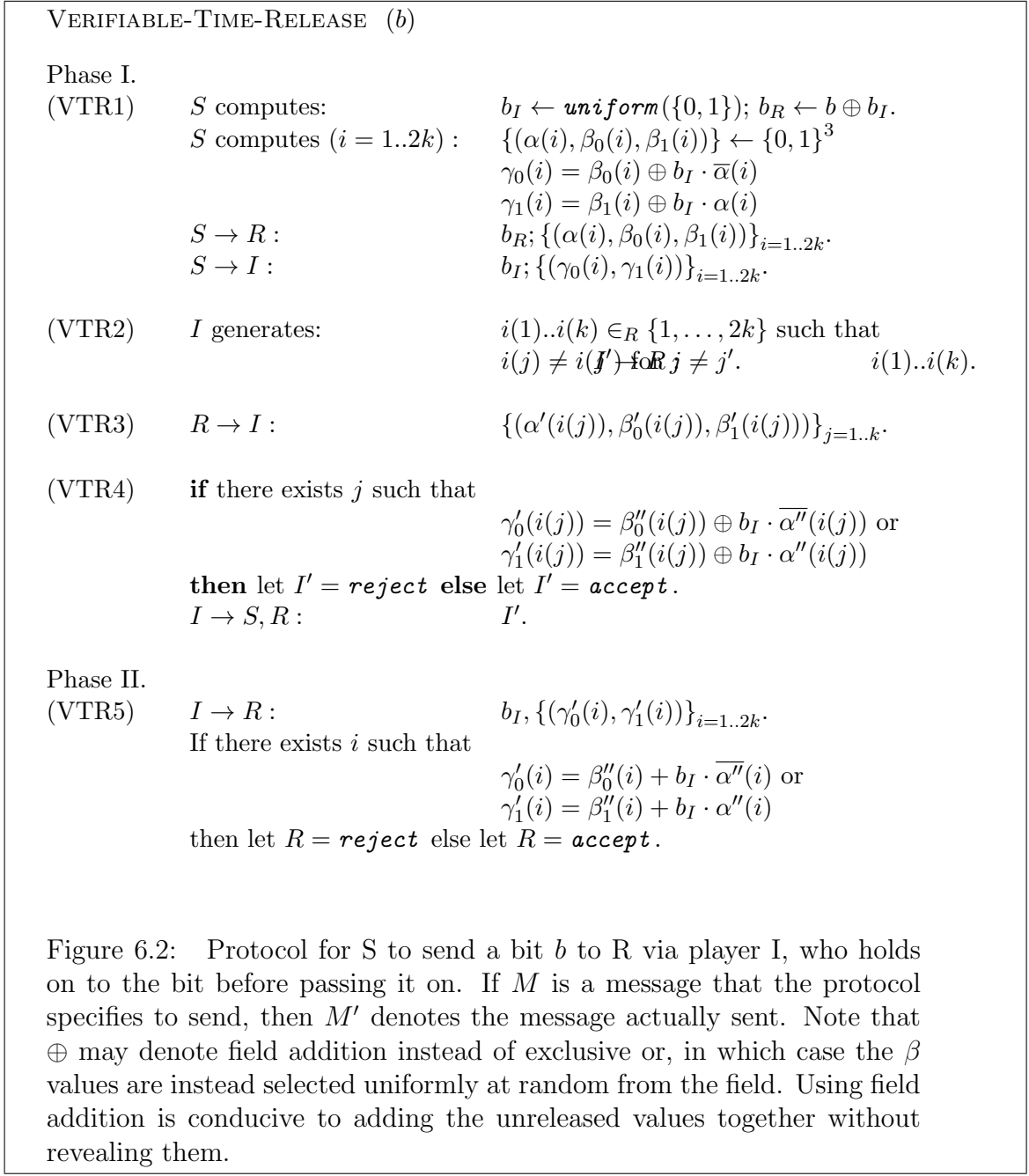

Verifiable-Time-Release  (*b*)

**Phase I.**
(VTR1)   $S$ computes:          $b_I \leftarrow \mathtt{uniform}(\{0,1\}); b_R \leftarrow b \oplus b_I.$
         $S$ computes $(i = 1..2k):$   $\{(\alpha(i), \beta_0(i), \beta_1(i))\} \leftarrow \{0,1\}^3$
                                        $\gamma_0(i) = \beta_0(i) \oplus b_I \cdot \overline{\alpha}(i)$
                                        $\gamma_1(i) = \beta_1(i) \oplus b_I \cdot \alpha(i)$
         $S \rightarrow R:$       $b_R; \{(\alpha(i), \beta_0(i), \beta_1(i))\}_{i=1..2k}.$
         $S \rightarrow I:$       $b_I; \{(\gamma_0(i), \gamma_1(i))\}_{i=1..2k}.$

(VTR2)   $I$ generates:        $i(1)..i(k) \in_R \{1, \ldots, 2k\}$ such that
                                $i(j) \neq i(j')$ for $j \neq j'.$          $i(1)..i(k).$

(VTR3)   $R \rightarrow I:$       $\{(\alpha'(i(j)), \beta_0'(i(j)), \beta_1'(i(j)))\}_{j=1..k}.$

(VTR4)   **if** there exists $j$ such that
                                $\gamma_0'(i(j)) = \beta_0''(i(j)) \oplus b_I \cdot \overline{\alpha''}(i(j))$ or
                                $\gamma_1'(i(j)) = \beta_1''(i(j)) \oplus b_I \cdot \alpha''(i(j))$
         **then** let $I' = \mathtt{reject}$ **else** let $I' = \mathtt{accept}.$
         $I \rightarrow S, R:$    $I'.$

**Phase II.**
(VTR5)   $I \rightarrow R:$       $b_I, \{(\gamma_0'(i), \gamma_1'(i))\}_{i=1..2k}.$
         If there exists $i$ such that
                                $\gamma_0'(i) = \beta_0''(i) + b_I \cdot \overline{\alpha''}(i)$ or
                                $\gamma_1'(i) = \beta_1''(i) + b_I \cdot \alpha''(i)$
         then let $R = \mathtt{reject}$ else let $R = \mathtt{accept}.$

Figure 6.2:   Protocol for S to send a bit $b$ to R via player I, who holds
on to the bit before passing it on. If $M$ is a message that the protocol
specifies to send, then $M'$ denotes the message actually sent. Note that
$\oplus$ may denote field addition instead of exclusive or, in which case the $\beta$
values are instead selected uniformly at random from the field. Using field
addition is conducive to adding the unreleased values together without
revealing them.

---

Linear-Combine **(a,b,c,u,v,w).**
Compute $w = au + bv + c$.

- Set
$$\text{PIECE}_i(w) = a \cdot \text{PIECE}_i(u) + b \cdot \text{PIECE}_i(v) + c.$$

- For each $j$, compute pieces of the pieces:
$$\text{PIECE}_i(\text{PIECE}_j(w)) = a \cdot \text{PIECE}_i(\text{PIECE}_j(u)) + b \cdot \text{PIECE}_i(\text{PIECE}_j(v)) + c.$$

- Choose new check vectors for $\text{PIECE}_j(\text{PIECE}_i(w))$ for all players $j \neq i$, and participate in their verification, as per the VSS protocol.

- Receive check vectors for other pieces $\text{PIECE}_i(\text{PIECE}_j(w))$ and participate in their verification, as per the VSS protocol.

Figure 6.3: Protocol for linear combinations of secrets. (Code for processor $i$.)

---

Multiply **(u,v,w).**

- Secretly share $h(i) = \text{PIECE}_i(u)\text{PIECE}_i(v)$.

- Run the Prove-Product protocol to prove that player $i$ shared this value.

- Receive pieces $\text{PIECE}_i(h(j))$ from each player $j \neq i$.

- Participate in the Prove-Product protocol for $2t < n$ to verify that player $j$ shared the correct value.

- Run the Truncate protocol using $h(i)$ to obtain a piece $\text{PIECE}_i(w)$ of a random, degree $t$ polynomial whose free term is $w$.

Figure 6.4: Protocol to multiply two secrets. (Code for processor $i$.)

PROVE-PRODUCT .

- Let $a$ and $b$ be verifiably shared.

- Alice verifiably shares $c = ab$.

- Alice shares secrets $r_1, \ldots, r_{2k_0}$ and $s_1, \ldots, s_{2k_0}$ chosen uniformly at random over the field $E$.

- For $j = 1, \ldots, 2k_0$, Alice computes $d_j = (a + r_j)(b + s_j)$, and shares $d_j$.

- The network confirms with high probability that each $d_j = (a + r_j)(b + s_j)$ :

    - For $i = 1, \ldots, k_0$, the network selects a random index $j_i$ by having each player select a random secret $\mod (2k_0 + 1 - i)$, sharing it (over fields of characteristic $p$ such that $p \mid 2k_0 + 1 - i$, so that the distribution is uniform), and then computing the sum.

    - Let $Y = \{j_1, \ldots, j_{k_0}\}$.

    - For all $j \in Y$, run the LINEAR-COMBINE protocol on $(a, r_j)$ to obtain the sum $a + r_j$.

    - For all $j \in Y$, run the LINEAR-COMBINE protocol on $(b, s_j)$ to obtain the sum $b + s_j$.

    - For all $j \in Y$, reconstruct $a + r_j$, $b + s_j$, and $d_j$. Disqualify Alice if any of the $d_j$ are not equal to $(a + r_j)(b + s_j)$.

- The network confirms that $c$ matches the product $ab$.

    - For all $j \notin Y$, reconstruct the values $r_j, s_j$.

    - For all $j \notin Y$, compute

    $$c_j = c - d_j + as_j + br_j + rs$$

    - For all $j \notin Y$, reconstruct $c_j$.

    - If any $c_j \neq 0$, disqualify Alice.

Figure 6.5:   Protocol to prove that the product of two secrets is a third secret. (Code for Alice and network.)

# Chapter 7

# Multiparty Zero-Knowledge Proof Systems

In this chapter, we generalize the concept of a zero-knowledge proof system in two ways. First, we consider proofs that some condition holds on secretly shared values. One player, the prover `Pro`, knows the values of a collection of secrets, and through an interaction with the rest of the network, it proves that a given predicate $\mathcal{P}$ holds on those secrets. Second, we consider efficient interactive proofs of general theorems, where secrecy is not the goal. Rather, the prover must prove a given theorem (*i.e.* prove membership in a language) to one or more verifiers. Instead of making unproven cryptographic assumptions to guarantee zero-knowledge, we utilize the presence of a majority of trusted players. No single player need be trusted.

**Assumptions made in this chapter.** The network is complete, with private lines, $n$ processors, and at most $t < \frac{n}{2}$ Byzantine faults, chosen dynamically. The protocols are unconditionally secure with high probability; the results are statistical zero-knowledge (with high probability *no* extra information is revealed). There are two distinguished parties, a *prover*, which is either polynomial-time or polynomial-space bounded, and a *verifier*, which is either unbounded or polynomial-time bounded.

## 7.1   Zero-Knowledge Proofs About Secrets

In Chapter 6 we considered the ABC Problem: Alice, the prover, must prove that $c = ab$, where $a, b$, and $c$ are secrets. Here, we extend Lemma 6.4 to demonstrate that a player may prove that a given predicate $\mathcal{P}$ holds on $m$ shared secrets $v_1, \ldots, v_m$, revealing no additional information. In a sense, this is analogous to the two-player problem of giving zero-knowledge proofs that a predicate holds on committed bits [Kil88a]. Here, we consider values shared among a network instead of committed bits, which have slightly different properties. For example, though committed bits are secret and unchangeable, they cannot be added together directly. Since most secure multiparty protocols are based on threshold schemes like secret sharing, proofs about shared values are an important tool.

Let $\mathcal{P}$ be a function on $m$ variables over the field $E$ and let $C_{\mathcal{P}}$ be an algebraic circuit over $E$ which computes $\mathcal{P}$; note that any boolean predicate on boolean inputs is easily modelled using arithmetic over $E$.

Let Pro and Ver be distinguished parties in the network. We consider Pro or Ver to be "cheating" if they are corrupted by an adversary $A$.

Let the prover's input be $x_{\text{Pro}} = (v_1, \ldots, v_m, w)$ and let all other $x_i = 0$. We define a function whose output is known only to the verifier: $F_i(x_1, \ldots, x_n) = 0$ if $i \neq \text{Ver}$, and

$$F_{\text{Ver}}(x_1, \ldots, x_n) = \begin{cases} 1 & \text{if } x_{\text{Pro}} = (v_1, \ldots, v_m, w) \text{ and } w = \mathcal{P}(v_1, \ldots, v_m) \\ 0 & \text{otherwise.} \end{cases}$$

Without regard to zero-knowledge, a simple proof system is trivial: simply reconstruct all the secrets and check if $w = \mathcal{P}(v_1, \ldots, v_m)$.

**Definition 7.1** *A protocol* $\Pi$ *is a* $(n, t)$ **multiparty zero-knowledge interactive proof system on secrets for** $\mathcal{P}$ *if it is a $t$-resilient $n$-player protocol for the function* $F_{\mathcal{P}}$ *defined above.*

The terms *statistical* and *computational* apply according to whether the protocol is statistically or computationally resilient. We call such a protocol a *MZKIPSS*, for short.

Theorem 5.7 tells us that there exists a *MZKIPSS* for any function $\mathcal{P}$ : simply compute $F_{\mathcal{P}}$ using the Evaluate protocol.

The purpose of this section, aside from defining the concept of proof systems with many parties, is to introduce a protocol which allows a direct proof system which is far more efficient than simulating $C_{\mathcal{P}}$. Simulating a circuit level by level (or even $(\log n)$-slice by slice) is an expensive undertaking. We can take advantage here, however, of the fact that the prover already knows the results of all of the gates. The prover shares the results of each and every gate, reducing the task of the verifier to checking that each shared result is correct. In other words, the system does not need to evaluate the circuit layer by layer, but rather only to verify that the collection of secrets representing the output values of each gate is correct. This verification requires only a local examination of the secret inputs and secret output of each gate, an operation which can be performed simultaneously for all the gates. Figure 7.1 details the protocol.

The results of Chapter 6 give the following:

**Theorem 7.2** *Let* $\mathcal{P}$ *be a family of predicates and let* $F_{\mathcal{P}}$ *and* $C_{\mathcal{P}}$ *be as described above. For $2t < n$, protocol* Prove-Secret-P *is a statistical $(n, t)$-MZKIPSS for* $\mathcal{P}$. *It runs in constant rounds and has message complexity polynomial in $(n, m, k, C_{\mathcal{P}})$. of* $\mathcal{P}$.

When $3t < n$, the results of [BGW88] ensure that the gatewise verification of $C_{\mathcal{P}}$ occurs without error, so that the simulation is perfect:

Prove-Secret-P ($\mathtt{Pro}, \mathtt{Ver}; \mathcal{P}; v_1, \ldots, v_m; w$).

> **For** $0 \le l \le \mathtt{depth}(C_{\mathcal{P}})$ **for** $1 \le j \le \mathtt{width}(C_{\mathcal{P}})$ **do in parallel :**
>> $\mathtt{Pro}$ shares the output of each gate $g_{lj}$ as a secret, $w_{lj}$.
>
> **For** $0 \le l \le \mathtt{depth}(C_{\mathcal{P}})$ **for** $1 \le j \le \mathtt{width}(C_{\mathcal{P}})$ **do in parallel :**
>> the network secretly computes $u_{lj}$ as follows:
>>> **if** $g_{lj} = (\times)$ **then** run Multiply to compute
>>>> $u_{lj} = g_{\mathtt{in}_1(g_{lj})} \cdot g_{\mathtt{in}_2(g_{lj})}.$
>>>
>>> **if** $g_{lj} = (+, a_0, \ldots, a_m)$ **then** run Linear-Combine to compute
>>>> $u_{lj} = a_0 + \sum_{j=1}^{m} g_{\mathtt{in}_m(g_{lj})}.$
>
> **For** $0 \le l \le \mathtt{depth}(C_{\mathcal{P}})$ **for** $1 \le j \le \mathtt{width}(C_{\mathcal{P}})$ **do in parallel :**
>> reveal $u_{lj}$ to $\mathtt{Ver}$ .
>
> If all $u_{lj} = 0$, $\mathtt{Ver}$ outputs *accept*, else it outputs *reject*.

Figure 7.1: Protocol to prove $w = \mathcal{P}(v_1, \ldots, v_m)$.

**Theorem 7.3** *Let $\mathcal{P}$ be a family of predicates and let $F_{\mathcal{P}}$ and $C_{\mathcal{P}}$ be as described above. For $3t < n$, protocol Prove-Secret-P is a perfect $(n,t)$-MZKIPSS for $\mathcal{P}$. It runs in constant rounds and has messages complexity polynomial in $(n, m, k, C_{\mathcal{P}})$.*

Protocols whose message complexity is polynomial in $n$, *independently* of the circuit complexity of $F_{\mathcal{P}}$, are presented in Chapter 13. They require the machinery of *locally random reductions*, and may require the players to compute time-consuming functions locally. Though the results of this section require polynomial local time if the circuit is of polynomial size, verifying huge circuits requires very little local time per gate but great cumulative local time. When the goal is to prove predicates regardless of their computational complexity, then the cost must be paid somewhere; better it be on the local computation than on the communication lines. The next section, however, describes a particular class of intractable functions for which only the prover itself need perform more than polynomial time computations.

## 7.2  Zero-Knowledge Proofs for IP

Any language $L \in$ IP certainly has a circuit $C_L$ which describes its characteristic function $\chi_L$; Theorems 7.2 and 7.3 imply there is a *MZKIPSS* for $L$ in the presence of a partly trusted network. The size of $C_L$, however, may be inordinately large, especially because IP = PSPACE [Sha89]. Protocol Prove-Secret-P would therefore require a great deal of time and very large messages to compute, and is unsuitable for proving intractable predicates in a network of polynomial time machines.

We present a protocol which achieves a zero-knowledge proof system for any language $L \in$ IP, yet uses only a polynomial number of rounds and polynomial message complexity. In particular, zero-knowledge proofs for any language in IP can be made without complexity theoretic assumptions, if a partly trusted network of probabilistic polynomial time Turing machines is available. In contrast, when there are only a prover and a verifier, perfect zero-knowledge proofs for NP-complete languages (which are in IP) are impossible unless the polynomial hierarchy collapses [For87].

The task at hand is distinguished from *MZKIPSS* in a few ways. First of all, there are no secretly shared values *a priori*. Second, the string $x$ and the language $L$ that must be proven to contain it are both public knowledge, or at least known to both prover and verifier. Third, the purpose of the network is not to act as a repository for secrets and committed values but to minimize the information leaked by the proof system while ensuring that the prover behaves.

We distinguish two members of the network as before, the prover `Pro` and the verifier `Ver`. In a two-party interactive proof system ([GMR89, BM88], and §2.2.3), the prover is a PSPACE machine and the verifier is a polynomial-time machine. The protocol is such that, if $x \in L$ and the prover is not corrupted, then an uncorrupted verifier will output *accept* with high probability. Otherwise, an uncorrupted verifier will output *reject*. Any language in PSPACE admits an interactive proof [Sha89]; the presence of a network of processors is irrelevant.

For the purpose of restricting information, however, a network of processors is invaluable. Zero-knowledge proof systems are often based on unproven complexity assumptions [GMR89]; perfect zero-knowledge proofs even for languages in NP cannot be achieved, unless the polynomial hierarchy collapses [For87]. Perfect zero-knowledge proof systems can be obtained given *two* provers that are not allowed to communicate [BGKW88, BGG88]. Here, we allow only one prover, and that prover is allowed to communicate with all of the participants and to collaborate with a constant fraction.

The idea of zero-knowledge proofs again corresponds to secure multiparty protocols. The protocol must provide the verifier with one of two outputs, *accept* and *reject*. If the prover and verifier are not corrupted, then the verifier accepts with high probability if $x \in L$, and otherwise rejects. If the verifier is not corrupted and $x \notin L$, then the verifier rejects with high probability. As in the definition of two-party zero-knowledge, a cheating verifier must be able to simulate the conversation. The definition of multiparty protocol resilience suffices to provide privacy not only with respect to the verifier but with respect to the prover as well.

We define a particular function $F_L$ that states whether a string $x$ agreed upon by two players is in a given language $L$: $F_i(x_1, \ldots, x_n) = 0$ if $i \neq$ `Ver`, and

$$F_{\texttt{Ver}}(x_1, \ldots, x_n) = \begin{cases} 1 & \text{if } x_{\texttt{Pro}} = x_{\texttt{Ver}} = x \neq \Lambda \text{ and } x \in L \\ 0 & \text{otherwise.} \end{cases}$$

Note that either player can refuse to participate by failing to supply a proper input ($x = \Lambda$). The only way for an honest verifier to be convinced that $x \in L$ is if the prover decides to participate and $x \in L$.

**Definition 7.4** *Let $L$ be a language. A $(t, n)$ **zero-knowledge network proof system** (ZKNPS) for $L$ is a $t$-resilient $n$-player protocol for $F_L$.*

As usual, the adjectives *perfect, statistical, computational* apply.

**Theorem 7.5** *For $2t < n$ and any $L \in$ PSPACE, Prove-P is a statistical ZKNPS for $L$. For $3t < n$, Prove-P is a perfect ZKNPS for $L$.*

**Proof:** By a result of [GS86], $L$ admits an interactive proof if and only if there is an Arthur-Merlin protocol [BM88] for $L$, in which the verifier takes the position of Arthur, and simply sends random coins to Merlin. Arthur accepts the proof based on a polynomial time computation using the set of messages sent by Merlin.

Without loss of generality, assume that for some polynomial $p_1(m, k)$, the AM protocol runs in $p_1(m, k)$ steps, and that at each step, the messages are of length $p_2(m, k)$. The program of Arthur is simple to specify: at round $r$, generate $p_2(m, k)$ random bits, send them to Merlin, receive a message $M_r$ from Merlin, and repeat $p_1(m, k)$ times; afterwards, compute a polynomial-time function $V$ on the messages $M_1, \ldots, M_{p_1(m,k)}$ from Merlin, and output the result. Merlin, however, must compute some presumably complicated function $P_r$ to generate the correct response $M_r$ at each round.

Thus we may take the interactive proof system to be the hidden composition of the functions

$$V \Diamond P_{p_1(m,k)} \Diamond V_{p_1(m,k)} \Diamond \cdots \Diamond P_1 \Diamond V_1.$$

The result, and only the result, is revealed to `Ver` .

Because each function $V_r$ is simply a set of random bits, protocol Rand-Secret suffices. The key is to construct a protocol that computes each $P_r$ in a $t$-resilient fashion. In general, the functions $P_r$ may be too complex to admit small circuits, and hence would be expensive in terms of interaction. In fact, however, there is no need to compute $P_r$ through circuit simulation. Instead, the network reveals $V_r$ to `Pro` , who computes $P_r$ on its own, and shares the result. Notice that the functions $V_r$ are private, since they are simply random bit sequences independent of all input values; hence revealing them to $P_r$ is perfectly tolerable.

The computation of function $V$ could be performed through circuit simulation, but at an expense of polynomially many rounds, since there is no guarantee that $V$, a polynomial-time calculation, admits sub-polynomial depth circuits. Instead, `Pro` computes $V$ and shares the result, $v$. Player `Pro` then provides a zero-knowledge proof on *secrets*, as per §7.2, to ensure that $v$ is correct. By verifying each step in the computation of $V$ locally yet in parallel, the number of rounds is kept down to a constant.

If `Pro` is corrupted and fails to share the correct value of $P_r$ for some $r$ or fails to supply a correct proof at the end, this corresponds to its choosing, in the ideal protocol, an input $x = \Lambda$. The prover has the inevitable right not to participate; but the protocol we present ensures that lack of participation is reflected properly in the final results. Notice that by the definition of two-party interactive proof systems, function $V$ is robust against incorrect computations of $P_1, \ldots, P_r$, so that a corrupt `Pro` who shares incorrect values corresponds to a prover in the ideal case who chooses not to participate.

---

PROVE-P

> **For** $u = 1..p_1(m,k)$ **do in parallel**
> > **For** $v = 1..p_2(m,k)$ **do in parallel**
> > > Run RAND-BIT to generate random secret $b_{uv}$.
> **For** $r = 1..p_1(m,k)$ **do**
> > Reconstruct $V_r = (b_{r,1}, \ldots, b_{r,p_1(m,k)})$ for Pro.
> > Pro computes $M_r = P_r$ and shares it.
> Pro computes $v = V(M_{p_1(m,k)}, V_r p_1(m,k), \ldots, M_1, V_1)$ and shares it.
> Run PROVE-SECRET-P ( Pro , Ver ;$V$; $M_{p_1(m,k)}, V_r p_1(m,k), \ldots, M_1, V_1$;$v$).
> > **if** output is *reject* or $v \neq$ *accept* **then**
> > > Ver outputs *reject.*
> > **else**
> > > Ver outputs *accept.*

Figure 7.2: Protocol for Pro to prove $x \in L$ to Ver.

---

Formally speaking, we perform the composition of the following *ideal* protocols. In the first protocol, each player supplies a sequence of $p_1(m,k)p_2(m,k)$ random bits, and the trusted host computes a string of $p_1(m,k)p_2(m,k)$ uniformly random bits by computing the parities of each group of $n$ bits. It divides these into $p_1(m,k)$ strings $V_r$ of length $p_2(m,k)$ and returns a robust and secret representation of them to the players. In each protocol $\mathrm{ID}^{2r}$, the trusted party gives $V_r$ to Pro. In each protocol $\mathrm{ID}^{2r+1}$, Pro computes $P_r$ and gives it to the trusted host, who computes a robust and secret representation of what Pro supplied, and returns pieces to the players. In the final protocol $(2p_1(m,k)+2)$, each player provides its pieces of the robust and private representation of all of the previous inputs, and the trusted host computes $V$ on the reconstructed values. It returns the result to Ver .

The composition of these protocols provides Ver with the closed composition

$$V \Diamond P_{p_1(m,k)} \Diamond V_{p_1(m,k)} \Diamond \cdots \Diamond P_1 \Diamond V_1$$

which reveals only the final result $V$, as desired. By Theorems 7.2, 7.3, 4.31, 5.7, and 6.1, the results are perfectly (statistically) $t$-resilient if $3t < n$ $(2t < n)$. ∎

# Chapter 8

# Privacy for Passive Adversaries

> "Notice all the computations, theoretical scribblings, and lab equipment,
> Norm.... Yes, curiosity killed these cats."
>
> — Gary Larson

Multiparty computations are generally impossible when the number of faulty players exceeds half the network, unless additional assumptions are made. Chapter 9 describes how noisy channels allow computations to proceed correctly and fairly, and demonstrates how to achieve the same results using a cryptographic protocol for oblivious transfer.

In order to elucidate the nature of privacy in distributed computations, we examine protocols which tolerate very large numbers of passive faults. We call this the "honest-but-curious" model: each player is honest and never sends a faulty message, but players may form coalitions and pool their knowledge to try to obtain extra information to which they are not entitled. We may more or less ignore the issues of correctness and fault-tolerance, focusing only on privacy.

When the bound $t$ on the number of curious parties satisfies $t < \lceil \frac{n}{2} \rceil$, any function can be computed privately in the "honest-but-curious" model (*cf.* [BGW88, CCD88] and Chapters 5 and 6).

When $t \geq \lceil \frac{n}{2} \rceil$, the class of privately computable functions is restricted. Chor and Kushilevitz [CK89] characterized the set of *boolean* functions that can be computed privately for $t \geq \lceil \frac{n}{2} \rceil$ as those of the form

$$f(x_1, x_2, \ldots, x_n) = f_1(x_1) \oplus f_2(x_2) \oplus \cdots f_n(x_n).$$

In other words, for the boolean case, the only functions privately computable when a majority of the parties are curious are exclusive-or's of $n$ functions, each depending on one input.

The results of [Yao86, GMW87, GHY87, BG89] demonstrate that when the participants are computationally bounded but still only curious, *any* function can be privately computed for $t \leq n-1$. Furthermore, even functions that have non-boolean outputs are privately computable. In an information-theoretic sense, however, coalitions of curious parties do hold more information than that to which they are entitled.

The complete characterization of privately computable functions when the participants are not bounded remains an open question:

> What *general* functions (say from $n$ inputs to $n$-bit outputs) can be computed privately by $n$ parties, allowing $t \geq \lceil \frac{n}{2} \rceil$, and maintaining privacy in an *information-theoretic* sense?

In this chapter, we take the penultimate step toward a complete characterization, by solving the following problem:

> What *general* functions can be computed privately by *two* parties?

We give a characterization of functions that are privately computable by two parties, based on a simple property of the table for the functions. Specifically, any function whose table can be *partitioned* in a certain manner can be privately computed by two parties; and any function that can be privately computed by two parties has a table which can be partitioned. (See §8.2.4 for a statement of the main theorem.)

In preliminary papers, Kushilevitz [Kus89] and Beaver [Bea89a] independently obtained this characterization, but the proofs appearing in those papers contain a subtle flaw. We present here a different proof, which uses a different approach (see §8.2.2).

Our characterization bears on the $n$-party problem in that any $n$-party protocol that allows $t \geq \lceil \frac{n}{2} \rceil$ can be adapted to a two-party protocol. Thus, functions that are privately computable in the $n$-party case must satisfy certain properties satisfied by functions that are private in the two-party case.

We also address a different version of the $n$-party problem, in which the output of the function is not revealed. Instead, it is maintained in a distributed form, using an arbitrary threshold scheme. In view of the recent development of multiparty computation protocols based on threshold schemes (cf. [BGW88, CCD88]), as well as the protocols described in the rest of this dissertation, all of which allow the output to be maintained in a shared form, a characterization of the functions that can be computed secretly is needed. We show a strongly negative result: the only functions that can be computed secretly, maintaining the result as a secret (for use in further protocols), are additive functions. This result fits neatly with that of [CK89] showing that privately computable boolean functions are addition functions ( mod 2).

**Assumptions made in this chapter.** The network is complete, with private lines, 2 or $n$ computationally unbounded processors. Protocols are perfectly resilient against passive $t$-adversaries where $t$ may be greater than $\frac{n}{2}$.

## 8.1 Definitions

For the two-party case, we assume that the two parties communicate using finite strings and for a finite number of exchanges. Neither party is computationally bounded; the privacy of the protocols is based on information-theoretic bounds. Because we consider only two parties, without an active adversary, we can make some simplifying observations.

A party can be described simply by a family of distributions on strings, parametrized by the input and the current transcript of the protocol. These distributions are induced by the formal transition functions, to which we shall not refer again. Thus, party 1 is a family of distributions $\{D^1_{x,t} \mid x \in X, t \in \Sigma^\star\}$, where $\Sigma = \{0, 1\}$, and each $D^1_{x,t}$ is a distribution on $\Sigma^\star$. The message that party 1 sends after round $r$ is a finite string selected at random according to distribution $D^1_{x,t_r}$, where $x$ is its input and $t_r$ is the transcript of messages through round $r$. Party 2 is defined similarly. We assume that the transcripts are delimited so that the messages are uniquely decodable from the transcripts.

Let the domain of $f$ be $X \times Y = \{1, \ldots, \alpha\} \times \{1, \ldots, \beta\}$. Let $T$ be a random variable which describes the transcript of message exchanges between parties 1 and 2. We say that $f$ is *private* if there is a protocol such that, when party 1 holds $x \in X$ and party 2 holds $y \in Y$:

- After any run of the protocol, each party knows $f(x, y)$.

- **(Privacy for party 1)** For any $x'$ such that $f(x', y) = f(x, y)$ and for any transcript $t$:
$$\Pr[t \mid x', y] = \Pr[t \mid x, y].$$

- **(Privacy for party 2)** For any $y'$ such that $f(x, y') = f(x, y)$ and for any transcript $t$:
$$\Pr[t \mid x, y'] = \Pr[t \mid x, y].$$

We shall also denote the prefix of a transcript $t$ after the $i^{th}$ round by $t_i$, and the corresponding random variable by $T_i$.

## 8.2 Two Party Privacy for Passive Adversaries

### 8.2.1 Partitions

For any $x \in X$ and any subset $P \subseteq Y$, we define the *row set* for row $x$ and columns in $P$ to be the range of values that $f$ takes over that row of its table:

$$R(x, P) = \{f(x, y) \mid y \in P\}.$$

Similarly, we define the *column set* for $y \in Y$ and a subset of values $P \subseteq X$:

$$C(P, y) = \{f(x, y) \mid x \in P\}.$$

We say that $f$ is *column-partitionable* into $f_P$ and $f_Q$ if if there exists a nontrivial partition of $Y$ into blocks $P$ and $Q$ such that

$$(\forall x \in X)R(x, P) \quad \cap R(x, Q) = \emptyset,$$

and $f_P$ and $f_Q$ are the restriction of $f$ to the sets $P$ and $Q$. In other words, if we look at a particular row in the table for $f$ and consider the range of values which $f$ takes on for $y \in P$, then no such value will appear as the output of $f$ for $y \in Q$. Similarly, $f$ is *row-partitionable* if there exists a nontrivial partition of $X$ into blocks $P$ and $Q$ such that

$$(\forall y \in Y)R(P, y) \quad \cap R(Q, y) = \emptyset,$$

and $f_P$ and $f_Q$ are the restriction of $f$ to the sets $P$ and $Q$.

Partitionability is recursively defined as follows. We say that $f$ is *partitionable* if:

1. $f$ is constant; or

2. $f$ is column-partitionable or row-partitionable into $f_P$ and $f_Q$, each of which are themselves partitionable.

For example, the AND function is not partitionable, while the function $f(x, y) = x + y \bmod 7$ is partitionable. Functions that are not partitionable need not be based on AND or OR, as demonstrated by the following table, which cannot be partitioned by rows or by columns:

| $f(x,y)$ | 0 | 1 | 2 |
|----------|---|---|---|
| 0 | 0 | 0 | 1 |
| 1 | 3 | 4 | 1 |
| 2 | 3 | 2 | 2 |

It is easy to see that any function that is insensitive to $x$ or to $y$ is partitionable.

It is also straightforward to determine from an $n \times n$ table for $f$ whether $f$ is partitionable, by using a transitive-closure algorithm, which runs in time polynomial in $n$, to determine the columns or rows that must fall in the same blocks of any allowable partition.

### 8.2.2 Two Parties Cannot Compute Unpartitionable Functions

We show the following result:

**Lemma 8.1** *If $f$ is not partitionable, then $f$ cannot be computed privately by two parties.*

Before proceeding to the proof, let us expose the flaw in the earlier proofs of Kushilevitz [Kus89] and Beaver [Bea89a].

Without loss of generality, let party 1 send messages $m_r$ to party 2 during odd-numbered rounds $r$; let party 2 send messages during even-numbered rounds. This does not restrict who "speaks" first, since null or completely random messages are allowed. We expand the probability of a transcript $t$ given $x$ and $y$ according to its prefixes $t_1, t_2, \ldots$, where $t_i = m_1 \circ m_2 \circ \cdots \circ m_i$ :

$$
\begin{aligned}
\Pr\left[t \mid x, y\right] &= \prod_r \Pr\left[t_r \mid x, y, t_{r-1}\right] \\
&= (\prod_j \Pr\left[t_{2j+1} \mid x, y, t_{2j}\right])(\prod_j \Pr\left[t_{2j} \mid x, y, t_{2j-1}\right]) \\
&= P_1(t, x) P_2(t, y)
\end{aligned}
$$

where $P_1$ and $P_2$ are defined as the parenthesized expressions. (This notation appears literally as $P_1(s \mid x), P_2(s \mid y)$ in [Kus89], and as $\gamma_t(x), \delta_t(y)$ in [Bea89a].)

The faulty proof appearing in [Kus89, Bea89a] runs along the following lines. Using the privacy of $f$, it is shown by induction that $\Pr\left[t \mid x, y\right]$ is constant over all $x$. The induction step fails, however, since it is based on the following statement, which intends to show that $P_1$ is insensitive to $x$ : If $\Pr\left[t \mid x_1, y\right] = \Pr\left[t \mid x_2, y\right]$ then $P_1(t, x_1) = P_1(t, x_2)$. Unfortunately, if $P_2(t, y) = 0$, this conclusion does not necessarily hold.

For example, consider the function described by

| $f(x, y)$ | $y_1$ | $y_2$ |
|---|---|---|
| $x_1$ | 1 | 2 |
| $x_2$ | 1 | 3 |

and the 4-round deterministic protocol where party 1 starts, and each party sends 1 bit, described by the table of transcripts:

| | $y_1$ | $y_2$ |
|---|---|---|
| $x_1$ | 0000 | 0100 |
| $x_2$ | 0000 | 0110 |

Then $\Pr\left[0000 \mid x_1, y_1\right] = P_1(0000, x_1) P_2(0000, y_1) = \Pr\left[0000 \mid x_2, y_1\right] = P_1(0000, x_2) P_2(0000, y_1)$, and we have $P_2(0000, y_1) = 1$ and $P_1(0000, x_1) = P_1(0000, x_2) = 1$. The claim that $P_1$ is insensitive to $x$ is satisfied for this particular transcript. On the other hand, $\Pr\left[0100 \mid x_1, y_1\right] = \Pr\left[0100 \mid x_2, y_1\right] = 0$, but $P_1(0100, x_1) = 1$ while $P_1(0100, x_2) = 0$. Informally, transcript 0100 is impossible given $y_1$, but it may or may not be possible for different $x$, a fact that cannot be determined simply by looking at the column corresponding to $y_1$. The claim that $P_1$ is insensitive to $x$ fails.

It turns out that $P_1$ is indeed insensitive to $x$ when $f$ cannot be partitioned into columns, but this is not a trivial observation, and the proof must use the unpartitionability of $f$ by columns. The earlier proofs did not use this property, and thereby fail on a counterexample like the one presented above (note that $f$ is not partitionable by rows, but it is partitionable by columns). Introducing the property of unpartitionability by columns into the earlier proof techniques seems to be an unwieldy approach, without an easy fix; instead, we present an alternative proof.

**Proof of Lemma 8.1:** Since $t_{2j+1} = t_{2j} \circ m_{2j+1}$, and $m_{2j+1}$ is selected by party 1 according to distribution $D^1_{x,t_{2j}}$, we may write:

$$\Pr\left[t_{2j+1} \mid x,y\right] = \Pr\left[t_{2j+1} \mid x,t_{2j}\right] \Pr\left[t_{2j} \mid x,y\right].$$

Similarly,

$$\Pr\left[t_{2j} \mid x,y\right] = \Pr\left[t_{2j} \mid y,t_{2j-1}\right] \Pr\left[t_{2j-1} \mid x,y\right].$$

First we shall show that if $f$ is privately computable but not partitionable, then party 1 "never speaks first." In other words, all conversations are independent of $x$ until party 2 gives away some information depending on $y$. Then we shall show that party 2 never speaks first, leading to a contradiction.

In order to formalize the notion of not speaking first, consider the following notations. Let $D \subseteq X \times Y$, and let $\tau(x,y) = \{t \mid \Pr\left[t \mid x,y\right] \neq 0\}$. Given a transcript $t$ and two pairs of inputs $(x,y)$ and $(u,v)$, let $\theta(t,x,y,u,v)$ be the value $r$ such that

$$\Pr\left[t_r \mid x,y\right] \neq \Pr\left[t_r \mid u,v\right] \text{ and } (\forall \rho < r)\Pr\left[t_\rho \mid x,y\right] = \Pr\left[t_\rho \mid u,v\right].$$

Clearly, such an $r$ is unique if it exists. If no such $r$ exists, let $\theta(t,x,y,u,v) = \infty$. The earliest round at which the probabilities of conversations on $(x,y)$ differ from those for *some* other input $(u,v)$ is denoted

$$\phi(t,D,x,y) = \min_{(u,v) \in D} \quad \theta(t,x,y,u,v).$$

Now let

$$\psi(D) = \{\phi(t,D,x,y) \mid (x,y) \in D, t \in \tau(x,y)\}.$$

We say that party 1 *does not speak first on $D$* if $\psi(D) \subset \{2j \mid j \in \mathbf{N}\} \cup \{\infty\}$, that is, if the earliest times at which conversations differ are always even-numbered rounds, corresponding to party 2 sending a message to party 1.

First, let us show an easy lemma:

**Lemma 8.2** *(Row Lemma) For any $x \in X$ and $y_1, y_2 \in Y$, and for any possible transcript $t \in \tau(x,y_1)$, if*

$$Pr[t_r \mid x,y_1] \neq Pr[t_r \mid x,y_2] \text{ and } (\forall \rho < r)Pr[t_\rho \mid x,y_1] = Pr[t_\rho \mid x,y_2],$$

*then $r$ is not odd.*

**Proof:** If $r$ is odd, party 1 sends the message at round $r$ :

$$
\begin{aligned}
\Pr\left[t_r \mid x,y_1\right] &= \Pr\left[t_r \mid x,t_{r-1}\right] \Pr\left[t_{r-1} \mid x,y_1\right] \\
&= \Pr\left[t_r \mid x,t_{r-1}\right] \Pr\left[t_{r-1} \mid x,y_2\right] \\
&= \Pr\left[t_r \mid x,y_2\right]
\end{aligned}
$$

which is a contradiction. □

The crucial lemma we need is the following:

**Lemma 8.3** *If $f$ is privately computable but not partitionable, then party 1 does not speak first on $X \times Y$.*

**Proof:** We show by induction that there exist $P_1 \subset P_2 \subset \cdots \subset P_{|X|} = X$ such that party 2 speaks first on each $P_i \times Y$.

Using lemma 8.2, we see that party 1 does not speak first on $P_1 \times Y$, where $P_1 = \{a_1\}$ and $a_1 \in X$ is arbitrary. Since party 1 only has one argument in $P_1$, this is intuitively clear.

Assume by way of induction that party 1 does not speak first on $P_i \times Y$. We must demonstrate a $P_{i+1}$ such that $P_i \subset P_{i+1}$ and party 1 does not speak first on $P_{i+1}$.

Since $f$ is not partitionable, there exist values $x_1 \in P_i, x_2 \in \overline{P_i}$, and $y_1 \in Y$ so that $f(x_1, y_1) = f(x_2, y_1)$. Let $P_{i+1} = P_i \cup \{x_2\}$. To show that party 2 speaks first on $P_{i+1} \times Y$, it suffices to show that for each $y \in Y$, $\phi(t, \{x_2\} \times Y, x_2, y)$ is never odd and $\phi(t, P_i \times Y, x_2, y)$ is never odd. In other words, we consider the earliest times that a meaningful message is sent for the new inputs in $\{x_2\} \times Y$ relative to the original set $P_i \times Y$ and relative to the new input set itself.

Lemma 8.2 shows that $\phi(t, \{x_2\} \times Y, x_2, y)$ is never odd.

To obtain the claim that $\phi(t, P_i \times Y, x_2, y)$ is never odd, let us divide the row $\{x_2\} \times Y$ into three sets, $A = \{(x_2, y_1)\}$, $B = \{(x_2, y) \mid (\exists a \in P_i) f(a, y) = f(x_2, y)\} \backslash A$, and $C = \{(x_2, y) \mid (\forall a \in P_i) f(a, y) \neq f(x_2, y)\}$.

First, we show that $\phi(t, P_i \times Y, x_2, y_1)$ is even for any possible transcript $t \in \tau(x_2, y_1)$. If $t \in \tau(x_2, y_1)$ then by the privacy of $f$,

$$(\forall r)\Pr[t_r \mid x_1, y_1] = \Pr[t_r \mid x_2, y_1]$$

Then it follows easily that $\phi(t, P_i \times Y, x_2, y_1) = \phi(t, P_i \times Y, x_1, y_1)$, and our first claim follows.

By a similar argument, $\phi(t, P_i \times Y, x_2, y)$ is never odd for any $y \in B$.

Thirdly, for any $(x_2, y) \in C$, we show $\phi(t, P_i \times Y, x_2, y)$ is never odd. We have two cases to consider: $\theta(t, x_2, y, x_1, y_1)$ is minimal or it is not. Say it is minimal. Again, by the privacy of $f$ we have that

$$(\forall r)\Pr[t_r \mid x_1, y_1] = \Pr[t_r \mid x_2, y_1]$$

and the smallest $r$ at which $\Pr[t_r \mid x_1, y_1] \neq \Pr[t_r \mid x_2, y]$ is identical to that for which $\Pr[t_r \mid x_2, y_1] \neq \Pr[t_r \mid x_2, y]$. Since $(x_2, y_1)$ and $(x_2, y)$ are in the same row, $r$ must be even.

If $\theta(t, x_2, y, x_1, y_1)$ is not minimal, let $(u, v)$ be such that $\theta(t, x_2, y, u, v)$ is minimal. For a given transcript $t$, consider the probabilities $\Pr[t_r \mid u, v]$, $\Pr[t_r \mid x_1, y_1]$, and $\Pr[t_r \mid x_2, y]$. At some earliest round $r$, two or three of these differ. Since $(u, v)$ gave the minimal value with respect to $(x_2, y)$ we see that $\Pr[t_r \mid x_2, y] \neq \Pr[t_r \mid u, v]$. Since $(x_1, y_1)$ did not give the minimal value, $\Pr[t_r \mid x_2, y] = \Pr[t_r \mid x_1, y_1]$. Therefore $\Pr[t_r \mid x_1, y_1] \neq \Pr[t_r \mid u, v]$. By the induction hypothesis, since $(x_1, y_1)$ and $(u, v)$ are in $P_i \times Y$, we deduce $r$ is even. □

A symmetric argument gives the corresponding lemma:

**Lemma 8.4** *If $f$ is privately computable but not partitionable, then party 2 does not speak first on $X \times Y$.*

Combining lemmas 8.3 and 8.4, we see that neither party speaks first on $X \times Y$, implying that $\psi(X \times Y) = \{\infty\}$. Let $a \in X$ and $b \in Y$ be arbitrary, and let $t \in \tau(a, b)$. Then for every $x \in X$ and $y \in Y$, we have $\theta(t, a, b, x, y) = \infty$, implying $\Pr[t \mid a, b] = \Pr[t \mid x, y]$. Since $f(x, y)$ is determined by the transcript $t$, $f$ must be constant and hence partitionable, giving a contradiction. □

**Remarks:** The proof of lemma 8.1 does not assume that the protocol is deterministic. In the next section we observe that a deterministic protocol suffices to compute any partitionable function. Furthermore, even if the protocol is only correct with probability exceeding $\frac{1}{2}$, the argument shows that the distributions on transcripts are identical regardless of the inputs, so that $f$ must be constant, giving a contradiction.

### 8.2.3   Two Parties Can Compute Partitionable Functions

We now show a converse result:

**Lemma 8.5** *If $f$ is partitionable, then there is a (deterministic) protocol whereby $f$ can be computed privately by two parties.*

**Proof:** By induction on $\alpha = |X|$.
**(Base Case.)** If $X = \{x_1\}$, then party 2 simply announces $f(x_1, y)$. This protocol is trivially private for party 1, and since for all $y, y' \in Y$ such that $f(x_1, y) = f(x_1, y')$ the transcript consists exactly of $f(x_1, y)$, the protocol is private for party 2.
**(Inductive Hypothesis.)** If $|X| \leq \alpha$, then if $f$ is partitionable then there is a deterministic protocol $\Pi_f$ to compute $f$.
**(Inductive Step.)** Let $|X| = \alpha + 1$. Say that $f$ is partitionable. If $f$ is constant there is a trivial private protocol. Otherwise, $f$ is row-partitionable or column-partitionable.

If $f$ is row-partitionable into $f_P$ and $f_Q$, then by the induction hypothesis there exist private protocols $\Pi_{f_P}$ and $\Pi_{f_Q}$ for $f_P$ and $f_Q$. Define the protocol $\Pi_f$ as follows. If $x \in P$, party 1 sends a 0; then the parties execute the protocol for $\Pi_{f_P}$. Otherwise if $x \in Q$, party 1 sends a 1, and the parties execute the protocol for $\Pi_{f_Q}$.

Since the protocols are deterministic, let $\Pi(x, y)$ denote the transcript of protocol $\Pi$ on inputs $x$ and $y$.

Protocol $\Pi_f$ is private with respect to party 1. If $f(x, y) = f(x', y)$, then either $x, x' \in P$ or $x, x' \in Q$; otherwise $f$ would not be row-partitionable. If $x, x' \in P$, then

$$\Pi_f(x, y) = 0 \circ \Pi_{f_P}(x, y) = 0 \circ \Pi_{f_P}(x', y) = \Pi_f(x', y).$$

Similarly, if $x, x' \in Q$, then $\Pi_f(x, y) = \Pi_f(x', y)$.

Protocol $\Pi_f$ is also private with respect to party 2. If $f(x, y) = f(x, y')$, then either $x \in P$, in which case

$$\Pi_f(x, y) = 0 \circ \Pi_{f_P}(x, y) = 0 \circ \Pi_{f_P}(x, y') = \Pi_f(x, y');$$

or $x \in Q$, which likewise gives $\Pi_f(x, y) = \Pi_f(x, y')$.

Now, if $f$ is not row-partitionable then it must be column-partitionable into $P$ and $Q$. We now prove by induction on $\beta = |Y|$ that $f$ is private. Say $\beta = 1$; then $f$ is trivially private. Now, assume that the hypothesis holds for $|Y| \le \beta$. Consider $|Y| = \beta + 1$. Then $f$ is partitionable into $f_P$ and $f_Q$, where $|P|, |Q| \le \beta$. Hence there exist private protocols $\Pi_{f_P}$ and $\Pi_{f_Q}$ for $f_P$ and $f_Q$. The protocol for $f$ is straightforward: If $y \in P$, party 2 sends a 0; then the parties execute the protocol for $\Pi_{f_P}$. Otherwise if $y \in Q$, party 2 sends a 1, and the parties execute the protocol for $\Pi_{f_Q}$. The proof that this $\Pi_f$ is private follows the same lines as above. □

### 8.2.4 Main Result

We can summarize the main result of this chapter in the following theorem:

**Theorem 8.6** *A finite function $f(x, y)$ is 1-private if and only if it is partitionable.*

**Proof:** Follows from lemmas 8.1 and 8.5. □

## 8.3 Multiparty Privacy for Modular Protocols

As described in Chapters 4, 5 and 6, Many of the methods for performing arbitrary secret computations follow the method of constructing a modular library of protocols that can be selected and combined to construct new protocols. A general library based on threshold (secret sharing) methods allows the output of one protocol to be maintained in a secret format suitable for input to another protocol, so that intermediate computations are not revealed.

In this section we consider what functions can be computed given an arbitrary threshold scheme that allows addition of secrets over the field used for sharing, when we also require that the output of the function be retained as a secretly shared value for potential use in further computations.

Let $E$ be a field and let $L_0$ be a set of function families mapping $E^n$ to $E$. Let $L$ be the closure of functions that can be written as a finite composition of functions in $L_0$. For example, with $g, f_1, \ldots, f_n \in L$, the following function is in $L$.

$$g(f_1(x_1, \ldots, x_n), \ldots, f_n(x_1, \ldots, x_n)).$$

A $t$-private *library* $\mathcal{L}$ for $L$ is a set of protocols satisfying:

- Each protocol $\Pi \in \mathcal{L}$ computes a function $f \in L$, $t$-privately.

- For every $f \in L$ there is a protocol $\Pi \in \mathcal{L}$ that computes it.

The protocols for multiplication and addition given in [BGW88, CCD88] satisfy these properties, and in fact are complete for any finite function when $t < \lceil \frac{n}{2} \rceil$. In other words, they form a $\lceil \frac{n}{2} \rceil$-private library for the class of all finite functions.

For $t \geq \lceil \frac{n}{2} \rceil$, however, the situation is less optimistic. We show that any library sufficiently powerful to compute affine functions over $E$ can *only* compute affine functions over the field $E$.

**Theorem 8.7** *If $\mathcal{L}$ is a $t$-private library of protocols that includes all affine functions over the field $E$, and if $t \geq \lceil \frac{n}{2} \rceil$, then any function $f(x_1, \ldots, x_n) \in L$ can be written*

$$f(x_1, \ldots, x_n) = f_1(x_1) + f_2(x_2) + \ldots + f_n(x_n).$$

**Proof:** Let $f(x_1, \ldots, x_n)$ be $t$-private for some $t \geq \lceil \frac{n}{2} \rceil$, and let $f$ depend on $k$ variables, $1 \leq k \leq n$. That is, for some $j_1, \ldots, j_k$ and $\hat{f}$,

$$f(x_1, \ldots, x_n) = \hat{f}(x_{j_1}, \ldots, x_{j_k}).$$

Then we show by induction on $k$ that

$$(\exists f_1, \ldots, f_n, \hat{f}) \ \ f(x_1, \ldots, x_n) = \sum_{i=1}^{n} f_i(x_i).$$

The statement is trivially true for $k = 1$. Assume that it holds for $k$.

Let $f$ be $t$-private and let $f$ depend on $k + 1$ variables:

$$(\exists j_1, \ldots, j_{k+1}) \ \ f(x_1, \ldots, x_n) = \hat{f}(x_{j_1}, \ldots, x_{j_{k+1}}).$$

Without loss of generality let $j_1 = 1, j_2 = 2, \ldots, j_{k+1} = k+1$. Fix arbitrary arguments $a_1, \ldots, a_{k+1}$, and let $l = \lceil \frac{k}{2} \rceil$. Suppose there exist $b_1, \ldots, b_{k+1}$ such that

$$\begin{aligned}
\hat{f}(b_1, \ldots, b_l, b_{l+1}, \ldots, b_{k+1}) + \hat{f}(a_1, \ldots, a_l, a_{l+1}, \ldots, a_{k+1}) \ &\neq \ \ \ \ \ \ \ \ \ (8.1) \\
\hat{f}(a_1, \ldots, a_l, b_{l+1}, \ldots, b_{k+1}) + \hat{f}(b_1, \ldots, b_l, a_{l+1}, \ldots, a_{k+1}). &
\end{aligned}$$

It is not hard to see that there exist $r$ and $s$ with $r \neq s$, such that $a_r \neq b_r$ and $a_s \neq b_s$.

Define the linear functions $p_r$ and $p_s$ as

$$\begin{aligned}
p_r(x_1, \ldots, x_{k+1}) \ &= \ (x_r - a_r)(b_r - a_r)^{-1}(\hat{f}(b_1, \ldots, b_l, a_{l+1}, \ldots, a_{k+1}) \\
&\quad - \hat{f}(a_1, \ldots, a_l, a_{l+1}, \ldots, a_{k+1})) \\
&\quad + \hat{f}(a_1, \ldots, a_l, a_{l+1}, \ldots, a_{k+1}) \\
p_s(x_1, \ldots, x_{k+1}) \ &= \ (x_s - a_s)(b_s - a_s)^{-1}(\hat{f}(a_1, \ldots, a_l, b_{l+1}, \ldots, b_{k+1}) \\
&\quad - \hat{f}(a_1, \ldots, a_l, a_{l+1}, \ldots, a_{k+1})) \\
&\quad + \hat{f}(a_1, \ldots, a_l, a_{l+1}, \ldots, a_{k+1})
\end{aligned}$$

Using the assumption that the library contains all affine functions, the function $G$ defined as follows is $t$-private:

$$G(x_1, \ldots, x_n) = f(x_1, \ldots, x_n) - p_r(x_1, \ldots, x_{k+1}) - p_s(x_1, \ldots, x_{k+1}) + \hat{f}(a_1, \ldots, a_{k+1}).$$

Clearly,

$$
\begin{aligned}
G(a_1, \ldots, a_l, a_{l+1}, \ldots, a_{k+1}, 0, 0, \ldots, 0) &= 0, \\
G(a_1, \ldots, a_l, b_{l+1}, \ldots, b_{k+1}, 0, 0, \ldots, 0) &= 0, \\
G(b_1, \ldots, b_l, a_{l+1}, \ldots, a_{k+1}, 0, 0, \ldots, 0) &= 0, \\
G(b_1, \ldots, b_l, b_{l+1}, \ldots, b_{k+1}, 0, 0, \ldots, 0) &\neq 0.
\end{aligned}
$$

Let $\Pi_G$ be a $t$-private protocol for $G$. We use $\Pi_G$ to construct a 1-private protocol for two parties to compute AND(x,y), by having party 1 simulate half the parties and party 2 simulate the other half. Party 1 holds an input $x \in \{0, 1\}$, and party 2 holds an input $y \in \{0, 1\}$, though when they simulate $\Pi_G$ they will "pretend" to hold general arguments instead. Let $m = \lfloor \frac{n-k-1}{2} \rfloor$. Party 1 follows $\Pi_G$, simulating the parties holding inputs $x_1, \ldots, x_l; x_{k+2}, \ldots, x_{k+1+m}$. Party 2 follows $\Pi_G$, simulating the parties holding inputs $x_{l+1}, \ldots, x_{k+1}; x_{k+m+2}, \ldots, x_n$. Notice that neither party 1 nor party 2 simulates more than $t$ parties.

If party 1 holds input $x = 0$, it selects $x_1 = a_1, \ldots, x_l = a_l$. If party 1 holds input $x = 1$, it selects $x_1 = b_1, \ldots, x_l = b_l$. It sets all other variables to 0. Similarly, if party 2 holds input $y = 0$, it selects $x_{l+1} = a_{l+1}, \ldots, x_{k+1} = a_{k+1}$. If party 2 holds input $y = 1$, it selects $x_{l+1} = b_{l+1}, \ldots, x_{k+1} = b_{k+1}$. It sets all other variables to 0.

Together, party 1 and party 2 simulate $\Pi_G$, sending messages to one another when $\Pi_G$ specifies that a player from the group simulated by party 1 interact with a player from the group simulated by party 2. Both party 1 and party 2 learn the value of $G$ from the simulated protocol. If $G = 0$, each outputs AND(x,y)= 0; if $G \neq 0$, each outputs AND(x,y)= 1. It is easy to see that their outputs are correct.

Since $\Pi_G$ is $t$-private, where $t \geq \lceil \frac{n}{2} \rceil$, this two-party protocol is also 1-private. But Theorem 8.6 implies that there is no two-party 1-private protocol for AND; hence supposition (8.1) is false. Thus,

$$
\begin{aligned}
f(x_1, \ldots, x_n) &= f(a_1, \ldots, a_l, x_{l+1}, \ldots, x_{k+1}, 0, \ldots, 0) \\
&\quad + f(x_1, \ldots, x_l, a_{l+1}, \ldots, a_{k+1}, 0, \ldots, 0) \\
&\quad - f(a_1, \ldots, a_l, a_{l+1}, \ldots, a_{k+1}, 0, \ldots, 0).
\end{aligned}
$$

Define

$$
\begin{aligned}
g(x_1, \ldots, x_n) &= f(a_1, \ldots, a_l, x_{l+1}, \ldots, x_{k+1}, 0, \ldots, 0), \\
h(x_1, \ldots, x_n) &= f(x_1, \ldots, x_l, a_{l+1}, \ldots, a_{k+1}, 0, \ldots, 0).
\end{aligned}
$$

Since $f$ is $t$-private, so are $g$ and $h$. But $g$ and $h$ each depend on at most $k$

variables. By the induction hypothesis, there are $g_1, \ldots, g_n, h_1, \ldots, h_n$ such that

$$
\begin{aligned}
g(x_1, \ldots, x_n) &= g_1(x_1) + \cdots + g_n(x_n), \\
h(x_1, \ldots, x_n) &= h_1(x_1) + \cdots + h_n(x_n).
\end{aligned}
$$

Let $d = f(a_1, \ldots, a_l, a_{l+1}, \ldots, a_{k+1}, 0, \ldots, 0)$. Then we have,

$$
\begin{aligned}
f(x_1, \ldots, x_n) &= g(x_1, \ldots, x_n) + h(x_1, \ldots, x_n) - d \\
&= g_1(x_1) + \cdots + g_n(x_n) + h_1(x_1) + \cdots + h_n(x_n) - d \\
&= f_1(x_1) + \cdots + f_n(x_n),
\end{aligned}
$$

where we define $f_1(x_1) = g_1(x_1) + h_1(x_1) - d$ and $f_i(x_i) = g_i(x_i) + h_i(x_i)$ for $i \geq 2$. This completes the induction step.

Applying the result for $k = n$, we conclude that for an arbitrary $t$-private function $f$ there exist $f_1, \ldots, f_n$ such that

$$
f(x_1, \ldots, x_n) = f_1(x_1) + \cdots + f_n(x_n),
$$

completing the proof of Theorem 8.7. $\square$

# Chapter 9

# Cryptographic Methods Tolerating a Faulty Majority

> In fact the incontinent person is like a city that votes for all the right decrees and has good laws, but does not apply them, as in Anaxandrides' taunt, "The city willed it, that cares nothing for laws."

> — Aristotle, *Nicomachean Ethics*

The public-key, complexity-based approach to security introduced by Diffie and Hellman [DH76] affords a greater range of resilience than the unconditional, information-theoretic approach discussed thus far. An adversary that has bounded resources will not be able to break encryptions or generate effective malicious messages if the encryptions and the protocols require tremendous resources to corrupt, even though they might leak information to an unbounded adversary in the pure, Shannon sense.

We shall consider protocols in which the players and the adversary are polynomial-time Turing machines. A tool common to complexity-based cryptography is the one-way function [Yao82b, BM84]. A one-way function is, informally, a function that is easy to compute but hard to invert. A wide variety of candidates exist, but none have been proven to be one-way; this is not surprising in view of the fact that the existence of a one-way function would imply P$\neq$ NP. For example, exponentiation modulo a prime $p$ is easy to perform, but no efficient solution is known for its inverse, the discrete logarithm problem (Definition 11.13). One-way *trapdoor* functions are functions that are difficult to compute but with a small amount of advice become easy. Computing quadratic residuosity modulo a product of two primes $n = pq$ is a common example; without the factorization of $n$, computing the residuosity is presumably difficult, whereas with the factors of $n$ there is a simple polynomial-time algorithm to compute residuosity.

The drawback to complexity-based cryptography is the problem that unproven assumptions are made. Security is not unconditional, but conditioned on *cryptographic* assumptions.[1] It is therefore essential to keep the assumptions to a minimum. Rather

---

[1]It may be observed that assuming the presence of a private channel is a *cryptographic* assumption,

than make a particular assumption like the intractability of factoring large integers, it is desirable to design protocols and encryption techniques based on the existence of an arbitrary one-way trapdoor function. Assuming a one-way function is preferable to assuming a one-way trapdoor function or a one-way permutation, for example.

Unfortunately, without assuming new primitives such as unproven complexity-theoretic conjectures or measurably noisy communication channels, it is impossible to achieve general multiparty computations when a majority of players may be faulty. Consider intuition, first. If a minority of players is able to determine the input of a given player, then certainly a faulty majority could do so. Therefore, a protocol cannot allow any minority the power to determine inputs. But if the group of nonfaulty players is a minority, the information it holds will be insufficient to determine the input of any faulty player (even in some oblivious, shared or distributed manner that preserves the privacy of the faulty player), and the joint computation cannot hope to depend on that player's input.[2] The old rule of thumb stating, "The majority rules," applies even when the majority is bad.

Even worse, faulty players may withdraw precisely after learning the output, gaining full benefit at the expense of honest players. Notice that when the faulty players hold only a minority, the majority of nonfaulty players always holds enough information to determine the result and cannot be denied the answer.

Fairness has been treated in the two-party, cryptographic setting by Yao [Yao86], and in the $n$-party case by Galil, Haber, and Yung [GHY87], primarily through techniques based on exchanging secret keys simultaneously. Luby, Micali and Rackoff [LMR83] use the Quadratic Residuosity Assumption to design a biased coin that two parties compute jointly; using this biased coin, the two parties exchange secret keys gradually. All of these results rely on strong and specific complexity theoretic assumptions. Cleve [Cle86] examines impossibility results independent of complexity theoretic assumptions.

This chapter presents techniques to attain fair, secure, and reliable computations even though the majority of processors may be malicious. Based on certain cryptographic assumptions, however, a minority of honest processors can restrict the power of a faulty majority to the ability only to withdraw; a faulty majority cannot cause the honest processors to adopt incorrect values.

We discuss different and stronger definitions for fairness in the $n$-party scenario, and present a learning-based approach for which fairness and secret computation are in fact achievable. Our solution uses a technique called *gradual disclosure*, in which each party learns the result slowly, so that if the procedure halts, each player has gained roughly equal knowledge. This technique has similarities to the methods of [LMR83] for secret exchange, but does not use strong assumptions.

For clarity, our exposition starts with a weakened adversary, who can corrupt

---

but the use of the term *cryptographic assumption* often refers to unproven complexity-theoretic conjectures.

[2]One cannot in general "penalize" a faulty player at any stage simply by ignoring its input, since an unfair bias may result (consider a faulty player that can withdraw its input of "1" when a parity computation isn't proceeding to its liking).

$t > n/2$ players but who is allowed only fail-stop corruptions. (Chapter 8 discusses an even weaker, *completely passive* adversary.) No incorrect messages may be sent. Our solution assumes two-player oblivious circuit evaluation, which we shall later replace by assuming either the presence of noisy ("oblivious transfer") channels or by assuming that a cryptographic protocol for two-party oblivious transfer exists. A result of Impagliazzo and Rudich [IR89] implies that weaker complexity theoretic assumptions are difficult to make without proving P$\neq$NP.

Finally, to investigate a fully Byzantine adversary that can corrupt a majority, we present methods that restrict the powers of the Byzantine adversary to those of a fail-stop adversary. Through zero-knowledge proofs using either noisy channels or one-way functions, we essentially compile our fail-stop protocol (in the sense of [GMW87]) into one resilient (with certain limitations) against Byzantine adversaries. That is, faced with a possibly cheating Byzantine adversary, nonfaulty players can detect cheating and disqualify faulty players, thus limiting the power of the Byzantine adversary. The use of noisy channels for this purpose is novel.

**Assumptions made in this chapter.** The network has private channels, a broadcast channel, $n$ processors, and any number $t$ of Byzantine faults, chosen statically. We assume that a protocol for oblivious transfer exists, and consider the goal of computing Boolean functions. We ultimately remove the assumptions of private channels. The protocols are computationally resilient.

## 9.1   Fairness

For $2t \geq n$, full resilience cannot be achieved, but we can measure how far each player progresses toward an answer in order to show that parity is maintained. As earlier, we would like a standard: a *fair, ideal protocol,* and we would like a means to compare an arbitrary protocol to it. For the latter we shall continue to use relative resilience.

The fair, ideal protocol should allow a stronger adversary, restricted to the ideal $t$-fault-class but given the ability to corrupt player $(n + 1)$ (the trusted host) in a fail-stop manner. We call this an *extended $t$-adversary class.* Thus, an adversary can see rushed messages in a given round and prevent the host from completing the round.

To measure how much each player learns in a protocol, consider the chance that a player guesses a particular output, for any value $y$ and state $q$ :

$$
\begin{aligned}
p_i(y, q_i) &= \Pr\left[\mathcal{O}_i(q_i) = y_i\right] \\
p_A(y, q_A) &= \Pr\left[\mathcal{O}_A(q_A) = y_A\right].
\end{aligned}
$$

The initial probability of player $i$ is $p_i(y, x_i \circ a_i)$. Motivated by definitions of likelihood and weight of evidence frequently used in learning theory, we define

$$
\mathtt{odds}(p) = \frac{p}{1-p}.
$$

The increase in odds that a player or adversary outputs the correct result serves as a measure of how much that party learns from a protocol.

A fair, ideal protocol computes some function $f$ while maintaining parity in knowledge as it reveals the result. In measuring the increase in knowledge of adversaries, we make two assumptions. The first is that the adversary is static. Defining fairness with respect to dynamic adversaries is more involved and fraught with subtleties. For example, it is hard to define an adversary's initial information, since a dynamic adversary may (intentionally) start off ignorant but later gain huge amounts of information, completely disparate with the minor gains of nonfaulty players, by obtaining the inputs of newly corrupted players. These difficulties suggest that a proper formulation of fairness for *dynamic* adversaries is an attractive open problem. Because our protocols are designed to withstand static adversaries, we need not treat it in this chapter.

The second assumption we make is that an adversary maximizes its initial chances to guess the result. As mentioned, an adversary might intentionally compute wrongly when given only the initial information, in order that a quantitative measure of fairness be broken. Here, it is not a question of ignorance, but of programming, since the adversary knows all the corrupt inputs at the start. Given an adversary class $\mathcal{A}$, we define the *maximal initial probability of coalition $T$*:

$$p_T(y, \vec{x}_T \circ \vec{a}_T \circ a) = \max \left\{ \Pr\left[ \mathcal{O}_A(\vec{x}_T \circ \vec{a}_T \circ a) = y_A \right] \mid A \in \mathcal{A}, T = T(A) \right\}$$

**Definition 9.1** *A $(\delta, t)$-**fair, ideal protocol for** $F$ is a protocol with the following properties, for any adversary $A$ from an extended, ideal, static $t$-adversary class:*

1. *The host must collect all inputs in the first round and compute $F(\vec{x}')$, where $\vec{x}'$ is the vector of inputs it collects;*

2. *For any $(\vec{x} \circ \vec{a} \circ a, k)$, for any execution, for all $r \leq R$, for all $i \notin T = T(q_A)$, and for all $\vec{x}'$ such that $\vec{x}_T = \vec{x}'_T$, either*

$$\frac{\mathsf{odds}(p_A(F(\vec{x}'), q_A^r))}{\mathsf{odds}(p_i(F(\vec{x}'), q_i^r))} \leq (1 + \delta) \cdot \frac{\mathsf{odds}(p_T(F(\vec{x}'), \vec{x}_T \circ \vec{a}_T \circ a))}{\mathsf{odds}(p_i(F(\vec{x}'), x_i \circ a_i))},$$

*or*

$$\mathsf{odds}(p_A(F(\vec{x}'), q_A^r)) = \infty \qquad and \qquad \mathsf{odds}(p_i(F(\vec{x}'), q_i^r)) \geq \frac{1}{\delta};$$

3. *The host must return $F(\vec{x}')$ in round $R$.*

*We say $\delta$-**fair, ideal** when $t$ is clear from context.*

Condition (3) seems pointless because the adversary ought to stop the host before this step, always learning $F(\vec{x}')$ while preventing nonfaulty players from learning $F(\vec{x}')$ with certainty. Condition (2), however, ensures that the nonfaulty players nevertheless have a very good idea of the output. Our protocols will satisfy an additional constraint, that in order to stop the trusted host, the adversary must identifiably corrupt at least $n - t$ players. In a litigious society, this is a high price to pay. At the very least, it prevents further infractions.

The ideal standard for fairness allows us to define fairness for a general protocol:

**Definition 9.2** *A protocol* $\Pi$ **computes** $F$ $t$**-fairly** *if, for some* $c > 0$, *there exists a* $O(k^{-c})$-*fair ideal protocol* $\mathtt{FID}(F)$ *for* $F$ *such that*

$$\Pi \succeq^s \mathtt{FID}(F).$$

We shall make use of protocols that would be resilient if the adversary could not halt the protocol entirely. Specifically, a *semi-ideal* protocol $\mathtt{id}(F)$ for $F$ is one that accommodates an extended adversary, but if the adversary halts the host at some round $r$, then all nonfaulty players output $\mathcal{O}(q_i^r) = (\textsc{Cheating}, y_i)$, where $y_i$ indicates a "best guess" for the output. Otherwise, as in the ideal protocol, each nonfaulty player receives the value of $F(\vec{x})$.

**Definition 9.3** *A protocol* $\Pi$ *is* $t$-**semi-resilient for** $F$ *if there exists a semi-ideal protocol* $\mathtt{id}(F)$ *such that, with respect to an extended ideal* $t$-*adversary class,*

$$\Pi \succeq \mathtt{id}(F)$$

### 9.1.1   Comparison to Previous Work

Galil, Haber, and Yung extended Yao's two-party protocol to an $n$-party protocol that is fair under a different, weaker definition [GHY87, Yao86]. Their definition allows nonfaulty players to run a recovery protocol *based on the programs of faulty players*. This is unrealistic in two ways. First, it requires that adversaries have less computational power. Second, the adversaries' programs cannot depend in any way on the nonfaulty players' programs.

Their solution involves encrypting the result using a trapdoor function and revealing the trapdoor bit by bit. At any point, it would seem that no player is more than one bit ahead of another. Unfortunately, if an adversary is a powerful investment firm with a supercomputer that is a thousand times faster than the personal computer of a small investor, the firm can always quit once it has enough of the trapdoor, leaving the small investor in the dust.

We make no such restrictions and allow a full range of adversaries that know the programs of nonfaulty players (though not, of course, their inputs).

Our protocols use a near optimal number of rounds relative to a two-party lower bound of Cleve [Cle89] that generalizes to the $n$-party, $2t \geq n$ case. If $p$ and $q$ are the *a priori* probabilities of each player to guess the result, there is a quitting strategy for one player allowing him to predict $F$ with probability $\frac{\min(1-p, 1-q)}{2k}$ better than the other, where $k$ is the number of rounds. A certain number of rounds are required to attain fairness.

## 9.2 Gradual Disclosure: Fail-Stop from Passive

We begin our analysis with a method to achieve fairness against fail-stop adversaries, given protocols to compute functions resiliently against passive adversaries. The fundamental idea is that the result should be revealed slowly; the particular method we employ uses a series of coin flips biased slightly toward the answer (*cf.* [LMR83]), each computed using a semi-resilient multiparty protocol.

The ideal coin-flip protocol is described in Figure 9.1. We show:

**Lemma 9.4** *Protocol* ID*(coin(F))is a* $(4k^{-1}, t)$-*fair, ideal protocol.*

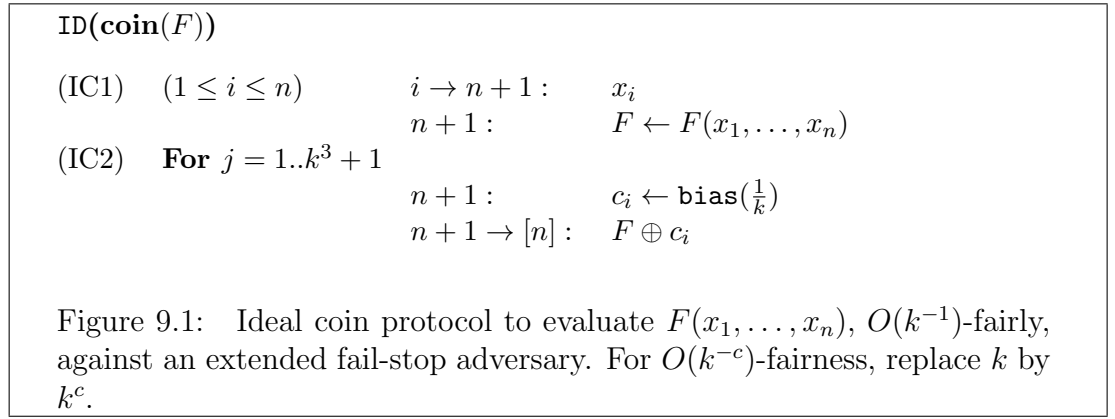

Figure 9.1: Ideal coin protocol to evaluate $F(x_1, \ldots, x_n)$, $O(k^{-1})$-fairly, against an extended fail-stop adversary. For $O(k^{-c})$-fairness, replace $k$ by $k^c$.

**Proof:** If $\mathrm{odds}(p_T(F(\vec{x}'), \vec{x}_T \circ \vec{a}_T \circ a)) = \infty$, there is nothing to prove. Otherwise, the adversary's odds become infinite only in the final round, when $F(\vec{x}')$ is revealed, but at that point the odds of nonfaulty players exceeds $k$. This can be seen from bounds on the tail of the binomial distribution proven by Chernoff and Angluin and Valiant [Che52, AV79]. A *binomial random variable* $X_{N,P}$ measures the number of successes in $N$ independent Bernoulli trials, where the probability of success is $P$.

**Lemma 9.5** *(Chernoff, Angluin, Valiant) If* $X_{N,P}$ *is a binomial random variable, then*

$$Pr[X \leq (1-\epsilon)NP] \leq \exp(-\tfrac{1}{2}\epsilon^2 NP).$$

Let $X$ represent the number of coins of value $F(\vec{x}')$, $N = k^3$, and $P = (\frac{1}{2} + \frac{1}{k})$. Then

$$\Pr\left[X \leq \tfrac{1}{2}k^3\right] \leq \exp(-\tfrac{1}{2}\left(\frac{2}{k+2}\right)^2 (\tfrac{1}{2}k^3 + k^2)) = \exp(-\frac{k^2}{k+2}) \leq \exp(-\frac{k}{2}),$$

for $k \geq 2$. Therefore

$$\mathrm{odds}(p_i(F(\vec{x}'), q_i^R)) \geq \frac{1 - \exp(-\frac{k}{2})}{\exp(-\frac{k}{2})} = \exp(\tfrac{1}{2}k) - 1 > k,$$

for $k \geq 3$. If the adversary allows a nonfaulty player to see the $(k+1)^{st}$ coin flip, the odds are clearly better.

Thus it suffices to show that the ratio of odds increases by a factor of no more than $(1+4k^{-1})$ in each round. First let us show that the best adversarial strategy for guessing the output is simple (choose the majority, *i.e.* the best Bayesian guess) and achievable. Let $A$ be an adversary that, for some sequence $\rho$ of coin flips for which $\Pr[F = 0 \mid \rho] > \frac{1}{2}$, but $A$ outputs 1 with probability $\alpha > 0$. Let $B$ be $A$ modified to output 1 always in this case. Then $B$ is more successful in guessing the output when the sequence $\rho$ comes up:

$$
\begin{aligned}
& (\Pr[B = 0, F = 0 \mid \rho] + \Pr[B = 1, F = 1 \mid \rho]) \\
& -(\Pr[A = 0, F = 0 \mid \rho] + \Pr[A = 1, F = 1 \mid \rho]) \\
= \quad & \Pr[B = 0 \mid \rho]\Pr[F = 0 \mid \rho] + \Pr[B = 1 \mid \rho]\Pr[F = 1 \mid \rho] \\
& -\Pr[A = 0 \mid \rho]\Pr[F = 0 \mid \rho] - \Pr[A = 1 \mid \rho]\Pr[F = 1 \mid \rho] \\
= \quad & (1 - (1 - \alpha))\Pr[F = 0 \mid \rho] - \alpha\Pr[F = 1 \mid \rho] \\
= \quad & \alpha(\Pr[F = 0 \mid \rho] - \Pr[F = 1 \mid \rho]) \\
> \quad & 0.
\end{aligned}
$$

Given the best possible adversary, we now show that the amount it learns from a round of the ideal coin protocol increases its odds by no more than a factor of $(1+4k^{-1})$. Let $b = \frac{1}{k}$, $\gamma_0 = (\frac{1}{2} + b)$, and $\gamma_1 = (\frac{1}{2} - b)$. Clearly, if $p$ is the number of 0's in $\rho$, and $q = |\rho| - p$,

$$
\Pr[\rho \mid F = d] = (\gamma_d)^p (\gamma_{\bar{d}})^q.
$$

Now,

$$
\begin{aligned}
\Pr[F = d \mid \rho] \quad = \quad & \Pr[\rho \mid F = d]\Pr[F = d] \\
& \cdot (\Pr[\rho \mid F = 0]\Pr[F = 0] + \Pr[\rho \mid F = 1]\Pr[F = 1])^{-1}.
\end{aligned}
$$

Consider what happens when the next coin toss is a 0. We must consider the odds of a correct guess given the sequence $\rho 0$. It suffices to show that the ratio $\Delta$ of these odds to those of the adversary when given only $\rho$ is no more than $(1+4k^{-1})$.

$$
\begin{aligned}
\Delta \quad = \quad & \frac{\mathtt{odds}(p_A(0, \rho 0))}{\mathtt{odds}(p_A(0, \rho))} \\
= \quad & \frac{\Pr[F = 0 \mid \rho 0]}{1 - \Pr[F = 0 \mid \rho 0]} \cdot \left( \frac{\Pr[F = 0 \mid \rho]}{1 - \Pr[F = 0 \mid \rho]} \right)^{-1}
\end{aligned}
$$

With a bit of manipulation, letting $L = \gamma_0 \gamma_1^{-1}$ and $G = \Pr[F = 0](\Pr[F = 1])^{-1}$,

$$
\Delta = \frac{1 + L^{q-p}G^{-1}}{1 + L^{q-p-1}G^{-1}} \cdot \frac{1 + L^{p+1-q}G}{1 + L^{p-q}G}
$$

The extremal cases occur when $p \approx q$ and $G \approx 1$;

$$\Delta \;<\; \frac{1+L}{1+L^{-1}} = L$$
$$(\frac{1}{2} + \frac{1}{k})(\frac{1}{2} - \frac{1}{k})^{-1} \leq 1 + 4k^{-1}.$$

∎

**Lemma 9.6** *Let $\mathtt{FID}(F)$ be a $\delta$-fair, ideal protocol for $F$ having $R(n,m)$ steps. If $\Pi = \{\Pi(i)\}$ is a family of protocols such that $\Pi(i)$ statistically $t$-semi-resiliently computes the $i^{th}$ step of $\mathtt{FID}(F)$, then the protocol $\hat{\Pi}$ defined as $\Pi(R(n,m)) \circ \cdots \circ \Pi(1)$ is a $t$-semi-resilient and $\delta$-fair protocol for $F$.*

**Proof:** We write $F$ as a composition of functions $\hat{F} = \hat{F}^{R(n,m)} \circ \cdots \circ \hat{F}^1$, where each can also take an input of the form $x_i = (\textsc{Cheating}, g)$, in which case it outputs $y_i = (\textsc{Cheating}, g)$. By Theorem 4.31, $\hat{\Pi} \succeq^s \; \circ_1^{R(n,m)} \mathtt{id}(\hat{F}^i) \succeq \mathtt{FID}(F)$. ∎

For any protocol $\Pi$, define protocol $\mathtt{fs}(\Pi)$ by modifying each player to broadcast \textsc{Cheating} if it fails to receive an expected message, receives an improper message (according to some local computation it can make), or receives a broadcast \textsc{Cheating} message. If any of these occasions occur, it halts in the next round, producing and output of the form $(\textsc{Cheating}, y)$.

**Lemma 9.7** *If $\Pi$ is $t$-resilient leaking $F$ against passive adversaries, then $\mathtt{fs}(\Pi)$ is $t$-semi-resilient leaking $F$ against fail-stop adversaries.*

**Proof:** Let $\mathcal{S}'$ be an interface for passive adversaries, from $\Pi$ to $\mathtt{ID}(F)$. Define interface $\mathcal{S}$ for fail-stop adversaries from $\mathtt{fs}(\Pi)$ to $\mathtt{id}(F)$ as follows. Interface $\mathcal{S}$ runs $\mathcal{S}'$ until adversary $A$ requests the failure of some subset of processors, by specifying they send no messages to some set $U$ of nonfaulty players. Interface $\mathcal{S}$ halts the host in $\mathtt{id}(F)$, causing all players to halt and output \textsc{Cheating}, just as in $\mathtt{fs}(\Pi)$. The interface also supplies $A$ with messages of \textsc{Cheating} from players in $U$, uses $\mathcal{S}'$ to generate the outgoing messages from the remaining nonfaulty players in the current round of $\mathtt{fs}(\Pi)$, and in the next round reports \textsc{Cheating} from all nonfaulty players that did not already broadcast \textsc{Cheating}.

If $A$ never causes a failure, then $\mathcal{S}$ runs $\mathcal{S}'$ to the end, never halting the trusted host. The distribution (conditioned on no halting) on outputs in $<A, \mathtt{fs}(\Pi)>$ is the same as in $<A, \Pi>$, which in turn is the same as in $<A, \mathcal{S}', \mathtt{ID}(F)>$, which (conditioned on no halting) matches that in $<A, \mathcal{S}, \mathtt{id}(F)>$. ∎

The semi-resilient general protocol, even though unfair, is easily utilized to accomplish what the ideal, fair protocol accomplishes. Figure 9.2 describes a $O(k^{-c})$-fair, semi-resilient protocol for any function $F$. The protocol is a concatenation of protocols that compute the result of each round of the ideal coin-flip protocol, namely a random coin biased toward the result, $F(x_1, \ldots, x_n)$, with probability $\frac{1}{2} + \frac{1}{k}$. Clearly, $k$ can be replaced by $k^c$ for any $c > 0$, providing greater fairness if desired.

**Theorem 9.8** *Let* $t \leq n$. *If for any $G$ there exists a protocol for $G$ that is $t$-resilient against passive adversaries, then for any $F$ and $c$ there exists a $(k^{-c}, t)$-fair protocol* EvalFairFS *for $F$ that is resilient against fail-stop adversaries. If $F$ is described by a circuit family $C_{F^{n,m}}$, the protocol requires $O(\mathtt{Size}(C_{F^{n,m}})^{c_0})$ bits and $O(\mathtt{Depth}(C_{F^{n,m}})^{c_0})$ rounds of interaction for some fixed $c_0$.*

**Proof:**

---

EvalFairFS

(EF0)  If any player halts, broadcast Cheating.
　　　　If any player broadcasts Cheating, halt and output Cheating.
(EF1)  Run EvalSemi to supply each player $i$ with $\mathrm{piece}_i(F(x_1, \ldots, x_n))$.
(EF2)  **For** $j = 1..k^3 + 1$ **do**
　　　　　　Run EvalSemi to compute $\mathrm{piece}_i(c_j)$ for player $i$,
　　　　　　　　where $c_j \leftarrow \mathtt{bias}(\frac{1}{k})$.
　　　　　　Run EvalSemi on $\{\mathrm{piece}_i(F), \mathrm{piece}_i(c_j)\}_{i \in [n]}$
　　　　　　　　to compute and reveal $F \oplus c_j$.

Figure 9.2: Protocol to evaluate $F(x_1, \ldots, x_n)$, $O(k^{-1})$-fairly, against a fail-stop adversary. For $O(k^{-c})$-fairness, replace $k$ by $k^c$. Biased coins $c_1, \ldots, c_{k^3+1}$ may also be computed beforehand and masked with $F$ all at once, so that the bulk of the protocol (Step (EF2)) becomes simply a gradual disclosure of $F$, secret by secret.

---

Let $F$ and $c$ be arbitrary. Lemma 9.4 states that there exists a $(k^{-c}, t)$-fair protocol $\mathtt{ID}(\mathrm{coin}(F))$ for $F$. The condition of the theorem states that there exist protocols $\Pi(1), \ldots, \Pi(R(n,m))$ for each step of $\mathtt{ID}(\mathrm{coin}(F))$ that are $t$-resilient against passive adversaries. By Lemma 9.7, protocols $\mathtt{fs}(\Pi(1)), \ldots, \mathtt{fs}(\Pi(R(n,m)))$ are $t$-semi-resilient against fail-stop adversaries, and Lemma 9.6 concludes that their composition is $(k^{-c}, t)$-fair and resilient against fail-stop adversaries. ∎

## 9.3　Degree Reduction: Passive from Two-Party Protocols

Let us assume either that a black box for two-party function evaluation exists (*i.e.,* in the formal model for protocol execution, two players write values on a special tape, and in the next round the result of a particular function $f(x_1, x_2)$ is written on another tape for them to read) or that some such protocol exists (*i.e.,* a 2-resilient protocol admitting an appropriate interface). Based on two-party function evaluation, we show how $n$ parties can achieve the same result: general function computation.

We consider secret sharing with a threshold of $n-1$. In this case, $n$ players are needed to determine a secret, and sharing reduces essentially to a form we call *sum-sharing,* which can be regarded as a generalization of parity-sharing (sum modulo 2) originally developed by Haber [HM87], and applied to improve message complexity and to protect information of faulty players in [GHY87]. In particular, a secret is represented as a sum of all pieces, where the pieces arise by selecting $n$ random field elements subject to their sum being the secret.[3] Addition remains easy (add pieces individually), but multiplication is again somewhat more difficult. Our solution, independently derived, generalizes the parity-based method of Haber and Micali.

Protocols for sum sharing and secret addition are listed in Figures 9.3 and 9.4. For multiplication of secrets $a$ and $b$, each player must receive a piece $\textsc{piece}_i(ab)$. Noting that

$$ab = (\sum_{i=1}^{n} \textsc{piece}_i(a))(\sum_{j=1}^{n} \textsc{piece}_j(b))$$

we see that for every combination of $i$ and $j$, player $i$'s piece of $a$ must be multiplied by player $j$'s piece of $b$; the sum of these products is $ab$. It would not do for player $i$ to know $\textsc{piece}_i(a) \cdot \textsc{piece}_j(b)$, for then it could easily calculate $\textsc{piece}_j(b)$. In fact, we use a two-party protocol for player $i$ and player $j$ to calculate the values $(\textsc{piece}_i(a)\textsc{piece}_j(b) + r_{ij})$ and $(-r_{ij})$, where $r_{ij}$ is a uniformly random field element. Each receives one of these values. No information is revealed, however, since each value by itself is uniformly random. On the other hand, the sum of the two values is the desired contribution to the final result, namely the pairwise product $\textsc{piece}_i(a)\textsc{piece}_j(b)$. Summing over all results of all two-party combinations gives the desired result, $ab$; thus the sum of each player's individual results represents a sum share of $ab$. That is, the newly constructed pieces $\textsc{piece}_i(ab)$ are such that any $n-1$ of them are distributed uniformly at random, while all $n$ pieces sum to $ab$. The protocol is given in Figure 9.3.

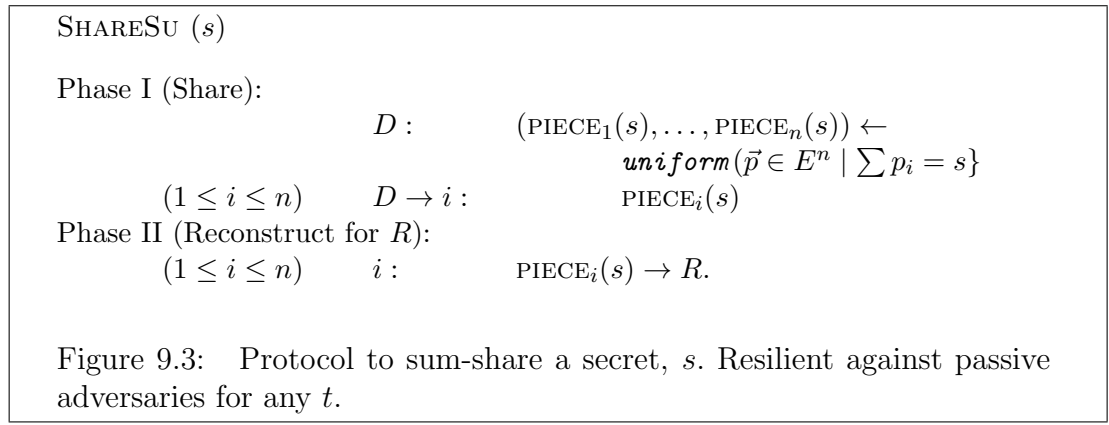

Figure 9.3: Protocol to sum-share a secret, $s$. Resilient against passive adversaries for any $t$.

---

[3]We leave it to the interested reader to identify the simple correspondence between polynomial secret sharing having threshold $n-1$ and sum sharing.

---

AᴅᴅSᴜ $(a, b)$

$\qquad (1 \leq i \leq n) \qquad i: \qquad \text{PIECE}_i(a + b) \leftarrow \text{PIECE}_i(a) + \text{PIECE}_i(b).$

Figure 9.4: Noninteractive protocol to add two secrets shared using the sum representation. Resilient against passive adversaries for any $t$. Linear combinations are a simple modification.

---

MᴜʟᴛSᴜ $(a, b)$

(MS0) Define $\alpha((x, r), (y, s)) = xy + r + s$,
$\qquad\qquad\quad \beta((x, r), (y, s)) = -r - s$.

(MS1) $(1 \leq i, j \leq n), i \neq j$:
$\qquad\qquad i: \qquad r_{ij}^i \leftarrow \textit{uniform}(E)$

(MS3) $(1 \leq i, j \leq n; i \neq j)$:
$\qquad\qquad$ Run TwᴏOᴄᴇ between players $i$ and $j$
$\qquad\qquad$ on inputs $((\text{PIECE}_i(a), r_{ij}^i), (\text{PIECE}_j(b), r_{ij}^j))$,
$\qquad\qquad$ to compute $\alpha_{ij} = (\text{PIECE}_i(a)\text{PIECE}_j(b) + r_{ij}^i + r_{ij}^j)$, $beta_{ij} = (-r_{ij}^i - r_{ij}^j)$.

(MS3) $(1 \leq i, j \leq n) \;\; i$:
$\qquad\qquad \text{PIECE}_i(ab) = \text{PIECE}_i(a)\text{PIECE}_i(b) + \sum_{j=1}^{n}(\alpha_{ij} + \beta_{ji}).$

---

Figure 9.5: Protocol to multiply two secrets shared using the sum representation. Resilient against passive adversaries for any $t$.

**Lemma 9.9** *Given a 2-semi-resilient protocol* TwᴏOᴄᴇ *for two-party circuit evaluation, protocol* MᴜʟᴛSᴜ *achieves $n$-semi-resilient secret multiplication against fail-stop adversaries.*

**Proof:** Clearly, the functions $\alpha$ and $\beta$ described in Figure 9.5 are $n$-private; if one of the pair of participants is corrupted, the messages from the nonfaulty processors are obtained from the assumed sub-interface for TwᴏOᴄᴇ run as a subroutine, supplied with a uniformly random output for the faulty player. For each $i$ and $j$, the protocol computes an $n$-private representation of $\text{PIECE}_i(a)\text{PIECE}_j(b)$, and then computes an $n$-private representation of their sum, which is the desired product. ∎

**Theorem 9.10** *Let $\{F^{n,m}\}$ be a family of functions described by circuit family $C_{F^{n,m}}$. Given a 2-semi-resilient protocol* TwᴏOᴄᴇ *for two-party circuit evaluation, for any $t$ there exists a $t$-semi-resilient protocol* EᴠᴀʟSᴇᴍɪ *leaking $F$ against fail-stop adversaries. The protocol requires $O(\texttt{Size}(C_{F^{n,m}})^{c_0})$ bits and $O(\texttt{Depth}(C_{F^{n,m}})^{c_0})$ rounds of interaction for some fixed $c_0$.*

**Proof:** The protocol EᴠᴀʟSᴇᴍɪ is almost identical to protocol Eᴠᴀʟᴜᴀᴛᴇ $-C_F$, described in Chapter 5, using SʜᴀʀᴇSᴜ , AᴅᴅSᴜ , and MᴜʟᴛSᴜ instead of the secret sharing, secret addition, and secret multiplication subprotocols. If any player detects cheating, *i.e.* if any nonfaulty player observes that another player halts, then it broadcasts CHEATING and each nonfaulty player outputs CHEATING. The proof of relative resilience applies here, *mutatis mutandis*. Note in particular that, when faced with a request from the adversary to halt a player in EVALSEMI , the interface simply requests that the trusted host in $\mathtt{id}(F)$ be halted. We conclude for a passive adversary class $\mathcal{A}$ that EVALSEMI $\mathrel{\dot{\succeq}} \mathtt{ID}(F)$, hence for fail-stop adversaries, EVALSEMI $\mathrel{\dot{\succeq}} \mathtt{id}(F)$.
∎

---

EVALSEMI -$C_F$

(ES0)  If any player halts, broadcast CHEATING.
       If any player broadcasts CHEATING, halt and output CHEATING.
(ES1)  Each player $i$ runs SHARESU $(x_i)$.
              Denote these secrets at level 0 of the circuit by $z_{01}, \ldots, z_{0w}$.
(ES2)  /* Evaluate addition and multiplication layers */
       **For** $l = 1..d/2$ **do**
              Run ADDSU with coefficients $\{c_{lv,j}\}$, secrets $\{z_{\mathtt{in}_j g_{lv}}\}$ to compute:
              $$z_{2l,v} = c_{2l,v,0} + \sum c_{2l,v,j} z_{\mathtt{in}_j(g_{2l,v})} \qquad (v = 1..w, \ j = 1..|\mathtt{in}(g_{lv})|)$$
              Run MULTSU with secrets $\{(z_{\mathtt{in}_1(g_{lv})}, z_{\mathtt{in}_2(g_{lv})})\}$ to compute:
              $$z_{2l,v} = z_{\mathtt{in}_1(g_{lv})} \cdot z_{\mathtt{in}_2(g_{lv})} \qquad (v = 1..w)$$
(ES3)  **For** $i = 1..n$ **do in parallel**
              **for** $j \in \mathtt{out}(i)$ **do in parallel**
                     Reveal $z_{lj}$ to player $i$.

Figure 9.6: Protocol to evaluate a circuit $C_F$ for $F(x_1, \ldots, x_n)$, with a fail-stop adversary.

---

### 9.3.1  Polynomial Sharing

The protocols just described are prone to a single error of omission. Polynomial sharing is not, and forces an adversary to corrupt more players in order to disrupt the protocol. The motivating factor for developing a scheme to multiply values shared via sum-sharing arose from designing a multiplication protocol for secrets shared using polynomials. We briefly discuss that scheme.

Degree reduction is again the problem. We reduce it to the problem of generating publicly known evaluation points of $h(x) = f(x)g(x) + xr(x)$, where $f(x)$ and $g(x)$ are the original polynomials of degree $t$ used to share secrets $u$ and $v$, and, as in §5.2.2, $r(x)$ is a random polynomial of degree $t$. Returning to LaGrange interpolation, we express $h(m)$ as the product of two sums, where the addenda in each sum are weighted pieces of $u$ and $v$, and may be regarded as sum-shares:

$$h(m) \;\;=\;\; f(m)g(m)$$

$$= (\sum_i L_i(m) f(i))(\sum_j L_j(m) g(i))$$

$$= (\sum_i L_i(m) \cdot \text{PIECE}_i(u))(\sum_j L_j(m) \cdot \text{PIECE}_i(v))$$

Define $\overline{h}(x) = h(x) \bmod x^{t+1}$. It is a matter of straightforward algebra for each player $i$ to calculate the value $\overline{h}(i)$ from $h(i)$ and the public values $h(-t), h(-t+1), \ldots, h(-1)$.

Figure 9.7 describes the protocol to synthesize values and reduce the degree of $h(x)$.

---

SYNTHESIZE

Run the RAND-SECRET protocol to provide each $i$ with $r_i$,
    a piece of a random polynomial $r(x)$ of degree $t$.
For $m = -1, \ldots, -n$:
    Run the MULTSU protocol with player $i$'s pieces being:
        $L_i(m) \cdot \text{PIECE}_i(u), L_i(m) \cdot \text{PIECE}_i(v)$.
    Each $i$ receives a piece $c_i^m$ of the result.
    Each $i$ shares $c_i^m$ as a secret,
        and receives $\text{PIECE}_i(c_j^m)$ for $1 \le j \le n$.
    Run ADDSU to compute $\text{PIECE}_i(\sum_{j=1}^n c_j^m)$.
    Broadcast $(i r_i L_i(m) + \text{PIECE}_i(\sum_{j=1}^n c_j^m))$,
        and interpolate the value, $h(m)$.
Player $i$ computes $\overline{h}(i)$ from $h(-1), \ldots, h(-n)$ and
$h(i) = \text{PIECE}_i(u) \cdot \text{PIECE}_i(v) + r_i$.

Figure 9.7: Protocol to compute and broadcast $h(-1), \ldots, h(-n)$, and to compute $\overline{h}(i)$ for player $i$.

---

## 9.4 Two-Party Protocols from Oblivious Transfer

### 9.4.1 Cryptographic Primitives

**Definition 9.11** *A* **one-way function** *$f$ is a family $\{f^n\}$ mapping $\Sigma^n$ to $\Sigma^{n^c}$ for some $c$ such that*

- *$f$ is polynomial-time computable;*

- *for all $c$, there exists $n_0$ such that for all $n > n_0$, for all polynomial time Turing machines $M$,*

$$Pr[x \leftarrow \texttt{uniform}(\Sigma^n) : f(x) = f(M(f(x)))] < n^{-c}.$$

**Definition 9.12** *A **pseudorandom number generator** $G$ is a family $\{G^n\}$ mapping $\Sigma^n$ to $\Sigma^{dn^c}$ for some $c, d$ such that $G$ is polynomial-time computable and*

$$\{x \leftarrow \mathtt{uniform}(\Sigma^n) : G(x)\} \stackrel{\approx}{\sim} \mathtt{uniform}(\Sigma^{dn^c}).$$

Impagliazzo, Levin, and Luby [ILL89] show how to produce a pseudorandom number generator (PRG) from any one-way function:

**Theorem 9.13** *(Impagliazzo* et al *(1989)) If there exists a one-way function in the non-uniform model of security then there exists a pseudorandom generator in the non-uniform model of security.*

Based on PRG's, we may construct all sorts of primitives, including simple private-key cryptosystems (if two parties share a seed, they generate a one-time pad of pseudorandom bits which they use to mask messages) and special encrypted forms of gates to evaluate circuits obliviously. It will be useful to note that, by Lemma 4.13, any simple subrange of the output of a PRG is itself indistinguishable from uniformly random strings.

### 9.4.2 Yao Gates

An essential tool to the construction of cryptographic multiparty protocols is that of two-party oblivious circuit evaluation, introduced by Yao [Yao86]. Two players wish to evaluate a circuit $C$ at private inputs. The output may go to one or both of the players, and may be different for each.

A Yao gate is a circuit gate that is encrypted in a special way to allow its evaluation without learning the result. Yao [Yao86] introduced a particular implementation and Goldreich *et al* improved upon it [GMW87]. Both approaches used particular cryptographic assumptions, such as the intractability of factoring or quadratic residuosity, or even the existence of any one-way trapdoor function. We present an implementation that requires only an arbitrary one-way function, not any particular function [BG89]. Our solution is based only on the assumption of a protocol for oblivious transfer. By [BCG89], the existence of a protocol for oblivious transfer implies the existence of a one-way function. The existence of a trapdoor function need not be assumed.

The Yao gate has two inputs, $X$ and $Y$, and one output $Z$. The gate computes some function $g(X, Y) = Z$. Rather than allowing the wire values $X$, $Y$, and $Z$ to be known, however, there are random *keys*, $X_0$, $X_1$, $Y_0$, $Y_1$, $Z_0$, and $Z_1$, that represent the *wire values* 0 and 1. The correspondence between $X_0/X_1$ and wire values 0/1 is random, however; the same holds true for $Y$ and $Z$. In other words, there are random bits $\omega_X$, $\omega_Y$, and $\omega_Z$ that encode the correspondence between the keys and the wire values:

$$X_0 \leftrightarrow \omega_X$$
$$X_1 \leftrightarrow \overline{\omega}_X$$

Thus, if $\omega_X = 1$, then key $X_0$ represents the wire value 1 and key $X_1$ represents the wire value 0. The *key translations* $\omega_X$, $\omega_Y$, and $\omega_Z$, are kept hidden.

The construction of the gate is such that knowing $X_\alpha$ and $Y_\beta$ allows one to compute $Z_{g(\alpha \oplus \omega_X, \beta \oplus \omega_Y) \oplus \omega_Z}$. That is, without knowing any of the key translations, one can compute the $Z$ key that corresponds to the output wire value. Furthermore, the only information one learns is the appropriate $Z$ key itself, not the wire value it represents, nor the other $Z$ key. One only need know two input keys $X_\alpha$ and $Y_\beta$. In fact, none of the other input keys are ever revealed; knowing both $X_0$ and $X_1$, for example, would allow one to perform two computations — and the results $[Z_{g(0, y \oplus \omega_Y) \oplus \omega_Z}, Z_{g(1, y \oplus \omega_Y) \oplus \omega_Z}]$ would be either two different $Z$ keys or the same $Z$ key, potentially revealing information about $Y$.

Thus, the encryption of the gate is public, and exactly one of each pair of input keys may be revealed, but all other values must remain secret. In a two-player protocol based on Yao gates [Yao86, GMW87], one of the players creates the encrypted circuit by encrypting each gate in a coordinated fashion, gives the circuit to the second, and obliviously allows the second player to learn exactly one input key for every input gate. The second player then evaluates the circuit by computing the output keys at each level, until it reaches the final output. The first player reveals the correspondence between the final output keys and the wire values. Note that this method requires little interaction; the computation of the new keys from the old ones is performed locally.

Yao's original method was based on assuming the intractability of factoring large numbers. We assume only that a protocol for oblivious transfer exists. Oblivious transfer, introduced by Rabin [HR83], is an important two-player protocol to compute the probabilistic function on $\emptyset \times \{0,1\}^2$ defined by $F(x_1, x_2) = \{b \leftarrow \textbf{\textit{uniform}}(\{0,1\}) : (\Lambda, (b, b \cdot x_1))\}$. The first player, Alice, has a bit, $x_1$, which she transfers to the second player, Bob, with probability $\frac{1}{2}$. Alice never learns, however, whether Bob received the bit; her output is always the same, 0. Bob knows whether he received the bit by checking whether $b = 1$ or not.

Bellare, Cowen, and Goldwasser [BCG89] show that the existence of an oblivious transfer (OT) protocol implies that one-way functions exist. On the other hand, even though it would be safer to assume only that one-way functions exist, Impagliazzo and Rudich [IR89] show that proving the existence of an OT protocol based on one-way functions is likely to be as hard as proving $P \neq NP$.

Fix a function family $F = \{F^m\}$ mapping $\{0,1\}^m \times \{0,1\}^m \rightarrow \{0,1\}^m$. Let $\texttt{Gen}(X)$ be a PRG that outputs $3|X|$ bits. Denote the first $|X|$ bits by $\texttt{Tag}(X)$, the next $|X|$ bits by $\texttt{Mask}(0, X)$, and the final $|X|$ bits by $\texttt{Mask}(1, X)$. These functions will be used to create identifying tags and masks for the keys. In the sequel, all keys have the same length $k$.

Our gate consists of four $3k$-bit entries as follows: for $a = 0, 1$ and for $b = 0, 1$,

$$\texttt{Ent}(a, b, X_0, X_1, \omega_X, Y_0, Y_1, \omega_Y, Z_0, Z_1, \omega_Z) =$$
$$\texttt{Tag}(X_a) \circ \texttt{Tag}(Y_b) \circ$$
$$[\texttt{Mask}(b, X_a) \oplus \texttt{Mask}(a, Y_b) \oplus Z_{g(a \oplus \omega_X, b \oplus \omega_Y) \oplus \omega_Z}]$$

Given a $12k$-bit string, we consider it as four $3k$-bit strings, and within each of those $3k$-bit strings we refer to segments $\mathtt{yg}_1(a,b)$, $\mathtt{yg}_2(a,b)$, and $\mathtt{yg}_3(a,b)$. The entire table is thus:

$$\mathtt{yg}(X_0, X_1, \omega_X, Y_0, Y_1, \omega_Y, Z_0, Z_1, \omega_Z) =$$
$$\mathtt{Ent}(0, 0, X_0, X_1, \omega_X, Y_0, Y_1, \omega_Y, Z_0, Z_1, \omega_Z) \circ$$
$$\mathtt{Ent}(0, 1, X_0, X_1, \omega_X, Y_0, Y_1, \omega_Y, Z_0, Z_1, \omega_Z) \circ$$
$$\mathtt{Ent}(1, 0, X_0, X_1, \omega_X, Y_0, Y_1, \omega_Y, Z_0, Z_1, \omega_Z) \circ$$
$$\mathtt{Ent}(1, 1, X_0, X_1, \omega_X, Y_0, Y_1, \omega_Y, Z_0, Z_1, \omega_Z)$$

Figure 9.8 describes the the *decoding* of a gate given input keys $X$ and $Y$.

---

Decode-2-Gate $(\mathtt{yg}, X, Y)$

(DY1) Compute $\mathtt{Tag}(X), \mathtt{Tag}(Y)$.
(DY2) **For** $a = 0..1$ **do**
$\qquad\qquad$ Compute $\mathtt{Mask}(a, X), \mathtt{Mask}(a, Y)$.
(DY3) Determine the least $\alpha, \beta$ such that
$\qquad\qquad \mathtt{Tag}(X_\alpha) = \mathtt{yg}_1(\alpha, \beta)$ and $\mathtt{Tag}(Y_\beta) = \mathtt{yg}_2(\alpha, \beta)$
(DY4) Set $Z \leftarrow \mathtt{yg}_3(\alpha, \beta) \oplus \mathtt{Mask}(\beta, X) \oplus \mathtt{Mask}(\alpha, Y)$.

Figure 9.8: Obtaining the output key $Z$ from a generalized Yao gate, given the encrypted table and two input keys $X$ and $Y$.

---

### 9.4.3 Two-Party Yao Circuit Construction

Given the means to construct gates, constructing a circuit is straightforward.

Let us fix a notation for the wires of a circuit. Given a circuit of depth $d$ and width $w$, assume without loss of generality that there is a gate $g_{i,j}$ at every location $(i, j)$ ($i \in [d], j \in [w]$), and assume that no wires skip from one level to the next. Dummy gates computing the identity function can be introduced for this purpose; without loss of generality, all gates have exactly two inputs (which could be the same wire).

We consider an array $W[i, j, l]$ of wire keys for $i \in \{0, \ldots, d\}, j \in \{1, \ldots, w\}$, $l \in \{0, 1\}$, $I \in [n]$. The keys $W[i, j, 0]$ and $W[i, j, 1]$) represent wire $(i, j)$, which carries the output of gate $g_{i,j}$. Keys $W[0, j, 0]$ and $W[0, j, 1]$) represent inputs to the circuit. Associated with each wire $(i, j)$ is a *key translation bit* $w[i, j]$.

Given $S$, a set of components to void, and $W$, the construction of the circuit is as follows. First, construct a table $\mathtt{GenTab}$ of $6dwnk$ pseudorandom bits, where

$$\mathtt{GenTab} = \mathtt{Gen}(W)$$

and where the function $\mathtt{Gen}(W)$ gives a table of $(2dwn)$ $3k$-bit strings defined by

$$\mathtt{Gen}(W)[i,j,l] = \mathtt{Gen}(W[i,j,l]).$$

Second, convert this table to a circuit in the same fashion as the gates are constructed. That is, with respect to table $\mathtt{GenTab}$, define the functions $\mathtt{tag}(\mathtt{GenTab}, i, j, l)$ and $\mathtt{mask}(c, \mathtt{GenTab}, i, j, l)$ $(c \in \{0,1\})$ as the three substrings of length $k$ of $\mathtt{GenTab}[i,j,l]$.

The Yao gate $\mathtt{YG}(W, \mathtt{GenTab}, w, i, j)$ is then constructed as in §9.4.2, using $\mathtt{tag}$ and $\mathtt{mask}$ as the tag and mask functions. If $\mathtt{in}_0(i,j)$ and $\mathtt{in}_1(i,j)$ are the indices of the left and right inputs to $g_{i,j}$, then the wire keys are as follows.

$$
\begin{aligned}
X_l &= W[\mathtt{in}_0(i,j), l] \\
Y_l &= W[\mathtt{in}_1(i,j), l] \\
Z_l &= W[i,j,l]
\end{aligned}
$$

The key translations used to construct the gate are $w_X = w[\mathtt{in}_0(i,j)]$, $w_Y = w[\mathtt{in}_1(i,j)]$, and $w_Z = w[i,j]$.

The circuit is denoted $\mathtt{YC}(W, \mathtt{GenTab}, w)$. Notice that the construction takes $\mathtt{GenTab}$ into account independently of $W$; it is well-defined whether or not $\mathtt{GenTab}$ arises from $W$.

---

Make-2-Circuit $(W, \mathtt{GenTab}, w)$

**For** $i = 1..d$ **do**
    **For** $j = 1..w$ **do**
        $\mathtt{YC}(W, \mathtt{GenTab}, w)[i,j] \leftarrow \mathtt{YG}(\mathtt{GenTab}, w, i, j)$;
        more precisely, $\mathtt{YC}(W, \mathtt{GenTab}, w)$ is the string
          $\mathtt{YG}(W, \mathtt{GenTab}, w, 1, 1) \circ \mathtt{YG}(W, \mathtt{GenTab}, w, 1, 2) \circ \cdots$
          $\circ \mathtt{YG}(W, \mathtt{GenTab}, w, d, w-1) \circ \mathtt{YG}(W, \mathtt{GenTab}, w, d, w)$

Figure 9.9: How to construct a Yao-type circuit from a table $W$ of keys, a table $\mathtt{GenTab}$ of tags and masks, and a set $w$ of key translations.

---

The inputs $x_1$ and $x_2$ are written as a set of input bits, $(x[1,1], \ldots, x[1,m], x[2,1], \ldots, x[2,m])$. The *value* $v[i,j]$ of wire $(i,j)$ is determined by the input bits and the gates $g_{i,j}$ (in the natural fashion in which the values are percolated through a circuit). We use a bijection $L_0(J) = (J \operatorname{div} m, J \bmod m)$ to map wire $(0, J)$ to input $x[L_0(J)]$. The set of keys necessary to evaluate the encrypted circuit is given by

$$\mathtt{InKeys}(W, w, x_1, x_2) = \{W[0, J, w[L_0(J)] \oplus x[L_0(J)]]\}_{J \in [2m]}$$

Finally, the output key translations that must be revealed in order to interpret the keys of the final level are

$$B(w) = (w[w, 1], w[w, 2], \ldots, w[w, d]).$$

The entire encrypted circuit is:

$$\texttt{EncCkt}(W, \texttt{GenTab}, w, x_1, x_2) = \texttt{YC}(W, \texttt{GenTab}, w) \circ B(w) \circ \texttt{InKeys}(W, w, x_1, x_2)$$

The initial portion corresponding to the circuit, the output keys, and the input keys for inputs from player 1 is denoted $\texttt{EncCkt}^1(W, \texttt{GenTab}, w, x_1, x_2)$, and does not depend on $x_2$. The suffix containing the input keys for player 2 is called $\texttt{EncCkt}^2(W, \texttt{GenTab}, w, x_1, x_2)$, and does not depend on $x_1$.

### 9.4.4   Two-Party Protocols for Passive Adversaries

The encrypted circuits we have just described allow us to specify a protocol for two players to evaluate some function $F(x_1, x_2)$. We examine the case where player 2 learns the output, and player 1 learns only whether player 2 cheats or not; the case where both players learn the output is covered by two applications of the unidirectional protocol.

Figure 9.10 describes the protocol. Player 1 constructs an encrypted circuit and gives it to player 2. Player 1 and player 2 also engage in a 1-out-of-2 oblivious transfer protocol in order for player 2 to obtain the needed wire keys corresponding to its inputs. Player 1 should not learn which keys player 2 obtained, and player 2 should obtain only one of each pair of keys.

One out of two oblivious transfer is like oblivious transfer except that instead of receiving a given secret with 50-50 probability, Bob is allowed to choose exactly one of two secrets to obtain. Alice, instead of not knowing whether Bob received anything, now fails to know which secret Bob chose. Crépeauand Kilian show how to implement 1-out-of-2 OT using standard OT. We denote their protocol, with the assumed OT protocol built in, as 1-2-OT $(b_1, b_2; c)$; Bob receives bit $b_c$ from Alice. The protocol is not limited to transferring bits; strings may be transferred as well.

To show this protocol secure, we need to show that the encrypted circuit $\texttt{EC}$, when generated properly, reveals nothing more than $F(x_1, x_2)$. We show that $\text{EVAL2}$ is as resilient as an ideal protocol that provides player 2 with $\texttt{EC}$ and we show the latter is as resilient as an ideal protocol to provide player 2 with $F(x_1, x_2)$.

**Theorem 9.14** *Protocol* $\text{EVAL2}$ *is 2-resilient against static, passive adversaries.*

**Proof:** Let $\texttt{ID2}(\texttt{EncCkt})$be an ideal protocol in which player 1 creates $W$, $\texttt{GenTab}$, and $w$ as above and sends them along with $x_1$ to the trusted host, and player 2 sends $x_2$ to the trusted host. The host computes $\texttt{EncCkt}(W, \texttt{GenTab}, w, x_1, x_2)$ and returns the value to player 2. It suffices to show $\text{EVAL2} \succeq \texttt{ID2}(\texttt{EncCkt})$ and $\texttt{ID2}(\texttt{EncCkt}) \overset{.}{\succeq} \text{ID}(F)$.

First let us show $\text{EVAL2} \succeq \texttt{ID2}(\texttt{EncCkt})$. If both players are corrupt, the interface $\mathcal{S}$ simply corrupts them both in the ideal protocol, obtains the inputs, operates the players as a subroutine, and returns the results to the adversary. If player 1 is corrupt, then $\mathcal{S}$ corrupts player 1 in the ideal protocol, obtains $x_1$, $W$, GenTab, and $w$, computes $\texttt{EC}^1$ and delivers it to $A$ as the view of player 1 in step (TE1). We have assumed 1-2-OT is resilient, so there is an interface that $\mathcal{S}$ can run as a subroutine for step (TE2) of EVAL2 (note that $\mathcal{S}$ has all keys that player 1 puts up in the protocol). If player 2 is corrupt, then $\mathcal{S}$ corrupts player 2 in the ideal protocol, obtains $x_2$ and $\texttt{EC} = \texttt{EncCkt}(W, \texttt{GenTab}, w, x_1, x_2)$, and gives to $A$ the message $\texttt{EC}^2$ from player 1 in step (TE1). For step (TE2), $\mathcal{S}$ runs a subinterface for 1-2-OT , providing it with $x_2$ when it requests the corruption of player 2 in the subprotocol. It should be clear that for this interface $\mathcal{S}$,

$$[A, \textsc{Eval2}] \stackrel{\cdot}{\asymp} [A, \mathcal{S}, \texttt{ID2}(\texttt{EncCkt})].$$

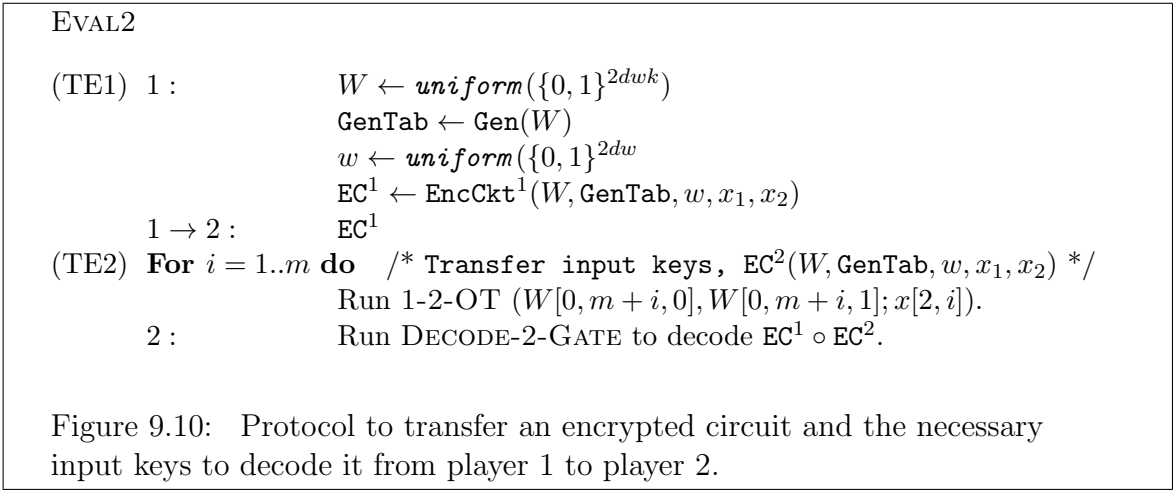

Figure 9.10: Protocol to transfer an encrypted circuit and the necessary input keys to decode it from player 1 to player 2.

The proof that $\texttt{ID2}(\texttt{EncCkt}) \succeq \texttt{ID}(F)$ is slightly more involved. If both player 1 and player 2 are corrupted, the interface $\mathcal{S}'$ has a trivial job. If player 1 is corrupted, $\mathcal{S}'$ need only corrupt player 1 in $\texttt{ID2}(\texttt{EncCkt})$ to obtain $x_1$ and supply it to $A$. The difficulty arises when only player 2 is corrupt, for then the interface needs to generate an encrypted circuit without knowing all the keys. Intuitively, the solution is to construct the part of the encrypted circuit that would be decrypted during DECODE-2-CIRCUIT , given the known subset of keys and the value of $F(x_1, x_2)$. In other words, each gate contains four entries, only one of which is used in the percolation. This entry must be generated accurately, but the known subset of keys suffice to generate it. The remaining entries are normally produced using the remaining keys, but $\mathcal{S}'$ does not have them. Instead, $\mathcal{S}'$ places uniformly random strings in their place. Because these entries are normally produced by a PRG, the resulting message that $\mathcal{S}'$ produces for $A$ is computationally indistinguishable from one generated in $\texttt{ID2}(\texttt{EncCkt})$, so the behavior of $A$, by Lemma 4.14, is essentially the same, and the corresponding final outputs in the two protocols are indistinguishable.

The proof is an easy corollary of the proof of Lemma 10.2. The proof of the lemma defines a set of ensembles which move progressively from one generated according to the real protocol to one generated by the interface. It is shown that, because each successive ensemble is indistinguishable, the first and last ensembles are indistinguishable. The encryption technique is more general, applying to $n$ parties, but the encryption allows a subset $S$ of the $n$ parties to be ignored (in case of misbehavior). If we take $S = [n] - \{1\}$, as though only player 1 behaves, the distributions are identical to the ones considered here, and the result follows directly. ∎

**Theorem 9.15** *Let $t \leq n$. If there exists a two-party protocol for oblivious transfer or oblivious transfer channels are available, then for any $F$ and $c$ there exists a $(k^{-c}, t)$-fair protocol for $F$ that is resilient against fail-stop adversaries. If $F$ is described by a circuit family $C_{F^{n,m}}$, the protocol requires $O(\mathtt{Size}(C_{F^{n,m}})^{c_0})$ bits and $O(\mathtt{Depth}(C_{F^{n,m}})^{c_0})$ rounds of interaction for some fixed $c_0$.*

**Proof:** The theorem follows directly from Theorems 9.8, 9.10, and 9.14. ∎

## 9.5 Byzantine Adversaries

### 9.5.1 Private and Public Channels

Before attacking the issue of Byzantine adversaries, we must consider how to eliminate private channels. In a cryptographic setting, *i.e.* when players are resource-bounded, we may employ encryption schemes between each pair of players. Before the protocol begins, each pair is given secret encryption and decryption keys $(E_{ij}, D_{ij})$, or each pair runs a protocol to generate and exchange them, such that each encryption scheme is independent of all the others whenever at least one player in the pair is nonfaulty. We convert a private channel protocol, $\Pi$, to a public channel protocol, $\mathtt{pub}(\Pi)$, by specifying that in $\mathtt{pub}(\Pi)$, every message $\mathtt{mesg}(i, j, r)$ that would otherwise be privately sent is encrypted and broadcast by $i$. That is, in $\mathtt{pub}(\Pi)$, $\mathcal{M}^{broad}(i, [n], r; \mathcal{M}(i, j, r)) = E_{ij}(\mathcal{M}(i, j, r))$.

An unpublished claim of Feldman states that private channels may be replaced by public, encrypted channels without loss of security [Fel88]. Indeed, a probability walk like the ones used to prove Theorems 9.14 and 10.4 (*cf.* §10.2.1) shows that the encryptions broadcast between nonfaulty players are indistinguishable to the adversary from uniformly random strings. Hence an interface need merely supply the adversary with uniformly random strings in their stead. It is important to note that this argument works only for static adversaries. We have:

**Proposition 9.16** *(After Feldman) For any protocol $\Pi$, $\mathtt{pub}(\Pi) \stackrel{\cdot}{\succeq} \Pi$.*

**Theorem 9.17** *Let $t \leq n$. If there exists a two-party protocol for oblivious transfer or oblivious transfer channels are available, then for any $F$ and $c$ there exists a $(k^{-c}, t)$-fair protocol $\mathtt{pub}(\textsc{EvalFairFS})$ for $F$ that is resilient against fail-stop adversaries and uses only broadcast channels. If $F$ is described by a circuit family $C_{F^{n,m}}$, the protocol requires $O(\mathtt{Size}(C_{F^{n,m}})^{c_0})$ bits and $O(\mathtt{Depth}(C_{F^{n,m}})^{c_0})$ rounds of interaction for some fixed $c_0$.*

### 9.5.2 Byzantine Adversaries

For technical reasons we consider only static adversaries. Recent work of Haber, Yung, and Beaver [BHY89] shows that, with additional techniques, dynamic adversaries can be withstood in the memoryless, cryptographic model, but we shall not delve into those issues here. The definition of fairness itself requires deeper examination with respect to dynamic adversaries.

We compile a public-channel cryptographic Turing machine protocol $\Pi$, resilient against passive or fail-stop adversaries, to a protocol $\texttt{byz}(\Pi)$, resilient against Byzantine adversaries, in the sense of [GMW86, GMW87]. Each player broadcasts an encryption of the machine $M_i$ it runs, its state, and the contents of all of its tapes. Later, each player gives zero-knowledge proofs that the new state and outgoing messages are computed correctly according to the encryptions and incoming messages. Let us consider this in more detail.

At each round of protocol $\Pi$, machine $M_i$ is described precisely by its *superstate* $s_i$ (Turing machine program, current state and position of tape heads, all tape contents). An *encrypted superstate* $e_i$ is the result of applying encryption $E_i$ to the string $s_i$. At each round of $\texttt{byz}(\Pi)$, every player broadcasts $e_i^r$, the encrypted superstate of $M_i$ at round $r$. The contents of $M_i's$ communication tapes, $c_i^r$, are all broadcast messages, hence public knowledge.

To prevent a Byzantine adversary from causing some player to simulate $M_i$ incorrectly, each player must prove publicly to every other player that the new, encrypted superstate and set of outgoing messages it broadcasts are correct with respect to the previous encryption and incoming messages. Define the predicate P1 by:

$$\text{P1}(e_i^r, c_i^r, e_i^{r+1}, c_i^{r+1}) = 1 \quad \Leftrightarrow \quad (\exists D_i)(D_i(e_i^r) = s_i^r)(D_i(e_i^{r+1}) = s_i^{r+1}) \quad \text{and}$$
$$s_i^r(c_i^r) = (s_i^{r+1}, c_i^{r+1})$$

where $s_i^r(c_i^r)$ indicates the local computation of a machine $M_i$ as specified by superstate $s_i^r$ with incoming messages $c_i^r$. The predicate claims that player $i$ can decode the encrypted superstate to demonstrate that the local computation is correct, something it cannot do with nonnegligible probability unless it actually encrypts a valid transition.

Each player $i$ conducts a zero-knowledge proof of $\text{P1}(e_i^r, c_i^r, e_i^{r+1}, c_i^{r+1})$ over broadcast channels with each other player as the verifier. As long as one player honestly plays the part of the verifier, with exponentially high probability no false statement will go undetected. Furthermore, all nonfaulty players concur as to the validity of the proof, having witnessed it over broadcast lines. Of course, proofs by nonfaulty players are zero-knowledge, and an interface easily creates their transcripts for the adversary without having to corrupt nonfaulty players to obtain their knowledge.

Another important predicate must be proved at the start of the protocol. Each player must prove it broadcast a proper description of the Turing machine $M_i$ in its initial state with zeros written over unused portions of the tapes (*i.e.* everything but the random tape (a sequence of some bounded number $p(m, n)$ of uniformly random bits), the $m$ bits of input, and the auxiliary tape). Call this predicate $\mathrm{P0}(e_i^0)$.

Should any proof fail, all players halt and output Cheating. In a certain sense, nothing better is achievable, since a faulty majority always has the power to halt the protocol. A slightly more robust treatment, in the spirit of using polynomial secret sharing instead of sum sharing, directs each player initially to polynomially share its initial superstate, $s_i^0$. If player $i$ fails, then $s_i^0$ is reconstructed and the current state computed using the public knowledge of broadcast messages. Better yet, if states are maintained as polynomially secretly shared secrets as the protocol progresses, then the reconstruction is immediate and no history need be recorded. Yet more involved treatments are possible: the shared portions of $s_i^r$ can be incorporated into the computation without revealing them. The protocol starts over with a new goal in mind, that of computing the appropriate result based on current states and the private pieces of $s_i^r$.

The protocol is described in Figure 9.11.

**Proposition 9.18** *(After Goldreich* et al*) Let $\mathcal{A}_1$ be a static, Byzantine, polynomial-time adversary class and let $\mathcal{A}_2$ be a static, fail-stop, polynomial-time adversary class. Then for any protocol $\Pi$,*

$$\mathtt{byz}(\Pi) \succeq_{(\mathcal{A}_1, \mathcal{A}_2)} \Pi.$$

---

**byz($\Pi$)**

(B0)    If any player fails to give a proper proof, halt and output Cheating.

(B1)    $(1 \leq i, j \leq n)$  Generate secret encryption and decryption keys $E_{ij}, D_{ij}$.

(B2)    $(1 \leq i \leq n)$     Generate secret encryption and decryption key $E_i, D_i$.

(B3)    **For** $i = 1..n$ **do**

        $i \to [n] :$      $e_i^0 = E_i(s_i^0)$

        **For** $j = 1..n$ **do**

        $(i, j) :$       $\mathtt{ZKP}(i, j, \mathrm{P0}(e_i^0))$

(B4)    **For** $r = 1..R(n, m)$ **do**  **For** $i = 1..n$ **do**

        $i :$        $(s_i^{r+1}, \mathtt{mesg}(i, [n], r+1)) \leftarrow s_i^r(\mathtt{mesg}([n], i, r)),$

                $e_i^{r+1} \leftarrow E_i(s_i^{r+1})$

        $i \to [n] :$     $e_i^{r+1}, \mathtt{mesg}(i, [n], r+1)$

        **For** $j = 1..n$ **do**

        $(i, j) :$       $\mathtt{ZKP}(i, j, \mathrm{P1}(e_i^r, \mathtt{mesg}([n], i, r), e_i^{r+1}, \mathtt{mesg}(i, [n], r+1)))$

Figure 9.11: Protocol to simulate protocol $\Pi$ in order to ensure resilience against Byzantine adversaries. Here, $\mathtt{ZKP}(i, j, P)$ is a zero-knowledge proof protocol with prover $i$, verifier $j$, predicate $P$, and all messages broadcast.

---

**Proof:** As discussed, the job of the interface $\mathcal{S}$ is relatively simple. For each step in $\Pi$, it participates with the adversary $A$ in $\mathtt{byz}(\Pi)$ to supply it with zero knowledge proofs from honest players or to supply it uniformly random strings substituted for encrypted messages between nonfaulty players. All encrypted messages known to the adversary are obtained from protocol $\Pi$. Should any proof from a corrupt player fail, $\mathcal{S}$ causes the corresponding player in $\Pi$ to halt. Protocol $\Pi$ deems what occurs thereafter (*e.g.*, all nonfaulty players may halt). It is not hard to see that all messages obtained by $A$ are computationally indistinguishable from those it sees in $<A, \texttt{byz}(\Pi)>$. Stripping away the zero-knowledge proofs, the messages processed by nonfaulty players are the same in $<A, \texttt{byz}(\Pi)>$ and $<A, \mathcal{S}, \Pi>$ — that is, the behavior of each nonfaulty internal $M_i$ in $\texttt{byz}(\Pi)$ matches that of each $M_i$ in $\Pi$. Hence all final outputs are distributed indistinguishably, by Lemma 4.14. ∎

**Theorem 9.19** *If there exists a protocol for two-party oblivious transfer, then for any $F$ and $c > 0$ there exists a protocol* EVALFAIR *for $F$ that is $(k^{-c})$-fair against Byzantine adversaries.*

**Proof:** Define EVALFAIR = $\texttt{byz}$(EVALFAIRFS), the compiled version of the fair, fail-stop protocol for $F$. By Proposition 9.18, EVALFAIR satisfies the claim. ∎

### 9.5.3 Restarting the Protocol

For most functions it is impossible to restart the protocol without allowing the adversary some degree of bias on the final result. For example, after learning the result of a parity computation with 80% likelihood, the adversary may halt the protocol in order to bias the result in the other direction. A certain amount of information about inputs is leaked from learning the result $F(x_1, \ldots, x_n)$; a second sample of $F$ after faulty players have changed their inputs may easily give away tremendous information. For example, $F(x_1, \ldots, x_n) = x_1 x_3 + x_2(1 - x_3)$ allows corrupt player 3 to learn $x_1$ with a reasonable degree of certainty by first using $x_3 = 1$, and in a restarted protocol to learn $x_2$ by setting $x_3 = 0$. Though inputs may be forced to be the same from one execution to another by using the encryptions as commitments, an astute adversary can glean information by a particular choice of inputs to *omit*, without having to change inputs.

The polynomial secret sharing method is advantageous in a setting where a correct answer must be had, even at the cost of allowing bias or loss of privacy. At least $n - t$ players must identifiably cheat in order for the protocol to halt. The maximal number of restarts is reduced by a factor of $(n - t)$ over the simple sum sharing methods. Eventually the number of nonfaulty players, $n - t$, becomes greater than half the number of remaining players, and a protocol for faulty minority (see Chapter 6) may be employed.

## 9.6 Simultaneous, Verifiable Secret Sharing

Fairness is precisely the most significant problem with secret sharing when the majority is faulty. An adversary can easily collect the pieces of nonfaulty players at reconstruction time but then refuse to reveal pieces held by faulty players.

A method for one player to share a secret that can later be revealed in a fair manner would be a useful tool. This is the problem of Simultaneous Verifiable Secret Sharing (SVSS): the secret must be verifiable when shared, but it also must later be revealed simultaneously to all players.

The ideal protocol utilizes a fair, ideal host who accepts the secret and in the next stage reveals it fairly. This is equivalent to a fair, ideal host who computes a trivial identity function.

**Definition 9.20** *A protocol is an $(\delta, t)$-**ideal SVSS protocol** with dealer $D$ if it is a $(\delta, t)$-fair, ideal protocol for the function $F_D(x_1, \ldots, x_n) = x_D$.*

The SVSS problem is to find a protocol that satisfies $\Pi \mathrel{\dot{\succeq}} \texttt{FID}(F_D)$.

Define the protocol $\text{SVSS}(D)$ to be the EVALFAIR protocol as run on $F_D(x_1, \ldots, x_n) = x_D$. Then by Theorem 9.19, $\text{SVSS}(D) \mathrel{\dot{\succeq}} \texttt{FID}(F)$. The solution to SVSS is clear, once we have the machinery of this chapter to construct fair protocols for arbitrary functions. The revelation of a secret boils down to of a sequence of unfair revelations of coin tosses biased toward the secret value.

# Chapter 10

# Cryptographic Methods for Constant Rounds

We have seen that cryptographic assumptions of one sort or another facilitate solutions where none are possible, according to information theory. The level of fault-tolerance is vastly improved by assuming the existence of an oblivious-transfer protocol; can other parameters be improved by making cryptographic assumptions?

As it turns out, the number of rounds for any protocol can be reduced to a constant, without any concomitant explosion in message size, if one is willing to make the weakest sort of cryptographic assumption, namely that there exists a one-way function. No details about the structure of such a function are needed for our protocols. Thus, in turning to the vulnerability of making unproven assumptions, we take the smallest possible gamble. Until such time as there is a proof that P $\neq$ NP and that one-way functions exist, our techniques provide all the advantages of complexity-based cryptography but at the least risk.

The solution is inspired by Yao's method for oblivious circuit evaluation, in which one player supplies the other with an encrypted circuit and a partial set of keys to decrypt it (*cf.* Chapter 9). The encrypted circuit is evaluated locally, without any interaction. Certainly in the unbounded-resource models of Chapters 5 and 6, such an encryption is not possible with perfect, information-theoretic security. On the other hand, if we limit the adversary to begin polynomial time, an encrypted circuit of the type introduced by Yao can be computed itself as a result of a multiparty protocol. If the construction of the encrypted circuit requires constant rounds, then the overall protocol will require only constant rounds, since the encrypted circuits can be evaluated locally, without interaction.

There are crucial differences between the two-party construction and the multiparty construction — most notably the fact that no single party actually knows how the circuit was constructed — but the important property is that, if the generalized Yao circuit can be constructed, then it requires no interaction to evaluate it. The encryption we describe here is more general and more involved than that in Chapter 9 but is quite similar in other ways.

The complication arises from the need to combine pseudorandom sequences gen-erated by *all* the players in order to ensure that no single player can fathom any part of the encrypted circuit not corresponding to the restricted path of percolated keys. Interestingly, this approach would not work if general secret computation were needed to construct the encrypted circuit, because the construction itself must use constant rounds, and generating pseudorandom bits is a polynomial time computation, not known to be in $NC^1$, and hence not known to be computable in constant rounds. The second technique that allows the solution to go through arises from the observation that each player can compute pseudorandom sequences locally and *prove* to the network, using methods of Chapter 7, that it shares these results properly. Proceeding from the pseudorandom outputs to the encrypted circuit is an extremely fast and simple computation.

## 10.1   Generalized Yao Gates

Fix $n$, $m$, and $F$. Let $\mathtt{Gen}(X)$ be a PRG that outputs $(1 + 2n)|X|$ bits (or to be pedantic, let $\mathtt{Gen}(X)$ be the first $(1 + 2n)|X|$ bits produced by a PRG that outputs $3|X|^2$ bits, with $|X| \geq n$). Denote the first $|X|$ bits by $\mathtt{Tag}(X)$, the next $n|X|$ bits by $\mathtt{Mask}(0, X)$, and the final $n|X|$ bits by $\mathtt{Mask}(1, X)$. In the sequel, all keys have the same length $k$. Let $S \subseteq [n]$, and define the string $S(i) = 0^{nk}$ if $i \notin S$ and $S(i) = 1^{nk}$ if $i \in S$. The set $S$ will ultimately be used to zero out components corresponding to faulty players. If $\wedge$ is the bitwise logical AND, define the following strings, for $a \in \{0, 1\}$ :

$$\mathtt{Tag}(S, \vec{X}_a) = [\mathtt{Tag}(X_a) \wedge S(1)] \circ [\mathtt{Tag}(X_a^2) \wedge S(2)] \circ \cdots \circ [\mathtt{Tag}(X_a^n) \wedge S(n)]$$
$$\mathtt{Tag}(S, \vec{Y}_a) = [\mathtt{Tag}(Y_a) \wedge S(1)] \circ [\mathtt{Tag}(Y_a^2) \wedge S(2)] \circ \cdots \circ [\mathtt{Tag}(Y_a^n) \wedge S(n)]$$

(The vector notation gives $X^i$ as the $i^{th}$ component of $\vec{X}$.) Define the following strings for $a, b \in \{0, 1\}$ :

$$\mathtt{Mask}(S, a, \vec{X}_b) = [\mathtt{Mask}(a, X_b^1) \wedge S(1)] \oplus \cdots \oplus [\mathtt{Mask}(a, X_b^n) \wedge S(n)]$$
$$\mathtt{Mask}(S, a, \vec{Y}_b) = [\mathtt{Mask}(a, Y_b^1) \wedge S(1)] \oplus \cdots \oplus [\mathtt{Mask}(a, Y_b^n) \wedge S(n)]$$

Knowing any one of the keys will allow a player to match it to a tag (by evaluating the PRG), but computing the masks requires knowledge of all the keys.

The generalized Yao gate is the table consisting of four $3nk$-bit entries as follows: for $a = 0, 1$ and for $b = 0, 1$,

$$\mathtt{Ent}(S, a, b, \vec{X}_0, \vec{X}_1, \omega_X, \vec{Y}_0, \vec{Y}_1, \omega_Y, \vec{Z}_0, \vec{Z}_1, \omega_Z) =$$
$$\mathtt{Tag}(S, \vec{X}_a) \circ \mathtt{Tag}(S, \vec{Y}_b) \circ$$
$$[\mathtt{Mask}(S, b, \vec{X}_a) \oplus \mathtt{Mask}(S, a, \vec{Y}_b) \oplus \vec{Z}_{g(a \oplus \omega_X, b \oplus \omega_Y) \oplus \omega_Z}]$$

We refer to the three $nk$-bit strings making up an entry of the table as segments $\mathtt{yg}_1(a, b)$, $\mathtt{yg}_2(a, b)$, and $\mathtt{yg}_3(a, b)$. The entire table is thus:

$$\texttt{yg}(S, \vec{X}_0, \vec{X}_1, \omega_X, \vec{Y}_0, \vec{Y}_1, \omega_Y, \vec{Z}_0, \vec{Z}_1, \omega_Z) =$$
$$\texttt{Ent}(S, 0, 0, \vec{X}_0, \vec{X}_1, \omega_X, \vec{Y}_0, \vec{Y}_1, \omega_Y, \vec{Z}_0, \vec{Z}_1, \omega_Z) \circ$$
$$\texttt{Ent}(S, 0, 1, \vec{X}_0, \vec{X}_1, \omega_X, \vec{Y}_0, \vec{Y}_1, \omega_Y, \vec{Z}_0, \vec{Z}_1, \omega_Z) \circ$$
$$\texttt{Ent}(S, 1, 0, \vec{X}_0, \vec{X}_1, \omega_X, \vec{Y}_0, \vec{Y}_1, \omega_Y, \vec{Z}_0, \vec{Z}_1, \omega_Z) \circ$$
$$\texttt{Ent}(S, 1, 1, \vec{X}_0, \vec{X}_1, \omega_X, \vec{Y}_0, \vec{Y}_1, \omega_Y, \vec{Z}_0, \vec{Z}_1, \omega_Z)$$

Figure 10.1 describes the the *decoding* of a gate given input keys $\vec{X}$ and $\vec{Y}$.

---

DECODE-N-GATE $(S, \texttt{yg}, \vec{X}, \vec{Y})$

(DC1) Compute $\texttt{Tag}(S, \vec{X}), \texttt{Tag}(S, \vec{Y})$.

(DC2) **For** $a = 0..1$ **do**

　　　　　Compute $\texttt{Mask}(S, a, \vec{X}), \texttt{Mask}(S, a, \vec{Y})$.

(DC3) Determine the least $\alpha, \beta$ such that

　　　　　$\texttt{Tag}(S, \vec{X}_\alpha) = \texttt{yg}_1(\alpha, \beta)$ and $\texttt{Tag}(S, \vec{Y}_\beta) = \texttt{yg}_2(\alpha, \beta)$

(DC4) Set $\vec{Z} \leftarrow \texttt{yg}_3(\alpha, \beta) \oplus \texttt{Mask}(S, \beta, \vec{X}) \oplus \texttt{Mask}(S, \alpha, \vec{Y})$.

Figure 10.1: Obtaining the output key $\vec{Z}$ from a generalized Yao gate, given the encrypted table and two input keys $\vec{X}$ and $\vec{Y}$. The set $S$ describes the indices of strings to consider as 0 when computing $\texttt{Tag}$, $\texttt{Mask}$.

---

### 10.1.1 Yao Circuit Construction

Given the means to construct gates, constructing a circuit is straightforward. We use the notation for gates $g_{i,j}$ and input wires $\texttt{in}_0(i, j)$ and $\texttt{in}_1(i, j)$ as described in §9.4.3.

We consider an array $W[i, j, l, I]$ of wire keys for $i \in \{0, \ldots, d\}, j \in \{1, \ldots, w\}$, $l \in \{0, 1\}, I \in [n]$. The keys $W[i, j, 0, I]$ and $W[i, j, 1, I]$) represent wire $(i, j)$, which carries the output of gate $g_{i,j}$. Keys $W[0, j, 0, I]$ and $W[0, j, 1, I]$) represent inputs to the circuit. Associated with each wire $(i, j)$ is a *key translation bit* $w[i, j]$.

Given $S$, a set of components to void, and $W$, the construction of the circuit is as follows. First, construct a table $\texttt{GenTab}$ of $2dwn(1 + 2n)k$ pseudorandom bits, where

$$\texttt{GenTab} = \texttt{Gen}(S, W)$$

and where the function $\texttt{Gen}(S, W)$ gives a table of $(2dwn)$ $((1 + 2n)k)$-bit strings defined by

$$\texttt{Gen}(S, W)[i, j, l, I] = \texttt{Gen}(W[i, j, l, I]) \wedge S(I).$$

Second, convert this table to a circuit in the same fashion as the gates are constructed. That is, with respect to table $\texttt{GenTab}$, define the functions $\texttt{tag}(\texttt{GenTab}, i, j, l, I)$ and $\texttt{mask}(c, \texttt{GenTab}, i, j, l, I)$ $(c \in \{0, 1\})$ as the substrings of length $k$, $nk$, and $nk$ of $\texttt{GenTab}[i, j, l, I]$.

The generalized Yao gate $\texttt{YG}(W, \texttt{GenTab}, w, S, i, j)$ is then constructed as in §10.1, using $\texttt{tag}$ and $\texttt{mask}$ as the tag and mask functions. The wire keys are as follows.

$$\begin{aligned}
\vec{X}_l &= (W[\texttt{in}_0(i,j), l, 1], W[\texttt{in}_0(i,j), l, 2]), \ldots, W[\texttt{in}_0(i,j), l, n]) \\
\vec{Y}_l &= (W[\texttt{in}_1(i,j), l, 1], W[\texttt{in}_1(i,j), l, 2]), \ldots, W[\texttt{in}_0(i,j), l, n]) \\
\vec{Z}_l &= (W[i, j, l, 1], W[i, j, l, 2]), \ldots, W[i, j, l, n])
\end{aligned}$$

The key translations used to construct the gate are $w_X = w[\texttt{in}_0(i,j)]$, $w_Y = w[\texttt{in}_1(i,j)]$, and $w_Z = w[i,j]$. As mentioned, a set $S$ of components to be voided is used to place 0's over appropriate components.

The circuit thus constructed is denoted $\texttt{YC}(W, \texttt{GenTab}, w, S)$. Notice that the construction takes $\texttt{GenTab}$ into account independently of $W$; it is well-defined whether or not $\texttt{GenTab}$ arises from $W$.

---

Make-N-Circuit $(W, \texttt{GenTab}, w, S)$

**For** $i = 1..d$ **do**
    **For** $j = 1..w$ **do**
        $\texttt{YC}(W, \texttt{GenTab}, w, S)[i, j] \leftarrow \texttt{YG}(\texttt{GenTab}, w, S, i, j)$;
        more precisely, $\texttt{YC}(W, \texttt{GenTab}, w, S)$ is the string
            $\texttt{YG}(W, \texttt{GenTab}, w, S, 1, 1) \circ \texttt{YG}(W, \texttt{GenTab}, w, S, 1, 2) \circ \cdots$
            $\circ \texttt{YG}(W, \texttt{GenTab}, w, S, d, w - 1) \circ \texttt{YG}(W, \texttt{GenTab}, w, S, d, w)$

Figure 10.2: How to construct a generalized Yao circuit from a table $W$ of keys, a table $\texttt{GenTab}$ of tags and masks, a set $w$ of key translations, and a set $S \subseteq [n]$ of components to be voided.

---

A set of inputs $\vec{x} = (x_1, \ldots, x_n)$ is written as a set of input bits, $(x[1,1], \ldots, x[1,m], x[2,1], \ldots, x[2,m], \ldots, x[n,1], \ldots, x[n,m])$. The *value* $v[i,j]$ of wire $(i, j)$ is determined by the input bits and the gates $g_{i,j}$ (in the natural fashion in which a circuit is evaluated) We use a bijection $L_0(J) = (J \text{ div } m, J \text{ mod } m)$ to map wire $(0, J)$ to input $x[L_0(J)]$.

The input keys $\texttt{InKeys}$ that must be revealed for the circuit to be evaluable through key percolation are given by

$$\begin{aligned}
\texttt{InKeys}_I(W, w, \vec{x}) &= \{W[0, J, w[L_0(J)] \oplus x[L_0(J)], I]\}_{J \in [nm]} \\
\texttt{InKeys}(W, w, \vec{x}) &= \{\texttt{InKeys}_I\}_{I \in [n]}
\end{aligned}$$

Finally, the output key translations that must be revealed in order to interpret the keys of the final level are

$$B(w) = (w[w, 1], w[w, 2], \ldots, w[w, d]).$$

The encrypted circuit is then

$$\texttt{EncCkt}(W, \texttt{GenTab}, w, S, \vec{x}) = \texttt{YC}(W, \texttt{GenTab}, w, S) \circ B(w) \circ \texttt{InKeys}(W, w, \vec{x})$$

## 10.2  Protocols in Constant Rounds

**Theorem 10.1** *Assume there exists a one-way function. Let $F$ be a polynomial-time function family. Consider a complete broadcast network with private channels, where each player is a probabilistic polynomial-time Turing machine. The adversary class consists of polynomial-size circuit families and a fault class allowing $2t < n$. Then there is a $t$-resilient protocol for $F$ that runs in* constant *rounds and has polynomial message complexity.*

**Proof:**

---

CRYPTO-CONST

(C1)   **For** $i = 1..n$ **do in parallel**
        Player $i$ shares input $x_i$ as secret bits $x[i1], \ldots, x[im]$.

(C2)   Run RAND-BIT to generate random bits $\{\rho[i,j]\}_{i \in [n], j \in [dwnk]}$.

(C3)   **For** $i = 1..n$ **do in parallel**
        Reveal $\rho[i,1], \ldots, \rho[i, dwnk]$ to player $i$.

(C4.1)  **For** $i = 1..n$ **do for** $j = 1..dwn$ **do**
        Player $i$ locally computes $G_{i,j} = \texttt{Gen}(\rho[i, jk+1] \cdots \rho[i, (j+1)k])$.

(C4.2)  **For** $i = 1..n$ **do in parallel for** $j = 1..dwn$ **do in parallel**
        Player $i$ shares $G_{i,j}$.

(C5)   **For** $i = 1..n$ **do in parallel for** $j = 1..dwn$ **do in parallel**
        Player $i$ proves $G_{i,j} = \texttt{Gen}(\rho_{i,jk+1} \cdots \rho_{i,(j+1)k})$ using PROVE-SECRET-P .

(C6.1)  Let $S$ be the set of players whose proof failed.

(C6.2)  Define $G_{i,j} = 0^{(1+2n)k}$ for $i \in S$ and $j \in [dwn]$.

(C6.3)  Compute $\texttt{YC}(S, W)$, where $W$ is defined by the $G_{i,j}$ values.

(C7)   Compute $\chi$ from $W, \vec{x}$.

(C8)   Reveal $\texttt{YC}(S, W), \chi, B$ to all players.
        (Note that secrets in $B$ are contained in $W$.)
        Each player $i$ uses DECODE-N-GATE to compute $F(\vec{x})$.

Figure 10.3:   Protocol to create a generalized Yao circuit $\texttt{YC}$ and the decrypting keys needed to evaluate it.

---

The protocol is given in Figure 10.3. Constructing the encrypted circuit requires constant rounds. Revealing the encrypted circuit reveals nothing more than the final outputs. The resilience of the protocol follows from Lemmas 10.2 (see §10.2.1), 10.3 (see §10.2.2), and 4.26.   ∎

### 10.2.1 Generalized Yao Circuits Are Private

Consider the following ideal protocol $\texttt{ID}(\texttt{EncCkt})$ to generate encrypted circuits. Each player supplies an input $(x_i, p_i)$ where $p_i \in \{participate, quit\}$. The trusted host sets $S = \{i \mid p_i = quit\}$ and selects keys $W$ and key translations $w$ uniformly at random for all the indices $i \notin S$. It computes $\texttt{GenTab}$ using $\texttt{Gen}$ and $W$. It then constructs the circuit and returns $\texttt{EncCkt}(W, \texttt{GenTab}, w, S, \vec{x})$. (By §2.7 we can assume that each player learns every output bit, so there is no need to return different subsets of output key translations.) Finally, each player computes its output by percolating the keys from $B$ through the gates, using DECODE-N-GATE .

**Lemma 10.2** *Let the adversary class $\mathcal{A}$ include all nonuniform polynomial-time Turing machines, consider only static adversaries, and let $t \leq n - 1$. Then*

$$\texttt{ID}(\texttt{EncCkt}) \overset{\cdot}{\succeq}_{\mathcal{A}} \texttt{ID}(F)$$

**Proof:** The bijection $L(m) = (m \text{ div } w, m \text{ mod } w)$ between $[dw]$ and $[d] \times [w]$ defines a natural ordering on pairs, namely $L(m) < L(m+1)$. This row-major ordering extends directly to longer tuples. In particular, we consider the bijection $L(m)$ between $[d \cdot w \cdot 2 \cdot n]$ and $[d] \times [w] \times \{0, 1\} \times [n]$. The tuple $(i, j, l, I)$ corresponds to a particular wire key and to one level of the tag and mask table, $\texttt{GenTab}$.

We construct a progressive obliteration of a generalized Yao circuit by stepping through the pseudorandom table row by row, replacing each string in the table by a uniformly random string. With respect to a particular input assignment, a set $S$ of voided components, and an additional parameter $T \subseteq [n]$, the strings that are actually used in the percolation and evaluation of the circuit, or that are voided or otherwise given special exception, are not replaced. This generates a sequence of hidden circuits that are indistinguishable.

For a given set of inputs and a given circuit for $F$, the values $v[i, j]$ on each wire are determined. The indices $u[i, j]$ of the keys that are percolated during evaluation are determined by the key translation bits $w$ and the wire values:

$$u[i, j] = w[i, j] \oplus v[i, j].$$

For a given $T$, $S \subseteq T$, and $M$ ($0 \leq M \leq 2dwn$), the obliterated table is as follows. Let $\texttt{RandTab}$ be a string of $2dwnk$ bits, let $L(\mu) = (i, j, l, I)$, and define the obliteration of entry $m$ :

$$\texttt{OblEnt}(W, \texttt{RandTab}, S, T, m) \;=\; \begin{cases} 0^{3k} & \text{if } I \in S \\ \texttt{RandTab}[L(m)] & \text{if } I \notin T, \, l \neq u[i, j] \\ \texttt{Gen}(S, W)[L(m)] & \text{otherwise} \end{cases}$$

Now, row $m$ of the $M^{th}$ obliterated table is:

$$\texttt{OblTab}(W, \texttt{RandTab}, S, T, M)[m] \;=\; \begin{cases} \texttt{OblEnt}(W, \texttt{RandTab}, S, T, m) & \text{if } m \leq M \\ \texttt{Gen}(S, W)[L(m)] & \text{otherwise} \end{cases}$$

The obliterated circuit is obtained by using the partly obliterated pseudorandom table:

$$\texttt{PYC}(W, \texttt{RandTab}, w, S, T, M) = \texttt{YC}(W, \texttt{OblTab}(W, \texttt{RandTab}, S, T, M), w, S)$$

Finally, the progressively obliterated encrypted circuits are defined by:

$$\texttt{OblEncCkt}(W, \texttt{RandTab}, w, S, T, \vec{x}, M) =$$
$$(\texttt{PYC}(W, \texttt{RandTab}, w, S, T, M), \texttt{InKeys}(W, w, \vec{x}), B(w))$$

Define ensembles $D(M)(z, k)$ for all $M$ as follows. If $z$ is not of the form $n \circ m \circ d \circ w \circ S \circ T \circ \vec{x} \circ \vec{a} \circ w$ or if $k < 2ndw$, then all probability weight is placed on the string 0. Otherwise, using the security parameter $k$ as the key length, set

$$
\begin{aligned}
D(M)(z, k) \;=\; \{ & W \leftarrow \{0, 1\}^{2dwn}; \\
& \texttt{RandTab} \leftarrow \{0, 1\}^{2dwn(1+2n)k} : \\
& \texttt{OblEncCkt}(W, \texttt{RandTab}, w, S, T, \vec{x}, M) \}
\end{aligned}
$$

Successive pairs $D(M-1)(z, k)$ and $D(M)(z, k)$ are indistinguishable because they differ at most in the generation of a single pseudorandom sequence. Let us argue formally.

Suppose $D(M-1)(z, k)$ and $D(M)(z, k)$ are $(k^{-c-1})$-distinguishable. In particular this means that $z = n \circ m \circ d \circ w \circ S \circ T \circ \vec{x} \circ \vec{a} \circ w$ and

$$\texttt{OblEnt}(W, \texttt{RandTab}, S, T, M) = \texttt{Gen}(W[L(M)]),$$

for otherwise $D(M-1)(z, k) = D(M)(z, k)$. Then for any $c$ there exists a distinguisher $\mathcal{M}$ such that

$$|\mathcal{M}_{D(m)(z,k)} - \mathcal{M}_{D(m-1)(z,k)}| > k^{-c}$$

for infinitely many $k$.

We overwrite the $(1 + 2n)k$-bit string in location $L(m)$ of $\texttt{RandTab}$ using the function

$$
\texttt{OverWriteCkt}(\texttt{RandTab}, \sigma)[L(m')] \;=\; \left\{
\begin{array}{ll}
\texttt{RandTab}[L(m')] & \text{if } m' \neq m \\
\sigma & \text{if } m' = m
\end{array}
\right.
$$

Consider the machine $\mathcal{M}'$ that on input $\sigma$, sets $k = |\sigma|$ and selects $W$ and $\texttt{RandTab}$ uniformly at random, It then sets $\texttt{RandTab}' = \texttt{OverWriteCkt}(\texttt{RandTab}, \sigma)$, computes $\texttt{OblTab}(W, \texttt{RandTab}', S, T, M)$, and constructs hidden circuit $\texttt{EncCkt}$. Finally, it runs $\mathcal{M}$ on $\texttt{EncCkt}$ and returns the output of $\mathcal{M}$.

Define the ensembles

$$
\begin{aligned}
\mathcal{G}(k) \;&=\; \{ X \leftarrow \boldsymbol{uniform}(\{0, 1\}^k) : \texttt{Gen}(X) \} \\
\mathcal{H}(k) \;&=\; \boldsymbol{uniform}(\{0, 1\}^{(1+2n)k}).
\end{aligned}
$$

Then

$$
\begin{aligned}
D(M-1)(z,k) = \{ &\texttt{RandTab} \leftarrow \{0,1\}^{2dwn(1+2n)k}; \\
&\sigma \leftarrow \mathcal{G}(k); \\
&\texttt{RandTab}' = \texttt{OverWriteCkt}(\texttt{RandTab}, \sigma): \\
&\texttt{OblEncCkt}(W, \texttt{RandTab}', w, S, T, \vec{x}, M)\} \\
D(M)(z,k) = \{ &\texttt{RandTab} \leftarrow \{0,1\}^{2dwn(1+2n)k}; \\
&\sigma \leftarrow \mathcal{H}(k); \\
&\texttt{RandTab}' = \texttt{OverWriteCkt}(\texttt{RandTab}, \sigma): \\
&\texttt{OblEncCkt}(W, \texttt{RandTab}', w, S, T, \vec{x}, M)\}
\end{aligned}
$$

so that by the construction of $\mathcal{M}'$,

$$
|\mathcal{M}'_{\mathcal{G}(k)} - \mathcal{M}'_{\mathcal{H}(k)}| = |\mathcal{M}_{D(m)(z,k)} - \mathcal{M}_{D(m-1)(z,k)}| > k^{-c}
$$

This holds for infinitely many $k$, and certainly infinitely many $k$ larger than $2dwn$, so $\mathcal{M}'$ distinguishes $\mathcal{G}$ from $\mathcal{H}$, contradicting the assumption that $\texttt{Gen}$ is a PRG.

Therefore, for all $M$, $D(M-1) \approx^{k^{-c-1}} D(M)$, so $D(0) \approx^{k^c} D(2dwn)$ (recall $k \geq 2dwn$).

With this in mind, we argue that the task of an interface is easy: obtain and return $x_i$ and $a_i$ for all players $i$ that $A$ wishes to corrupt, by corrupting $i_{id}$ in $\texttt{ID}(F)$; use the input values shared by corrupted players $i$ to replace those of corrupted players $i_{id}$ in $\texttt{ID}(F)$, or have $i_{id}$ send $\Lambda$ if $i$ does not share a value; determine the set $S$ of non-participating faulty players from the interaction with $A$ ($S$ is the list of players who are disqualified); and return an *obliterated* circuit that is constructed according to distribution $D(2dwn)$, namely according to the circuit for $F$ and the output returned by the trusted host.

It is clear that distribution $D(2dwn)$ is easy to sample given the list of gates in the circuit for $F$, the value returned by the trusted host, and the list $S$ of voided components. Now, in protocol $<A, \texttt{ID}(\texttt{EncCkt})>$,

$$
\texttt{view}_A^f = a_A \circ \vec{a}_T \circ \vec{x}_T \circ \texttt{EC}
$$

where $T = T(q_A^f)$ and $\texttt{EC}$ is returned by the host. Note that the host samples

$$
\texttt{EC} \leftarrow D(0)(n \circ m \circ d \circ w \circ S \circ T \circ \vec{x} \circ \vec{a}, k).
$$

On the other hand, in protocol $<A, \mathcal{S}, \texttt{ID}(F)>$,

$$
\texttt{view}_A^f = a_A \circ \vec{a}_T \circ \vec{x}_T \circ \texttt{EC}
$$

where $\texttt{EC}$ is returned by $\mathcal{S}$. Note that $\mathcal{S}$ samples

$$
\texttt{EC} \leftarrow D(2dwn)(n \circ m \circ d \circ w \circ S \circ T \circ \vec{x} \circ \vec{a}, k).
$$

We have already shown that the ensembles $D(0)$ and $D(2dwn)$ are computationally indistinguishable, so the families of ensembles satisfy

$[A, \mathtt{ID}(\mathtt{EncCkt})]^{Y_A} \mathrel{\dot{\asymp}} [A, \mathcal{S}, \mathtt{ID}(F)]^{Y_A}$. Now, by the construction of $D(0)$ and $D(2dwn)$, the value of $F$ that is percolated through an encrypted circuit is the same regardless of which method is used to generate it, so $[A, \mathtt{ID}(\mathtt{EncCkt})]^{\vec{Y}} \mathrel{\dot{\asymp}} [A, \mathcal{S}, \mathtt{ID}(F)]^{\vec{Y}}$. Hence $\mathtt{ID}(\mathtt{EncCkt}) \mathrel{\dot{\succeq}} \mathtt{ID}(F)$. ∎

### 10.2.2 Protocol Crypto-Const Resiliently Computes $F$

**Lemma 10.3** *Let the adversary class $\mathcal{A}$ be static and restricted to nonuniform polynomial-time Turing machines, and let $t < n$. Then:*

$$\textsc{Crypto-Const} \mathrel{\dot{\succeq}}_{\mathcal{A}} \mathtt{ID}(\mathtt{EncCkt})$$

**Proof:** The Crypto-Const protocol is a composition of resilient computations. The computation of random bits is a private function and $t$-resilient. The computation of each secret PRG output is $t$-resilient, using Theorem 7.1. Revealing $S$ is private. The Evaluate $^{(2t<n)}$ protocol evaluates $\mathtt{YC}(W, G, w, S)$ in a constant number of rounds since $\mathtt{YC}$ is of constant depth. In fact, the circuit construction requires an exclusive-or of bit streams, some of which are voided according to $S$. Because $S$ is public, the indices of the appropriate pseudorandom sequences to include are public and no secret AND computations need be performed. Therefore, constructing the encrypted circuit secretly requires no interaction. Secretly computing the input keys to reveal requires an execution of a multiplication protocol (an AND must be secretly computed). The encrypted circuit itself is a robust (all players learn it, so a minority of alterations are not effective) and private (*cf.* Lemma 10.2) representation of the result, $F(x_1, \ldots, x_n)$. The reconstruction stage is $t$-resilient.

By Theorem 4.28, the composition is as resilient as the composition of the corresponding ideal vacuous protocols with $\mathtt{ID}(\mathtt{EncCkt})$, the which composition is in turn as resilient as $\mathtt{ID}(\mathtt{EncCkt})$. The adversary must be static in order that the subprotocols be post-protocol corruptible and hence composable. Each subprotocol of Crypto-Const requires only a constant number of rounds, so Crypto-Const requires only a constant number of rounds. ∎

## 10.3 Without Private Channels

By an unpublished claim of Feldman [Fel88] (see §9.5.1), private channels in a protocol can be replaced by public channels over which encrypted messages are broadcast. Lemma 4.14 along with a standard probability walk as in §10.2.1 would provide a proof of this claim.

**Theorem 10.4** *Assume there exists a one-way function. Let $F$ be a polynomial-time function family. Consider a complete broadcast network of public channels, where each player is a probabilistic polynomial-time Turing machine. The adversary class consists of polynomial-size circuit families and a fault class allowing $2t < n$. Then* *there is a t-resilient protocol for F that runs in* constant *rounds and has polynomial message complexity.*

# Chapter 11

# Instance Hiding Schemes

An *instance-hiding scheme* is a method by which a weak (polynomial-time) processor can obtain the value $f(x)$ of a function it cannot compute on its own, by querying more powerful processors, without having to reveal the instance $x$ at which it would like to compute $f$.

Our research into instance-hiding schemes motivated the development of a new tool, called a *locally random reduction,* which inspired a broad range of results in cryptography and complexity theory, including a theory of program testing [Lip89] and a recent line of research leading to the proof that IP = PSPACE [Nis89, LFKN89, Sha89]. These applications are discussed in Chapter 12, which investigates locally random reductions in depth. In this chapter we provide some of the motivations for our development of locally random reductions and investigate the first of many applications which locally random reductions solve, namely instance-hiding schemes.

Abadi, Feigenbaum, and Kilian were the first to investigate the problem of using a powerful, public resource to solve a problem without revealing the instance of the problem [AFK89]. They developed a formal model to measure the information which is hidden or leaked during the interaction between querier and oracle. They were motivated by the practical question of whether a weak, private computing device, such as a smart card or terminal, can take advantage of a powerful, shared computing device while keeping private some important aspects of its user's data. They were also motivated by the theoretical question of whether several well-studied yet intractable number-theoretic functions, such as discrete logarithm and quadratic residuosity, whose instances *can* be hidden significantly when querying an oracle for the solution, are examples of a more general phenomenon; that is, do other seemingly intractable problems, such as SAT, also have instance-hiding schemes?

The main result of Abadi, Feigenbaum, and Kilian is negative: if $f$ is an NP-hard function, a weak querier $A$ cannot query a single oracle $B$ while hiding all but the size of the instance, assuming that the polynomial hierarchy does not collapse. Their proof draws on a connection between single-oracle instance-hiding schemes and the nonuniform complexity classes NP/poly and CoNP/poly, and it is related to other complexity-theoretic notions, such as random-self-reducibility. This negative result holds even if $B$ is modelled as an oracle Turing Machine and is given access to an arbitrary (non-r.e.) oracle set. We refer the interested reader to [AFK89] for full details.

**Theorem 11.1** *[Abadi, Feigenbaum, Kilian 1989] If language L admits an instance-hiding scheme hiding all but $|x|$, then $L \in NP/poly \cap coNP/poly$.*

Following a question originally posed by Rivest [Riv86], we generalize instance-hiding schemes to handle several, physically separate oracles, $B_1, \ldots, B_m$ and show that, given sufficiently many oracles, any function admits a multioracle instance-hiding scheme. This contrasts with the single oracle case, where not only is one oracle insufficient for some functions [AFK89] but, as we shall see, one oracle is insufficient for most functions.

We consider two models for $m$-oracle instance-hiding schemes:

- Oracles $B_1$ through $B_m$ may collude before the start of the protocol, but they are kept physically separate during the protocol. In this model, $m = |x|$ oracles suffice for any function $f$.

- Oracles $B_1$ through $B_m$ may not collude at all, either before or during the protocol. In this model, $m = 2$ oracles suffice for any function $f$. Conversely, for most boolean functions $f$, two oracles are necessary.

In the first model, our proof of sufficiency demonstrates an unintuitive connection between instance-hiding schemes and secure multiparty protocols [BGW88, CCD88]. The connection is unintuitive for many reasons, the most obvious of which is that, in the first problem, we require explicitly that the oracles not communicate at all and, in the second problem, we require that they communicate extensively. More fundamentally, the two problems seem at first to exemplify two incompatible views of distributed computations with secret data. The instance-hiding problem first defined in [AFK89] and generalized in Section 11.1 below formalizes the following view. A weak processor A requires interaction with powerful processors $B_1$ through $B_m$, because A does *not* have enough computational resources to compute $f$; A does not want to reveal more than necessary about its private input $x$ because the $B_i$'s are public resources. Secure multiparty protocols address an alternative view: mutually untrusting, *equally powerful* processors $B_1$ through $B_m$ must interact in a computation because each of them has a private input $x_i$ *without which* the computation cannot proceed.

## 11.1 Preliminaries

Following Abadi, Feigenbaum, and Kilian, we take A to be a probabilistic polynomial-time Turing Machine transducer [AFK89]. Let $f$ be a boolean function on $\{0,1\}$ for which no probabilistic polynomial-time algorithm is known. In order to compute $f(x)$, A consults players $B_1, \ldots, B_m$, where $m$ is (necessarily) bounded by a polynomial in $|x|$. Each $B_i$ is an interactive oracle Turing machine that can use an unbounded amount of time and space. It is convenient to think of the oracle tape of $B_i$ as a random variable $O_i$. Player $B_i$ is completely specified by its finite control and the value of $O_i$.

An *m-oracle instance-hiding scheme* for $f$ is a $R$-round, synchronous protocol executed by players $A$, $B_1$, ..., $B_m$. Player $A$ draws input $x$ according to a distribution $X$. The round-complexity $R$ is bounded by a polynomial in $|x|$. Let $\texttt{view}_A^{1\cdot r}$ denote the sequence of messages sent and received by $A$ in rounds 1 through $r$, let $\delta_A$ denote its finite control, and let $R_A$ denote its random tape.

At round $r + 1$ of the protocol, $A$ performs a probabilistic polynomial-time computation producing $m$ messages $y_{r,1}, \ldots, y_{r,m}$ for $B_1, \ldots, B_m$, based on $x$, $\texttt{view}_A^{1\cdot r}$, $\delta_A$, and $R_A$. Each $B_i$ computes a response $z_{r+1,i}$ based on $y_{1,i}, \ldots, y_{r+1,i}$ and sends it to player $A$.

In the simple case, which corresponds to the motivating idea of a collection of public servers supplying answers to queries, this response is a function $g$ applied to $y_{r+1,i}$; player $B_i$ can perform an unbounded amount of local computation, and replies with $z_{r+1,i} = g(y_{r+1,i})$. After round $R$ of the protocol, $A$ computes $f(x)$ based on $x$, $\texttt{view}_A^{1\cdot r}$, $\delta_A$, and $R_A$. Note that we do not allow the output of $A$ to be incorrect.

Let $Y_i$ be the induced distribution on the sequence $<y_{1,i}, \ldots, y_{R,i}>$ of messages $A$ sends to player $B_i$.

We wish to make precise the statements that an instance-hiding scheme "leaks at most" some function $L(x)$ to player $B_i$ or that it "hides at least" some function $H(x)$ from player $B_i$. Note that $L(X)$ and $H(X)$ are also induced random variables.

We consider two generalizations of the definitions in [AFK89]. In both models, all players "know" the plaintext distribution $X$ and the contents of the finite controls $\delta_A, \delta_{B_1}, \ldots, \delta_{B_m}$. Furthermore, in both models, each player $B_i$ "does not know" the content of the random tape $R_A$, and, for $i \neq j$, player $B_i$ "does not see" the messages $y_{r,j}$ that A sends to $B_j$ or the responses that $B_j$ sends back. The difference between the two models lies in whether or not $B_i$ "knows" the content of oracle tape $O_j$ for $j \neq i$.

**Model 1:**

- An instance-hiding scheme *leaks* at most $L$ to oracle $B_i$ if, for all plaintext distributions $X$, for all $u \in Range(L)$, the random variables $X$ and $<Y_i, O_1, \ldots, O_m>$ are independent given $L(X) = u$.

- An instance-hiding scheme *hides* at least $H$ from oracle $B_i$ if, for all plaintext distributions $X$, the random variables $H(X)$ and $<Y_i, O_1, \ldots, O_m>$ are independent.

Intuitively, a player "knows" nothing about the outcome of a random variable $X$ if the observations of random variables to which it has access are independent of $X$. Thus, in Model 1, we allow $B_i$ to "know" the contents of every oracle tape by virtue of measuring its knowledge with respect not only to the query sequence $Y_i$ and oracle tape $O_i$ but to the entire set of oracle tapes.

Informally, this corresponds to the case in which the powerful players can collude before, but not during, the execution of the protocol. In Model 2, however, we allow no collusion at all, before or during the execution of the protocol:

**Model 2:**

- An instance-hiding scheme *leaks* at most $L$ to oracle $B_i$ if, for all plaintext distributions $X$, for all $u \in Range(L)$, the random variables $X$ and $<Y_i, O_i>$ are independent given $L(X) = u$.

- An instance-hiding scheme *hides* at least $H$ from oracle $B_i$ if, for all plaintext distributions $X$, the random variables $H(X)$ and $<Y_i, O_i>$ are independent.

Specifically, in Model 2, player $B_i$ has access only to $Y_i$ and $O_i$; it is possible that the additional knowledge of other oracle tapes might lead to a dependence among random variables, revealing some additional information.

In either model, we say that the scheme "leaks $L$" if it leaks at most $L$ to each $B_i$, and we say that it "hides $H$" if it hides at least $H$ from each oracle $B_i$. Throughout this paper, we are primarily concerned with schemes that leak $|x|$. In either model, we say that "$f$ has an $m$-oracle instance-hiding scheme" or that "$m$ oracles suffice for $f$" to mean that $f$ has an $m$-oracle instance-hiding scheme that leaks $|x|$. As in [AFK89], we define leaking and hiding in terms of independence of random variables. Complexity-based cryptography is irrelevant, because A is time-bounded and the $B_i$'s are time-unbounded.

**Observation 11.2** *Models 1 and 2 are equivalent if $m = 1$.*

We end this section with the proposition that if the querier A is limited in space, but not in time or in number of rounds of interaction with the $B_i$'s, then the instance-hiding problem is trivial.

**Theorem 11.3** *Every function has a 1-oracle instance-hiding scheme that leaks at most $|x|$ in which the querier A is limited to (deterministic) constant space.*

The oracle simply sends every instance of size $|x|$ along with its answer, and the querier checks the instance against $x$, copying the answer to the output when the instance matches $x$. We refer the interested reader to [Con88, Cle89, DS88, Kil88c], for example, for a discussion of the related topic of (zero-knowledge) interactive proof systems with space-bounded verifiers.

## 11.2 Model 1: $\mid x \mid$ Oracles Suffice

Our main result on model-1 instance-hiding schemes uses Shamir's method for secret sharing. For notational convenience, we shall refer to a value $p(\alpha_i)$ of a polynomial of degree $t$ with free term $s$ as a *$t$-point* of $s$. The difference between a *$t$-point* and a *piece* of $s$ is that the latter indicates that the polynomial has been selected uniformly at random (subject to the constraints on degree and free term), whereas the former concerns an arbitrary polynomial. The randomness property will be essential only for the initial queries of $A$.

The following lemma is an easy consequence of Lemma 5.21.

**Lemma 11.4** *Any boolean function $f(x)$ on inputs $x$ of length $n$ can be represented as a polynomial $c_f(x_1, \ldots, x_n)$ over an arbitrary field $E$, such that when the values $x_1, \ldots, x_n$ match the bits of $x$, $f(x) = c_f(x_1, \ldots, x_n)$.*

With a few straightforward observations, we shall be able to construct efficient instance-hiding schemes for arbitrary boolean functions. First, note that if elements $\gamma_1, \ldots, \gamma_k$ of $E$ are $t$-points of $s_1, \ldots, s_k$, respectively, then $\gamma_1 + \cdots + \gamma_k$ is a $t$-point of $s_1 + \cdots + s_k$. In fact, any this holds for any linear combination: if $\beta_1, \ldots, \beta_k$ are fixed constants in $E$, then $\beta_1 \gamma_1 + \cdots + \beta_k \gamma_k$ is a $t$-point of $\beta_1 s_1 + \cdots \beta_k s_k$. Furthermore, $\gamma_1 \times \cdots \times \gamma_k$ is a $kt$-point of $s_1 \times \cdots \times s_k$. (By comparison, recall that in the MULTIPLY protocol (Chapter 5), the product of two $t$-shares is a $2t$-point, which later is converted to a $t$-share.)

Specifically, if $c_f(x_1, \ldots, x_n)$ has degree $n$, and $\gamma_1, \ldots, \gamma_k$ are $t$-points of $x_1, \ldots, x_n$, then $c_f(\gamma_1, \ldots, \gamma_n)$ is a $nt$-point of $c_f(x_1, \ldots, x_n)$. A total of $(nt + 1)$ $nt$-points suffice to determine $c_f(x_1, \ldots, x_n)$.

**Theorem 11.5** *Let $f(x)$ be any function whose output $|f(x)|$ is polynomially bounded. Then for any positive constant $c$, $f(x)$ has a model-1 $(|x| - c \log |x|)$-oracle instance-hiding scheme that leaks at most $|x|$.*

**Proof:** Assume WLOG that $f$ is boolean. The result for general functions then follows by regarding each output bit as a boolean function of the input. We first show that $|x| + 1$ oracles suffice and then improve upon the construction.

For concreteness, let $E$ be $\mathbf{Z}_p$ for the smallest prime exceeding $n + 2$. By lemma 5.21, there is a polynomial $c_f(x_1, \ldots, x_n)$ over $E$ which is equal to $f(x)$ at 0/1-valued inputs. Let each $B_i$ have access to an oracle for $c_f$.

The protocol begins with the querier secretly sharing $x_1, \ldots, x_n$ among $n+1$ oracles using $t = 1$ as the bound on coalition sizes. Each oracle evaluates $c_f$ on the collection of 1-points it receives, giving a $n$-point of the final value. Finally, $A$ interpolates the $(n + 1)$ $n$-points to determine the $n^{th}$-degree polynomial $c_f$. Figure 11.1 lists the steps exactly. The `interpolate` function outputs the coefficients of the minimal-degree polynomial running through its arguments.

Intuitively, basing the protocol on ideas from secret-sharing ensures that each $B_i$ learns nothing about $x$. Even though $B_i$ can use its unbounded resources to attempt to recover $x$, it cannot communicate with any of the other $B_j$, so the only physically possible coalitions are trivial ones of size 1. Because player $B_i$ sees only the 1-*shares* of $x_1, \ldots, x_n$ and does not see the later collection of $n$-points, he receives no information about $x$.

To prove that each $B_i$ learns nothing but $n$, we use a simple lemma, akin to the statement that secret sharing is 1-private:

---

| | |
|---|---|
| $A$ : | $n \leftarrow |x|$, where $x = x_1 x_2 \cdots x_n$. |
| $A$ : | select $p_1(u), \ldots, p_n(u)$ randomly of degree 1 with $p_i(0) = x_i$. |
| $A$ : | $\textsc{piece}_i(x_j) \leftarrow p_i(j)$.                    (for $1 \le i \le n+1$.) |
| $A$ : | $y_i \leftarrow {<}n, \textsc{piece}_i(x_1), \ldots, \textsc{piece}_i(x_n){>}$    (for $1 \le i \le n+1$.) |
| $A \rightarrow B_i$ : | $y_i$                                          (for $1 \le i \le n+1$.) |
| $B_i$ : | $z_i \leftarrow c_f(\textsc{piece}_i(x_1), \ldots, \textsc{piece}_i(x_n))$    (for $1 \le i \le n+1$.) |
| $B_i \rightarrow A$ : | $z_i$                                          (for $1 \le i \le n+1$.) |
| $A$ : | $p(u) \leftarrow \texttt{interpolate}(z_1, z_2, \ldots, z_{n+1})$ |
| $A$ : | $f(x) \leftarrow p(0)$. |

Figure 11.1: Instance-hiding scheme for $f(x)$ with $n+1$ oracles. Each $B_i$ has an oracle $O_i$ for $c_f$.

**Lemma 11.6** *For any $x_1, \ldots, x_n \in E$ and for any $i \ne 0$, the following distribution is the same as* $\texttt{uniform}(E^n)$ :

$$\{(a_1, \ldots, a_n) \leftarrow \texttt{uniform}(E^n) : (a_1 i + x_1, \ldots, a_n i + x_n)\}$$

**Proof:** Clearly, for fixed $x_1, \ldots, x_n$ and $i \ne 0$, the mapping $(a_1, \ldots, a_n) \mapsto (a_1 i + x_1, \ldots, a_n i + x_n)$ is a bijection, so if $(a_1, \ldots, a_n)$ is uniform over $E^n$, so is $(a_1 i + x_1, \ldots, a_n i + x_n)$. ∎

Thus, the distribution on messages $y_i$ seen by $B_i$ is the same for any $x$ : uniform over $\{n\} \times E^n$. Therefore the random variables $X$ and ${<}Y_i, O_i{>}$ are independent. In fact, since each $B_j$ ($j \ne i$) computes the same polynomial $c_f(x_1, \ldots, x_n)$, each oracle tape is the same, and hence the random variables $X$ and ${<}Y_i, O_1, \ldots, O_{n+1}{>}$ are independent.

To reduce the number of oracles queried in the protocol from $n+1$ to $n$, we can "instantiate" the input bit $x_1$ and treat $f$ as though it were a function of $n-1$ inputs. The querier $A$ first executes the basic protocol on input $0x_2 \cdots x_n$, then executes it on input $1x_2 \cdots x_n$, and uses the result that corresponds to the original input $x_1 x_2 \cdots x_n$. More generally, by instantiating $O(\log n)$ input bits (thus generating $n^{O(1)}$ queries) the number of $B_i$'s queried can be reduced to $n - O(\log n)$. ∎

## 11.3   Model 2: Two Oracles Suffice

In this section, it is convenient to regard a boolean function $f$ as the characteristic function $\chi_S$ of a set $S \subseteq \{0,1\}^*$ or as a language $L_f = \{x \mid f(x) = 1\}$. By *a random set* $S$, we mean one in which $\chi_S(x)$ is the outcome of a fair coin toss, for each $x \in \{0,1\}^*$. The expression $S_1 \triangle S_2$ denotes the symmetric difference of sets $S_1$ and $S_2$; that is,

$$\chi_{S_1 \triangle S_2}(x) \equiv \chi_{S_1}(x) \oplus \chi_{S_2}(x).$$

If $s_1$ and $s_2$ are both $n$-bit strings, then $s_1 \oplus s_2$ is the $n$-bit string whose $i^{\text{th}}$ bit is the exclusive-or of the $i^{th}$ bits of $s_1$ and $s_2$.

We also use the following nonstandard terminology and notation. A *singleton sequence* is a subset of $\{0,1\}^*$ that contains exactly one string of each length. A *random singleton sequence* is one in which the length-$n$ string is chosen u.a.r. from $\{0,1\}^n$. If $S$ is an arbitrary set and $V = \{v_1, v_2, \ldots\}$ is an arbitrary singleton sequence, then $S \circ V$ denotes the set with characteristic function

$$\chi_{S \circ V}(x) \equiv \chi_S(x \oplus v_{|x|}).$$

Note that each $v_n$ in $V$ effects a permutation of the bits in the characteristic vector of $S \cap \{0,1\}^n$.

**Theorem 11.7** *Every function[1] has a model-2 2-oracle instance-hiding scheme that leaks at most $|x|$. Conversely, two oracles are necessary for most boolean functions.*

**Proof:** Necessity follows from Observation 11.2 and Lemma 11.8, which is proved in the next section. We show sufficiency by demonstrating the existence of a model-2 2-oracle instance-hiding scheme for an arbitrary function $\chi_S$ that leaks at most $|x|$. As in the proof of Theorem 11.5, we assume WLOG that $f$ is boolean.

Let us illustrate the basic idea of the proof through the following simpler argument. Assume first that the conditional plaintext distribution $P(X \mid |X|)$ is uniform. Under this simplifying assumption, we can see that the characteristic function of a random set $S$ has a model-2 2-oracle instance-hiding scheme that, with probability 1, leaks at most $|x|$ to $B_1$ and at most $\chi_S(x)$ to $B_2$: Let $B_1$ have an oracle for a random singleton sequence $V = \{v_n\}_{n=1}^{\infty}$, and let $B_2$ have an oracle for $S \circ V$. On cleartext input $x$, A first sends $|x|$ to $B_1$, who sends back $v_{|x|}$. A then sends $x \oplus v_{|x|}$ to $B_2$, who sends back $\chi_{S \circ V}(x \oplus v_{|x|}) = \chi_S(x)$. Clearly, $B_1$ learns only $|x|$. Intuitively, $B_2$ learns only $\chi_S(x)$ because, for random $S$ and $V$, $S \circ V$ is also random, and the encrypted input $x \oplus v_{|x|}$ is a uniformly distributed $n$-bit string. To make this idea work for the theorem as stated, we must make an arbitrary $S$ "look random," and we must show how to avoid leaking $\chi_S(x)$ to $B_2$. We accomplish this by using the symmetric difference of $S$ with a random set and by "splitting" the singleton sequence $V$ into two halves, one of which is given to each of $B_1$ and $B_2$.

Suppose that $\chi_S$ is the (arbitrary) boolean function for which we seek an instance-hiding scheme. Let $R$ be a random set, $V = \{v_n\}_{n=1}^{\infty}$ and $U = \{b_{1,n}\}_{n=1}^{\infty}$ be random singleton sequences, and let $T = \{b_{2,n}\}_{n=1}^{\infty}$ be such that $b_{2,n} = v_n \oplus b_{1,n}$. Let the oracle $O_1$ for $B_1$ encode both $R \circ V$ and $U$ in a standard way, and let the oracle $O_2$ for $B_2$ encode both $(S \triangle R) \circ V$ and $T$. The instance-hiding scheme is described in Figure 11.2.

The querier obtains the result:

$$\chi_{R \circ V}(y) \oplus \chi_{(S \triangle R) \circ V}(y) \;=\; \chi_R(x \oplus b_{1,n} \oplus b_{2,n} \oplus v_n) \oplus \chi_{S \triangle R}(x \oplus b_{1,n} \oplus b_{2,n} \oplus v_n)$$

---

[1] As before, we assume the size of the output, $|f(x)|$, is polynomially bounded.

| A: | $n \leftarrow |x|$. |
|---|---|
| A $\to B_1, B_2$: | $n$. |
| $B_1 \to$ A: | $b_{1,n}$. |
| $B_2 \to$ A: | $b_{2,n}$. |
| A: | $y \leftarrow x \oplus b_{1,n} \oplus b_{2,n}$. |
| A $\to B_1, B_2$: | $y$. |
| $B_1 \to$ A: | $\chi_{R \circ V}(y)$. |
| $B_2 \to$ A: | $\chi_{(S \triangle R) \circ V}(y)$. |
| A: | $\chi_S(x) \leftarrow \chi_{R \circ V}(y) \oplus \chi_{(S \triangle R) \circ V}(y)$. |

Figure 11.2: Instance-hiding scheme for $f(x)$ with two oracles. Oracle $O_1$ encodes $(R \circ V, U)$; $O_2$ encodes $((S \triangle R) \circ V, T)$.

$$= \chi_R(x) \oplus \chi_{S \triangle R}(x)$$
$$= \chi_S(x).$$

The messages seen by $B_1$ are $y_{1,1} = n$ and $y_{2,1} = x \oplus b_{1,n} \oplus b_{2,n}$. First let us show that for every $x$ of size $n$, the distribution on $((y_{1,1}, y_{2,1}), (R \circ V, U))$ is the same.

The distribution on $(R, V, U)$ given $n$ is uniform over the $2^{2^n} 2^n 2^n$ choices for $(R, V, U)$. For any particular $x$ of size $n$, the protocol induces a bijection with $(y_{2,1}, R \circ V, U)$. Thus, for any particular $x$ of size $n$, the distribution on $(y_{2,1}, R \circ V, U)$ is uniform over all possibilities. Since $y_{1,1}$ is always $n$, the distribution on $((y_{1,1}, y_{2,1}), (R \circ V, U))$ is uniform over all possible values, for each $x$ of size $n$. Hence the random variable $<Y_1, O_1>$ is independent of $x$, given $n$.

Now consider $B_2$. Any fixed $S$ induces a bijection on the set of possible $R$. Hence there is a bijection between $(R, V, U)$ and $(S \triangle R, V, U)$. By definition, there is a bijection between the latter set and $(S \triangle R, V, T)$. Using the argument of the previous paragraph, there is a bijection between $(S \triangle R, V, T)$ and $(n, y_{2,2}, (S \triangle R) \circ V, T)$. Hence the distribution on $((y_{2,1}, y_{2,2}), ((S \triangle R) \circ V, T))$ is uniform, for any $x$ of size $n$. Thus $<Y_2, O_2>$ is independent of $x$, given $n$. ∎

## 11.4 Other Results

### 11.4.1 Unconditional Negative Results

The main theorem of [AFK89] is that NP-hard functions have no 1-oracle instance-hiding schemes that leak $|x|$, unless the polynomial hierarchy collapses at the third level. This is a conditional negative result. No unconditional negative results about instance-hiding schemes that leak $|x|$ are provided in [AFK89]. We give one here.

**Lemma 11.8** *Measure one of boolean functions do not have 1-oracle instance-hiding schemes that leak at most $|x|$.*

**Proof:** Let $\vec{c}_n(f)$ denote the characteristic vector of $f$ for inputs of $n$ bits (e.g., $\vec{c}_2(f)$ is the concatenation of $f(00)$, $f(01)$, $f(10)$, and $f(11)$). Let $C(f)$ denote the set of characteristic vectors for $f$, $\{\vec{c}_n(f) \mid n \in \mathbf{N}\}$. We refer to a set containing exactly one string of length $2^n$ for each $n$ as a characteristic set.

If $L_f$, the set of $x$ such that $f(x) = 1$, is a language in NP/*poly*, then it is not hard to see that for some constant $d$, the Kolmogorov complexity of $\vec{c}_n(f)$ is at most $n^d$ for all $n$. We shall show that for any $d$, the class of characteristic sets consisting of strings with Kolmogorov complexity $n^d$ is contained in a class having measure zero, using a natural measure.

The measure we use is the natural one defined on the class of boolean functions: any boolean function $f$ is equivalent to a real number $R_f$ between 0 and 1, where the $i^{th}$ bit of the number is $f(i)$. The measure of a set of boolean functions is the Lebesgue measure on the corresponding set of reals. Since the representations of $f$ as a boolean function, real number, language, or characteristic set are equivalent, we loosely apply "measure" to classes of all of them, without loss of generality.

For any $n$, a straightforward counting argument shows that at least $2^{2^n}/2$ strings of length $2^n$ have Kolmogorov complexity at least $2^n - 1$. Let $K_n$ be the set of the lexicographically first $2^{2^n}/2$ of these strings, and let $K = \cup_n K_n$. If $\mathcal{C}$ is the class of characteristic sets, then let $J = \{C \in \mathcal{C} \mid (\forall n)|K_n \cap C| < \infty\}$; each characteristic set in $J$ contains at most a finite number of vectors appearing in $K$. The class $J$ is composed of a countable number of subclasses ($J_l^m$, the class of sets having $l$ vectors in common with $K$ but each of length at most $2^{2^m}$), each similar to the Cantor set and having measure 0 by the standard argument. The measure of $J$ is thus 0.

Let us argue that the characteristic set for any language in NP/*poly* is a member of $J$. Let $L_f \in$ NP/*poly*, and let $L_f$ be recognized by a nondeterministic Turing machine with a program of length $l$ and with advice of length $n^e$. For some $d$, $l + n^e \leq n^d$, and the Kolmogorov complexity of any characteristic vector $\vec{c}_n(f)$ is therefore at most $n^d$. Now, for some constant $n_d$, $n \geq n_d \Rightarrow n^d < 2^n - 1$, and therefore $n \geq n_d \Rightarrow \vec{c}_n(f) \notin K_n$. Hence a finite number of characteristic vectors are in $K$, and $C(f) \in J$.

The class of characteristic sets for languages in NP/*poly* is thus contained in $J$ and has measure 0. For every function $f$ having a 1-oracle instance-hiding scheme, $L_f \in$ NP/*poly*, so the measure of functions having 1-oracle instance-hiding schemes is 0. ∎

## 11.4.2 Random-Self-Reducing Circuits

Intuitively, a random-self-reduction of a set $S$ is an expected-polynomial-time, randomized algorithm that maps $S$ to $S$, maps $\overline{S}$ to $\overline{S}$, and, on each input $x$ in $S$, outputs, with equal probability, each $y$ in $S \cap \{0,1\}^{|x|}$. Abadi, Feigenbaum, and Kilian defined random-self-reducing algorithms rigorously and related them to 1-oracle instance-hiding schemes [AFK89]. In this section, we consider random-self-reducing *circuits* in the same context.

A *randomized* PSIZE *circuit family* is a set $\{C_n\}_{n=1}^{\infty}$ of circuits in which the number of gates in $C_n$ is at most $p(n)$ for some polynomial $p$, and $C_n$ has $n$ inputs $x_1, \ldots, x_n$, random inputs $r_1, \ldots, r_{q(n)}$, and outputs $y_1, \ldots, y_{m(n)}$, for polynomials $q(n)$ and $m(n)$. We say that $\{C_n\}$ *generates* a collection of distributions $\{D_n(x)\}_{n \in \mathbf{N}, x \in \{0,1\}^*}$ if for all $x$, $C_{|x|}$ outputs a sample point from $D_n(x)$ with probability at least $1/2$, and otherwise outputs a distinguished string $\Lambda$.

**Definition 11.9** *A set $S$ has random-self-reducing circuits if there is a randomized* PSIZE *circuit family $\{C_n\}$ that generates the distribution $D_n(x)$ defined as the uniform distribution on $S \cap \{0,1\}^n$ for $x \in S$, or an arbitrary distribution on $\overline{S} \cap \{0,1\}^n$ for $x \in \overline{S}$. The family $\{C_n\}$ is called a* random-self-reducing circuit family for $S$.

**Definition 11.10** *A set $S$ has two-sided random-self-reducing circuits if there is a randomized* PSIZE *circuit family $\{C_n\}$ that generates the distribution $D_n(x)$ defined as the uniform distribution on $S \cap \{0,1\}^n$ for $x \in S$, or the uniform distribution on $\overline{S} \cap \{0,1\}^n$ for $x \in \overline{S}$. The family $\{C_n\}$ is called a* two-sided random-self-reducing circuit family for $S$.

**Lemma 11.11** *If $S$ has two-sided random-self-reducing circuits, then $\chi_S$ has a 1-oracle instance-hiding scheme that leaks at most $|x|$ and $\chi_S(x)$.*

Lemma 11.11 provides a way of showing that certain boolean functions have model-1 instance-hiding schemes that use many fewer oracles than the schemes given by Theorem 11.5. Unfortunately, as the following theorem indicates, neither SAT nor $\overline{\text{SAT}}$ is likely to have random-self-reducing circuits. (It is even less likely that SAT has two-sided random-self-reducing circuits, as it would need for Lemma 11.11 to apply.)

**Theorem 11.12**

  a. *If $\overline{\text{SAT}}$ has random-self-reducing circuits, then the polynomial hierarchy collapses at the third level.*

  b. *If SAT has random-self-reducing circuits, then the polynomial hierarchy collapses at the third level.*

**Proof:** By a proof of Nisan [Nis88] that, if SAT had a random-self-reducing algorithm, then $\overline{\text{SAT}}$ would be in IP[2]. Then, we use the facts that IP[2] $\subseteq$ AM[4] $\subseteq$ AM[2] $\subseteq$ NP/$poly$ (see [BM88, GMR89, GS86]) and that $\overline{\text{SAT}} \subseteq$ NP/$poly \Rightarrow$ PH $\subseteq \Sigma_3^p$ (see [Yap83]). These results extend *mutatis mutandis* to the case of random-self-reducing circuits for SAT and proof-systems with circuit-verifiers. ∎

## 11.4.3  Multiple Queries

The instance-hiding schemes we give are easily extended to allow the querier to ask an unlimited number of questions. The scheme for Model 1 given in the proof of Theorem 11.5 can be repeated directly. The scheme for Model 2 can be used only once, but by generating an exponential number of (non-reusable) schemes for each $n$ and encoding them in a direct way into the two oracles, the scheme can be modified to support many queries.

### 11.4.4 Nontrivial One-Oracle Schemes

Abadi, Feigenbaum, and Kilian show that certain well-known number-theoretic functions, such as discrete logarithm and quadratic residuosity, have one-oracle instance-hiding schemes that hide some significant information about the input [AFK89]. However, those schemes *do* leak more than the size of the input $x$, and [AFK89] provides no nontrivial examples of functions with instance-hiding schemes that leak at most $|x|$.[2] Let us present a nontrivial example of a presumably intractable function with a one-oracle instance-hiding scheme.

**Definition 11.13** *The* discrete logarithm *to the base $g$ modulo $p$ of a number $x$, $DLOG_{g,p}(x)$, is the exponent $e$ such that $g^e \equiv x \bmod p$. Here, $p$ is prime, and $g$ is a generator of the nonzero integers mod $p$. No efficient method for computing DLOG is known.*

**Example 11.14** The instance-hiding scheme for DLOG given in [AFK89] is as follows. On input $g\#p\#x$, choose $r \in \{1, \ldots, p-1\}$ uniformly at random, and let $y = xg^r \bmod p$ (if $x = 0$, pretend $x = 1$ and note that the answer is $\Lambda$.). Clearly, $y$ is distributed uniformly over $\mathbf{Z}_p^*$ for any $g\#p\#x$ of size $n$, given $g$ and $p$. On receiving the answer $z = \mathrm{DLOG}_{g,p}(y)$, return $z - r$. The reduction is correct:

$$\mathrm{DLOG}_{g,p}(y) - r = \mathrm{DLOG}_{g,p}(xg^r) - r = \mathrm{DLOG}_{g,p}(x) + r - r = \mathrm{DLOG}_{g,p}(x).$$

Thus this scheme leaks $|g\#p\#x|$, $g$, and $p$. We now define a function that has a scheme which leaks only $|g\#p\#x|$.

Let $p_n$ denote the smallest prime exceeding $2^n$, and $g_n$ the smallest generator of $\mathbf{Z}_{p_n}$. Define the function $f(x) = \mathrm{DLOG}_{g_{|x|}, p_{|x|}}(x)$. (If $x \equiv 0 \bmod p_{|x|}$, then say $f(x) = \Lambda$.) As in the general case of the discrete logarithm, no efficient method for computing $f(x)$ is known.

On the other hand, given $g_n$ and $p_n$, the reduction described above is easy to perform. Thus, an instance-hiding scheme for $f(x)$ can be constructed as follows. First, the querier sends $n = |x|$, and receives $g_n$ and $p_n$ in return. Then the querier performs the random-self-reduction described above, and obtains the result as described above. The scheme hides everything but $|x|$.

---

[2]Examples of trivial instance-hiding schemes that leak at most $|x|$ are the obvious schemes for $\chi_S$, where $S$ is a set in P/*poly*; the oracle simply sends the polynomial advice string, and the querier computes on his own.

# Chapter 12

# Locally Random Reductions

> However, this bottle was *not* marked "poison," so Alice ventured to taste it, and finding it very nice (it had, in fact, a sort of mixed flavour, of cherry-tart, custard, pine-apple, roast turkey, toffee, and hot buttered toast), she very soon finished it off.

> "What a curious feeling!" said Alice. "I must be shutting up like a telescope."

> And so it was indeed: she was now only ten inches high, and her face brightened up at the thought that she was now the right size for going through the little door into that lovely garden.

> — Lewis Carroll, *Alice's Adventures in Wonderland*

Locally random reductions are a new class of reductions from one computational problem to another. Inspired by the problem of *instance-hiding schemes* (presented in Chapter 11), locally random reductions have found a broad variety of applications since their introduction by Beaver and Feigenbaum. These results include methods for program testing [Lip89, BLR89a] and the surprising result that IP = PSPACE [Sha89]. We describe some of these important subsequent developments in §12.5, and show how LRR's drastically reduce the communication complexity of zero-knowledge proof systems and secure multiparty protocols in Chapters 13 and 14.

Reductions from one language to another are fundamental to complexity theory. One can reduce the problem "is $x \in L_1$?" to solving the problem "is $y \in L_2$?" if there is an efficiently computable function $P$ such that $x \in L_1 \Leftrightarrow P(x) \in L_2$. A reduction from a *function* $f(x)$ to another function $g(y)$ similarly maps an instance $x$ in the domain of $f$ to an instance $y = P(x)$ in the domain of $g$. It also requires that the value of $g(y)$ be *interpolated* to obtain the answer: $f(x) = Q(g(y))$.

In general, a reduction from $f$ to $g$ can produce several instances $y_1, \ldots, y_m$ at which to evaluate $g$, and need not generate these instances deterministically. Let $X$ be the domain of $f$ and $Y$ be the domain of $g$.

**Definition 12.1** *A random reduction from $f(x)$ to $g(y)$ is a pair of probabilistic, polynomial time algorithms $(P, Q)$ satisfying the following properties:*

**(Reduction)** $P : X \to \mathtt{dist}(Y^m \times \Sigma^\star)$, *that is, the querying algorithm $P$ produces a sample $(y_1, \ldots, y_m, \sigma)$ according to distribution $P(x)$, where each $y_i$ is in the domain of $g$, and $\sigma$ is a string.*

**(Interpolation)** *If $(y_1, \ldots, y_m, \sigma)$ has nonzero weight in the distribution $P(x)$, then the interpolation algorithm gives $Q(g(y_1), \ldots, g(y_m), \sigma) = f(x)$.*

The term *locally random* refers to the distributions on subsets of the queries $y_1, \ldots, y_m$ produced by $P(x)$. In a certain sense, it generalizes the idea of secret-sharing, in that secret-sharing ensures a uniform distribution on sufficiently small subsets of pieces, regardless of the secret. Here, we should like that the distributions on subsets of the pieces must be the same for any $x$ of a given size $n$, even though it need not be *uniform,* as it is in the case of secret sharing.

**Definition 12.2** *Let $B \subseteq \{1, \ldots, n\}$. Then $D_x^B$ denotes the distribution on $\{y_i \mid i \in B\}$ induced by $P(x)$.*

**Definition 12.3** *A $(k(n), m(n))$-locally random reduction (LRR) from $f(x)$ to $g(y)$ is a random reduction $(P, Q)$ satisfying:*

- *$P(x)$ produces $m(n)$ queries on inputs of size $n$;*

- *For all $n$, for all subsets $B \subseteq \{1, \ldots, n\}$ of size $\leq k(n)$, the distribution $D_x^B$ is the same for all $x$ of size $n$.*

**Example 12.4** The *discrete logarithm* function, $\mathrm{DLOG}_{g,p}(x)$, was defined in Definition 11.13. No efficient algorithm for computing DLOG is known. The random-self-reduction given for *DLOG* is in fact a $(1, 1)$-locally random reduction: it produces one query $y = xg^r$, and the distribution on query sets of size $\leq 1$ is the same for every $x$.

**Example 12.5** For illustration, let us demonstrate a $(10, 21)$-locally random reduction from the multiplication function to itself. Let $f(x_1, x_2) = x_1 x_2$, over $\mathbf{Z}_{p_n}$ for concreteness. Let $g(x_1, x_2) = x_1 x_2$. We reduce $f$ to $g$ via $(P, Q)$ as follows.

The querying algorithm $P(x_1, x_2)$ chooses $a_1, \ldots, a_{10}, b_1, \ldots, b_{10} \in \mathbf{Z}_{p_n}$ uniformly at random, and sets $p_1(u) = a_{10}u^{10} + \cdots + a_1 u + x_1$ and $p_2(u) = b_{10}u^{10} + \cdots + b_1 u + x_2$. It computes $y_i = (p_1(i), p_2(i))$ for each $1 \leq i \leq 21$.

The interpolation algorithm $Q(z_1, \ldots, z_{21}, \sigma)$ does just that: it interpolates the polynomial $q(u)$ of maximal degree 20 passing through the points $(i, z_i)$ for $1 \leq i \leq 21$. It returns the value $q(0)$.

To see that the reduction is correct, observe that $g(p_1(i), p_2(i)) = p_1(i)p_2(i) = q(i)$ for $1 \leq i \leq 21$, and both $g(p_1(u), p_2(u))$ and $q(u)$ are polynomials of maximal degree 20. By elementary algebra, $g(p_1(u), p_2(u)) = q(u)$, hence $g(p_1(0), p_2(0)) = q(0)$.

To see that the reduction is (10,21)-locally random, note that because the coefficients of $p_1(u)$ and $p_2(u)$ are chosen uniformly at random, the distribution on any 10-set of values of $p_1(u)$ or $p_2(u)$ is uniform over $(\mathbf{Z}_{p_n})^{10}$ regardless of $x_1$ and $x_2$. Thus for any subset $B$ of 10 or fewer queries, $D_{x_1,x_2}^B$ is the same for any $(x_1, x_2)$.

The main results of this chapter are twofold: the first shows the existence of locally random reductions for polynomials $f(x_1, \ldots, x_m)$ and the second for arbitrary boolean functions. These reductions inspire a variety of applications from program testing to interactive proofs to secure distributed protocols.

**Theorem 12.6**

a. *For any polynomial $f(x_1, \ldots, x_n)$ of degree at most $n$ over some field $E$, there is a $(1, n+1)$ locally random reduction from $f$ to itself.*

b. *For any boolean function $f(x)$, there exists a $(1, n+1)$ locally random reduction from $f(x)$ to some function $g(y)$. In particular, for any field of size exceeding $n + 1$, $g(y)$ can be taken to be a polynomial $c_f(x_1, \ldots, x_m)$ of degree $n$ over that field.*

**Proof:** See Lemmas 12.8 and 12.9, proved below in §12.1 and §12.2. ∎

In fact, a stronger statement is achievable:

**Theorem 12.7**

a. *For any polynomial $f(x_1, \ldots, x_n)$ of degree at most $n$ over some field $E$, for any $d(n)$, and for any constant $c > 0$, there is a $(d, dn/(c \log n))$ locally random reduction from $f$ to a polynomial $h_f(w_1, \ldots, w_{n^{c+1}/(c \log n)})$ of degree $n/(c \log n)$ over that field.*

b. *For any boolean function $f(x)$, for any $d(n)$, and for any constant $c > 0$, there exists a $(d, dn/(c \log n))$ locally random reduction from $f(x)$ to some function $g(y)$. In particular, for any field of size exceeding $(dn/(c \log n))$, $g(y)$ can be taken to be a polynomial $h_f(w_1, \ldots, w_{n^{c+1}/(c \log n)})$ of degree $n/(c \log n)$ over that field, where each $w_i$ itself is a product of at most $O(\log n)$ variables $x_i$.*

**Proof:** See Lemmas 12.11 and 12.12. ∎

## 12.1 LRR's for Multivariate Polynomials

**Lemma 12.8** *For any polynomial $f(x_1, \ldots, x_n)$ of degree at most $n$ over some field $E$, there is a $(1, n+1)$ locally random reduction from $f$ to itself.*

---

Generate-Query$^{(1,n+1)}$

- Choose $a_1, \ldots, a_n \in E$ uniformly at random, and set $p_1(u) = a_1 u + x_1, p_2(u) = a_2 u + x_2, \ldots, p_n(u) = a_n u + x_n$.

- Set $y_i \leftarrow (p_1(i), \ldots, p_n(i))$ for $1 \le i \le n+1$.

Figure 12.1: Query function $P(x_1, \ldots, x_n)$ for self-reducing the multivariate polynomial $f(x_1, \ldots, x_n)$.

---

**Proof:** We assume without loss of generality that $|E| > n+1$ (otherwise, we may use an extension field). The query function $P(x_1, \ldots, x_n)$ is described in Figure 12.1, and the interpolation function $Q(z_1, \ldots, z_{n+1})$ is described in Figure 12.2. By Lemma 11.6, the distribution $D^i_{x_1, \ldots, x_n}$ is uniform on $E^n$ for all assignments to $x_1, \ldots, x_n$.

The interpolation function $Q$ is easy to specify: given $n+1$ values $z_1, \ldots, z_n$, interpolate the polynomial $q(u)$ of degree $n$ running through the $n+1$ points. Return $q(0)$. That this reduction is correct can be verified by simple algebra: $f(p_1(u), \ldots, p_n(u))$ and $q(u)$ are both polynomials in $u$ of degree $n$. They agree at $n+1$ points ($q(i) = f(p_1(i), \ldots, p_n(i))$, $1 \le i \le n+1$). They must therefore be identical, so $q(0) = f(p_1(0), \ldots, p_n(0)) = f(x_1, \ldots, x_n)$.

---

Interpolate$^{(1,n+1)}$

- Interpolate the polynomial $q(u)$ passing through $(i, z_i)$ for $1 \le i \le n+1$.

- Return $q(0)$.

Figure 12.2: Interpolation function $Q(z_1, \ldots, z_{n+1})$ for self-reducing the multivariate polynomial $f(x_1, \ldots, x_n)$.

---

## 12.2  LRR's for Boolean Functions

**Lemma 12.9** *For any boolean function $f(x)$, there exists a $(1, n+1)$ locally random reduction on inputs of size $n$ from $f(x)$ to some function $g(y)$. In particular, for any field of size exceeding $n + 1$, $g(y)$ can be taken to be a polynomial $c_f(x_1, \ldots, x_n)$ of degree $n$ over that field.*

**Proof:** By Lemma 11.4, every function $f$ on inputs of length $n$ can be expressed as a polynomial $c_f$ of degree $n$ over an arbitrary field $E$, such that when the variables $x_1, \ldots, x_n$ are assigned values in $\{0, 1\}$ that match $x$, $c_f(x_1, \ldots, x_n) = f(x)$.

By Lemma 12.8, then, $c_f$ is $(1, n + 1)$-locally-random reducible to itself, which implies that $f$ is $(1, n + 1)$-locally-random reducible to $c_f$.    ∎

## 12.3    Reducing the Number of Queries

By an appropriate change of variables, we can decrease the number of queries needed to reduce a polynomial to itself by a logarithmic factor (and, correspondingly, to reduce a boolean function to a polynomial).

**Lemma 12.10** *Every polynomial $f(x_1, \ldots, x_n)$ of degree $n$ is expressible as a polynomial $h_f(w_1, \ldots, w_N)$ of degree $\frac{n}{\log n}$ over new variables $w_1, \ldots, w_N$, where $N = n^2/\log\ n$, and each variable $w_i$ is a product of at most $\log n$ of the $x$-variables.*

**Proof:** Express $f$ as a weighted sum of its monomials:

$$f(x_1, \ldots, x_n) = \sum_\epsilon c_\epsilon \prod_{i=1}^n x_i^{\epsilon(i)}$$

where $\epsilon : \{1, \ldots, n\} \to \{0, 1\}$ indicates whether variable $x_i$ is in a given monomial of $f$, and $c_\epsilon$ is 0 or 1 according to whether the monomial represented by $\epsilon$ appears in $f$. The sum is taken over all possible $\epsilon$. We group the $x$'s into blocks of size $\log n$ and assign a variable $w$ to every product of a combination of $x$'s taken from a single block. For example, if $n = 16$, then we take blocks $\{x_1, x_2, x_3, x_4\}, \{x_5, x_6, x_7, x_8\}, \{x_9, x_{10}, x_{11}, x_{12}\}, \{x_{13}, x_{14}, x_{15}, x_{16}\}$, and we take $w_1 = 1, w_2 = x_1, w_3 = x_2, w_4 = x_2 x_1, w_5 = x_3, \ldots, w_{256} = x_{13} x_{14} x_{15} x_{16}$. Defining

$$\psi(b, \epsilon) = bn + \sum_{i=1}^{\log n} 2^{i-1}\epsilon(i + (b - 1)\log n),$$

for block number $b$ in the range $\{1, \ldots, \frac{n}{\log n}\}$, we have the correspondence

$$w_{\psi(b,\epsilon)} = \prod_{i=(b \log n)+1}^{(b+1)\log n} x_i^{\epsilon(i)}.$$

Then $f$ is expressed in terms of the new variables as

$$f(x_1, \ldots, x_n) = h_f(w_1, \ldots, w_N) \equiv \sum_\epsilon c_\epsilon \prod_{b=1}^{n/\log\ n} w_{\psi(b,\epsilon)}$$

It is not hard to see that $h$ has degree $\frac{n}{\log n}$ over $N = n^2/\log\ n$ variables.    ∎

**Lemma 12.11** *For any polynomial $f(x_1, \ldots, x_n)$ of degree at most $n$ over some field $E$, for any $d(n)$, and for any constant $c > 0$, there is a $(d, dn/(c \log n))$ locally random* *reduction from $f$ to a polynomial $h_f(w_1, \ldots, w_{n^{c+1}/(c\log\ n)})$ of degree $n/(c\log\ n)$ over that field. In particular, a polynomial $h_f(w_1, \ldots, w_N)$ of degree $n/(c\log\ n)$ suffices, where $N = \frac{n^{c+1}}{c\log\ n}$.*

**Proof:** The idea behind the reduction is to decrease the degree of $f$ to $n/(c\log\ n)$, and then to use random polynomials of degree $d$ to hide the variables. Assume WLOG that $c\log\ n$ divides $n$.

By an easy extension of Lemma 12.10, there exists a polynomial $h_f(w_1, \ldots, w_N)$ of degree $n/(c\log\ n)$ in $N = n^{c+1}/(c\log\ n)$ variables. Given $x_1, \ldots, x_n$, computing $w_1, \ldots, w_N$ is an easy $NC^1$ computation: simply multiply together all the $x_j$ appearing in the monomial corresponding to $w_i$, for each $w_i$.

Instead of choosing linear polynomials as in the proof of Lemma 12.8, the query algorithm GENERATE-QUERY$^{(d, dn/\log n)}$ selects uniformly random polynomials $p_1(u), \ldots, p_n(u)$ of degree $d$ subject to $p_1(0) = w_0, \ldots, p_N(0) = w_N$. It then generates $y_i = (p_1(i), \ldots, p_N(i))$ for $1 \leq i \leq dn/(c\log\ n)$.

On input $z_1, \ldots, z_{1+dn/(c\log\ n)}$, INTERPOLATE$^{(d, dn/\log n)}$ interpolates the points $(i, z_i)$ to a polynomial $q(u)$ of degree $dn/(c\log\ n)$ and returns $q(0)$.

Correctness is satisfied through straightforward algebra: $q(u)$ and $h_f(p_1(u), \ldots, p_N(u))$ have degree $dn/(c\log\ n)$ and agree at $1 + dn/(c\log\ n)$ points, so they are identical;

$$q(0) = h_f(p_1(0), \ldots, p_N(0)) = h_f(w_0, \ldots, w_N) = f(x_1, \ldots, x_n).$$

Any $d$-subset of the $y_i$'s contains $d$ values on each polynomial $p_1(u), \ldots, p_N(u)$. By Lemma 5.1 these values are distributed uniformly over $E^d$ for any $w_1, \ldots, w_N$; that is, for any $x_1, \ldots, x_n$.

The extra "+1" in the number of queries is removed by instantiating $w_0$ to 0 and then to 1 (see the proof of Theorem 11.5 in Chapter 11), generating queries that are twice as long. ∎

**Lemma 12.12** *For any boolean function $f(x)$, for any $d(n)$, and for any constant $c > 0$, there exists a $(d, dn/(c\log n))$ locally random reduction from $f(x)$ to some function $g(y)$. In particular, for any field of size exceeding $(dn/(c\log n))$, $g(y)$ can be taken to be a polynomial $h_f(w_1, \ldots, w_{n^{c+1}/(c\log n)})$ of degree $n/(c\log\ n)$ over that field, where each $w_i$ itself is a product of at most $O(\log\ n)$ variables $x_i$.*

**Proof:** The proof of Lemma 12.12 is virtually the same as that of Lemma 12.9: $f(x)$ is equivalent to a polynomial $g(x_1, \ldots, x_n)$ of degree $n$ over an arbitrary field $E$. The only new observation we must make is that $g$ is $(d, dn/(c\log\ n))$ locally random reducible to some polynomial $h_f$, by virtue of Lemma 12.11. ∎

## 12.4 Instance Hiding Schemes

The construction of the scheme given in the proof of Theorem 11.5 used secret-sharing with maximal coalition size $t = 1$ to generate a a $(1, n+1)$-locally-random reduction, which gave rise to a $(n + 1)$-oracle instance hiding schemes.

In fact, it is not hard to show that any function $f(x)$ admitting a $(1, m)$-locally random reduction to some function $g(y)$ also has an $m$-oracle instance hiding scheme. This gives a direct way to generate instance hiding schemes for arbitrary functions.

**Theorem 12.13** *If there exists a $(1, m(n))$-locally random reduction from $f(x)$ to $g(y)$, then there exists an $m(n)$-oracle instance hiding scheme for $f(x)$.*

**Proof:** The protocol is similar to the one given in the proof of Theorem 11.5, except that instead of computing $y_i$ as a list of 1-shares of $x_1, \ldots, x_n$, we let $y_i$ be the result of computing the reduction $P(x)$. In other words, the querier computes $(y_1, \ldots, y_m) \leftarrow P(x)$, and sends $y_i$ to $B_i$, who computes $g(y_i)$. Then $A$ determines $f(x)$ by computing $Q(g(y_1), \ldots, g(y_m))$. Since $(P, Q)$ is a $(1, m)$ locally-random reduction, the distribution $D_n^{\{y_i\}}$ on query $y_i$ is the same for any $x$ of size $n$, so the random variables $<Y_i, O_1, \ldots, O_m>$ and $X$ are independent given $|x| = n$. (Note that the oracles are identical, as before, and each computes $g(y)$.) $\blacksquare$

## 12.5   IP=PSPACE and Other Subsequent Work

### 12.5.1   Program Testing and the Permanent

Lipton [Lip89] has designed a theory of program testing based on locally-random self reductions. Given a program which is presumed to work with reasonably high probability on most inputs, he investigates means to check the answer of the program on a particular input by calling the program on several random but related inputs. Based on Theorem 12.6, which gives a $(1, n+1)$ locally-random self reduction for any polynomial of degree $n$ in $n$ variables, his main theorem states that any program for a function expressible as a polynomial of degree $n$ is testable using $n + 1$ calls to the program. (See [BK89, BLR89a] for a related, earlier approach to program checking and correcting.)

In particular, Lipton observes the following:

**Observation 12.14** *[Lipton 1989] The* Permanent *function is testable of order $n + 1$, i.e. it has a $(1, n+1)$-locally random reduction.*

The Permanent of a $n \times n$ matrix $A$ is given by the following expression, similar to the determinant:

$$\text{Perm } A \equiv \sum_{\sigma} \prod_{1}^{n} A_{i, \sigma(i)},$$

where the sum is over all permutations of $\{1, \ldots, n\}$.

The interesting and important property to note is that the Permanent is not known to be efficiently computable. In fact, it is complete for the class $\#P$ of counting problems [Val79], presumed to be extremely intractable. The polynomial hierarchy (PH), which contains NP and coNP, is contained in $P^{\#P}$ by a result of Toda [Tod89];

the PERMANENT function is certainly as hard and presumably harder than any problem in NP or coNP.

## 12.5.2   Interactive Proofs: IP=PSPACE

Beaver, Feigenbaum, Kilian, and Rogaway [BFKR89] applied locally random reductions to show that zero-knowledge proofs of properties about bits which are committed but remain secret require only a constant number of rounds of interaction and a small polynomial message size. In other words, one processor places several bits in envelopes and seals them, and then claims and proves to another processor that a certain property holds on the bits in the envelopes. The bits are not revealed. Previous methods required a message complexity proportional to the size and depth of a circuit describing the property to be shown. Chapter 13 describes and proves this result in more detail.

Nisan [Nis89] used Observation 12.14 to show that there is a multiprover interactive proof system for $\#P$. That is, two physically separate provers can convince a polynomial-time verifier of the value of the permanent of some matrix, despite the difficulty of computing the permanent. Based on Nisan's solution, Fortnow, Lund, Karloff, and Nisan [LFKN89] showed that there is a *single*-prover proof system for $\#P$, namely that one prover can convince a verifier of the value of a permanent. Since the polynomial hierarchy is contained in $P^{\#P}$ by [Tod89], this implies interactive proof systems for languages in coNP, which was a large open question.

Shamir showed that in fact IP = PSPACE, namely that the class of languages that are interactively provable is the same as the class of languages computable by a Turing machine using a polynomial amount of memory [Sha89]. His proof, based on the line of research inspired by locally random reductions, uses simple algebraic techniques.

Interestingly, the proof that IP = PSPACE connects a class based on interaction and communication with a class based on computational complexity. The line of research sparked by the algebraic methods of locally random reductions joins communication complexity with computational complexity. On the other hand, the results of [BFKR89], described in Chapters 13 and 14, show how to make communication complexity and computational complexity *independent* in protocols for security and reliability.

Thus, locally random reductions elucidate the connection between communication complexity and computational complexity. Their straightforward, algebraic nature and the concise and simple solutions they provide and inspire suggest that more applications are waiting to be found.

# Chapter 13

# Zero Knowledge Proofs on Committed Bits

Notarized envelope schemes are a natural extension of the zero-knowledge proof systems introduced by Goldwasser, Micali, and Rackoff [GMR89] (see §2.2.4). An envelope scheme is a means to commit to a string $x$ (the contents of the envelope) in such a way that $x$ cannot later be changed. Until the envelope is opened, the string remains secret. It is similar to secret-sharing in these respects: a secretly shared value remains hidden until it is explicitly reconstructed, and its value cannot be changed in the meantime. In fact, threshold schemes provide a natural means to implement envelopes in the presence of a network, an idea that will be especially useful in Chapter 14.

A *notarized* envelope provides some additional information about the contents of the envelope. Namely, it is an envelope associated with a predicate $\mathcal{P}$ that holds true when applied to the contents. The truth of the predicate may reveal information about the contents; but nothing more is compromised. Like zero-knowledge proofs, the bearer of the envelope sees the simple fact that $x \in L$ — or here, the fact that $\mathcal{P}(x) = 1$ — and whatever else it can compute given the notarized envelope, it can compute given simply that $\mathcal{P}(x) = 1$. (We shall consider an even stronger statement: for some *function $F$*, the committer commits to $x$ and also provides a value $y$ such that $F(x) = y$. The example of a predicate is one-sided in that it does not apply when $\mathcal{P}(x) = 0$.)

Using cryptographic assumptions, notarized envelope schemes have been developed by, among others, Goldreich, Micali, Wigderson [GMW86], Brassard, Chaum, Crépeau [BCC88], and Impagliazzo, Yung [IY87]. In this model, at least one of the two players (prover and verifier) is limited in computational power, and bit commitment is implemented using cryptographic tools such as one-way functions.

We shall consider the *ideal envelope model*, in which both prover and verifier have unlimited computational power, no cryptographic assumptions are made, and bit commitment is assumed as a primitive. Any primitive that implements bit commitment in this model is called an *envelope scheme*. A natural question to address is whether notarized envelope schemes exist in this model; that is, can notarized envelopes be constructed from generic envelopes? This question was answered in the affirmative by Bennett, Brassard, Crépeau [BBC], Ben-Or *et al.* [BGG88], and Rudich [Rud89].

The resources measures for notarized envelope schemes are similar to those of multiparty protocols, or for that of any interaction, for that matter. Local computing time and space, number of rounds of interaction, message sizes, and number of generic envelopes ("envelope complexity") are of primary interest.

The notarized envelope schemes of [BBC, BCC88, BGG88, GMW86, IY87, Rud89] are not directly comparable, because they consider a wide variety of models, but they have one feature in common: All have bit complexity proportional to the circuit complexity of $\mathcal{P}$. Here, we achieve a more communication-efficient general construction of notarized envelope schemes. We consider the execution of an ideal generic envelope scheme to require $O(1)$ rounds.

**Theorem 13.1** *(Constant Rounds Notarization) In the ideal envelope model, every function family $F$ (and thus every predicate family $\mathcal{P}$) has an exponentially-secure notarized envelope scheme that uses a constant number of rounds and a small polynomial number (in $(n, M, k)$) of envelopes and bits, regardless of the computational complexity of $F$.*

By comparison, zero-knowledge proof systems are limited to languages in IP = PSPACE. Though it is unlikely that there are small circuits describing PSPACE languages, the class of provable statements is limited, and there are no proof systems requiring only a bounded number of rounds. Our construction for notarized envelopes makes no restriction on the predicates, and uses a bounded number of rounds.

## 13.1 Definitions

An *ideal (generic) envelope scheme* is the following two-stage ideal protocol for three players, one of whom is a trusted host. We refer to the players as the Sender S, the Recipient R, and the trusted Intermediary I. In the first stage, S sends either the message *refuse* or it sends a bit $b$ to I. If it sends a bit $b$, then I sends the message *accept* to R, to indicate having accepted a bit; otherwise I sends *reject* to indicate refusal or misbehavior. In the second stage, S sends one of two messages, *open* or *retain*, to I. If S sends *open*, then I sends $b$ to R. If S sends *retain*, then I sends $\Lambda$ to R. This protocol is similar to the Verifiable Time-Release Schemes of Chapter 6, though here we are not concerned whether the intermediary knows the bit.

Committing to a string $x$ simply means performing $|x|$ repetitions of the ideal envelope scheme, one for each bit of $x$. We assume a different trusted intermediary for each committed bit. We say that S "opens an envelope" to mean that S sends *open* in stage 2.

An *ideal notarized envelope scheme* for (deterministic) function family $F$ requires a few simple modifications. The sender sends strings $x$ and $y$ to I, who computes $F(x)$. If $F(x) = y$, then I returns (*accept,y*) to R; otherwise, I sends (*reject,y*). In the second stage, I sends either $x$ or $\Lambda$ to R, according to the request of S.

Where clear from context, we in fact refer to a notarized envelope for $F$ as one of two cases: as above, the sender S has provided a value $y$ for $F(x)$; or secondly, the sender has also committed $y$ and is "really" using a notarized envelope for the function $\chi_F(x,y)$ defined as 1 iff $F(x) = y$, and otherwise 0. In the latter case we implicitly assume the sender is claiming $\chi_F(x,y) = 1$.

A notarized envelope for a predicate $\mathcal{P}$ has classically been defined as the special modification in which $y = 1$ always. This weak notion does not allow the sender to claim that the predicate does *not* hold. It corresponds to zero-knowledge in the sense that zero-knowledge proof systems are one-sided; they show that a string $x$ is in a language, but do not need to show $x$ is not in a language. Thus, notarized envelopes for predicates in NP are distinguished from notarized envelopes for coNP, or even from notarized envelopes for NP∩coNP — which corresponds to the more powerful definition in terms of *functions*. That is, given an NP predicate $\mathcal{P}$, if the sender is allowed to *choose* $y = 0$ or $y = 1$ then the function $F$ is an NP∩coNP decision question; if the sender were allowed only to claim $y = 1$, then the function $F$ would be an NP decision problem, corresponding to the case of proof systems for NP.

No efficient notarized envelope schemes built on generic envelope schemes have previously been discovered for predicates outside of NP, even with the weaker one-sided definition. We observe that the techniques employed to show that IP = PSPACE provide an unbounded-round method for notarized envelopes, even with the stronger two-sided definition; and they provide an implementation for a sender and receiver of different computational power. In this chapter, however, we consider equally powerful parties and present results that are not only more general but more efficient than past methods and derivations of those methods.

## 13.2    Notarized Envelopes from Generic Envelopes

**Proof of Theorem 13.1:** Our goal is to construct a protocol that uses generic envelope schemes to implement a notarized envelope scheme. In order to construct *notarized* envelopes for an arbitrary function family $F$, we use a $(1, m)$ locally random reduction from $F$ to a family $G$ of polynomials.

We shall make use of the following important result for functions that can be computed by logarithmic depth, polynomial size circuits. It certainly applies to the reduction (P) and interpolation (Q) functions of the locally random reductions of Chapters 11 and 12, which are randomized $NC^1$ computations.

**Theorem 13.2** *(Bennett et al. [BBC], Ben-Or et al. [BGG88], Rudich [Rud89]) If $F^{n,M}$ is computable by a boolean or arithmetic circuit of depth $O(\log nM)$ and size polynomial in $nM$, then it has a notarized envelope scheme* Notarized-Envelope-$NC^1$ *with constant round complexity, polynomial bit complexity, and polynomial envelope complexity.*

The protocol Notarized-Envelope (F,x,y) for notarized envelopes for arbitrary function $F$ is described in Figure 13.1. Essentially, the sender repeats the following protocol $kn^2$ times in parallel. It first commits to two sequences of random bits $r_1$ and $r_2$ that the reduction and interpolation algorithms, P and Q, will use. (Assume that the two randomized algorithms require $p(n)$ bits on inputs of length $n$; with these bits, the algorithms are essentially deterministic.) The sender uses a notarized envelope to commit to a string containing $m$ outputs $y_1', y_2', \ldots, y_m'$ that it claims to be the outputs of $P(r_1, x)$. The sender also uses a notarized envelope to commit values $z_1, z_2', \ldots, z_m$ that it claims satisfy $Q(r_2, z_1', z_2', \ldots, z_m') = y$. These two claims are satisfied using the notarization schemes for polynomial size circuits [BBC, BGG88, Rud89].

The crucial claim is the following. The sender claims that for each $i \in [m]$, $z_i' = G(y_i')$. Of course, the computation of $G(y_i')$ might require a large circuit, and previous techniques will not apply without incurring tremendous expense in terms of bits, rounds, and commitments. The sender "proves" his claim by allowing the receiver to open *one* of the pairs $(y_i', z_i')$. If S has behaved then indeed $z_i' = G(y_i')$, and R will accept it. If S has not behaved, namely if $F(x) \neq y$, then given that $P(r_1, x) = (y_1', y_2', \ldots, y_n')$ and $Q(r_2, z_1', z_2', \ldots, z_n') = y$, there must be some pair $(y_i', z_i')$ for which $z_i' \neq y_i'$. With probability at least $1/m$, then, R will detect cheating and reject. The probability is at least $1/n$ if a $(1, n)$ LRR is used, and through amplification by repetition, the probability is at least $1 - (1 - 1/n)^{kn^2}$ or asymptotically $1 - e^{-kn}$ that R detects cheating if it exists. (Locally random reductions with sharper parameters will decrease the number of repetitions needed to amplify the probabilities.)

More formally, the overall protocol computes the composition of three functions, $F^1$, $F^2$, and $F^3$. The first computes a robust and private representation of $x$ and $\{r_1^K, y_1^K, y_2^K, \ldots, y_m^K\}_{K \in [nk^2]}$, as well as the appropriate output *accept* or *reject* for receiver R. The output is clearly private: a reliable sender will always provide inputs producing *accept* for R, and the output induced by a corrupted sender depends only on the sender's information. The second computes a robust and private representation of $y$ and $\{r_2^K, z_1^K, z_2^K, \ldots, z_m^K\}_{K \in [nk^2]}$ as well as the appropriate output *accept* or *reject* for receiver R. The third computes the function that computes *reject* if R has rejected anything so far, and computes *reject* if any $i_j$ selects a pair $(y_i, z_i)$ that was computed incorrectly, and otherwise computes *accept* for R. Function $F^3$ is private, since if S is reliable then the output is always *accept*. It is a robust representation of the statement that $r_1, r_2, i_1, i_2, \ldots, i_{nk^2}$ select a random reduction and a set of pairs that are consistent.

The overall protocol runs three subprotocols in sequence, each of which computes one of these robust and private representations of intermediate values. It uses a security parameter of $k' = k/4$ to ensure that the concatenation is sufficiently resilient.

The ideal notarized envelope protocol allows no chance to cause R to output *accept* improperly. The composite interface for the three functions $F^1$, $F^2$, and $F^3$ can simulate all outcomes accurately given that cheating does not go undetected, namely that the executions of NOTARIZED-ENVELOPE-NC$^1$ for the first two phases do not run into an undetected inconsistency, and that $r_1, r_2, i_1, i_2, \ldots, i_{nk^2}$ do not lead to an undetected inconsistency. The probability of such runs is at most $2^{-k'} 2^{-k'} 2^{-k'} < 2^{-k}$. Hence the output distributions induced by the simulator are exponentially close to

---

NOTARIZED-ENVELOPE (F,x,y)

**Stage I (Commit):**

    S computes $y = F(x)$ and sends $y$ to R.

    **For $K = 1..nk^2$ do in parallel** :

        **begin**

        S commits two uniformly random strings $\hat{r}_1^K, \hat{r}_2^K$ of length $p(n)$.

        R sends to S two uniformly random strings $\hat{s}_1^K, \hat{s}_2^K$ of length $p(n)$.

        S sets $r_1^K = \hat{r}_1^K \oplus \hat{s}_1^K, r_2^K = \hat{r}_2^K \oplus \hat{s}_2^K$.

        S computes $P(r_1^K, x) = (y_1^K, y_2^K, \ldots, y_m^K)$.

        **For** $i = 1..m$ : S computes $z_i^K = G(y_i^K)$.

        S runs NOTARIZED-ENVELOPE-NC$^1$ $(P;x, r_1^K;y_1^K, y_2^K, \ldots, y_m^K)$.

        R assigns $P_K$ to be *accept* or *reject* accordingly.

        S runs NOTARIZED-ENVELOPE-NC$^1$ $(P;r_2^K, z_1^K, z_2^K, \ldots, z_m^K;y)$.

        R assigns $Q_K$ to be *accept* or *reject* accordingly.

        R randomly chooses $(i_1, \ldots, i_K) \in_R [m]^K$ and sends it to S.

        S opens the generic envelopes containing $y_i$ and $z_i$ for each $i$.

        R receives the opened values $y_i'$ and $z_i'$.

        R computes $G(z_i')$.

        **end**

    **if** for all $K \in [nk^2]$, $y_i' = G(z_i')$ and $P_K = Q_K = accept$,

        **then** R outputs $(accept,y)$

        **else** R outputs $(reject,y)$.

**Stage II (Open):**

    S opens the generic envelopes containing $x$.

Figure 13.1: Protocol for sender S to place $x$ in a notarized envelope claiming $F(x) = y$. $(P, Q)$ is a $(1, m)$ locally random reduction from $F$ to $G$, and $P$ and $Q$ are $NC^1$ computations.

---

that of the ideal notarized envelope protocol. ▮

# 13.3 Zero-Knowledge Proofs About Secrets

In §7.1 we saw that any function $F_{\mathcal{P}}$ for a predicate $\mathcal{P}$ admits a multiparty zero-knowledge interactive proof system having constant round complexity and having message complexity polynomial in the size of $C_{\mathcal{P}}$, a circuit for $\mathcal{P}$.

Using locally random reductions, we can show that the message complexity is indeed independent of the circuit complexity, namely that the message complexity is polynomial in $n$, $m$, and $k$, for any $\mathcal{P}$.

Secret-sharing is indeed a method for committal, given one technical restriction: since the committer has the option of refusing to decommit, the reconstruction pro-tocol must be adapted so that reliable players give up their pieces only if the dealer requests.

A multiparty zero-knowledge proof system on secrets (definition 7.1) is then a special case of a notarized envelope scheme where envelopes are replaced by secret-sharing. As a direct application of Theorem 13.1, we have the following:

**Theorem 13.3** *(Constant Rounds ZKMIPSS) In the ideal envelope model, every function family F (and thus every predicate family $\mathcal{P}$) has an exponentially-resilient notarized envelope scheme that uses a* constant *number of rounds and a small polynomial number of envelopes and bits.*

# Chapter 14

# Efficient Multiparty Protocols for Any Function

> She was pale when she talked with the chaplain. She said that the war instead of ennobling soldiers made beasts of them. Downstairs the patients had stuck their tongues out at her and told her she was a scarecrow and a frightful skinny old frump. "Oh, how dreadful, chaplain," she said in German. "The people have been corrupted."
>
> — Jaroslav Hašek, *The Good Soldier Švejk*

Chapter 5 gives secure and reliable protocols for any function in $NC^1$ using a constant number of rounds and tolerating faults in a third of the network. The construction also gives constant round protocols for any function but at a potential exponential blowup in message sizes. Here we present secure and reliable protocols to compute *arbitrary* functions in *constant* rounds, using small, polynomial-size messages, and tolerating $t = O(\log n)$ faults. Thus, instead of incurring a price in message size in order to be able to compute any function, we tolerate a smaller yet still reasonable number of faults.

Locally random reductions appear twice in our protocol. First, they serve to to reduce a highly interactive network computation of $F$ to an $n$-fold *local* computation of a polynomial for $F$. Second, they ensure that Byzantine faults are detected and corrected.

The power of locally random reductions is evident in the results: *any* function admits an unconditionally $t$-resilient protocol having *constant* round complexity and *small* message complexity, *regardless* of its computational complexity. Through simple algebraic approaches, we separate the *communication complexity of secure computing* from the *computational complexity* of the function to be computed.

Let us now consider the inputs as a single sequence of bits instead of a set of $n$ different inputs each having length $m$. That is, for $a = bm + c$, define $\hat{x}_a$ to be the $c^{th}$ bit of input $x_b$. Each player shares each bit of its input separately. Since we do have a bound on the number of input bits in terms of the number of players, we shall consider the number of output bits, $N(n)$, as a separate parameter.

**Theorem 14.1** *Let $t(n)$ and $m(n)$ satisfy $t(n)m(n) = O(\log n)$. Let $F$ be a family of functions $\{F^n\}$ mapping $(\Sigma^{m(n)})^n \to \Sigma^{N(n)}$, where $N(n)$ is polynomial in $n$. Then there exists a $t$-resilient protocol $\text{EVALUATE}^{\,t<\log n}$ for $F$ that runs in a constant number of rounds and uses a small polynomial number of message bits.*

**Proof:** Let $\alpha_1, \ldots, \alpha_n$ be fixed nonzero field elements over a field $E$ of size exceeding $n$. Let $q = \lceil \log n \rceil$ and let $M$ be the least multiple of $q$ exceeding $nm$. Let $r = (2^q)\frac{M}{q}$, which is clearly polynomially bounded in $n$. The function $F^n$ maps $\hat{x}_1, \ldots, \hat{x}_{nm}$ to an $N$-bit string; let $F^{nb}(\hat{x}_1, \ldots, \hat{x}_{nm})$ be the $b^{th}$ bit of $F^n(\hat{x}_1, \ldots, \hat{x}_{nm})$. Without loss of generality, consider $F^{nb}$ to be an $M$-variate function that is insensitive to the extra $M - nm$ variables. By Theorem 12.7, there is a polynomial $h_F^{nb}(w_1, \ldots, w_r)$ of degree $(m/q)$ for $F^{nb}$, such that each $w_i$ is a product of at most $2^q$ variables $\hat{x}_j$.

Figure 14.1 describes the protocol. We have expanded on the locally-random reduction so that the steps of this protocol are more concrete; note that any locally-random reduction computable in $NC^1$, not simply the one based on polynomial evaluation that sufficed for the proof of Theorem 12.7, will do. The particular locally-random reduction of Theorem 12.7 is easy to implement and very efficient, so we use it directly.

Because the locally random reduction is in $NC^1$ (each $w_i$ is the product of at most it can be secretly evaluated in constant rounds using Theorem 5.16. The system generates $N(n)$ sets of $n$ queries, one set for each output bit. The queries from each set are given to the players individually. The crux of the solution is that, though computing the reduced functions $h^{nb}$ may be complicated, it is a local computation requiring no interaction.

Locally random reductions, however, are not necessarily robust against errors in the reduced computations. Reducing the overall computation to locally independent computations suffices for privacy but not for robustness. Thus we need to verify that the results of step 5 are accurate before using them to interpolate the final answer.

Let us expand upon step 6. Each player $i$ gives a ZKMIPS proof (§7.1) to each player $j$ that the secret value of $v_i^b$ it shared satisfies $v_i^b = h_F^{nb}(w_1^i, \ldots, w_r^i)$, using the methods of Chapter 13. This is not quite sufficient *per se*, since a faulty player may cause some good players to believe the proof while others find fault. Thus, after the proving phase, every player $j$ broadcasts a vote $V_{ji}$ as to whether it believes the proof of player $i$. If a majority vote to accept $i$, then player $i$ is accepted. A majority vote in favor of player $i$'s honesty implies that at least one reliable player found player $i$ correct, which means that player $i$ was able to prove its correctness to someone, and its results are therefore correct. If player $i$ is rejected by the vote, it failed to satisfy any reliable player of its behavior, and the query $(w_1^i, \ldots, w_r^i)$ with which it is supposed to compute is revealed so that the reliable players may incorporate it properly.

Thus, we have specified eight function families corresponding to the eight modules of the protocol. Each of these computes a robust and private representation of its results (small lists of questions from the reductions are independent of the inputs and hence private; the $v$ values are robust given that the zero-knowledge proofs succeed, or else they are broadcast implicitly through the revelation of $w_1^i, \ldots, w_r^i$; pieces of the various secrets are robust and private representations). By Theorem 4.31, the composition is $t$-resilient.

The number of rounds is constant, by the following observations. Step 2 is an evaluation of several polynomials $w_i$ of degree $2^q$, which is an $NC^1$ computation. By Theorem 5.16, this takes constant rounds and polynomial message sizes.

Step 3, the selection of $n$ random polynomials of degree $t$ with free terms $p_1^1, \ldots, p_1^t; \ldots; p_r^1, \ldots, p_r^t$, requires the secret choice of $rt$ coefficients. The selection of these random secret values is performed in parallel and requires constant rounds. Applying polynomials $p_1, \ldots, p_r$ to the known, fixed values $\alpha_1, \ldots, \alpha_n$ is a linear combination and hence requires no interaction.

Each player $i$ computes $h_F^{nb}$ locally and non-interactively, and its resharing requires polynomial bits and constant rounds. The zero-knowledge proofs of step 6 are slightly more complicated yet still require constant rounds and polynomial message sizes. The interpolation of $y^b$ from $v_1^b, \ldots, v_n^b$, step 8, uses an $NC^1$ circuit. ∎

Evaluate $^{t<\log n}(F)$

1. **(Share** $\hat{x}_1, \ldots, \hat{x}_m$.**)** Each player $i$ shares its input as $\hat{x}_{m(i-1)+1}, \ldots, \hat{x}_{mi}$. Variables $\hat{x}_{mn+1}, \ldots, \hat{x}_m$ are taken to be 0.

2. **(Expand variables.)** Compute new secrets $w_1, \ldots, w_r$ by running protocol Evaluate-NC [1] to evaluate $w_i(\hat{x}_1, \ldots, \hat{x}_m)$.

3. **(Select LRR.)** Select $rt$ random secrets $p_1^1, \ldots, p_1^t; \ldots; p_r^1, \ldots, p_r^t$, using protocol Rand-Secret . These denote the higher coefficients of polynomials $p_i(u) = w_i + \sum_j p_i^j u^j$.

4. **(Apply LRR.)** Using protocol Evaluate-NC [1], compute new secrets $w_j^i = p_j(\alpha_i)$ for $i = 1..n$, $j = 1..r$, in parallel. Reveal $(w_1^i, \ldots, w_r^i)$ to player $i$.

5. **(Compute reduced problem locally.)** Each player $i$ computes

$$v_i^b = h_F^{nb}(w_1^i, \ldots, w_r^i) \qquad 1 \le b \le N(n)$$

and secretly shares $(v_i^1, \ldots, v_i^{N(n)})$ among the network.

6. **(Protect against Byzantine failures.)** Run Protocol Prove-Behavior $(i, j)$ for each $1 \le i, j \le n$; thus each player $i$ proves to the network in zero-knowledge that

$$(v_i^1, \ldots, v_i^{N(n)}) = (F^{n1}(w_1^i, \ldots, w_r^i), \ldots, F^{n,N(n)}(w_1^i, \ldots, w_r^i))$$

7. Each player $j$ broadcasts a vote as to whether each player $i$ gave him a satisfactory proof in step 6. Disqualify each player $i$ who is impeached by a majority; reveal its query $(w_1^i, \ldots, w_r^i)$; and use

$$v_i^b = h_F^{nb}(w_1^i, \ldots, w_r^i) \qquad 1 \le b \le N(n)$$

as publicly known constants for the remainder of the protocol.

8. **(Reconstruct** $f$.**)** Run protocol Evaluate-NC [1] to secretly interpolate each set $(v_1^b, v_2^b, \ldots, v_n^b)$, for $b = 1..N(n)$ in parallel. For each set, obtain the free term $y^b$ of the polynomial of degree $rt$ passing through the points.

Figure 14.1: Protocol to evaluate any $F$ in constant rounds and low fault-tolerance.

# Chapter 15

# Applications

> The artist possesses the ability to breathe soul into the lifeless product of the machine.
>
> — Walter Gropius

The bulk of this dissertation has addressed formal specifications of protocols for networks of players to compute arbitrary functions in a distributed manner. In this chapter we present efficient and practical solutions for useful and important computations.

It would certainly be impractical to require multiparty protocols around each distributed computation a collection of processors must make. Some computations, especially administrative and system-security oriented computations, are of greater importance than others, but at the same time need to be performed less often. For example, authenticating a user or a remote process must be correct and secure but are performed with lower frequency than other computations. Authorizing access to a file is another example.

We propose very fast protocols for authentication, authorization, and anonymous mail. These protocols are designed with specific goals in mind and capitalize on properties that a general-purpose protocol-compiler (based only on circuit specifications) ignores.

For the purposes of ensuring distributed system reliability, we envision a *security kernel*, in terms of processors and of processes. That is, not all the processors in a network need be involved in ensuring the resilience of every computation. A specific committee of, say, seven processors would suffice, with the assurance that no more than three are compromised. These processors are tightly coupled with high-speed secure communication lines.

With respect to the operations themselves, we propose that authentication and authorization are infrequent yet essential computations, and that it is reasonable to take special care to ensure the correct and secure operation of a small body of desired processes. Once a process has been authenticated and authorized to read a file, the authorization need not be repeated and the authentication can be omitted during repeated file access requests.

The Secure UNIX of Reeds and McElroy [MR87, MR88a, MR88b, Ree89] adopts this sort of approach, allowing two remote processes to communicate over a single pipe that excludes other processes and that fails as soon as one process detaches. While this does allow certain attacks to succeed — a corrupt operating system might capture an authorized line to a remote system by running an authenticated process and substituting its own process after the authentication with the remote host has occurred — the attacks are more difficult and complex. Furthermore, a repeated authentication of the *host* operating system, not each individual process at each individual request, would serve as a more efficient safety check. If the host is continually authenticated then presumably its operations are valid and the processes it runs cannot be substituted after their initial authentication. This is a reasonable trade-off of absolute security for practicality.

Our efficient algorithms for specific tasks support the approach of infrequent yet highly efficient and secure kernel operations. We focus not simply on achieving a constant number of rounds of interaction but on achieving a *small* constant number of rounds, on the order of 2 or 3.

## 15.1   On Strong Assumptions

In general, we assume a completely connected, synchronous network with private lines. The robustness of polynomial interpolation suggests the techniques developed here should be adaptable to asynchronous models. The fault-tolerance levels would be lower, and many technical issues must be considered (such as the availability of broadcast channels *vs.* Byzantine Agreement subprotocols). Because our protocols are robust against fail-stop faults, their extension to problems of asynchronicity is natural.

A network without private lines can still use the protocols developed here if suitable encryption of the lines is available. Care must be taken to ensure that this approach of treating sub-issues in a modular fashion does in fact treat each module independently — for example, to avoid unintended interaction between the design of the overall protocol and the design of a communication line protection scheme, the same encryption methods should not be used at all levels. That is, if a protocol assumes a private channel but also requires an encryption for other reasons, the same encryption should not be used to replace the private channel with an encrypted channel. Proofs of security remain necessary.

We have also assumed a complete network. This assumption is highly unlikely to hold except on a local-area network. In practical terms, however, it is not hard to achieve for a processor security kernel with a small number of tightly-coupled processors. In theoretical terms, the issue of simulating a complete network using a $t+1$-connected network is an interesting separate subproblem.

The modular treatment of the various subproblems provides optimism for the design of very general protocols that are robust under very harsh circumstances. A word of caution must be interjected: as always, intuitive claims to security require formal treatment. But the modular approach makes the analysis easier.

## 15.2   Choice of Field

For the purpose of implementation, we suggest a few natural finite fields.

The first is GF(257), the set of integers modulo 257. This field easily encodes bytes using integers in the range 0—255. Modular arithmetic is fast and standard, though individual bit operations on the eight bits are not quite so facile. Two-byte words are supported by a modulus of 32771.

The other fields we suggest have characteristic 2: $GF(256) = GF(2^8)$, $GF(65536) = GF(2^{16})$, $GF(2^{32})$, and $GF(2^{64})$, Field elements are represented as polynomials over GF(2) modulo an irreducible polynomial (for example, $x^8+x^4+x^3+x^2+1$ for $GF(2^8)$, $x^{16}+x^{12}+x^3+x+1$ for $GF(2^{16})$, and $x^{32}+x^{22}+x^2+x+1$ for $GF(2^{32})$). Each coefficient is a single bit, so that the field elements themselves correspond naturally to common word-sizes. Addition over these fields is quite simple and normally requires a single machine instruction: take the bitwise exclusive-or. Multiplication is somewhat more complicated, requiring one to reduce the product modulo the appropriate irreducible polynomial, but fast algorithms exist.

Generating random *bits* as the exclusive-or of random bits supplied by many players is much easier using a field of characteristic 2 than one of characteristic 257, since addition of random secret 0/1 values will suffice. In general, bitwise operations are facilitated by using arithmetic over a field of characteristic 2.

## 15.3   A Fast and Practical Password Scheme

In the UNIX$^{tm}$ password scheme for user authentication, the passwords are not stored in a file on the system. Rather, each password is encrypted (using a DES-like encryption function) and the result is stored in a password file. Even if the password file is compromised, an attacker cannot learn the passwords directly from the file. When a user would like to log in to the system, his attempted password is encrypted and compared against the stored encryption. If they match, the user is authenticated.

One can imagine a generalization to the distributed environment in which each host authenticates the user and the system as a whole considers the user acceptable if a majority of the hosts have authenticated him. An immediate problem surfaces with password schemes: if the same password is used for all the hosts, then a single corrupt host will learn the user's global password when the user requests authentication. This can be circumvented by having a different password for each host (requiring some nontrivial and probably security-prone bookkeeping by the user). Nevertheless, there are many disadvantages, including the fundamental problem that the encryption function itself is only *assumed* to be unbreakable.

We propose a simple and efficient password scheme [1] discovered independently

---

[1]This work was done while the author was on leave of absence from Harvard University in the

by Michael Rabin [Rab89] that has some similar properties but does not depend on the security of an encryption function like DES. That is, passwords cannot be obtained by capturing the files on some fraction of the systems, and are certainly not compromised during an execution in which *one* of the hosts is corrupt. In order to break the password directly, an attacker must corrupt a majority of the hosts (we ignore dictionary searches and repeated attempts, which are certainly valid issues for any password scheme but are orthogonal to the problem of breaking the system directly).

The basic idea is as follows: a user secretly shares his password `password` among the system at start-up. Later, when he requests authentication, he demonstrates his knowledge of the password by sharing it again. Let us call the second secret his attempt, `attempt`. The system must compare `password` to `attempt` *without revealing* `password`.

---

INITIALIZE-PASSWORD

Let $E = \mathrm{GF}(2^{64})$. Simultaneously:

       User shares `password`.

       Each host $i$ selects two uniformly random 64-bit elements $r_i, s_i$, and shares them.

       Run the LINEAR-COMBINE protocol (non-interactively) to secretly compute

$$r = \sum r_j, s = \sum s_j :$$

      — More precisely, each host $i$ sets

$$\mathrm{PIECE}_i(r) = \sum\nolimits_j \mathrm{PIECE}_i(r_j)$$
$$\mathrm{PIECE}_i(s) = \sum\nolimits_j \mathrm{PIECE}_i(s_j)$$

Figure 15.1: Protocol to initialize a user's password in a distributed system of $n$ hosts.

---

It does not suffice to compute (`password` − `attempt`) and reveal the result, since knowing `attempt` would then reveal `password`. This remains true even if the *user* is not supplied with the result; if the user cooperates with just one corrupted host involved in the authentication, then that host can reveal the result when it learns it.

One solution is to use Theorem 5.16 to secretly compute $\|\texttt{password} - \texttt{attempt}\|$ in a constant number of rounds. The protocol most certainly is correct and uses only a few rounds, but we propose an even simpler and more efficient protocol that is designed with the particular password problem in mind.

The essential information that the system must compute is whether or not `password` = `attempt`. The result of the computation should specify whether or not they are equal, but it need not be a deterministic result; that is, the result need

---

spring and summer of 1988.

not be either 0 or 1. What matters is that if the values are not equal, the result of the computation should have the same distribution regardless of what attempt is made and of what the password is. The problem of revealing information about the password through a failed attempt is then solved.

The trick is the following: compute and then reveal $(\texttt{password} - \texttt{attempt}) \cdot r$, where $r$ is secret and uniformly random over $E - \{0\}$. If the attempt is correct, then the result is always 0. But if the attempt is invalid, then the result will be a uniformly distributed nonzero random field element, for any $\texttt{password}$ and $\texttt{attempt}$. Thus, no information about $\texttt{password}$ or $\texttt{attempt}$ is revealed by the result, apart from whether they are identical or not. The system accepts the user if and only if the result is 0; each host outputs *accept* if the result is 0 and otherwise outputs *reject*.

The only minor inefficiency is to generate a nonzero random field element. The RAND-SECRET protocol will generate a uniformly random secret $r$ in $E$, but we should then have to normalize it in order to test if $r = 0$, returning us to the original problem. One simple and acceptable approach is simply to finesse the problem by using a large field $E$ so that the chance that $r = 0$ is negligible. This would allow some very low probability that a failed attempt will succeed. In fact, the chance of such an event is the same as guessing a valid password. See Figure 15.2.

---

AUTHENTICATE-FAST

(A1) User shares $\texttt{attempt}$.
Run the LINEAR-COMBINE protocol (non-interactively) to secretly compute $v = \texttt{password} - \texttt{attempt}$.
— More precisely, each host $i$ sets
$$\text{PIECE}_i(v) = \text{PIECE}_i(\texttt{password}) - \text{PIECE}_i(\texttt{attempt})$$
(A2) To be ready for the *next* authentication, each host $i$ selects two uniformly random 64-bit elements $r_i^{new}, s_i^{new}$, and shares them. Each host $i$ sets
$$\text{PIECE}_i(r^{new}) = \sum_j \text{PIECE}_i(r_j^{new})$$
$$\text{PIECE}_i(s^{new}) = \sum_j \text{PIECE}_i(s_j^{new})$$
Run the MULTIPLY protocol to secretly compute $w = v \cdot r$.
**if** $w = 0$ **then** *accept* the user, **else** *reject* the user.

Figure 15.2: Protocol to authenticate user who has already shared password $\texttt{password}$.

---

The chance that the scheme is broken through this error is less than that of randomly selecting a password, and it is neither repeatable (the user has no control over $r$) nor assisted by ulterior means (such as dictionary lookups). It therefore doubles at most the inherent chance of failure. Figure 15.2 describes the stripped-down scheme. One round suffices to share the attempt, and one to check it (with only fail-stop errors).

A completely accurate solution is the following: generate a second random secret, $s$, and multiply it by $r$. If the result is nonzero, then the system has a proof that $r \neq 0$ without revealing $r$.

The chance that randomly choosing $r$ and $s$ gives $r = 0$ or that $r \neq 0$ and $s = 0$ is small: $\frac{2|E|-1}{|E|^2}$. Otherwise, with high probability the system has generated a uniformly random nonzero secret $r$, along with a ("zero-knowledge") certificate that $r \neq 0$.

For concreteness, let us use the field GF($2^{64}$) to handle passwords of 8 bytes. The password set-up protocol is described in Figure 15.1. The authentication protocol is described in Figure 15.3. Note that in the authentication protocol, a failure to obtain a nonzero secret $r$ requires that the network generate one; this occurs with probability $2^{-64}$ and can essentially be ignored.

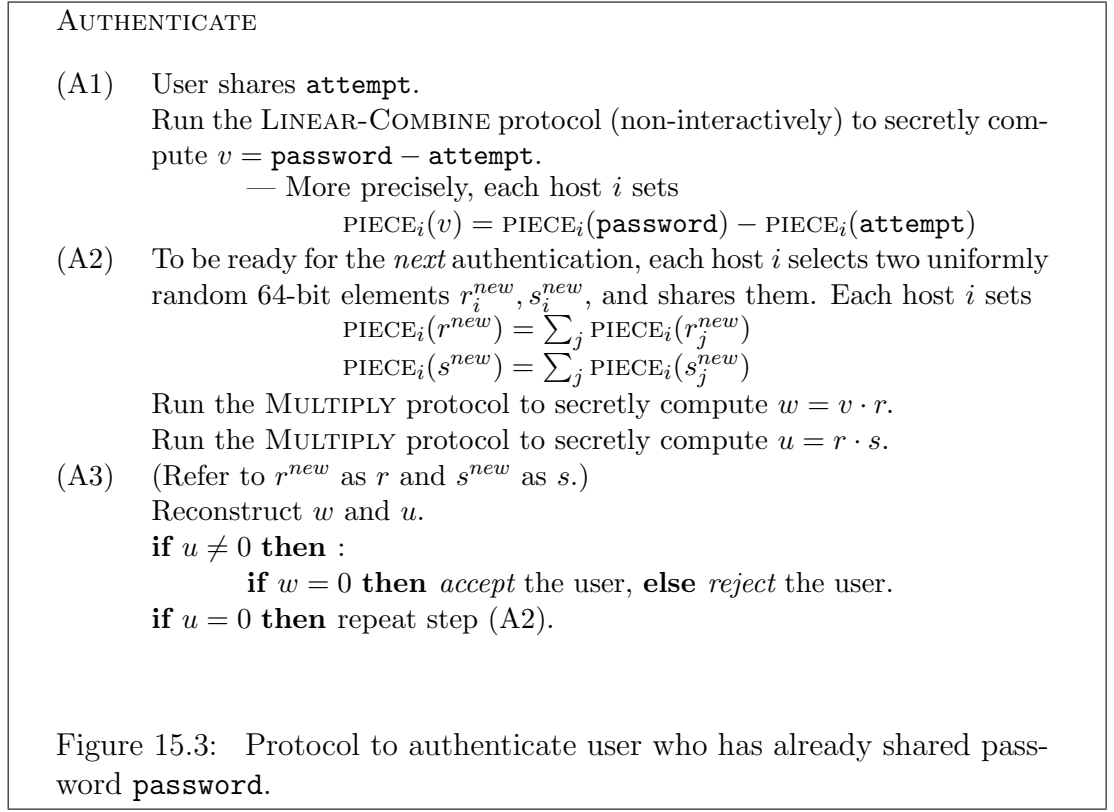

Figure 15.3: Protocol to authenticate user who has already shared password `password`.

**Theorem 15.1** *For $2t < n$, protocol* Authenticate *is statistically $t$-resilient and requires a constant expected number of rounds.*

**Proof:** Clearly, host $i$ outputs *accept* iff $r \neq 0$ and (`password` − `attempt`) $\cdot r = 0$, *i.e.* iff `password` = `attempt`. If `password` = `attempt`, the distribution on $(w, u)$ is

$$\{r \leftarrow \boldsymbol{uniform}(E); s \leftarrow \boldsymbol{uniform}(E) : (0, rs)\},$$

regardless of the values of `password` and `attempt`. If `password` $\neq$ `attempt`, the distribution on $(w, u)$ is the distribution

$$\{r \leftarrow \boldsymbol{uniform}(E); s \leftarrow \boldsymbol{uniform}(E) : ((\texttt{password} - \texttt{attempt}) \cdot r, rs)\},$$

which for any (unequal) values of `password` and `attempt`, is fixed and identical to

$$\{r \leftarrow \mathit{uniform}(E); s \leftarrow \mathit{uniform}(E); w \leftarrow \mathit{uniform}(E - \{0\}) : (w, rs)\}.$$

∎

**Corollary 15.2** *For $2t < n$, Protocol* AUTHENTICATE *is $t$-resilient against passive (or even fail-stop) adversaries and requires only $(3 + \epsilon)$ expected rounds, where $\epsilon = 4/|E|$ is small.*

**Proof:** Follows from Theorem 15.1 and the observation that, when messages are always correct, the number of rounds required for sharing and for multiplying secrets is 1. Step (A2) is repeated with probability $\frac{2|E|-1}{|E|^2}$, which gives an expected time less than $4/|E|$ when $|E| \geq 7$. ∎

**Remark:** The secure authentication schemes we have presented take the same amount of time as the simple method of having a user authenticate itself to each processor separately and then taking a vote. They provide, on the other hand, a good deal more robustness. For example, once an intruder has recorded the message from a valid user to one host in the voting scheme, that password can be re-used; the invader need only accumulate enough attacks over time. In our scheme, however, at least half of the communications must be tapped *simultaneously* to obtain *any* information at all.

### 15.3.1   Unanimous Secret Ballots

Secret ballots are fast and simple (§5.2.5), and by Theorem 5.16, checking whether a secret vote is *unanimous* or not, without revealing the tally, also admits a protocol in constant rounds. The authorization techniques listed above suggest a faster and more direct method: first compute $v = \sum_i x_i$, where each $x_1$ is a verified 0/1 secret vote, generate a random (nonzero) secret $r$, and secretly compute $vr$. Reveal the result. If the ballot is unanimous, the result will be 0; otherwise it will be a uniformly distributed nonzero number.

## 15.4   Anonymous Mail

A mail system ought to provide privacy on two levels. First, it should hide the contents of each message from the general public, and certainly from those involved in its delivery. Second, it should not need to identify the sender of the message, either to those involved in the delivery or to the recipient. It does not suffice simply to deliver a message without attaching a sender's name in the header, since the name of the originating host or workstation may be enough to reveal the sender. In fact, the delivery of anonymous messages in a network is powerful enough to support general function computation [BBR88], though we shall not go into the details here.

Delivering mail anonymously even when hosts are identified is thus a powerful and a useful tool.

To be practical, however, a mail scheme must be efficient. If the message is an undivided unit, then it traverses some path from the sender S to the receiver R. If every host on that path is corrupt, then the receiver can identify the origin. Thus the path must be longer than the number $t$ of corruptible hosts. Already this is a very inefficient solution. Furthermore, it requires that the message be encrypted in some form so that the intermediate hosts cannot read the contents.

We adopt a different approach: divide the message in an appropriate manner and send each portion via a different path. If sufficiently few pieces are captured, then nothing is learned about the contents. Secretly sharing the message and reconstructing it for the recipient suffices. This is trivial in a complete network with private channels, but in an incomplete network, when two players may be forced to communicate via intermediate players, secret sharing solves the problem of privacy robustly.

Preserving the privacy of the sender, however, is a slightly more difficult goal. We must ensure that the paths to R do not identify S. Thus, the direct approach of as secretly sharing the message and reconstructing it for R is suggestive, but not sufficient. Every host knows *who* shared the secret.

Consider the following situation: there is some secret permutation $\sigma$ on the elements $\{1, \ldots, n\}$, and each player $i$ knows $\sigma(i)$. Each player $i$ would like to send a message $M(i)$ to player $\sigma(i)$ without identifying itself. We would like to compute a function $F$ such that $F_{\sigma(i)}(M(1), \ldots, M(n), \sigma(1), \ldots, \sigma(n)) = M(i)$.

One method to perform this computation would be to compute

$$F_j(M(1), \ldots, M(n), \sigma(1), \ldots, \sigma(n)) = \sum_{i=1}^{n} M(i)\delta(\sigma(i), j)$$

Because the delta function is in $NC^1$, Theorem 5.16 implies there is a constant-rounds protocol to compute $F$. Note that we have omitted a verification that the secrets $\sigma(1), \ldots, \sigma(n)$ do indeed define a permutation; malicious players could upset the protocol. An additional function that compares each pair of destinations and makes sure no collisions occur is a necessary prerequisite.

A more efficient way is the following. Each player, instead of sharing $\sigma(i)$, shares a set of $n$ secrets $d(i, 1), \ldots, d(i, n)$, such that $d(i, \sigma(i)) = 1$ and the remaining secrets are 0. It is trivial to check privately that every $d(i, j)$ is 0 or 1 by computing and revealing $d(i, j)(d(i, j) - 1)$. It is also trivial to ensure that each player shares a valid vector, by computing $\sum_j d(i, j)$ for each $i$ and revealing the sums. Thirdly, it is trivial to check for collisions by taking $\sum_i d(i, j)$ for each $j$ and revealing the sums. (Collisions give a sum greater than 1, which reveals no information to the faulty processors that changed their destinations.) Thus there are three simple verifications to perform in parallel, requiring the time of one multiplication and reconstruction.

Given verified vectors, the final outputs are simply

$$F_j(M(1), \ldots, M(n), d(1,1), \ldots, d(n,n)) = \sum_{i=1}^{n} M(i)d(i,j).$$

This construction gives an easy solution to the converse problem, *limited espionage*.[2] Like 1-out-of-2 Oblivious Transfer, in which the recipient chooses one of two values to learn but the holder does not find out which, limited espionage allows each player in the network to learn one secret from a list of $N$ secrets without revealing which one it chose. The protocol requires each player to share a 0/1 vector specifying a 1 in the component corresponding to the secret it desires, and the verifications are performed as above before computing the simple weighted sum. Note that in this case, collisions need not be avoided; two players may look at the same secret.

The general problem of supporting arbitrary maps from sources to destinations and multiple messages from a single player is somewhat more involved. Allow each player to send some number $N$ of messages to any other player. The set $\{M(i,j,k) \mid i,j \in [n], k \in [N]\}$ describes the list of all messages. Player $i$ secretly shares each $M(i,j,k)$ as a secret $s(j, (i-1)nN+k)$. The goal is to construct a random secret permutation $\sigma \in \mathcal{S}_{nN}$ for each $j$, and apply it to the secrets $s(j,1), \ldots, s(j,nN)$ destined for player $j$. The permuted list of results hides the origins and is revealed directly to player $j$. To generate a random secret permutation $X(j)$ on $nN$ indices, protocol FULL-RANK-RANDOM-MATRIX of §5.3.3 suffices, using 0/1 matrices to represent permutations. The permuted list of results for player $j$ is computed by secretly multiplying the matrix $X(j)$ by the vector of secrets $(s(j,1), \ldots, s(j,nN))$.

While mathematically correct, the size of the matrices makes this solution prohibitive. In a more practical vein, a different sort of approach is preferable. The approach used in delivering mail whose destinations are a permutation of the sources is useful. We provide each player $j$ with a set of $n$ "mailboxes," namely a fixed set of secrets $b(j,l)$. Each mailbox has a corresponding set of flags $c(j,l)$ that indicate whether one or more players have attempted to place a message in the mailbox. The protocol is simple: player $i$ sends a message $M(i,j,k)$ to player $j$ by choosing a mailbox $(i,l)$ at random and sharing $M(i,j,k)$ as a secret value $B(j,l,i)$ and sharing $C(j,l,i) = 1$. Other $C(j,l,i)$ values are shared as 0. The 0/1 verification is easy, and it is easy to bound the number of messages sent by each player.

Finally, the players secretly compute $b(j,l) = \sum_i B(j,l,i)$ and $c(j,l) = \sum_i C(j,l,i)$. Each $b(j,l)$ and $c(j,l)$ is revealed to player $j$, who detects if there is a collision by virtue of $c(j,l) > 1$. Each player announces the boxes containing collisions and the protocol is repeated so that the senders can try again. By choosing the number of mailboxes sufficiently large for the expected traffic — a practical issue — the expected number of repetitions is small.

---

[2]The terms *espionage* and *anonymous mail* were suggested by M. Rabin.

# Chapter 16

# Conclusions

> Forsan et haec olim meminisse iuvabit. [1]

> — Virgil, the *Aeneid*

The issues of this dissertation are three-fold. First, reliability and security are best ensured by avoiding assumptions that particular processors are reliable. Second, proving the security and reliability of distributed protocols requires clear and concise definitions. Our definition of resilience captures security and reliability simultaneously and *a priori* captures all the intuitive properties one expects of a robust system. Third, theoretical protocols are interesting, but to be practical a higher standard of efficiency must be met. Protocols requiring more than a dozen rounds of interaction are likely to be too complex or too inefficient to be implementable; low communication complexity is of utmost importance.

In general, we have made some strong simplifying assumptions, such as the presence of a complete, synchronous network with private channels. These issues are of deep practical and theoretical significance, but we place them aside in the hope that a modular approach provides a deeper understanding of the fundamental nature of security and reliability in interactive protocols. Chapter 15 (§15.1) discusses the incorporation of solutions for underlying network problems. The robustness of our protocols against message omissions and fail-stop faults suggests a cautious optimism that, with appropriate formal proofs, they can be provided with modular subroutines to withstand a weakening of the simplifying assumptions.

The understanding of security and reliability has been hampered by a lack of formal proofs and definitions. Much current research develops techniques that are intuitively secure but which have not been proven secure. We provide a definition for resilience that solidifies the foundations of research into security. Precise definitions of standard properties such as correctness and privacy fall out as an easy consequence. We believe that all the desirable properties of security and reliability are captured *a priori* by considering the *ideal* protocol as our standard of computational resilience. Privacy, correctness, and other properties are aspects of the same definition, tied

---

[1]Perhaps this will be a pleasure to look back on one day.

together by the idea of an ideal, trusted party, and measurable by the idea of relative resilience.

Our definition for resilience provides not just a means to rate the security of a protocol with respect to an *ideal* situation but a means to compare arbitrary protocols. The ability to measure relative resilience provides a greater flexibility and modularity in proof techniques and protocol design. Rather than directly prove a protocol is secure, one can compare it to an intermediate protocol that is itself proven secure. Relative resilience also provides a conceptually easier means to show when protocol concatenation preserves the security of the protocols, a very important issue that is often difficult to analyze.

It remains to be seen if unconditional security, high fault-tolerance, highly complex function computation, and low communication complexity can be achieved simultaneously. For functions of arbitrarily high complexity, the locally random reductions of Chapter 14 provide low communication complexity, but at a cost of lower ($O(\log n)$) fault tolerance. Chapter 10 achieves low communication complexity and high fault-tolerance, but the results are conditioned on the existence of a one-way function. If the computational complexity of the function is restricted, Chapter 5 demonstrates that low communication complexity and low local computational complexity are simultaneously achievable.

Locally random reductions with better parameters — *i.e.* a higher ratio of the independence-set size $k$ to the number $m$ of queries — are one path to achieving higher fault-tolerance for arbitrarily complex functions while using a constant number of rounds. A characterization of the achievable parameters for locally random reductions would be of general interest beyond serving simply to provide more secure protocols.

We have separated the communication complexity of secure computation from the computational complexity of the desired result. This bodes well for practical implementations, but is more striking from a theoretical standpoint. The crucial tool of *locally random reductions* vastly reduces upper bounds on the number of rounds of communication for a variety of cryptographic protocols. In a deeper sense, it has triggered other research leading to a result, IP = PSPACE, which relates a complexity class associated with interaction to a complexity class associated with computational complexity. The algebraic nature of the locally random reductions we develop suggests not only wide applications but a deeper understanding of the nature of interaction, computational complexity, and reliable computation.

# Bibliography

[AFK89]  M. Abadi, J. Feigenbaum, J. Kilian. "On Hiding Information from an Oracle." *J. Comput. Systems Sci.* **39** (1989), 21–50.

[AV79]   D. Angluin, L. Valiant. "Fast Probabilistic Algorithms for Hamiltonian Paths and Matchings." *J. Comput. Systems Sci.* **18** (1979), 155–193.

[BFL90]  L. Babai, L. Fortnow, C. Lund. "Non-Deterministic Exponential Time has Two-Prover Interactive Proofs," Submitted to FOCS, IEEE, 1990.

[BM88]   L. Babai, S. Moran. "Arthur-Merlin Games: A Randomized Proof System, and a Hierarchy of Complexity Classes." *J. Comput. System Sci.* **36** (1988), 254–276.

[BB88]   J. Bar-Ilan, D. Beaver. "Non-Cryptographic Fault-Tolerant Computing in a Constant Expected Number of Rounds of Interaction." *Proceedings of PODC,* ACM, 1989, 201–209.

[BBR88]  J. Bar-Ilan, D. Beaver, M. Rabin. Personal communication, 1988.

[Bar86]  D. Barrington, "Bounded Width Polynomial Size Branching Programs Recognize Exactly those Languages in $NC^1$. " *Proceedings of the $18^{th}$ STOC,* ACM, 1986, 1–5.

[Bea88]  D. Beaver. "Distributed Computations Tolerating a Faulty Minority, and Multiparty Zero-Knowledge Proof Systems." *J. Cryptology,* 1990. A preliminary version appeared as, "Secure Multiparty Protocols Tolerating Half Faulty Processors," in *Proceedings of Crypto,* ACM, 1989, and in Technical Report TR-19-88, Harvard University, September, 1988.

[Bea89a] D. Beaver. "Perfect Privacy for Two-Party Protocols." *Proceedings of the DIMACS Workshop on Distributed Computing and Cryptography,* Princeton, NJ, October, 1989, J. Feigenbaum, M. Merritt (eds.). Preliminary version in TR-11-89, Harvard University.

[Bea89b] D. Beaver. "Formal Definitions for Secure Distributed Protocols." *Proceedings of the DIMACS Workshop on Distributed Computing and Cryptography,* Princeton, NJ, October, 1989, J. Feigenbaum, M. Merritt (eds.).

[BCG89]   D. Beaver, R. Cleve, S. Goldwasser. In preparation, 1989.

[BF89a]   D. Beaver, J. Feigenbaum. "Hiding Information from Several Oracles." Harvard University Technical Report TR-10-89, May 1, 1989.

[BF89b]   D. Beaver, J. Feigenbaum. "Encrypted Queries to Multiple Oracles." AT&T Bell Laboratories Technical Memorandum, August 14, 1989.

[BF90]    D. Beaver, J. Feigenbaum. "Hiding Instances in Multioracle Queries." *Proceedings of the the $7^{th}$ STACS,* Springer–Verlag LNCS **415**, 1990, 37–48.

[BFKR89]  D. Beaver, J. Feigenbaum, J. Kilian, P. Rogaway. "Cryptographic Applications of Locally Random Reductions." *Proceedings of Crypto 1990.* Also appeared as AT&T Bell Laboratories Technical Memorandum, November 15, 1989.

[BFS90]   D. Beaver, J. Feigenbaum, V. Shoup. "Hiding Instances in Zero-Knowledge Proof Systems." *Proceedings of Crypto 1990.*

[BG89]    D. Beaver, S. Goldwasser. "Multiparty Computation with Faulty Majority." *Proceedings of the $30^{th}$ FOCS,* IEEE, 1989, 468–473.

[BGM89]   D. Beaver, S. Goldwasser, Y. Mansour. Personal communication, 1989.

[BHY89]   D. Beaver, S. Haber, M. Yung. In preparation, 1989.

[BMR90]   D. Beaver, S. Micali, P. Rogaway. "The Round Complexity of Secure Protocols." *Proceedings of the $22^{st}$ STOC,* ACM, 1990, 503–513.

[BBC]     C. Bennett, G. Brassard, C. Crépeau. Personal communication, 1989.

[BCG89]   M. Bellare, L. Cowen, S. Goldwasser. "On the Power of Secret Key Exchange." In preparation, 1989.

[Ben87]   J. Benaloh. "Verifiable Secret Ballot Elections" PhD Thesis, Yale University, 1987.

[BC88]    M. Ben-Or, R. Cleve, "Computing Algebraic Formulas Using a Constant Number of Registers." *Proceedings of the $20^{th}$ STOC,* ACM, 1988, 254–257.

[BGG88]   M. Ben-Or, O. Goldreich, S. Goldwasser, J. Hastad, J. Kilian, S. Micali, P. Rogaway. "Everything Provable is Provable in Zero-Knowledge." *Proceedings of Crypto 1988,* Springer–Verlag, 1990.

[BGMR85]  M. Ben-Or, O. Goldreich, S. Micali, R. Rivest. "Fair Contract Signing." Proceedings of the ICALP (1985).

[BGKW88]  M. Ben-Or, S. Goldwasser, J. Kilian, A. Wigderson. "Multi-Prover Interactive Proofs: How to Remove Intractability." *Proceedings of the $20^{th}$ STOC,* ACM, 1988, 113–131.

[BGW88]  M. Ben-Or, S. Goldwasser, A. Wigderson. "Completeness Theorems for Non-Cryptographic Fault-Tolerant Distributed Computation." *Proceedings of the 20$^{th}$ STOC,* ACM, 1988, 1–10.

[BE88]  M. Ben-Or, R. El-Yaniv, "Interactive Consistency in Constant Expected Time." Unpublished Manuscript, 1988.

[Ber84]  S. Berkowitz. "On Computing Determinant in Small Parallel Time Using a Small Number of Processors." *Info. Proc. Letters* **18**:3 (1984), 147–150.

[Bla81]  Blakley, "Security Proofs for Information Protection Systems." *Proceedings of the the 1980 Symposium on Security and Privacy,* IEEE Computer Society Press, NY (1981), 79–88.

[BK89]  M. Blum, S. Kannan. "Designing Programs that Check Their Work." *Proceedings of the 21$^{st}$ STOC,* ACM, 1989, 86–97.

[BLL89]  M. Blum, L. Levin, M. Luby. "If Permanent with the Uniform Distribution is in Average Polynomial Time then #P $\subseteq$ ZPP." Unpublished Manuscript, October 31, 1989.

[BLR89a]  M. Blum, M. Luby, R. Rubinfeld. "Self-Testing/Correcting with Applications to Numerical Problems." Preprint, November, 1989.

[BLR89b]  M. Blum, M. Luby, R. Rubinfeld. "Program Result Checking Against Adaptive Programs and in Cryptographic Settings." *Proceedings of the DI-MACS Workshop on Distributed Computing and Cryptography,* Princeton, October, 1989.

[BLR90]  M. Blum, M. Luby, R. Rubinfeld. "Stronger Checkers and General Techniques for Numerical Problems, *Proceedings of the 22$^{nd}$ STOC,* ACM, 1990, 73–83.

[BM84]  M. Blum, S. Micali. "How to Generate Cryptographically Strong Sequences of Pseudo-Random Bits." *SIAM J. Comput.* **13** (1984), 850–864.

[BCC88]  G. Brassard, D. Chaum, C. Crépeau. "Minimum Disclosure Proofs of Knowledge." *J. Comput. System Sci.* **37** (1988), 156–189.

[CCD88]  D. Chaum, C. Crépeau, I. Damgaard. "Multiparty Unconditionally Secure Protocols." *Proceedings of the 20$^{th}$ STOC,* ACM, 1988, 11–19.

[CDG87]  D. Chaum, I. Damgaard, J. van de Graaf. "Multiparty Computations Ensuring Secrecy of Each Party's Input and Correctness of the Output." *Proceedings of Crypto 1987,* Springer–Verlag, 1988.

[Che52]  H. Chernoff. "A Measure of Asymptotic Efficiency for Tests of a Hypothesis Based on the Sum of Observations." *Annals of Math. Statistics* **23** (1952), 493–507.

[CGMA85]  B. Chor, S. Goldwasser, S. Micali, B. Awerbuch. "Verifiable Secret Sharing and Achieving Simultaneity in the Presence of Faults." *Proceedings of the $17^{th}$ STOC,* ACM, 1985, 383–395.

[CK89]  B. Chor, E. Kushilevitz. "A Zero-One Law for Boolean Privacy," *Proceedings of the $21^{st}$ STOC,* ACM, 1989, 62–72.

[CR86]  B. Chor, M. Rabin. "Achieving Independence in a Logarithmic Number of Rounds." *Proceedings of the $6^{th}$ PODC,* ACM, 1987.

[Cle86]  R. Cleve. "Limits on the Security of Coin Flips when Half the Processors are Faulty." *Proceedings of the $18^{th}$ STOC,* ACM, 1986, 364–370.

[Cle89]  R. Cleve. Personal communication, 1989.

[CF85]  J. Cohen, M. Fischer. "A Robust and Verifiable Cryptographically Secure Election."

[Con88]  A. Condon. "Space Bounded Probabilistic Game Automata." *Proceedings of the $3^{rd}$ Structure in Complexity Theory Conf.,* IEEE, 1988, 162–174.

[CL89]  A. Condon, R. J. Lipton. "On the Complexity of Space-Bounded Interactive Proofs." *Proceedings of the $30^{th}$ FOCS,* IEEE, 1989, 462–467.

[Coo85]  S. Cook. "A Taxonomy of Problems with Fast Parallel Algorithms." *Info. and Control* **64** (1985), 2–22.

[CrK88]  C. Crépeau, J. Kilian. "Achieving Oblivious Transfer Using Weakened Security Assumptions." *Proceedings of the $29^{th}$ FOCS,* IEEE, 1988, 42–52.

[Den82]  D. Denning, *Cryptography and Data Security.* Addison-Wesley, Reading, MA (1982).

[DOD83]  *Department of Defense Trusted Computer System Evaluation Criteria.* US Department of Defense, Fort Meade, MD (August 15, 1983).

[DH76]  W. Diffie, M. Hellman. "New Directions in Cryptography." *IEEE Transactions of Information Theory* **IT-22** (November 1976), 644–654.

[DS88]  C. Dwork, L. Stockmeyer. "Interactive Proof Systems with Finite State Verifiers." IBM Research Report RJ 6262 (61659), May 26, 1988. Extended Abstract in *Proceedings of Crypto 1988,* Springer–Verlag, 1990, 71–75.

[Duf89]  T. Duff. "Experience with Viruses on UNIX Systems." *Computing Systems* **2**:2 (1989), 155–171.

[EGL82]  S. Even, O. Goldreich, A. Lempel. "A Randomized Protocol for Signing Contracts." *Proceedings of Crypto 1982,* Springer–Verlag, 1983, 205–210.

[Fel87]   P. Feldman. "A practical scheme for Noninteractive Verifiable Secret Sharing." *Proceedings of the* $28^{th}$ *FOCS,* IEEE, 1987, 427–437.

[Fel88]   P. Feldman. "One Can Always Assume Private Channels." Unpublished Manuscript, 1988.

[FM88]   P. Feldman, S. Micali. "Optimal Algorithms for Byzantine Agreement." *Proceedings of the* $20^{th}$ *STOC,* ACM, 1988, 148–161.

[FKN89]   J. Feigenbaum, S. Kannan, N. Nisan. "Lower Bounds on Random-Self-Reducibility." AT&T Bell Laboratories Technical Memorandum, December 4, 1989.

[For87]   L. Fortnow. "The Complexity of Perfect Zero-Knowledge." *Proceedings of the* $19^{th}$ *STOC,* ACM, 1987, 204–209.

[FRS88]   L. Fortnow, J. Rompel, M. Sipser. "On the Power of Multi-Prover Interactive Protocols." *Proceedings of the* $3^{rd}$ *Structure in Complexity Theory Conf.,* IEEE, 1988, 156–161.

[GHY87]   Z. Galil, S. Haber, M. Yung. "Cryptographic Computation: Secure Fault-Tolerant Protocols and the Public-Key Model." *Proceedings of Crypto 1987,* Springer–Verlag, 1988, 135–155.

[GHY89]   Z. Galil, S. Haber, and M. Yung. "Minimum-Knowledge Interactive Proofs for Decision Problems." *SIAM J. Comput.* **18** (1989), 711–739.

[GGM86]   O. Goldreich, S. Goldwasser, S. Micali. "How to Construct Random Functions." *JACM* **33**:4 (1986), 792–807.

[GK89]   O. Goldreich and H. Krawczyk. "On the Composition of Zero-Knowledge Proofs." Technical Report 570, Technion, Israel, June, 1989.

[GL89]   O. Goldreich, L. Levin. "A Hard-Core Predicate for All One-Way Functions." *Proceedings of the* $21^{st}$ *STOC,* ACM, 1989, 25–32.

[GMW86]   O. Goldreich, S. Micali, A. Wigderson. "Proofs that Yield Nothing but Their Validity and a Methodology of Cryptographic Protocol Design." *Proceedings of the* $27^{th}$ *FOCS,* IEEE, 1986, 174–187.

[GMW87]   O. Goldreich, S. Micali, A. Wigderson. "How to Play Any Mental Game, or A Completeness Theorem for Protocols with Honest Majority." *Proceedings of the* $19^{th}$ *STOC,* ACM, 1987, 218–229.

[GV87]   O. Goldreich, R. Vainish. "How to Solve any Protocol Problem – An Efficiency Improvement." *Proceedings of Crypto 1987,* Springer–Verlag, 1988, 73–86.

[GM84]   S. Goldwasser, S. Micali. "Probabilistic Encryption." *J. Comput. System Sci.* **28** (1984), 270–299.

[GMR89]  S. Goldwasser, S. Micali, C. Rackoff. "The Knowledge Complexity of Interactive Proof Systems." *SIAM J. Comput.* **18**:1 (1989), 186–208.

[GS86]   S. Goldwasser, M. Sipser. "Private Coins vs. Public Coins in Interactive Proof Systems." *Proceedings of the 18th STOC,* ACM, 1986, 59–68.

[Hab88]  S. Haber. *Multi-Party Cryptographic Computation: Techniques and Applications*, PhD Thesis, Columbia University, 1988.

[HM87]   S. Haber, S. Micali. Personal communication, 1987.

[HR83]   J. Halpern, M. Rabin. "A Logic to Reason about Likelihood." *Proceedings of the 15th STOC,* ACM, 1983, 310–319.

[IL89]   N. Immerman, S. Landau. "The Complexity of Iterated Multiplication." *Proceedings of the 4th Structure in Complexity Theory Conf.,* IEEE, 1989, 104–111.

[ILL89]  R. Impagliazzo, L. Levin, M. Luby. "Pseudo-Random Generation from One-Way Functions." *Proceedings of the 21st STOC,* ACM, 1989, 12–24.

[IR89]   R. Impagliazzo, S. Rudich. "Limits on The Provable Consequences of One-Way Permutations." *Proceedings of the 21st STOC,* ACM, 1989, 44–62.

[IY87]   R. Impagliazzo, M. Yung. "Direct Minimum-Knowledge Computation." *Proceedings of Crypto 1987,* Springer–Verlag, 1988, 40–51.

[Kil88a] J. Kilian. "Founding Cryptography on Oblivious Transfer." *Proceedings of the 20th STOC,* ACM, 1988, 20–29.

[Kil88b] J. Kilian. Personal communication, 1988.

[Kil88c] J. Kilian. "Zero-Knowledge with Log-Space Verifiers." *Proceedings of the 29th FOCS,* IEEE, 1988, 25–35.

[KMR90]  J. Kilian, S. Micali, P. Rogaway. "The Notion of Secure Computation." Unpublished Manuscript, 1990.

[Kus89]  E. Kushilevitz. "Privacy and Communication Complexity." *Proceedings of the 30th FOCS,* IEEE, 1989, 416–421.

[Lip89]  R. Lipton. "New Directions in Testing." Preprint, October, 1989.

[LMR83]  M. Luby, S. Micali, C. Rackoff. "How to Simultaneously Exchange a Secret Bit by Flipping a Symmetrically Biased Coin." *Proceedings of the 24th FOCS,* IEEE, 1983, 11–21.

[LFKN89] K. Lund, L. Fortnow, H. Karloff, N. Nisan. "The Polynomial Time Hierarchy has Interactive Proofs." Electronic mail announcement, December 13, 1989.

[MR87] D. McIlroy, J. Reeds. "A Security Model for Files and Processes in the UNIX System." AT&T Bell Laboratories Technical Report, April 14, 1987.

[MR88a] D. McIlroy, J. Reeds. "Multilevel Security with Fewer Fetters." *Proceedings of the the Spring 1988 EUUG Conference, European UNIX Users' Group,* London.

[MR88b] D. McIlroy, J. Reeds. "Multilevel Windows on a Single-Level Terminal." *Proceedings of the UNIX Security Workshop (1988),* USENIX, Portland, Oregon.

[Mac89] D. McIlroy. "Virology 101." *Computing Systems* **2**:2 (1989), 173–181.

[Nis88] N. Nisan. Personal communication, 1988.

[Nis89] N. Nisan. "Co-SAT has Multiprover Interactive Proofs." Preliminary draft, November, 1989.

[NZ72] I. Niven and H. Zuckerman. *An Introduction to the Theory of Numbers.* Wiley, New York, 1972.

[Ore87] Y. Oren. "On the Cunning Power of Cheating Verifiers: Some Observations about Zero Knowledge Proofs." *Proceedings of the* 28th *FOCS,* IEEE, 1987, 462–471.

[PW72] W. Peterson and E. Weldon. *Error Correcting Codes.* Second Ed., MIT Press (1972).

[Rab78] M. Rabin. "Digitalized Signatures." *Foundations of Secure Computations,* R. Demillo *et al*, Ed. Academic Press (1978), 155–165.

[Rab79] M. Rabin. "Digitalized Signatures and Public-Key Functions as Intractable as Factorization." Technical Report LCS/TR-212, MIT, January, 1979.

[Rab88] M. Rabin. Personal communication, 1988.

[Rab89] M. Rabin. Personal communication, 1989.

[RT88] T. Rabin. "Robust Sharing of Secrets When the Dealer is Honest or Cheating." Masters Thesis, Hebrew University, 1988.

[RB89] T. Rabin, M. Ben-Or. "Verifiable Secret Sharing and Multiparty Protocols with Honest Majority." *Proceedings of the* 21st *STOC,* ACM, 1989, 73–85.

[Ree89]   J. Reeds. "Secure IX Network." Proceedings of the DIMACS Workshop on Distributed Computing and Cryptography, Princeton, NJ, October, 1989, J. Feigenbaum, M. Merritt (eds.).

[Riv86]   R. Rivest. Workshop on Communication and Computing, MIT, October, 1986.

[RSA78]   R. Rivest, A. Shamir, L. Adleman. "A Method for Obtaining Digital Signatures and Public Key Cryptosystems." *Communications of the ACM* **21**:2 (1978), 120–126.

[Rog89]   P. Rogaway, Personal Communication, 1989.

[Rog90]   P. Rogaway. "The Round Complexity of Secure Protocols." PhD Thesis, Massachusetts Institute of Technology, 1990.

[Rud89]   S. Rudich. Personal communication, 1989.

[Sha79]   A. Shamir. "How to Share a Secret." *Communications of the ACM,* **22** (1979), 612–613.

[Sha89]   A. Shamir. "IP = PSPACE." Electronic mail announcement, December, 1989.

[Tod89]   S. Toda. "On the Computational Power of PP and $\oplus P$." *Proceedings of the $30^{th}$ FOCS,* IEEE, 1989, 514–519.

[TW87]    M. Tompa and H. Woll. "Random Self-Reducibility and Zero-Knowledge Proofs of Possession of Information." *Proceedings of the $28^{th}$ FOCS,* IEEE, 1987, 472–482.

[Val79]   L. Valiant. "The Complexity of Computing the Permanent." *Theor. Comput. Sci.* **8** (1979), 189–201.

[Yao82a]  A. C. Yao. "Protocols for Secure Computations." *Proceedings of the $23^{rd}$ FOCS,* IEEE, 1982, 160–164.

[Yao82b]  A. Yao, "Theory and Applications of Trapdoor Functions," *Proceedings of the $23^{rd}$ FOCS,* IEEE, 1982, 80–91.

[Yao86]   A. Yao. "How to Generate and Exchange Secrets." *Proceedings of the $27^{th}$ FOCS,* IEEE, 1986, 162–167.

[Yap83]   C. Yap. "Some Consequences of Nonuniform Conditions on Uniform Classes." *Theor. Comput. Sci.* **26** (1983), 287–300.